\documentclass[11pt]{article}
\usepackage{times}

\usepackage[margin=1in]{geometry}
\usepackage{graphicx}
\usepackage{amsmath, amsthm, amssymb}
\usepackage{braket}
\usepackage{subcaption}
\usepackage{comment}
\usepackage{url}
\usepackage{algorithm}
\usepackage{algpseudocode}
\usepackage{tikz}
\usepackage{diagbox}
\usepackage[colorlinks=true,citecolor=green!70!black,linkcolor=red,urlcolor=blue,bookmarks,breaklinks]{hyperref}
\usepackage{mathabx}
\usepackage{natbib}
\usepackage{bm}
\usepackage{ulem}
\setcitestyle{numbers}

\allowdisplaybreaks

\makeatletter
\newtheorem*{rep@theorem}{\rep@title}
\newcommand{\newreptheorem}[2]{%
\newenvironment{rep#1}[1]{%
 \def\rep@title{#2 \ref{##1} (restated)}%
 \begin{rep@theorem}}%
 {\end{rep@theorem}}}
\makeatother

\makeatletter
\newcommand{\subalign}[1]{%
  \vcenter{%
    \Let@ \restore@math@cr \default@tag
    \baselineskip\fontdimen10 \scriptfont\tw@
    \advance\baselineskip\fontdimen12 \scriptfont\tw@
    \lineskip\thr@@\fontdimen8 \scriptfont\thr@@
    \lineskiplimit\lineskip
    \ialign{\hfil$\m@th\scriptstyle##$&$\m@th\scriptstyle{}##$\hfil\crcr
      #1\crcr
    }%
  }%
}
\makeatother

\newtheorem{thm}{Theorem}
\newtheorem*{thm*}{Theorem}
\newtheorem{cor}[thm]{Corollary}

\newtheorem{lem}[thm]{Lemma}
\newtheorem*{lem*}{Lemma}

\newtheorem{prop}[thm]{Proposition}
\newtheorem{defn}[thm]{Definition}
\newtheorem{rem}[thm]{Remark}
\newtheorem{eg}[thm]{Example}

\newtheorem{assp}[thm]{Assumption}

\newreptheorem{thm}{Theorem}
\newreptheorem{lem}{Lemma}
\newreptheorem{prop}{Proposition}
\newreptheorem{cor}{Corollary}

\definecolor{darkgreen}{rgb}{0,0.6,0}

\newcommand{\FIXME}[1]{}

\newcommand{\eqpoly}[0]{\stackrel{\textrm{poly}(n)}{\approx}}

\makeatletter
\newcommand{\ALOOP}[1]{\ALC@it\algorithmicloop\ #1%
  \begin{ALC@loop}}
\newcommand{\ENDALOOP}{\end{ALC@loop}\ALC@it\algorithmicendloop}

\newcommand{\algorithmicbreak}{\textbf{break}}

\makeatother

\title{Solving boolean satisfiability problems with the quantum approximate optimization algorithm}
\date{\today}
\author{Sami Boulebnane\thanks{University College London and Phasecraft Ltd.}\ \ and Ashley Montanaro\thanks{Phasecraft Ltd.}}

\begin{document}

\maketitle

\begin{abstract}
The quantum approximate optimization algorithm (QAOA) is one of the most prominent proposed applications for near-term quantum computing. Here we study the ability of QAOA to solve hard constraint satisfaction problems, as opposed to optimization problems. We focus on the fundamental boolean satisfiability problem, in the form of random $k$-SAT. We develop analytic bounds on the average success probability of QAOA over random boolean formulae at the satisfiability threshold, as the number of variables $n$ goes to infinity. The bounds hold for fixed parameters and when $k$ is a power of 2. We complement these theoretical results with numerical results on the performance of QAOA for small $n$, showing that these match the limiting theoretical bounds closely. We then use these results to compare QAOA with leading classical solvers. In the case of random 8-SAT, we find that for around $14$ ansatz layers, QAOA matches the scaling performance of the highest-performance classical solver we tested, WalkSATlm. For larger numbers of layers, QAOA outperforms WalkSATlm, with an ultimate level of advantage that is still to be determined. Our methods provide a framework for analysing the performance of QAOA for hard constraint satisfaction problems and finding further speedups over classical algorithms.
\end{abstract}

\section{Introduction}

One of the most prominent application areas for quantum computers is solving constraint satisfaction and optimization problems. Grover's algorithm famously achieves a quadratic speedup over classical unstructured search~\cite{grover97}, and can be applied to solve unstructured optimization problems~\cite{durr96}. However, this family of algorithms requires a fault-tolerant quantum computer, with very significant overheads for error-correction~\cite{campbell19,babbush21}, and can be outperformed by classical methods tailored to the problem being solved. In the setting of near-term quantum computing, the most well-studied approach to solving optimization problems is the Quantum Approximate Optimization Algorithm (QAOA)~\cite{hogg_2000,1411.4028}. In a very influential pair of studies, Farhi, Goldstone and Gutmann~\cite{1411.4028,farhi14a} found provable bounds on the performance of this algorithm for instances of the optimization problems Max-Cut and Max-E3Lin2. In the latter case, for certain families of instances, QAOA outperformed the best classical algorithm known at the time. However, a classical algorithm was then found which outperformed QAOA~\cite{barak15}. Although there have been many subsequent works on the theoretical and empirical performance of QAOA for optimization problems (see~\cite{bharti22} for a review), none has yet shown an unambiguous advantage over the best classical algorithms.

Here, we study the performance of QAOA for hard constraint satisfaction problems. For problems of this form, we seek to find a solution that exactly satisfies all constraints, and expect the algorithm's running time to scale exponentially with the number of variables $n$. We focus on the fundamental boolean satisfiability problem, in the form of random $k$-SAT, where one is given a randomly generated boolean formula with $k$ variables per clause, and aims to find an assignment to the variables that satisfies all clauses. We define this problem formally in Definition \ref{def:random_ksat} below. QAOA was already applied to random $k$-SAT by Hogg in a pioneering work in 2000~\cite{hogg_2000}. (Hogg's ``quantum heuristic'' is essentially identical to QAOA, the chief difference being a prescribed ansatz (linear schedule) for the variational parameters.) Hogg applied his algorithm to hard random 3-SAT instances, using analytic arguments relying on a mean-field approximation which, while non-rigorous, seemed to be largely confirmed by small-scale numerical simulations. He did not find an improvement in running time compared with the best classical algorithms for 3-SAT.

In this work, we consider the performance of QAOA on random $k$-SAT at constant depth and \textit{fixed angles}. The last assumption means variational parameters are not allowed to depend on the problem instance; this feature is desirable as it means the quantum circuit ansatz need not be retrained for each instance. Fixed angles have been shown to achieve a non-trivial approximation ratio for a typical instance of the MAXCUT problem on random graphs, see e.g.\ \cite{1812.04170}. Here, we aim to maximise the probability $p_{succ}$ that QAOA outputs a satisfying assignment. Repeatedly running QAOA immediately translates into an algorithm for determining satisfiability whose expected running time on that instance is $1/p_{succ}$.

In the constant-depth regime, by a simple light-cone argument, see e.g.\ \cite{1411.4028}, QAOA outputs any optimal solutions with probability decaying exponentially with the problem size. This limitation is a generic feature of constraint satisfaction problems with a constant number of variables per constraint and a constant clauses-to-variables ratio. Nevertheless, in the context of random $k$-SAT close to the satisfiability threshold, this result should not necessarily lead to pessimism regarding the performance of QAOA: the reason is that state-of-the-art classical algorithms also empirically require exponential running time to solve this constraint satisfaction problem, see e.g.\ benchmarks from \cite{2110.10685}. The question then becomes which of QAOA or classical solvers has the smallest empirical or theoretical running time exponent.

\subsection{Summary of our results}

In this work we provide a theoretical and empirical analysis of the performance of QAOA on random $k$-SAT.
First, we propose an analytic method to estimate the mean QAOA success probability over instances in the infinite-size limit, together with a concrete algorithmic implementation. The correctness of the algorithm is rigorously proven for sufficiently small variational parameters. We underline that in this context, ``small" allows for constant parameters as the problem size goes to infinity, a regime in which QAOA remains hard to classically simulate. The analytic algorithm can be used in practice for relatively large numbers of layers $p$ (up to $p = 10$ shown in this work) to evaluate or even train QAOA on random $k$-SAT. However, we also empirically show that when it comes to finding near-optimal average-instance variational parameters, the analytic method essentially coincides with a much easier one, namely, estimating the expected success probability from an empirical average over a limited set of modest-size instances. In particular, full state vector simulation and training of QAOA even for large $p$ is very efficient on a classical computer at this size.

Encouraged by the agreement between analytic and numerical results, we then benchmark constant-depth QAOA, trained with the ``easier" method just described, against many classical solvers for random $k$-SAT. We find that the WalkSATlm solver~\cite{cai_luo_su_2014} is consistently the most efficient classical solver. We focus on relatively large $k$, as this is the regime in which we find that QAOA achieves the highest performance relative to classical algorithms. Based on both our analytic and numerical results, we estimate that for random $k=8$ instances at the satisfiability threshold, QAOA with $\approx 14$ ansatz layers would match the performance of WalkSATlm, with a running time of at most $\approx 2^{0.33n}$ to find a satisfying assignment. Notably, this is significantly faster than na\"ive use of Grover's algorithm, and with a far lower-depth quantum circuit. For larger numbers of layers, we predict that QAOA will start to outperform WalkSATlm. The extent of the advantage is unclear. For 60 layers, for example, numerical estimates of the median running time for small instances suggest a $\approx 2^{0.30n}$ scaling, whereas based on theoretical results on the average success probability, the scaling could be as low as $\approx 2^{0.19n}$. We also tested a combination of QAOA and the classical WalkSAT algorithm, but found the improvement in performance over standard QAOA to be modest.  We remark that, given a fault-tolerant quantum computer, amplitude amplification can be used to reduce all of these exponents by a factor of 2.

The main theoretical contribution of this work is a technique to estimate a certain family of ``generalized multinomial sums", extending the standard binomial and multinomial theorems. A similar goal was very recently achieved in \cite{2204.10306}, leading to an estimate of the performance of QAOA on spin-glass models. Their analysis relied on a sophisticated combinatorial analysis of generalized multinomial sums combined with complex analysis techniques, and involved the new and non-trivial concept of \textit{well-played polynomial}. This work is similar in that it considers generalized multinomial sums involving a certain family of (exponentiated) polynomials; however, instead of the ``well-played" property, it rather requires the polynomial to be expressible as a sum of perfect powers. Despite its simplicity, this assumption covers the case of QAOA applied to random $k$-SAT where $k$ is a power of 2, which is the focus of this work. More precisely, the success probability of this quantum algorithm can be expressed as a generalized multinomial sum satisfying all required properties. We then estimate generalized multinomial sums by recasting them as integrals. The asymptotic scaling of these integrals (in the limit where the problem size $n$ goes to infinity) can in turn be rigorously estimated using the \textit{saddle-point method}. Unfortunately, the method is only fully justified if certain parameters defining the multinomial sum are sufficiently small. In the context of QAOA, this requirement translates to sufficiently small variational angles; however, they may be held constant as $n \to \infty$, a parameter regime where classical simulation or even prediction of QAOA performance remains non-trivial in general. This additional requirement is a shortcoming compared to the method developed in \cite{2204.10306}, which remains operational unconditional on the magnitude of the QAOA angles. However, the two approaches rely on very different assumptions and presumably do not apply to the same problems.

This work is organized as follows. In section \ref{sec:preliminaries}, we introduce the required elements of background on random $k$-SAT and QAOA for the statement of our results. Analytic and numerical results are then discussed in section \ref{sec:results}, including the analytic method of evaluationg the expected success probability of random $k$-SAT QAOA, exposed in proposition \ref{prop:success_probability_random_ksat_qaoa_saddle_point}. Section \ref{sec:derivation_analytic_formulae} is technical and dedicated to the proof of the last result. It starts (\ref{sec:qaoa_multinomial_sums}) by recasting the expected success probability of QAOA on random $k$-SAT as a \textit{generalized multinomial sum} (definition \ref{defn:generalized_binomial_sum}). The rest of the work is dedicated to analyzing generalized multinomial sums ---its application is therefore not necessarily limited to random $k$-SAT QAOA. In section \ref{sec:hamming_weight_squared_example}, which is self-contained, we introduce a trivial yet instructive toy-model example (optimizing the Hamming weight squared cost function with QAOA) which gives an accurate flavour of the general method. In fact, the analysis of random $2$-SAT (among other examples) almost immediately follows from this example, as discussed in section \ref{sec:quadratic_case}. In \ref{sec:power_of_two_case}, we outline the analysis of the general case, applying in particular to random $k$-SAT QAOA with $k$ a power of $2$, is given; for clarity, the proofs of the most technical results are deferred to sections \ref{sec:iterated_gaussian_integration} and \ref{sec:quadratic_form_power_two_case}. The algorithmic implementation of our method for estimating multinomial sums, hence the success probability of random $k$-SAT QAOA, is made explicit in section \ref{sec:algorithmic_implementation}.

\section{Other background and related work}

\textbf{Random $k$-SAT}. We refer to \cite{0911.2322} for a thorough and accessible review of recent progress in the field, recalling only a few salient facts here. Instances of random $k$-SAT are generated from a random ensemble of constraint satisfaction problems which is parametrized by a positive integer $k$; the precise description of this ensemble used in this work is given in definition \ref{def:random_ksat}. An important fact is, the existence of solutions to random $k$-SAT and the complexity of algorithmically finding them are related to the ratio between the number of constraints $m$ and the number of variables $n$. For each integer $k \geq 1$ and problem size $n$, there exists a threshold $r_k(n)$, known as \textit{satisfiability ratio}, such that for all $\varepsilon > 0$, a randomly generated instance admits solutions with high probability if $\frac{m}{n} < r_k(n) - \varepsilon$, while it is almost surely unsatisfiable for $\frac{m}{n} > r_k(n)$. The thresholds $r_k(n)$ are believed to admit a limit $r_k$ as $n \to \infty$ ($k$ fixed). This quantity can be practically estimated to good precision, either through numerical simulations or non-rigorous analytic arguments from statistical physics \cite{mertens_mezard_zecchina_2006}. It is rigorously known that for sufficiently large $k$, $r_k(n) \sim 2^k\log(2)$ to leading order in $k$ \cite{0911.2322}. Remarkably, exact algorithms or heuristics are at least empirically known to solve random $k$-SAT efficient for a ratio $\frac{m}{n} < 2^k\frac{\log(k)}{k}$, but not any further beyond this ratio. The last ratio is known as \textit{algorithmic ratio} and is strictly smaller than the \textit{satisfiability ratio}, with a discrepancy increasing with $k$. It is therefore an outstanding problem to understand the ultimate limitations of different types of algorithms in the region between the algorithmic and the satisfiability threshold. This possibly involves considering non-conventional computational paradigms such as quantum computing.

\textbf{Related work on QAOA.} Recently, a numerical study \cite{2109.13346} considered the complexity of training QAOA for a parametrized constraint satisfaction problem: exact $k$-cover, which is distinct from random $k$-SAT but also admits a threshold. The authors showed that the difficulty of training the variational circuit with the goal of producing a solution with high probability dramatically increased as one approached the threshold. This led them to conjecture that QAOA, similar to classical algorithms, underwent a phase transition when approaching the satisfiability threshold.

A distinctive feature of our work compared with earlier theoretical work on MAXCUT-QAOA is that we optimize the expected success probability in the average-instance case rather than the instance-wise or average-instance energy. A similar idea was explored e.g.\ in \cite{li_fan_coram_riley_leichenauer_2020}, which proposed to optimize the Gibbs free energy instance-wise, leading to an empirical improvement on the probability of finding a high-quality solution; we recall that depending on the temperature parameter the Gibbs free energy interpolates continuously between the expected energy of a sampled solution and the probability of sampling an optimal solution. 

\section{Definitions and preliminaries}
\label{sec:preliminaries}

\subsection{Notation}
Given an integer variable $n$, we shall denote by $\eqpoly$ equality up to a factor which is at most polynomial in $n$. Intuitively, if one considers exponential scalings, a polynomial factor is irrelevant and this approximate equality therefore signifies the exponential scalings are the same.

Given an integer $n \geq 1$ and $r$ integers $n_0, \ldots, n_{r - 1}$ summing to $n$, we denote by
\begin{align}
	\binom{n}{n_0, \ldots, n_{r - 1}} = \binom{n}{\left(n_j\right)_{j \in [r]}} := \frac{n!}{\prod_{j \in [r]}n_j!}
\end{align}
multinomial coefficients, generalizing binomial coefficients and obeying a generalization of the binomial theorem (known as \textit{multinomial theorem}):
\begin{align}
	\sum_{\substack{n_0, \ldots, n_{r - 1}\\n_0 + \ldots + n_{r - 1} = n}}\binom{n}{n_0, \ldots, n_{r - 1}}\prod_{j \in [r]}x_j^{n_j} & = \left(\sum_{j \in [r]}x_j\right)^n.
\end{align}

\subsection{The Quantum Approximate Optimization Algorithm}

In this section, we recall the principle of the Quantum Approximate Optimization Algorithm (QAOA) as described by Farhi et al.\ in \cite{1411.4028} (see also Hogg's prior work~\cite{hogg_2000}). QAOA is a quantum algorithm designed to find approximate solutions to combinatorial optimization problems; for the purpose of this work, it is sufficient to think of such a problem as the task of minimizing a cost function of $n$ bits $H(x)$ ($x \in \{0, 1\}^n$). Finding approximate minimizers of this cost function can be rephrased as finding low-energy eigenstates of the corresponding $n$-qubit classical Hamiltonian:
\begin{align}
    H_C & := \sum_{x \in \{0, 1\}^n}H(x)\ket{x}\bra{x}.
\end{align}
QAOA attempts to achieve this task by starting with a product state corresponding to a uniform superposition of bitstrings: \begin{align}
    \ket{+}^{\otimes n} & = \frac{1}{\sqrt{2^n}}\sum_{x \in \{0, 1\}^n}\ket{x},
\end{align}
and alternating Hamiltonian evolution under $H_C$ and the transverse field Hamiltonian
\begin{align}
    H_B & := \sum_{j \in [n]}X_j.
\end{align}
The evolution times under $H_B$ and $H_C$ are hyperparameters of the algorithm to be optimized. Explicitly, the variational state prepared by a $p$-layer QAOA ansatz can be expressed:
\begin{align}
\label{eq:qaoa_state}
    \ket{\Psi_{\textrm{QAOA}}(\bm\beta, \bm\gamma)} & = e^{-\frac{i\beta_{p - 1}}{2}H_B}e^{-\frac{i\gamma_{p - 1}}{2}H_C}\ldots e^{-\frac{i\beta_1}{2}H_B}e^{-\frac{i\gamma_1}{2}H_C}e^{-\frac{i\beta_0}{2}H_B}e^{-\frac{i\gamma_0}{2}H_C}\ket{+}^{\otimes n},
\end{align}
where the variational parameters $\bm{\beta}, \bm{\gamma} \in \mathbf{R}^p$ are often referred to as ``QAOA angles". They are optimized in order to minimize an empirical cost function, estimated by repeatedly preparing the quantum state and measuring it in the computational basis. The expected energy achieved by the state: $\braket{\Psi_{\textrm{QAOA}}(\bm\beta, \bm\gamma)|H_C|\Psi_{\textrm{QAOA}}(\bm\beta, \bm\gamma)}$ is the most frequently used such function, but other candidates have been reported, including the ``CVar" \cite{barkoutsos_nannicini_robert_tavernelli_woerner_2020} (average over energies after discarding samples with energy above a certain threshold) or the Gibbs free energy \cite{li_fan_coram_riley_leichenauer_2020}. Once satisfying variational parameters have been determined, preparing the corresponding QAOA state in equation \ref{eq:qaoa_state} and measuring it in the computational basis (ideally) provides good approximate solutions to the original combinatorial problem.

\subsection{Random $k$-SAT and QAOA}

This work considers the performance of the Quantum Approximate Optimization Algorithm on the random $k$-SAT combinatorial optimization problem. An instance of $k$-SAT is a formula on $n$ Boolean variables $x_0, \ldots, x_{n - 1}$ which is a \textit{conjunction} of $m$ \textit{clauses}; conjunction means the formula is satisfied iff. all clauses are. Besides, each clause is a \textit{disjunction} of $k$ \textit{literals}, where a literal is a Boolean variable or its negation; disjunction means the clause is satisfied iff. at least one of its literals is. Such a formula is said to be in \textit{conjunctive normal form} (abbreviated CNF), meaning it is expressed as a conjunction of disjunctions. An example of a $k$-SAT formula with $k = 2$ and $n = 4$ is:
\begin{align}
    \left(x_0 \vee \overline{x_1}\right) \wedge \left(x_1 \vee x_2\right) \wedge \left(x_1 \vee \overline{x_3}\right),
\end{align}
where  the bar over a Boolean variable denotes negation and $\vee$ (resp. $\wedge$) denotes disjunction (resp. conjunction). The formula above is for instance satisfied by the assignment $x_0 = x_1 = 1, x_2 = x_3 = 0$. An algorithmically interesting setting for $k$-SAT is when instances are generated at random with a number of clauses $m$ proportional to the number of variables $n$, where the ratio $r := \frac{m}{n}$ is known as \textit{clauses-to-variables ratio}. In random $k$-SAT, the existence of solutions to a random $k$-SAT instance and the hardness of finding it are determined by $r$ \cite{0911.2322}. In our case, rather than fixing $m$ to be a constant multiple $\left\lfloor rn \right\rfloor$ of $n$, we sample it from a distribution of expectation $rn$ peaked around this mean, namely $\textrm{Poisson}(rn)$. This technical choice allows to write the success probability of random $k$-SAT QAOA as a generalized multinomial sum of the form of equation \ref{eq:generalized_binomial_sum}, making its analysis possible via our main technical result proposition \ref{prop:exp_scaling_generalized_binomial_sum}.
\begin{defn}[Random $k$-SAT problem]
\label{def:random_ksat}
Let $k \geq 1$ an integer and $r > 0$. The random $k$-SAT problem is a constraint satisfaction problem on $n$ variables, temporarily denoted by $x_{0}, \ldots, x_{n - 1}$ for convenience. A random instance of such a problem is defined as follows:
\begin{itemize}
    \item Sample $m \sim \textrm{Poisson}(rn)$.
    \item Generate $m$ random OR-clauses $\bm{\sigma} = \left(\sigma_0, \ldots, \sigma_{m - 1}\right)$. Each clause consists of $k$ literals chosen uniformly (with replacement) from $\left\{x_0, \overline{x_0}, x_1, \overline{x_1}, \ldots, x_{n - 1}, \overline{x_{n - 1}}\right\}$. The OR-clause is satisfied iff.\ at least one of its literals is.
    \item The random instance thereby generated, characterized by $\bm{\sigma} = \left(\sigma_0, \ldots, \sigma_{m - 1}\right)$, is satisfied iff. all OR-clauses $\sigma_j$ are.
\end{itemize}
A problem instance generated from this random ensemble will be denoted by $\bm{\sigma} \sim \mathrm{CNF}\left(n, k, r\right)$. Besides, for $y = \left(y_0, \ldots, y_{n - 1}\right) \in \{0, 1\}^n$, one denotes
\begin{align}
    y & \vdash \bm{\sigma}.
\end{align}
to signify that assignment $y$ of the literals satisfies all clauses in $\sigma$.
\end{defn}

\begin{defn}[Random $k$-SAT QAOA]
\label{def:random_ksat_qaoa}
Let $k \geq 1$ an integer, $r > 0$ and $n$ a positive integer. Given a random $k$-SAT instance $\bm{\sigma} = \left(\sigma_0, \ldots, \sigma_{m - 1}\right) \sim \mathrm{CNF}\left(n, k, r\right)$ generated according to definition \ref{def:random_ksat}, we denote by
\begin{align}
    H[\bm{\sigma}] & := \sum_{y \in \left\{0, 1\right\}^n}\left|\left\{j \in [m]\,:\,y \not\vdash \sigma_j\right\}\right|\ket{y}\bra{y}
\end{align}
the diagonal quantum Hamiltonian corresponding to the classical cost function counting the number of unsatisfied clauses in $\bm\sigma$. The diagonal elements of this Hamiltonian are in $\left\{0, \ldots, m\right\}$. For each $m' \in \left\{0, \ldots, m\right\}$, we then denote by
\begin{align}
    \left\{H[\bm{\sigma}] = m'\right\} & := \sum_{\substack{y \in \left\{0, 1\right\}^n\\\left|\left\{j \in [m]\,:\,y \not\vdash \sigma_j\right\}\right| = m'}}\ket{y}\bra{y}
\end{align}
the orthogonal projector onto the eigenspace of $H[\bm\sigma]$ of eigenvalue $m'$. In particular, $\left\{H[\bm\sigma] = 0\right\}$ is the orthogonal projector onto the space of satisfying assignments. Besides, for $\bm\beta, \bm\gamma \in \mathbf{R}^p$, we denote by
\begin{align}
    \ket{\Psi\left(\bm\sigma, \bm\beta, \bm\gamma\right)} & := e^{-\frac{i\beta_{p - 1}}{2}\sum\limits_{j \in [n]}X_j}e^{-\frac{i\gamma_{p - 1}}{2}H[\bm\sigma]}\ldots e^{-\frac{i\beta_1}{2}\sum\limits_{j \in [n]}X_j}e^{-\frac{i\gamma_1}{2}H[\bm\sigma]}e^{-\frac{i\beta_0}{2}\sum\limits_{j \in [n]}X_j}e^{-\frac{i\gamma_0}{2}H[\bm\sigma]}\ket{+}^{\otimes n}
\end{align}
the state prepared by level-$p$ QAOA for the optimization problem defined by Hamiltonian $H[\bm\sigma]$.
\end{defn}

\section{Results}
\label{sec:results}
\subsection{Theoretical results}
\label{sec:theoretical_results}

The main technical result of this work, proposition \ref{prop:exp_scaling_generalized_binomial_sum}, allows to estimate the leading exponential contribution of ``generalized multinomial sums" (precisely defined in definition \ref{defn:generalized_binomial_sum}), extending the standard multinomial theorem
\begin{align}
    \sum_{\substack{n_0, \ldots, n_{r - 1}\\n_0 + \ldots + n_{r - 1} = n}}\binom{n}{n_0, \ldots, n_{r - 1}}\prod_{j \in [r]}x_j^{n_j} & = \left(\sum_{j \in [r]}x_j\right)^n.
\end{align}
The proof of proposition \ref{prop:exp_scaling_generalized_binomial_sum} uses the saddle-point method, whereby the generalized multinomial sum is expressed as an integral, whose exponential scaling is controlled by the unique critical point of the integrand; with our methods, the existence and uniqueness of the critical point requires certain parameters in the sum to be sufficiently small. Now, as we show in proposition \ref{prop:qaoa_ksat_expected_success_probability}, the expected success probability of QAOA on random $k$-SAT for fixed variational parameters $\bm\beta, \bm\gamma$ can be cast as a generalized multinomial sum. Combining proposition \ref{prop:exp_scaling_generalized_binomial_sum} and \ref{prop:qaoa_ksat_expected_success_probability} then leads to proposition \ref{prop:success_probability_random_ksat_qaoa_saddle_point} below. In the context of QAOA, the ``small parameters" assumption required by proposition \ref{prop:exp_scaling_generalized_binomial_sum} to estimate the generalized multinomial sum translates to small $\bm\gamma$ angles; however, $\bm\beta$ is allowed to take any finite value.
\begin{prop}[Average-case success probability of random $k$-SAT QAOA by saddle-point method]
\label{prop:success_probability_random_ksat_qaoa_saddle_point}
Let $q \geq 1$ an integer, $p \geq 1$ an integer and $\bm\beta, \bm\gamma \in \mathbf{R}^p$. For $\bm\gamma$ sufficiently small (i.e. smaller than a constant independent of the problem size $n$), the expected success probability of random-$2^q$-SAT QAOA admits the following scaling exponent in the infinite-size limit:
\begin{align}
    & \lim_{n \to \infty}\frac{1}{n}\log\mathbf{E}_{\bm\sigma \sim \mathrm{CNF}(n, k, r)}\left[\left\langle\Psi_{\mathrm{QAOA}}(\bm\sigma, \bm\beta, \bm\gamma)|\mathbf{1}\left\{H[\bm\sigma] = 0\right\}|\Psi_{\mathrm{QAOA}}(\bm\sigma, \bm\beta, \bm\gamma)\right\rangle\right]\nonumber\\
    & = F\left(z^*\right) + \left(2^q - 1\right)\sum_{\alpha \subset [2p + 1]}\left(\frac{\partial F}{\partial z_{\alpha}}\left(z^*\right)\right)^{2^q},
\end{align}
where $z^* \in \mathbf{C}^{2^{2p + 1}}$ is the unique fixed point of the function
\begin{align}
    F: \left\{\begin{array}{ccc}
        \mathbf{C}^{2^{2p + 1}} & \longrightarrow & \mathbf{C}^{2^{2p + 1}}\\
        \left(z_{\alpha}\right)_{\alpha \subset [2p + 1]} & \longmapsto & \log\sum\limits_{s \in \{0, 1\}^{2p + 1}}b_s\exp\left(\frac{1}{2}\sum\limits_{\substack{\alpha \subset [2p + 1]\\\forall j, j' \in \alpha,\,s_j = s_{j'}}}(-c_{\alpha})^{1/2^q}z_{\alpha}\right)
    \end{array}\right.,
\end{align}
where:
\begin{align}
    b_s & := \frac{(-1)^{\mathbf{1}[s_0 \neq s_p]}}{2}\prod_{j \in [p]}\left(\cos\frac{\beta_j}{2}\right)^{\mathbf{1}\left[s_j = s_{j + 1}\right] + \mathbf{1}[s_{2p - j} = s_{2p - j - 1}]}\left(i\sin\frac{\beta_j}{2}\right)^{\mathbf{1}[s_j \neq s_{j + 1}] + \mathbf{1}[s_{2p - j} \neq s_{2p - j - 1}]},\\
    c_{\alpha} & := (-1)^{\mathbf{1}\left[p \in \alpha\right]}\prod_{j \in \alpha\,:\,j < p}\left(e^{-\frac{i\gamma_j}{2}} - 1\right)\prod_{j \in \alpha\,:\,j > p}\left(e^{\frac{i\gamma_{2p - j}}{2}} - 1\right).
\end{align}
The existence and uniqueness of the fixed point are guaranteed for sufficiently small $\bm\gamma$.
\end{prop}
A generalization of the multinomial theorem was already derived in \cite{2204.10306} in the context of QAOA applied to MAX-$k$-XOR, though with distinct assumptions and a very different method. Besides, while the results from \cite{2204.10306} apply to arbitrary variational angles in the context of MAX-$k$-XOR, our analysis for random-$k$-SAT is (in principle) limited to sufficiently small $\bm\gamma$ angles. However, as will be extensively discussed in section \ref{sec:numerical_methods}, the method empirically appears to be quantitatively accurate for a sufficiently wide range of angles, most interestingly for the optimal ones.

While proposition \ref{prop:success_probability_random_ksat_qaoa_saddle_point} establishes a rigorous scaling for the expected success probability of QAOA on random $k$-SAT under certain assumptions, this quantity may not the most natural to consider to benchmark QAOA against other algorithms. In fact, it may be more natural to consider the \textit{median running time} of the algorithm, which is a common method to benchmark classical SAT solvers, see e.g. \cite{campbell19}. The choice of the median running time as a benchmark, as opposed for instance to the expected or maximum running time, addresses an important difficulty: since QAOA is based on sampling bitstrings from the quantum state until one finds a satisfying assignment, the algorithm will never terminate if the problem instance is unsatisfiable. This would lead to an infinite expected running time as soon as a randomly generated instance has a finite probability of being unsatisfiable, which is the case for the random ensemble introduced in definition \ref{def:random_ksat}. In contrast, the median running time will remain finite and is besides less sensitive to outliers with finite, yet unusually large, running time.

In addition, the expected success probability gives a lower bound on the median running time. First, the expected success probability can be related to the median success probability $\mu$ via the following straightforward argument.
\begin{align}\mathbf{E}_{\bm\sigma}\left[p_{\textrm{succ}(\bm\sigma)}\right] &= \sum_{\bm\sigma} P(\bm\sigma) p_{\textrm{succ}}(\bm\sigma) = \sum_{\bm\sigma,p_{\textrm{succ}}(\bm\sigma) < \mu} P(\bm\sigma) p_{\textrm{succ}}(\bm\sigma) + \sum_{\bm\sigma,p_{\textrm{succ}}(\bm\sigma) \ge \mu} P(\bm\sigma) p_{\textrm{succ}}(\bm\sigma)\\
&\ge \sum_{\bm\sigma,p_{\textrm{succ}}(\bm\sigma) \ge \mu} P(\bm\sigma) p_{\textrm{succ}}(\bm\sigma) \ge \frac{\mu}{2}.
\end{align}
Hence
\begin{align} \frac{1}{\mathbf{E}_{\bm\sigma}\left[p_{\textrm{succ}(\bm\sigma)}\right]} \le \frac{2}{\mu} \end{align}
and the latter quantity is upper-bounded by twice the median running time via Jensen's inequality for medians.

\subsection{Validation of the analytic algorithm}
\label{sec:validation_analytic_formula}

Next we validate the theoretical formula given by proposition \ref{prop:success_probability_random_ksat_qaoa_saddle_point} for the expected success probability of QAOA on a random $k$-SAT instance.

\begin{figure}
    \centering
    \begin{subfigure}{0.49\textwidth}
    \includegraphics[width=\textwidth]{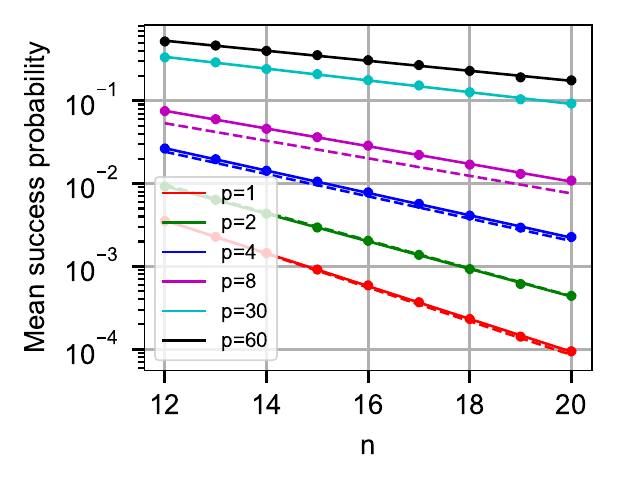}
    \caption{$k=8$, varying $p$}
    \end{subfigure}
    \begin{subfigure}{0.49\textwidth}
    \includegraphics[width=\textwidth]{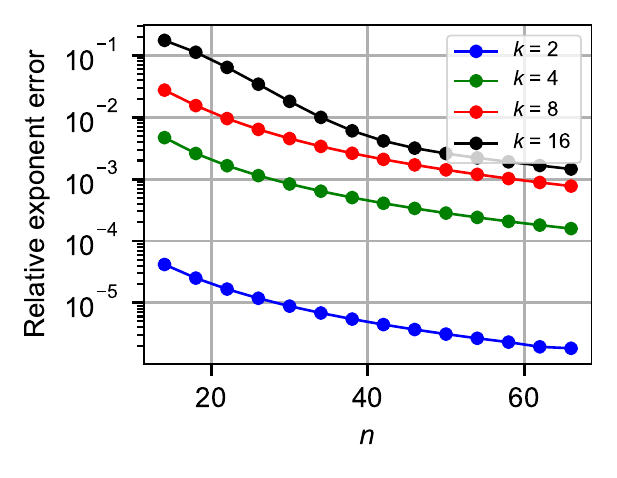}
    \caption{$p=1$, varying $k$}
    \end{subfigure}
    \caption{Comparison of numerical results to limiting theoretical predictions. LHS: points are empirical averages. Solid lines are fits to empirical averages, dashed lines are scaling predicted by theory (assuming unknown constant factor is 1). Error bars are too small to be seen. RHS: points are relative differences in excess scaling exponents between numerical and theoretical results, solid lines are added to guide the eye.}
    \label{fig:numerics_vs_theory}
\end{figure}

First, we compare the limiting average success probabilities for random $k$-SAT predicted by proposition \ref{prop:success_probability_random_ksat_qaoa_saddle_point} with actual average success probabilities determined by numerical experiments for small $n$. We sample up to $10000$ instances from $\mathrm{CNF}(n, k, r)$ for each value of $k$ and problem size $n \in \{12, \ldots, 20\}$ and retain only satisfiable instances. QAOA is then \textit{evaluated} (and not trained) on each of these instances using angles previously determined to achieve a good average success probability, as detailed in section \ref{sec:numerical_methods} below. Note that the instances used to train QAOA are much less numerous (100) and smaller (12) than the ones used to validate the performance here. For each $k$ and $n$, we compute the average success probability and median running time on the relevant set of random instances. For each problem size $n$, the instances generated for evaluation achieve an empirical uncertainty of order $< 0.5\%$ on the expected success probability at size $n$. This translates to an error of order $10\%$ with the $100$ instances used for training, confirming the latter provide a rather coarse approximation of the success probability. 
The results are shown in Figure \ref{fig:numerics_vs_theory} for the case $k=8$, where we perform a linear least-squares fit on the experimental data and compare against the scaling predicted from the theoretical results. As the constant factor in this scaling is unknown, we assume that this is equal to 1 in the plot.

Second, we exploit the fact, established in proposition \ref{prop:p1_qaoa_ksat_expected_success_probability}, that for $p = 1$ the expected success probability of random $k$-SAT QAOA at finite size $n$ can be computed in time $O(n^3)$, allowing for a practical evaluation, and even optimization of the expected QAOA success probability for large instance sizes of order $100$. Unlike the analytic prediction for the infinite-size scaling exponent, the finite-size calculation at $p = 1$ applies to all $k$ (not only $k$ a power of $2$) and arbitrary angles (not only sufficiently small $\bm\gamma$). We may therefore extract the empirical scaling exponent of this expected success probability by an exponential fit 
and compare it with the infinite-size scaling exponent predicted by proposition \ref{prop:success_probability_random_ksat_qaoa_saddle_point}. Here, the empirical scaling of the success probability at $n$ is defined as the ratio between success probabilities at size $n + 1$ and $n$, taken to the logarithm:
\begin{align}
    \log\frac{\mathbf{E}_{\bm\sigma \sim \mathrm{CNF}(n + 1, k, r)}\left[\left\langle\Psi_{\mathrm{QAOA}}(\bm\sigma, \beta, \gamma)|\mathbf{1}\left\{H[\bm\sigma] = 0\right\}|\Psi_{\mathrm{QAOA}}(\bm\sigma, \beta, \gamma)\right\rangle\right]}{\mathbf{E}_{\bm\sigma \sim \mathrm{CNF}(n, k, r)}\left[\left\langle\Psi_{\mathrm{QAOA}}(\bm\sigma, \beta, \gamma)|\mathbf{1}\left\{H[\bm\sigma] = 0\right\}|\Psi_{\mathrm{QAOA}}(\bm\sigma, \beta, \gamma)\right\rangle\right]}.
\end{align}
Although less robust than an exponential fit, this metric is usable in this context as the expectations can be evaluated exactly. Besides, it has the advantage of sharply capturing the \textit{local} scaling of the success probability around each $n$. The result of this comparison (between \textit{excess scaling exponents} rather than exponents themselves, see section \ref{sec:numerical_methods}) is represented for $k$-SAT, $k \in \left\{2, 4, 8, 16\right\}$, with instance sizes ranging from $10$ to $70$, in figure \ref{fig:numerics_vs_theory}. The plot shows that, as expected, the error decreases as $n$ increases. For a fixed size, the relative error incurred by the finite-size approximation worsens as $k$ increases. However, the asymptotic decay rate of the error seems comparable between the different $k$ values considered.

While these comparisons were performed for (empirically) optimal average-instance parameters, it is also instructive to consider the full optimization landscape of $p = 1$ QAOA for the same values of $k$, to determine how well the analytic and empirical results match. In addition, one may wonder whether the empirically optimal parameters are close to the limiting optimal parameters. Our experimental results to address these questions are included in Appendix \ref{app:further_figures}.

\subsection{Algorithm scaling}
\begin{figure}
    \centering
    \begin{subfigure}{0.49\textwidth}
    \includegraphics[width=\textwidth]{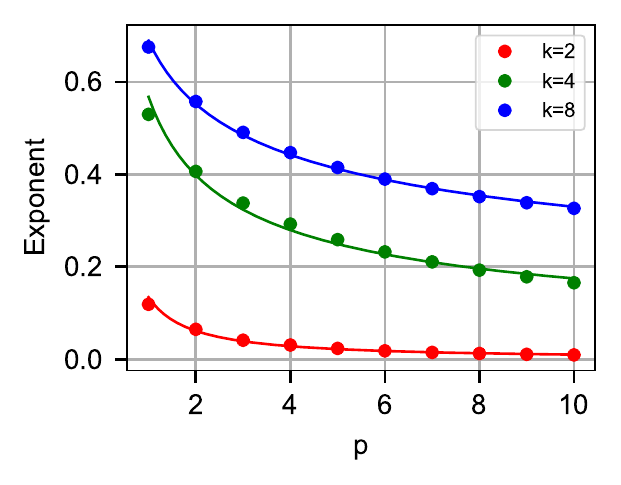}
    \end{subfigure}
    \begin{subfigure}{0.49\textwidth}
    \includegraphics[width=\textwidth]{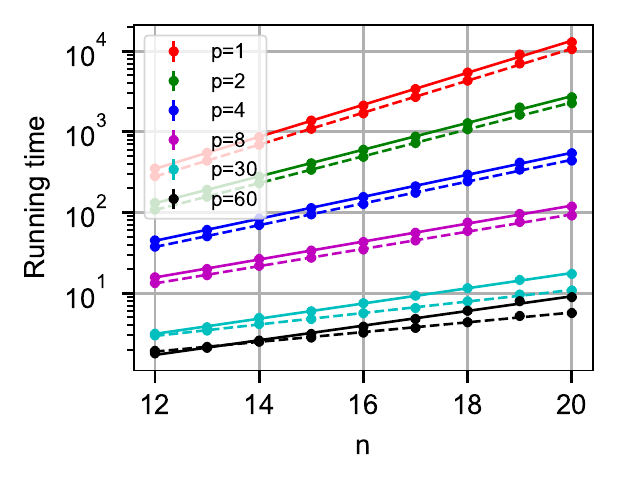}
    \end{subfigure}
    \caption{Scaling behaviour of QAOA on random $k$-SAT. LHS: Analytic scaling exponents $c$ in terms of $p$, such that success probability is predicted to be $2^{-cn}$ up to lower-order terms. Fit to a power law for each $k$. Fits are $c \approx 0.13 p^{-1.12}$ ($k=2$), $c\approx 0.57p^{-0.51}$ ($k=4$), $c\approx 0.69p^{-0.32}$ ($k=8$). RHS: Median running time (solid line) compared with running time estimated from average success probability for random 8-SAT instances (dashed line). Lines are linear fits. Error bars are too small to be seen.}
    \label{fig:pscaling}
\end{figure}

Having developed confidence that the analytic and empirical scaling exponents are close, we use our analytic formulae to determine the behaviour of the exponent in terms of $p$ (for a fixed set of parameters, determined for each $p$ using a small-scale experiment). This behaviour strongly suggests power-law decay; performing a fit to the data allows us to extrapolate the performance of QAOA to larger values of $p$ than are accessible to our algorithm. The results are shown in Figure \ref{fig:pscaling}.

We also studied whether the inverse of the expected success probability provides an accurate reflection of the median running time, by comparing these two quantities in numerical experiments. Using an exponential fit, we extract a scaling exponent for both the success probability and the median running time as functions of $n$. Observe that while the success probability was shown to admit a scaling exponent for sufficiently small variational angles in proposition \ref{prop:success_probability_random_ksat_qaoa_saddle_point}, no such rigorous statement exists for the median running time; the exponent for the latter quantity should therefore be regarded as purely empirical. The results are also shown in Figure \ref{fig:pscaling} for various choices of $p$. We observe that for small $p$, the two complexity measures are well-aligned, while for large $p$, their slope appears to differ. One possible explanation for this divergence is that training to maximise the average success probability does not necessarily minimise the median running time. This may be a particular issue in the scenario where $p$ is large and $n$ is small, because the QAOA success probability may be close to 1 for many ``easy'' instances. Optimising the average success probability may lead to finding parameters that are good for these easy instances, while performing poorly for harder instances. For example, taking $p=60$, $k=8$, $n=12$, the median success probability was $\sim 0.56$.

All in all, these results seem to provide theoretical backing for the approach described in section \ref{sec:numerical_methods} below of obtaining good average-case parameters for QAOA by estimating averages empirically on a small dataset of modest size instances.

\subsection{Comparison of fixed-parameters QAOA with classical SAT solvers}
\label{sec:classical_comparison}

Having built up confidence that our limiting theoretical results are well-aligned with numerical benchmarks for small $n$, we compare the performance of QAOA to a variety of classical solvers for $k$-SAT. We choose to focus on the case $k=8$, motivated by a trade-off between the need for a sufficiently large $k$ (making the problem hard enough for classical solvers) and the practical requirement to store all generated instances (recalling that the clauses-to-variables ratio at satisfiability threshold increases exponentially with $k$ \cite{mertens_mezard_zecchina_2006}), together with our theoretical results only being available for $k$ a power of 2.

\begin{figure}
    \centering
    \begin{tabular}{cc}
    \raisebox{-2cm}{
    \includegraphics[width=0.5\textwidth]{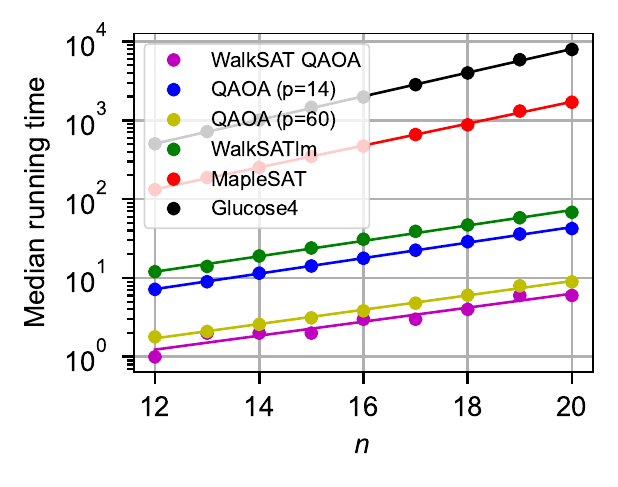}}&
	\begin{tabular}{c|c|c}
		Solver & Fit & Error\\
		\hline
		WalkSAT QAOA & $-3.232 + 0.295n$ & $0.011$\\
		\hline
		QAOA ($p=14$) & $-1.064 + 0.326n$ & $0.008$\\
		\hline
		QAOA ($p=60$) & $-2.842 + 0.302n$ & $0.007$\\
		\hline
		\verb|walksatlm| & $-0.309 + 0.325n$ &  $0.008$\\
		\hline
		\verb|maplesat| & $1.531 + 0.461n$ &  $0.004$\\
		\hline
		\verb|glucose4| & $2.998 + 0.498n$ & $0.005$\\
	\end{tabular}
	\end{tabular}
    \caption{Scaling behaviour of median running times of selected classical and quantum algorithms for 8-SAT. WalkSAT QAOA uses $p=60$.}
    \label{fig:fits}
\end{figure}

We benchmarked QAOA against the simple local search algorithm WalkSAT~\cite{papadimitriou91,selman93,schoning}, the optimized local search algorithm WalkSATlm \cite{cai_luo_su_2014} and the suite of state-of-the-art SAT solvers \verb|pySAT| \cite{imms-sat18}. WalkSATlm has demonstrated leading performance on random 5-SAT and 7-SAT instances with clause/variable ratio close to the satisfiability threshold~\cite{cai_luo_su_2014}. In the most recent SAT Competition which included a track for randomly generated instances~\cite{sat2018}, although WalkSATlm did not compete, the winning solver was based on the Sparrow solver, which performed significantly less well than WalkSATlm in previous experiments~\cite{cai_luo_su_2014}. We additionally considered comparing QAOA to the well-known \textit{survey propagation algorithm} introduced in \cite{marino_parisi_ricci_tersenghi_2016}. This work empirically showed this message passing algorithm to outperform competitors close to the satisfiability threshold. However, these conclusions were only reported for relatively small $k$ ($3$ or $4$), and initial experiments carried out at $k = 8$ using a publicly available implementation \cite{SurveyPropagationGithub} revealed the algorithm to be impractical to run at $k = 8$ due to the high number of constraints per variable and the large degree of the constraint graph. We therefore did not include this contender in our comparisons.

We also studied a combination of QAOA and WalkSAT, whereby assignments sampled from QAOA (hence, not necessarily satisfying) are given as a starting guess to WalkSAT. More precisely, the algorithm consists of sampling an assignment from QAOA, apply a single iteration of WalkSAT (consisting of a $\Theta(n)$ steps walk), and declare success or failure according to whether the walk updated the initial assignment to a satisfying one. Similarly to WalkSAT, this process is repeated until success. Each given instantiation of the algorithm has a success probability, and the expected number of instantiations is the inverse of this probability.

Similarly to results presented in section \ref{sec:validation_analytic_formula}, we extracted the median running time of the solver, with respect to our randomly generated instances, for each instance size $n \in \{12,\dots,20\}$ before determining an estimated scaling exponent based on a fit to this data. We illustrate this method for four of the most efficient classical solvers, together with some quantum ones, in figure \ref{fig:fits}; a more complete list of empirical scaling exponents is given in table \ref{tab:8sat_scaling_exponents} in Appendix \ref{app:k8results}. We found that, in each case, a simple exponential fit was highly accurate. Consistent with previous experiments~\cite{cai_luo_su_2014}, WalkSATlm's median running time, as measured by the number of input formula evaluations, was the lowest among solvers considered. In addition, an instance-by-instance analysis determined that WalkSATlm was the most efficient algorithm on all but a few percent of the instances considered. In the case of quantum algorithms, we found that the combination of WalkSAT and QAOA was not substantially more efficient than the use of QAOA alone: for example, as seen in figure \ref{fig:fits}, the two scaling exponents are within error bars for $p=60$. We therefore focused on the comparison between WalkSATlm and QAOA.

\begin{figure}
    \centering
    \includegraphics[width=0.49\textwidth]{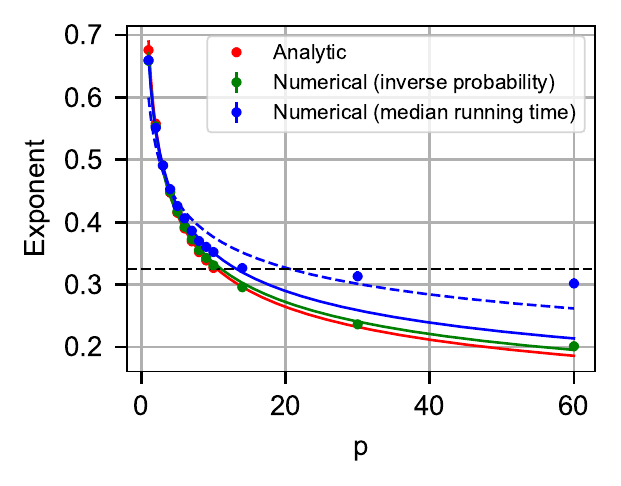}
    \includegraphics[width=0.49\textwidth]{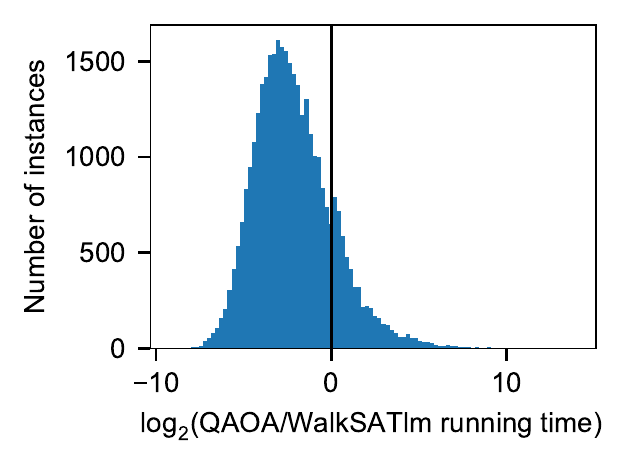}
    \caption{Running times of QAOA compared with WalkSATlm for random 8-SAT. LHS: Scaling exponent $\alpha$ in running time $\sim 2^{\alpha n}$ for QAOA estimated by inverting analytic ($p \le 10$) and numerical ($p \le 60$) results on average probabilities, and from numerical results on median running times for $n \in \{12,\dots,20\}$. Horizontal line is experimentally estimated WalkSATlm median running time scaling exponent. Other lines are fits. Blue dashed line is fitting based on all $p$, blue solid line is using $p\le 10$. Error bars are too small to be seen. RHS: histogram of ratios of running times of QAOA ($p=60$) and WalkSAT for $n=20$ instances.}
    \label{fig:qaoa_walksatlm}
\end{figure}

Figure \ref{fig:qaoa_walksatlm} shows the result of this comparison. We found that for $p\approx 14$, QAOA matches the performance of WalkSATlm. For larger $p$, the predicted running time scaling based on our theoretical bounds on success probability well matches the predicted scaling from empirical success probabilities. However, the median running time diverges from this prediction, and outperforms WalkSATlm more modestly. As discussed in the previous section, one possible cause of this is that the QAOA parameters used, which optimise the average success probability for small $n$, are less effective at optimising the median running time for larger $n$. Further experiments would be required to determine whether this is the case, or if there truly is a gap between the running time predicted by the average success probability, and the median running time. We also show in Figure \ref{fig:qaoa_walksatlm} a histogram of the ratios between running times of QAOA ($p=60$) and WalkSATlm for $n=20$ instances. We observe that there are rather substantial tails, corresponding to instances where one solver or the other has a large advantage over its counterpart.

\subsection{Methods}
\label{sec:numerical_methods}

\paragraph{Random $k$-SAT instances}\mbox{}\\\\
Random $k$-SAT instances are generated from the random ensemble described in definition \ref{def:random_ksat}. When not explicitly specified, the clauses-to-variables ratio $r$ is set to (an approximation of) the satisfiability threshold. For $k \leq 15$, we use the same reference values for this threshold as \cite{campbell19}, while for $k = 16$, we estimate the threshold using the third-order expansion from \cite[appendix]{mertens_mezard_zecchina_2006}, in agreement with the method used by the former work for lower $k$. For reference, we report in table \ref{tab:sat_thresholds} the relevant threshold values used in our numerical study.

\paragraph{Comparison with classical algorithms}\mbox{}\\\\
WalkSAT is an algorithm based on a randomised local search approach. Given an assignment which does not satisfy the formula, the algorithm picks a clause which is not satisfied, picks one of the variables within that clause, and flips it. Various strategies have been proposed for choosing clauses and variables; here we simply pick the clause at random (among unsatisfied clauses) and pick a random variable within that clause to flip. 
WalkSATlm is a modification of WalkSAT with a more complex cost function. For WalkSATlm, trial and error led to the choice of hyperparameters $p = 0.15, w_1 = 6, w_2 = 5$, see \cite{cai_luo_su_2014} for their detailed meaning. For QAOA, we used pseudo-optimal parameters obtained from $100$ size $12$ random problem instances as described in section \ref{sec:numerical_methods}.

\begin{table}
    \centering
    \begin{tabular}{c|c}
        $k$ & \textrm{Satisfiability threshold}\\
        \hline
        2 & 1.0\\
        \hline
        4 & 9.93\\
        \hline
        8 & 176.54\\
        \hline
        10 & 708.92\\
        \hline
        16 & 45425.2
    \end{tabular}
    \caption{Random $k$-SAT satisfiability thresholds for values of $k$ considered in this work}
    \label{tab:sat_thresholds}
\end{table}

\paragraph{Parameter optimization}\mbox{}\\\\
Since QAOA is a variational algorithm depending on angles $\bm\beta$ and $\bm\gamma$, all numerical experiments require either optimization of these parameters or evaluation at well-chosen guessed parameters. In this work, we choose the second option to demonstrate the potential of using QAOA without classical optimization loop. We therefore look for variational parameters achieving good success probability on an average instance. More precisely, this means identifying parameters that maximize the expected success probability of QAOA on a random instance of $k$-SAT.

Given mixer angles $\bm\beta$, this probability is evaluated analytically for small enough $\bm\gamma$ by proposition \ref{prop:p1_qaoa_ksat_expected_success_probability}. Besides, under this small angle assumption, the fixed point finding procedure is numerically efficient, requiring only time $\mathcal{O}\left(\log\frac{1}{\varepsilon}\right)$ for a tolerated error $\varepsilon$ (see discussion of the Hamming weight squared toy example in section \ref{sec:hamming_weight_squared_example} for an explicit illustration of this fact). This suggests to use proposition \ref{prop:p1_qaoa_ksat_expected_success_probability} and its implementation detailed in algorithms \ref{alg:find_fixed_point_small_c}, \ref{alg:find_fixed_point_improved} to perform this average-instance optimization. Unfortunately, this approach presents two difficulties. First, we do not know whether optimal parameters are sufficiently small that proposition \ref{prop:p1_qaoa_ksat_expected_success_probability} applies. Second, even if this assumption was satisfied, we empirically observed that for moderately large $p$, e.g. $p \geq 5$, variational optimization could not be performed in reasonable time (with gradient descent, allowing hundreds of iterations and a time budget of order hour) using the analytic algorithm to evaluate the success probability.

We therefore chose to estimate the expected success probability of QAOA empirically, using sets of randomly generated instances. More specifically, given a number of variables per clause $k$, a clauses-to-variables ratio $r$ and a finite instance size $n$, one generates a set of $100$ random instances sampled from $\textrm{CNF}(n, k, r)$. Then, for each set of angles $\bm\beta, \bm\gamma$, the fixed-parameters, average-instance success probability of QAOA is estimated by empirically averaging the success probability of QAOA over the set of instances; angles $\bm\beta, \bm\gamma$ are then updated accordingly. The set of instances does not change between iterations of the classical optimization algorithm. In this study, we initialize the optimization by setting all $\bm\beta$ to $0.01$ and $\bm\gamma$ to $-0.01$, which we conjecture correspond to the correct signs of the optimal angles based on many optimization trials from randomly starting points. The small magnitude $0.01$ was chosen following the observation that excessively large angles, e.g.\ $1$, led to false maxima and barren plateaus, notably for large $p$. The classical optimal algorithm  we used is a simple gradient descent, which we conjecture always converges to the optimum from this initial guess. Besides, for each $k$ and $r$, $100$ random instances of size $12$ were generated to evaluate the empirical average success probability.

A limitation of this empirical average method is that the average success probability is only approximated rather than exactly calculated. In fact, the very limited number of samples: 100 used for this approximation is not even sufficient to achieve acceptable statistical significance on the estimation of the success probability itself, as discussed in section \ref{sec:validation_analytic_formula}. Besides, individual instance sizes must be limited for the method to remain practical. Despite these apparent weaknesses, we find that the empirical technique provides near-optimal angles  (at least for small $p$) when compared to the analytic one introduced in proposition \ref{prop:success_probability_random_ksat_qaoa_saddle_point}. This justifies a posteriori the approach of determining near-optimal angles from averages over few sample instances. Therefore, we always use angles optimized by this technique in the rest of the study.

The evaluation and optimization of QAOA were carried out using the Yao.jl quantum circuit simulation framework \cite{Luo2020yaojlextensible}. Among other advantages, this library combines execution speed and seamless integration of differentiable programming, making it particularly suitable for the study of variational quantum circuits.

\paragraph{Comparison metrics}\mbox{}\\\\
In the presentation of our numerical results, we have made comparisons between different analytic and empirical methods to estimate the performance of QAOA on random $k$-SAT. We will also contrast QAOA against several classical algorithms tackling the $k$-SAT problem. We discuss here some aspects of the metrics used for these comparisons.

The main figure of merit we use to characterize QAOA is the scaling exponent of its success probability, averaged over problem instances. Such a scaling exponent is rigorously known to exist for sufficiently small $\bm\gamma$ according to proposition \ref{prop:success_probability_random_ksat_qaoa_saddle_point}. It can also be estimated through an exponential fit of the success probability against the instance size. In fact, rather than the scaling exponent itself, we occasionally consider its \textit{excess} over the value it would take for random assignment. Recalling that random assignment is the special case of QAOA where $\bm\beta = \bm{0}$ or $\bm\gamma = \bm{0}$, equation \ref{eq:qaoa_ksat_expected_success_probability} gives a scaling exponent $-2^{-k}r$ in this case; therefore, we systematically subtract this quantity from all scaling exponents and call the resulting values \textit{excess scaling exponents}. This adjustment prevents excessive optimism when comparing different algorithms to estimate scaling exponents. Precisely, in the case that QAOA does little better than random assignment (for instance, for small angles) and two exponent estimation algorithms capture this feature, these methods will simultaneously return a value close to $-2^{-k}r$, leading to a small relative error between the methods. However, should they differ more substantially when it comes to the excess scaling exponents, this would go relatively unnoticed if only comparing exponents.

We now discuss how the running time on a problem instance is quantified for QAOA and classical algorithms. In both cases, we used the median running time over randomly generated instances as a figure of merit. However, the running time of the algorithm on a single instance is defined differently for QAOA and classical solvers. In the case of QAOA, the instance running time is simply defined as the inverse $\frac{1}{p_{\mathrm{succ}}}$ of the success probability $p_{\mathrm{succ}}$ (probability of sampling a satisfying assignment) of QAOA on this instance. This is indeed the expected number of samples one needs to draw to obtain a satisfying assignment, and corresponds to the ``time to solution'' (TTS) metric used in experimental comparison of algorithms, taking a success probability of $1/2$. As for classical solvers, the instance running time is understood as the number of evaluations of the Boolean formula defining the SAT problem. For algorithms from the \verb|pySAT| suite \cite{imms-sat18}, this number is determined according to the information returned by the library after execution of the solver.

\paragraph{Exponential fits}\mbox{}\\\\
Our main analytic result for random $k$-SAT QAOA, proposition \ref{prop:success_probability_random_ksat_qaoa_saddle_point}, predicts an exponential scaling in the infinite size limit for the average-instance fixed-parameters success probability. To compare this result to numerical experiments, empirical scaling exponents need to be extracted using an exponential fit. In practice, in this work, the exponential fit is a least-squares linear regression on the logarithm of the quantity to fit. For each problem size, we obtain an empirical average success probability by averaging over instances. The fit is then performed on these empirical average success probabilities as a function of problem size. To estimate the error on the parameters returned by the fit (in particular, on the scaling exponent), one uses resampling. Precisely, one recalculates empirical average success probabilities using only half of the sample instances, where the half is chosen uniformly at random. This resampling process is repeated several times (typically $100$), leading to a probability distribution for the fit parameters. The error on each fit parameter is then estimated as the standard deviation of the distribution of this parameter.

\section{Derivation of analytic formulae}
\label{sec:derivation_analytic_formulae}
In this section, we derive the saddle-point formula for the average-instance success probability of $k$-SAT QAOA given in proposition \ref{prop:success_probability_random_ksat_qaoa_saddle_point}. We start with a derivation of an analogous result for a toy example (QAOA applied to the Hamming weight squared) which gives an accurate flavour of the general method, at least for the $2$-SAT case.

\subsection{Random $k$-SAT QAOA expectations as generalized multinomial sums}
\label{sec:qaoa_multinomial_sums}

We establish an expression for the expected success probability of random $k$-SAT QAOA (definition \ref{def:random_ksat_qaoa}) using a slight generalization of proposition 28 from \cite{2110.10685}. According to this proposition, recalled here with adapted notations:
\begin{prop}
\label{prop:permutation_invariant_csp_qaoa_average}
Let a random constraint satisfaction problem be defined by set of clauses $\bm\sigma$ and a diagonal quantum Hamiltonian $H[\bm\sigma]$; one temporarily denotes $\bm\sigma \sim \mathrm{CSP}(n)$ for a set of clauses on $n$ variables sampled from the random ensemble. For $\bm\beta, \bm\gamma \in \mathbf{R}^p$, let $\ket{\Psi\left(\bm\beta, \bm\gamma\right)}$ the state prepared by level-$p$ QAOA for this combinatorial optimization problem. Assume that for all $y^{(0)}, \ldots, y^{(2p)} \in \{0, 1\}^{n}$,
\begin{align}
\label{eq:expected_hamiltonian_applied_tensor_product}
    \mathbf{E}_{\bm\sigma \sim \mathrm{CSP}(n)}\left[\bigotimes_{j \in [p]}e^{\frac{i\gamma_j}{2}H[\bm\sigma]} \otimes \left\{H[\bm\sigma] = 0\right\} \otimes \bigotimes_{j \in [p]} e^{-\frac{i\gamma_{p - 1 - j}}{2}H[\bm\sigma]}\right]\bigotimes_{j \in [2p + 1]}\ket{y^{(2p - j)}}
\end{align}
only depends on the numbers:
\begin{align}
\label{eq:def_ns}
    n_s & := \left|\left\{i \in [n]\,:\,\forall j \in [2p + 1],\,y^{(j)}_i = s_j \right\}\right|, \quad s \in \{0, 1\}^{2p + 1}.
\end{align}
[Note that the quantity in equation \ref{eq:expected_hamiltonian_applied_tensor_product} is always colinear to $\bigotimes\limits_{j \in [2p + 1]}\ket{y^{(2p - j)}}$ since $H[\bm\sigma]$ is diagonal.] In this case, we introduce the notation:
\begin{align}
    & E\left(\left(n_s\right)_{s \in \{0, 1\}}\right)\nonumber\\
    & := \bigotimes_{j \in [2p + 1]}\bra{y^{(2p - j)}}\mathbf{E}_{\bm\sigma \sim \mathrm{CSP}(n)}\left[\bigotimes_{j \in [p]}e^{\frac{i\gamma_j}{2}H[\bm\sigma]} \otimes \left\{H[\bm\sigma] = 0\right\} \otimes \bigotimes_{j \in [p]} e^{-\frac{i\gamma_{p - 1 - j}}{2}H[\bm\sigma]}\right]\bigotimes_{j \in [2p + 1]}\ket{y^{(2p - j)}}.
\end{align}
Then:
\begin{align}
    & \mathbf{E}_{\bm\sigma \sim \mathrm{CSP}(n)}\braket{\Psi\left(\bm\sigma, \bm\beta, \bm\gamma\right)|\left\{H[\bm\sigma] = 0\right\}|\Psi\left(\bm\sigma, \bm\beta, \bm\gamma\right)} & = \frac{1}{2^n}\sum_{\substack{\left(n_s\right)_{s \in \{0, 1\}^{2p + 1}}\\\sum_sn_s = n}}\binom{n}{\left(n_s\right)_{s}}\prod_{s \in \{0, 1\}^{2p + 1}}B_{\bm\beta, s}^{n_s}E\left(\left(n_s\right)_{s}\right),
\end{align}
where
\begin{align}
    B_{\bm\beta, s} & := (-1)^{\mathbf{1}[s_0 \neq s_p]}\prod_{j \in [p]}\left(\cos\frac{\beta_j}{2}\right)^{\mathbf{1}\left[s_j = s_{j + 1}\right] + \mathbf{1}[s_{2p - j} = s_{2p - j - 1}]}\left(i\sin\frac{\beta_j}{2}\right)^{\mathbf{1}[s_j \neq s_{j + 1}] + \mathbf{1}[s_{2p - j} \neq s_{2p - j - 1}]}
\end{align}
depend only on the mixing angles $\bm\beta$ (in particular, not on the dephasing angles $\bm\gamma$ or the constraint satisfaction problem) and
\begin{align}
    \binom{n}{\left(n_s\right)_{s \in \{0, 1\}^{2p + 1}}} & = \frac{n!}{\prod_{s \in \{0, 1\}^{2p + 1}}n_s!}
\end{align}
is a multinomial coefficient.
\end{prop}
We will need the following lemma, which is a slight adaptation of the reasoning in \cite[section 3.1]{Achlioptas2006} to average over random $k$-SAT instances:
\begin{lem}[Averaging over random $k$-SAT clauses]
\label{lemma:average_clauses}
Let $k \geq 1$ and $p \geq 1$ integers and let $\sigma$ an OR-clause on $n$ variables $x_0, \ldots, x_{n - 1}$ sampled as defined in definition \ref{def:random_ksat}, i.e. by choosing $k$ literals uniformly at random among $\left\{x_0, \overline{x_0}, \ldots, x_{n - 1}, \overline{x_{n - 1}}\right\}$. Let $J$ bitstrings (representing literal assignment) $y^{(j)} \in \{0, 1\}^n,\,j \in [J]$.
\begin{align}
    \forall y \in \{0, 1\}^n,\quad\mathbf{E}_{\sigma}\left[\prod_{j \in [J]}\mathbf{1}\left[y^{(j)} \not\vdash \sigma\right]\right] & = \left(\frac{\left|y^{(0)} \cap \ldots \cap y^{(J - 1)}\right|}{2n}\right)^k,
\end{align}
where $y^{(0)} \cap \ldots \cap y^{(J - 1)} := \left\{i \in [n]\,:\,y^{(0)}_i = y^{(1)}_i = \ldots = y^{\left(J - 1\right)}_0\right\}$ is the set of indices where bitstrings $y^{(j)}$ all coincide. It is easily seen that in the degenerate case $J = 0$, it suffices to replace $\left|y^{(0)} \cap \ldots \cap y^{(J - 1)}\right| \longrightarrow 2n$ in the equation above for it to remain correct. 
\begin{proof}
It suffices to observe that for all $k$-literal OR-clause $\sigma$, $\sigma$ is simultaneously \textit{unsatisfied} by $y^{(0)}, \ldots, y^{(J - 1)}$ iff.:
\begin{itemize}
    \item all literals from $\sigma$ have variables in $y^{(0)} \cap \ldots \cap y^{(J - 1)}$;
    \item a variable appearing in the clause is negated iff. it is set to $1$ in $y^{(0)}, \ldots, y^{(J - 1)}$ (the value must be common between these bitstrings by definition of $y^{(0)} \cap \ldots \cap y^{(J - 1)}$).
\end{itemize}
Using this fact, and since the probability of choosing one such literal among $2n$ possible literals is $\frac{\left|y^{(0)} \cap \ldots \cap y^{(J - 1)}\right|}{2n}$, the probability of choosing $k$ such literals is $\left(\frac{\left|y^{(0)} \cap \ldots \cap y^{(J - 1)}\right|}{2n}\right)^k$ by independence of literal choices.
\end{proof}
\end{lem}

\begin{prop}
\label{prop:qaoa_ksat_expected_success_probability}
Let $k \geq 1$, $p \geq 1$ integers and let $r > 0$. Let $\bm\beta, \bm\gamma \in \mathbf{R}^p$. The success probability of level-$p$ QAOA on random $k$-SAT with $n$ variables and expected clauses-to-variables ratio $r$ (see definitions \ref{def:random_ksat} and \ref{def:random_ksat_qaoa}) is given by:
\begin{align}
\label{eq:qaoa_ksat_expected_success_probability}
    & \mathbf{E}_{\bm\sigma \sim \mathrm{CNF}(n, k, r)}\left[\left\langle\Psi_{\mathrm{QAOA}}(\bm\sigma, \bm\beta, \bm\gamma)|\mathbf{1}\left\{H[\bm\sigma] = 0\right\}|\Psi_{\mathrm{QAOA}}(\bm\sigma, \bm\beta, \bm\gamma)\right\rangle\right]\nonumber\\
    & = \frac{1}{2^n}e^{-2^{-k}rn\left(1 + 4\sum\limits_{j \in [p]}\sin^2\frac{\gamma_j}{4}\right)}\sum_{\substack{\left(n_s\right)_{s \in \{0, 1\}^{2p + 1}}\\\sum_sn_s = n}}\binom{n}{\left(n_s\right)_s}\prod_sB_{\bm\beta, s}^{n_s}\exp\left(rn\sum_{\substack{J \subset [2p + 1]\\|J| \geq 2}}c_J\left(\frac{1}{2n}\sum_{\substack{s \in \{0, 1\}^{2p + 1}\\\forall j, j' \in J,\,s_j = s_{j'}}}n_s\right)^k\right),
\end{align}
where
\begin{align}
\label{eq:def_cJ}
    c_J & := (-1)^{\mathbf{1}\left[p \in J\right]}\prod_{j \in J\,:\,j < p}\left(e^{-\frac{i\gamma_j}{2}} - 1\right)\prod_{j \in J\,:\,j > p}\left(e^{\frac{i\gamma_{2p - j}}{2}} - 1\right),
\end{align}
and $B_{\bm\beta, s}$ is defined in proposition \ref{prop:permutation_invariant_csp_qaoa_average}.
\begin{proof}
In order to apply proposition \ref{prop:permutation_invariant_csp_qaoa_average}, we start by computing:
\begin{align*}
    & \mathbf{E}_{\bm\sigma \sim \mathrm{CNF}(n, k, r)}\left[\bigotimes_{j \in [p]}e^{\frac{i\gamma_j}{2}H[\bm\sigma]} \otimes \left\{H[\bm\sigma] = 0\right\} \otimes \bigotimes_{j \in [p]}e^{-\frac{i\gamma_{p - 1 - j}}{2}H[\bm\sigma]}\right]\bigotimes_{j \in [2p + 1]}\ket{y^{(2p - j)}}\\
    & = \sum_{m\geq  0}\frac{e^{-rn}(rn)^m}{m!}\mathbf{E}_{\bm\sigma = \left(\sigma_0, \ldots, \sigma_{m - 1}\right)}\left[\bigotimes_{j \in [p]}e^{\frac{i\gamma_j}{2}H[\bm\sigma]} \otimes \left\{H[\bm\sigma] = 0\right\} \otimes \bigotimes_{j \in [p]}e^{-\frac{i\gamma_{p - 1 - j}}{2}H[\bm\sigma]}\right]\bigotimes_{j \in [2p + 1]}\ket{y^{(2p - j)}}.
\end{align*}
We analyze the terms in the sum for fixed $m$.
\begin{align*}
    & \bigotimes_{j \in [p]}e^{\frac{i\gamma_j}{2}H[\bm\sigma]} \otimes \left\{H[\bm\sigma] = 0\right\} \otimes \bigotimes_{j \in [p]}e^{-\frac{i\gamma_{p - 1 - j}}{2}H[\bm\sigma]}\bigotimes_{j \in [2p + 1]}\ket{y^{(2p - j)}}\\
    & = \left(\bigotimes_{j \in [p]}e^{\frac{i\gamma_j}{2}\sum\limits_{z^{(2p - j)} \in \{0, 1\}^n}\sum\limits_{l \in [m]}\mathbf{1}\left[z^{(2p - j)} \not\vdash \sigma_l\right]\ket{z^{(2p - j)}}\bra{z^{(2p - j)}}} \otimes \sum_{z^{(p)} \in \{0, 1\}^n}\prod_{l \in [m]}\mathbf{1}\left[z^{(p)} \vdash \sigma_l\right]\ket{z^{(p)}}\bra{z^{(p)}} \otimes \right.\\
    & \left. \hspace*{30px} \bigotimes_{j \in [p]}e^{-\frac{i\gamma_{p - 1 - j}}{2}\sum\limits_{z^{(p - 1 - j)} \in \{0, 1\}^n}\sum\limits_{l \in [m]}\mathbf{1}\left[z^{(p - 1 - j)} \not\vdash \sigma_l\right]\ket{z^{(p - 1 - j)}}\bra{z^{(p - 1 - j)}}}\right)\bigotimes_{j \in [2p + 1]}\ket{y^{(2p - j)}}\\
    & = \prod_{j \in [p]}e^{\frac{i\gamma_j}{2}\sum\limits_{l \in [m]}\mathbf{1}\left[y^{(2p - j)} \not\vdash \sigma_l\right]}\prod_{l \in [m]}\mathbf{1}\left[y^{(p)} \vdash \sigma_l\right]\prod_{j \in [p]}e^{-\frac{i\gamma_{p - 1 - j}}{2}\sum\limits_{l \in [m]}\mathbf{1}\left[y^{(p - 1 - j)} \not\vdash \sigma_l\right]}\bigotimes_{j \in [2p + 1]}\ket{y^{(2p - j)}}\\
    & = \prod_{l \in [m]}\left\{\mathbf{1}\left[y^{(p)} \not\vdash \sigma_l\right]\exp\left(\sum_{j \in [p]}\frac{i\gamma_j}{2}\mathbf{1}\left[y^{(2p - j)} \not\vdash \sigma_l\right] - \sum_{j \in [p]}\frac{i\gamma_{p - 1 - j}}{2}\mathbf{1}\left[y^{(p - 1 - j)} \not\vdash \sigma_l\right]\right)\right\}\bigotimes_{j \in [2p + 1]}\ket{y^{(2p - j)}}.
\end{align*}
By the factorization in $l \in [m]$ just obtained and independence of clause choices, it suffices to average independently over each random clause $\sigma_0, \ldots, \sigma_{m - 1}$. We then compute
\begin{align*}
    & \mathbf{E}_{\sigma_0}\left\{\mathbf{1}\left[y^{(p)} \vdash \sigma_0\right]\exp\left(\sum_{j \in [p]}\frac{i\gamma_j}{2}\mathbf{1}\left[y^{(2p - j)} \not\vdash \sigma_0\right] - \sum_{j \in [p]}\frac{i\gamma_{p - 1 - j}}{2}\mathbf{1}\left[y^{(p - 1 - j)} \not\vdash \sigma_0\right]\right)\right\}\\
    & = \mathbf{E}_{\sigma_0}\left\{\mathbf{1}\left[y^{(p)} \vdash \sigma_0\right]\prod_{j \in [p]}\exp\left(\frac{i\gamma_j}{2}\mathbf{1}\left[y^{(2p - j)} \not\vdash \sigma_0\right]\right)\prod_{j \in [p]}\exp\left(-\frac{i\gamma_{p - 1 - j}}{2}\mathbf{1}\left[y^{(p - 1 - j)} \not\vdash \sigma_0\right]\right)\right\}\\
    & = \mathbf{E}_{\sigma_0}\left\{\prod_{j \in [p]}\left(1 + \left(e^{i\gamma_j/2} - 1\right)\mathbf{1}\left[y^{(2p - j)} \not\vdash \sigma_0\right]\right)\prod_{j \in [p]}\left(1 + \left(e^{-i\gamma_{p - 1 - j}/2} - 1\right)\mathbf{1}\left[y^{(p - 1 - j)} \not\vdash \sigma_0\right]\right)\right.\\
    & \hspace*{60px} \times \left(1 - \mathbf{1}\left[y^{(p)} \not\vdash \sigma_0\right]\right) \Bigg\}\\
    & = \mathbf{E}_{\sigma_0}\left\{\sum_{J \subset [2p + 1]}c_J\prod_{j \in J}\mathbf{1}\left[y^{(j)} \not\vdash \sigma_0\right]\right\},
\end{align*}
where $J$ records the $j$ where we chose the term with $\mathbf{1}\left[y^{(j)} \not\vdash \sigma_0\right]$ when expanding the parenthesis, and
\begin{align*}
    c_J & := (-1)^{\mathbf{1}\left[p \in J\right]}\prod_{j \in J\,:\,j < p}\left(e^{-\frac{i\gamma_j}{2}} - 1\right)\prod_{j \in J\,:\,j > p}\left(e^{\frac{i\gamma_{2p - j}}{2}} - 1\right).
\end{align*}
Using lemma \ref{lemma:average_clauses} to average over $\sigma_0$ then gives
\begin{align*}
    \mathbf{E}_{\sigma_0}\left\{\sum_{J \subset [2p + 1]}c_J\prod_{j \in J}\mathbf{1}\left[y^{(j)} \not\vdash \sigma_0\right]\right\} & = 1 + \sum_{\varnothing \subsetneq J \subset [2p + 1]}c_J\left(\frac{\left|\bigcap_{j \in J}y^{(j)}\right|}{2n}\right)^k.
\end{align*}
Therefore, averaging the original $m$-clause expression over $\sigma_0, \ldots, \sigma_{m - 1}$ yields
\begin{align*}
    & \mathbf{E}_{\bm\sigma = \left(\sigma_0, \ldots, \sigma_{m - 1}\right)}\left\{\prod_{l \in [m]}\mathbf{1}\left[y^{(p)} \vdash \sigma_l\right]\exp\left(\sum_{j \in [p]}\frac{i\gamma_j}{2}\mathbf{1}\left[y^{(2p - j)} \not\vdash \sigma_l\right] - \sum_{j \in [p]}\frac{i\gamma_{p - 1 - j}}{2}\mathbf{1}\left[y^{(p - 1 - j)} \not\vdash \sigma_l\right]\right)\right\}\\
    & = \left(1 + \sum_{\varnothing \subsetneq J \subset [2p + 1]}c_J\left(\frac{\left|\bigcap_{j \in J}y^{(j)}\right|}{2n}\right)^k\right)^m,
\end{align*}
which, after averaging over $m \sim \textrm{Poisson}(rn)$, becomes
\begin{align*}
    & \sum_{m \geq 0}\frac{e^{-rn}(rn)^m}{m!}\nonumber\\
    & \hspace*{20px} \times \mathbf{E}_{\bm\sigma = \left(\sigma_0, \ldots, \sigma_{m - 1}\right)}\left\{\prod_{l \in [m]}\mathbf{1}\left[y^{(p)} \vdash \sigma_l\right]\exp\left(\sum_{j \in [p]}\frac{i\gamma_j}{2}\mathbf{1}\left[y^{(2p - j)} \not\vdash \sigma_l\right] - \sum_{j \in [p]}\frac{i\gamma_{p - 1 - j}}{2}\mathbf{1}\left[y^{(p - 1 - j)} \not\vdash \sigma_l\right]\right)\right\}\\
    & = \exp\left(rn\sum_{\varnothing \subsetneq J \subset [2p + 1]}c_J\left(\frac{\left|\bigcap_{j \in J}y^{(j)}\right|}{2n}\right)^k\right).
\end{align*}
Defining $\left(n_s\right)_{s \in \{0, 1\}}$ as in equation \ref{eq:def_ns} for bitstrings $y^{(0)}, \ldots, y^{(2p)}$, the above can be rewritten as
\begin{align*}
    & \exp\left(rn\sum_{\varnothing \subsetneq J \subset [2p + 1]}c_J\left(\frac{1}{2n}\sum\limits_{\substack{s \in \{0, 1\}^{2p + 1}\\\forall j, j' \in J,\,s_{j} = s_{j'}}}n_s\right)^k\right).
\end{align*}
This shows that random $k$-SAT satisfies the permutation invariance assumption from proposition \ref{prop:permutation_invariant_csp_qaoa_average}. One can slightly simplify the expression above by distinguishing the singleton $J = \left\{j\right\},\,j \in [2p + 1]$ from other $J$. Indeed, for these $J$,
\begin{align*}
    \sum_{\substack{s \in \{0, 1\}^{2p + 1}\\\forall j', j'' \in J, s_{j'} = s_{j''}}}n_s & = \sum_{\substack{s \in \{0, 1\}^{2p + 1}\\}}n_s\\
    & = n,
\end{align*}
giving total contribution in the exponential
\begin{align*}
    rn\sum_{j \in [2p + 1]}c_{\{j\}}\left(\frac{n}{2n}\right)^k & = r\sum_{j \in [2p + 1]}2^{-k}\left\{\begin{array}{cc}
        e^{-\frac{i\gamma_j}{2}} - 1 & \textrm{if } j < p\\
        -1 & \textrm{if } j = p\\
        e^{\frac{i\gamma_{2p - j}}{2}} - 1 & \textrm{if } j > p
    \end{array}\right.\\
    & = 2^{-k}rn\left(\sum_{j \in [p]}\left(2\cos\frac{\gamma_j}{2} - 2\right) - 1\right)\\
    & = -2^{-k}rn\left(4\sum_{j \in [p]}\sin^2\frac{\gamma_j}{4} + 1\right).
\end{align*}
\end{proof}
\end{prop}

\begin{prop}
For single-layer ($p = 1$) QAOA, equation \ref{eq:qaoa_ksat_expected_success_probability} for the expected success probability of random $k$-SAT QAOA specializes as follows:
\begin{align}
\label{prop:p1_qaoa_ksat_expected_success_probability}
    & \mathbf{E}_{\bm\sigma \sim \mathrm{CNF}(n, k, r)}\left[\left\langle\Psi_{\mathrm{QAOA}}(\bm\sigma, \beta, \gamma)|\mathbf{1}\left\{H[\bm\sigma] = 0\right\}|\Psi_{\mathrm{QAOA}}(\bm\sigma, \beta, \gamma)\right\rangle\right]\nonumber\\
    & = e^{-2^{-k}rn\left(1 + 4\sin^2\frac{\gamma}{4}\right)}\hspace*{-30px}\sum_{\substack{n_{a}, n_b, n_c, n_d\\n_{a} + n_b + n_c + n_d = n}}\hspace*{-10px}\binom{n}{n_a, n_b, n_c, n_d}\left(\cos^2\frac{\beta}{2}\right)^{n_a}\left(\sin^2\frac{\beta}{2}\right)^{n_b}\left(\frac{i\sin\beta}{2}\right)^{n_c}\left(-\frac{i\sin\beta}{2}\right)^{n_d}\nonumber\\
    & \hspace*{30px} \times \exp\left\{rn\left[4\sin^2\frac{\gamma}{4}\left(\left(\frac{n_b + n_a}{2n}\right)^k - \left(\frac{n_a}{2n}\right)^k\right) + \left(1 - e^{-i\gamma/2}\right)\left(\frac{n_c + n_a}{2n}\right)^k\right.\right.\nonumber\\
    & \left.\left. \hspace*{90px} + \left(1 - e^{i\gamma/2}\right)\left(\frac{n_d + n_a}{2n}\right)^k\right]\right\}.
\end{align}
\begin{proof}
In the $p = 1$ case, it is easy to enumerate the terms in the $J$ sum of the exponential in equation \ref{eq:qaoa_ksat_expected_success_probability}. There are 4 such terms given by $J = \{0, 1\}, J = \{0, 2\}, J = \{1, 2\}$ and $J = \{0, 1, 2\}$. We therefore explicit
\begin{align*}
    c_{\{0, 1\}} & = 1 - e^{-i\gamma/2},\\
    c_{\{0, 2\}} & = 4\sin^2\frac{\gamma}{4},\\
    c_{\{1, 2\}} & = 1 - e^{i\gamma/2},\\
    c_{\{0, 1, 2\}} & = -4\sin^2\frac{\gamma}{4}.
\end{align*}
Also,
\begin{align*}
    B_{\beta, 000} & = B_{\beta, 111} = \cos^2\frac{\beta}{2},\\
    B_{\beta, 001} & = B_{\beta, 110} = -\frac{i}{2}\sin\beta,\\
    B_{\beta, 010} & = B_{\beta, 101} = \sin^2\frac{\beta}{2},\\
    B_{\beta, 011} & = B_{\beta, 100} = \frac{i}{2}\sin\beta.
\end{align*}
Plugging it into the formula from proposition \ref{prop:qaoa_ksat_expected_success_probability},
\begin{align*}
    & \mathbf{E}_{\bm\sigma \sim \mathrm{CNF}(n, k, r)}\left[\left\langle\Psi_{\mathrm{QAOA}}(\bm\sigma, \beta, \gamma)|\mathbf{1}\left\{H[\bm\sigma] = 0\right\}|\Psi_{\mathrm{QAOA}}(\bm\sigma, \beta, \gamma)\right\rangle\right]\\
    & = \frac{e^{-2^{-k}rn\left(1 + 4\sin^2\frac{\gamma}{4}\right)}}{2^n}\sum_{\substack{\left(n_s\right)_{s \in \{0, 1\}^3}\\\sum_sn_s = n}}\binom{n}{\left(n_s\right)_s}\left(\cos^2\frac{\beta}{2}\right)^{n_{000} + n_{111}}\left(\sin^2\frac{\beta}{2}\right)^{n_{010} + n_{101}}\left(\frac{i\sin\beta}{2}\right)^{n_{011} + n_{100}}\\
    & \hspace*{20px} \times \left(-\frac{i\sin\beta}{2}\right)^{n_{001} + n_{110}}\exp\left\{rn\left[4\sin^2\frac{\gamma}{4}\left(\left(\frac{n_{000} + n_{010} + n_{101} + n_{111}}{2n}\right)^k - \left(\frac{n_{000} + n_{111}}{2n}\right)^k\right)\right.\right.\\
    & \left.\left. \hspace*{30px} + \left(1 - e^{-i\gamma/2}\right)\left(\frac{n_{000} + n_{100} + n_{011} + n_{111}}{2n}\right)^k + \left(1 - e^{i\gamma/2}\right)\left(\frac{n_{000} + n_{001} + n_{110} + n_{111}}{2n}\right)^k\right]\right\}.
\end{align*}
We now observe that for all $s \in \{0, 1\}^3$, variable $n_s$ always appears in the form $n_s + n_{\overline{s}}$, where $\overline{s}$ denotes $s$ with all bits flipped. (For instance, $n_{000}$ always appears as part of $n_{000} + n_{111}$. Therefore, we may apply the standard multinomial theorem to each pair of variables $\left(n_s, n_{\overline{s}}\right)$, reducing the summed variables to $4$ instead of $8$. Renaming these variables as follows:
\begin{align*}
    n_{000} + n_{111} & \longrightarrow n_a,\\
    n_{010} + n_{101} & \longrightarrow n_b,\\
    n_{100} + n_{011} & \longrightarrow n_c,\\
    n_{001} + n_{110} & \longrightarrow n_d,
\end{align*}
the expression above becomes
\begin{align*}
    & \mathbf{E}_{\bm\sigma \sim \mathrm{CNF}(n, k, r)}\left[\left\langle\Psi_{\mathrm{QAOA}}(\bm\sigma, \beta, \gamma)|\mathbf{1}\left\{H[\bm\sigma] = 0\right\}|\Psi_{\mathrm{QAOA}}(\bm\sigma, \beta, \gamma)\right\rangle\right]\\
    & = e^{-2^{-k}rn\left(1 + 4\sin^2\frac{\gamma}{4}\right)}\hspace*{-30px}\sum_{\substack{n_a, n_b, n_c, n_d\\n_a + n_b + n_c + n_d = n}}\hspace*{-10px}\binom{n}{n_a, n_b, n_c, n_d}\left(\cos^2\frac{\beta}{2}\right)^{n_a}\left(\sin^2\frac{\beta}{2}\right)^{n_b}\left(\frac{i\sin\beta}{2}\right)^{n_c}\left(-\frac{i\sin\beta}{2}\right)^{n_d}\\
    & \hspace*{30px} \times \exp\left\{rn\left[4\sin^2\frac{\gamma}{4}\left(\left(\frac{n_b + n_a}{2n}\right)^k - \left(\frac{n_a}{2n}\right)^k\right) + \left(1 - e^{-i\gamma/2}\right)\left(\frac{n_c + n_a}{2n}\right)^k\right.\right.\nonumber\\
    & \left.\left. \hspace*{90px} + \left(1 - e^{i\gamma/2}\right)\left(\frac{n_d + n_a}{2n}\right)^k\right]\right\}.
\end{align*}
\end{proof}
\end{prop}

\subsection{Generalized multinomial sums: a warm-up example}
\label{sec:hamming_weight_squared_example}
In this section, we estimate the leading exponential contribution of the success probability of QAOA applied to the Hamming weight squared Hamiltonian:
\begin{align}
    H_C & := \left(\sum_{j \in [n]}\frac{1 - Z_j}{2}\right)^2.
\end{align}
The success probability is defined as the probability of sampling the all-zero string from the QAOA state:
\begin{align}
    p_{\mathrm{succ}}(n) & := \left|\braket{0|e^{-\frac{i\beta H_B}{2}}e^{-\frac{i\gamma H_C}{2}}|+}\right|^2.
\end{align}
Applying all operators in the computational basis, one can easily derive:
\begin{align}
    \braket{0|e^{-\frac{i\beta H_B}{2}}e^{-\frac{i\gamma H_C}{3}}|+} & = \frac{1}{\sqrt{2^n}}\sum_{x \in \{0, 1\}^{n}}\left(\cos\frac{\beta}{2}\right)^{n - \sum\limits_{j \in [n]}x_j}\left(-i\sin\frac{\beta}{2}\right)^{\sum\limits_{j \in [n]}x_j}e^{-\frac{i\gamma}{2}\left(\sum\limits_{j \in [n]}x_j\right)^2}.
\end{align}
Note that the general term of the sum over bitstrings $x \in \{0, 1\}^n$ only depends on the Hamming weight on $x$ and the above can therefore be rewritten
\begin{align}
\label{eq:hamming_weight_squared_binomial_sum}
    \braket{0|e^{-\frac{i\beta H_B}{2}}e^{-\frac{i\gamma H_C}{2}}|+} & = \frac{1}{\sqrt{2^n}}\sum_{0 \leq k \leq n}\binom{n}{k}\left(\cos\frac{\beta}{2}\right)^{n - k}\left(-i\sin\frac{\beta}{2}\right)^ke^{-\frac{i\gamma k^2}{2}}
\end{align}
and evaluated in time $\mathcal{O}(n)$. Unfortunately, the infinite-size limit is not immediate due to the exponential of square $e^{-\frac{i\gamma k^2}{2}}$ preventing from applying the usual binomial theorem. However, this difficulty can be remedied at the expense of introducing an additional integral. Indeed, using
\begin{align}
    \int_{\mathbf{R}}\mathrm{d}\theta\,\frac{e^{-\theta^2/4}}{\sqrt{4\pi}}e^{\theta x} = e^{x^2},
\end{align}
\begin{align*}
    \braket{0|e^{-\frac{i\beta H_B}{2}}e^{-\frac{i\gamma H_C}{2}}|+} & = \frac{1}{\sqrt{2^n}}\sum_{0 \leq k \leq n}\binom{n}{k}\left(\cos\frac{\beta}{2}\right)^{n - k}\left(-i\sin\frac{\beta}{2}\right)^k\int_{\mathbf{R}}\mathrm{d}\theta\,\frac{e^{-\theta^2/4}}{\sqrt{4\pi}}e^{\theta\sqrt{-\frac{i\gamma}{2}}k}\\
    & = \frac{1}{\sqrt{2^n}}\int_{\mathbf{R}}\mathrm{d}\theta\,\frac{e^{-\theta^2/4}}{\sqrt{4\pi}}\frac{1}{\sqrt{2^n}}\sum_{0 \leq k \leq n}\binom{n}{k}\left(\cos\frac{\beta}{2}\right)^{n - k}\left(-i\sin\frac{\beta}{2}\right)^ke^{\theta\sqrt{-\frac{i\gamma}{2}}k}\\
    & = \frac{1}{\sqrt{2^n}}\int_{\mathbf{R}}\mathrm{d}\theta\,\frac{e^{-\theta^2/4}}{\sqrt{4\pi}}\left(\cos\frac{\beta}{2} - i\sin\frac{\beta}{2}e^{\theta\sqrt{-\frac{i\gamma}{2}}}\right)^n\\
    & = \frac{1}{\sqrt{2^n}}\int_{\mathbf{R}}\mathrm{d}\theta\,\frac{e^{-\theta^2/4}}{\sqrt{4\pi}}\left(\cos\frac{\beta}{2} - i\sin\frac{\beta}{2}e^{\theta\sqrt{-\frac{i\widetilde\gamma}{2n}}}\right)^n \qquad \gamma \longleftarrow \frac{\widetilde\gamma}{n}\\
    & = \sqrt{\frac{n}{2^n}}\int_{\mathbf{R}}\mathrm{d}\widetilde\theta\,\frac{e^{-n\widetilde\theta^2/4}}{\sqrt{4\pi}}\left(\cos\frac{\beta}{2} - i\sin\frac{\beta}{2}e^{\widetilde\theta\sqrt{-\frac{i\widetilde\gamma}{2}}}\right)^n \qquad \widetilde\theta \longleftarrow \theta\sqrt{n}\\
    & = \sqrt{n}\frac{1}{\sqrt{2^n}}e^{-\frac{i\beta n}{2}}\int_{\mathbf{R}}\mathrm{d}\widetilde\theta\,\frac{e^{-n\widetilde\theta^2/4}}{\sqrt{4\pi}}\left(e^{\frac{i\beta}{2}}\left(\cos\frac{\beta}{2} - i\sin\frac{\beta}{2}e^{\widetilde\theta\sqrt{-\frac{i\widetilde\gamma}{2}}}\right)\right)^n\\
    & = \sqrt{\frac{n}{4\pi}}\int_{\mathbf{R}}\mathrm{d}\widetilde\theta\,\exp\left\{n\left[-\frac{\log 2}{2} - \frac{i\beta}{2} - \frac{\widetilde\theta^2}{4} + \log\left(e^{\frac{i\beta}{2}}\left(\cos\frac{\beta}{2} - i\sin\frac{\beta}{2}e^{\widetilde\theta\sqrt{-\frac{i\widetilde\gamma}{2}}}\right)\right)\right]\right\}\\
    & =: \sqrt{\frac{n}{4\pi}}\int_{\mathbf{R}}\mathrm{d}\widetilde\theta\,\exp\left(n\Phi_{\beta, \widetilde\gamma}(\widetilde\theta)\right),
\end{align*}
where in the fourth line we prescribe the scaling $\frac{\widetilde\gamma}{n}$, $\widetilde\gamma$ constant, for $\gamma$, allowing for a change of integration variable $\theta \longrightarrow \widetilde\theta$ such that the argument of the integrated exponential in the last line is the product of $n$ and an $n$-independent function of $\widetilde\theta$; this is precisely the setting in which the saddle-point method applies. The sixth line introduces uses the dummy identity $1 = e^{-\frac{i\beta n}{2}}e^{\frac{i\beta n}{2}}$, which allows the argument of the logarithm to be $1$ when $\widetilde\theta = 0$ or $\widetilde\gamma = 0$. In the latter case, the logarithm vanishes for all $\theta$ and the integral is the last line is simply Gaussian with value $\frac{e^{-\frac{i\beta n}{2}}}{\sqrt{2^n}}$, as could have been more directly found using $H_B\ket{+} = n\ket{+}$: $\braket{0|e^{-\frac{i\beta H_B}{n}}e^{-\frac{i\gamma H_C}{2}}|+} = \braket{0|e^{-\frac{i\beta H_B}{2}}|+} = e^{-\frac{i\beta n}{2}}\braket{0|+} = \frac{e^{-\frac{i\beta n}{2}}}{\sqrt{2^n}}$. The logarithm is understood as the principal determination of the complex logarithm, i.e. $\log(\rho e^{i\varphi}) := \log\rho + i\varphi$ for $\rho > 0, \varphi \in [-\pi, \pi)$; it has a discontinuity across the negative real axis.

\begin{rem}
It is easily verified numerically that for large enough real $\theta$, the argument of the logarithm crosses the negative real axis; however, the exponential of the logarithm, hence $\exp\left(n\Phi_{\beta, \widetilde\gamma}(\theta)\right)$ is still analytic in $\theta$ on the whole complex plane. Concretely, only the analyticity of the logarithm around $\theta = 0$ will be relevant to apply the saddle-point method, while the integrand will only require a crude bound (not relying on the analyticity of the $\log$) for large $\theta$.
\end{rem}

We now estimate the integral in the limit $n \to \infty$ using the saddle-point method. The results we show only apply to small enough (but still constant) $\widetilde\gamma$; we will not dedicate much effort to accurately estimating this upper bound as numerical experiments suggest that our estimate for $\braket{0|e^{e^{-\frac{i\beta H_B}{2}}}e^{-\frac{i\gamma H_C}{2}}|+}$ holds beyond the assumptions required for the proofs anyway\footnote{Notably, the prediction for the leading exponential contribution to the scaling of the amplitude still holds numerically even when $\Phi_{\beta, \widetilde\gamma}$ has two critical points, which we conjecture happens for large enough $\widetilde\gamma$. In this case, we still conjecture that $\Phi_{\beta, \widetilde\gamma}$ has smaller real value when evaluated at the secondary critical point, constituting a plausible explanation for the robustness of the result. However, a rigorous application of the saddle-point method remains challenging since the large $\widetilde\gamma$ regime also requires to consider more complicated integration contours in the complex plane than the straight line which suffices for small $\widetilde\gamma$.}. In the following, we always assume $\widetilde\gamma$ is positive without loss of generality since $(\beta, \gamma) \longrightarrow (-\beta, -\gamma)$ is equivalent to conjugating the amplitude. We first show that for small enough $\widetilde\gamma$, $\Phi_{\beta, \widetilde\gamma}$ has a single critical-point and provide an accurate estimate for it. To achieve that, it will be convenient to look at critical points of $\Phi_{\beta, \widetilde\gamma}$ as the fixed points of a certain function. Namely, by simple differentiation,
\begin{align}
    \Phi_{\beta, \widetilde\gamma}'(\theta) = 0 \iff \theta = -\frac{i\sqrt{-2i\widetilde\gamma}\sin\frac{\beta}{2}e^{\theta\sqrt{-\frac{i\widetilde\gamma}{2}}}}{\cos\frac{\beta}{2} - i\sin\frac{\beta}{2}e^{\theta\sqrt{-\frac{i\widetilde\gamma}{2}}}}.
\end{align}
To justify the existence and uniqueness of the critical point as well as estimate it, we only need to show the function on the right-hand side is contractive with a sufficiently small constant.
\begin{lem}
\label{lemma:hamming_weight_squared_fixed_point}
Let $F(\theta) := -\frac{i\sqrt{-2i\widetilde\gamma}\sin\frac{\beta}{2}e^{\theta\sqrt{-\frac{i\widetilde\gamma}{2}}}}{\cos\frac{\beta}{2} - i\sin\frac{\beta}{2}e^{\theta\sqrt{-\frac{i\widetilde\gamma}{2}}}}$. There exists a universal constant $c > 0$ such that for all $\widetilde\gamma < c$ and all $\theta, \theta' \in \mathbf{C}$ with $|\theta|, |\theta'| \leq 2\sqrt{\widetilde\gamma}$, the following holds:
\begin{itemize}
\item $\left|F(\theta)\right|, |F(\theta')| \leq 2\sqrt{\widetilde\gamma}$.
\item $\left|F(\theta) - F(\theta')\right| \leq \gamma|\theta - \theta'|$.
\end{itemize}
\begin{proof}
Let's prove the first statement:
\begin{align*}
    \left|F(\theta)\right| & = \left|-\frac{i\sqrt{-2i\widetilde\gamma}\sin\frac{\beta}{2}e^{\theta\sqrt{-\frac{i\widetilde\gamma}{2}}}}{\cos\frac{\beta}{2} - i\sin\frac{\beta}{2}e^{\theta\sqrt{-\frac{i\widetilde\gamma}{2}}}}\right|\\
    & \leq \sqrt{2\widetilde\gamma}\frac{\left|e^{\theta\sqrt{-\frac{i\widetilde\gamma}{2}}}\right|}{\left|\cos\frac{\beta}{2} - i\sin\frac{\beta}{2}e^{\theta\sqrt{-\frac{i\widetilde\gamma}{2}}}\right|}\\
    & = \sqrt{2\widetilde\gamma}\frac{\left|e^{\theta\sqrt{-\frac{i\widetilde\gamma}{2}}}\right|}{\left|e^{-\frac{i\beta}{2}} - i\sin\frac{\beta}{2}\left(e^{\theta\sqrt{-\frac{i\widetilde\gamma}{2}}} - 1\right)\right|}\\
    & \leq \sqrt{2\widetilde{\gamma}}\frac{e^{\sqrt{2\widetilde\gamma}}}{1 - \left(e^{\sqrt{2\widetilde\gamma}} - 1\right)} \qquad (\left|\theta\right| \leq 2\sqrt{\widetilde\gamma})\\
    & \leq 2\sqrt{\widetilde\gamma} \qquad \textrm{(for small enough } \gamma \textrm{)}.
\end{align*}
As for the second statement, we consider the derivative:
\begin{align}
    F'(\eta) & = -\frac{e^{\eta\sqrt{-\frac{i\widetilde\gamma}{2}}}\sin\beta\widetilde\gamma}{2\left(\cos\frac{\beta}{2} - i\sin\frac{\beta}{2}e^{\eta\sqrt{-\frac{i\widetilde\gamma}{2}}}\right)^2}.
\end{align}
Similarly to the previous calculation, this is $\leq \widetilde\gamma$ for small enough $\widetilde\gamma$ and $|\eta| \leq 2\sqrt{\widetilde\gamma}$. This completes the proof.
\end{proof}
\end{lem}
The existence and uniqueness of the fixed point of $F$ (corresponding to the critical point of $\Phi_{\beta, \widetilde\gamma}$) then follows from the Banach fixed point theorem \cite{banach_1922}:
\begin{thm}[Banach fixed point theorem]
\label{thm:banach_fixed_point}
Let $(X, d)$ a non-empty complete metric space. Let $T: X \longrightarrow X$ a contraction mapping; that is, there exists $k \in [0, 1)$ such that for all $x, x' \in X$, $d(T(x), T(x')) \leq kd(x, x')$. Then $T$ has a unique fixed point $x^*$ in $X$. Besides, $x^*$ can be obtained by iteratively applying $T$ to an arbitrary element of $X$: for $x_0 \in X$, $x_n := T(x_{n - 1})\,\,(n \geq 1)$, $x_n \longrightarrow x^*$.
\end{thm}

\begin{prop}
\label{prop:hamming_weight_square_critical_point}
There exists a universal constant $c > 0$ such that for $\widetilde\gamma < c$, $\Phi_{\beta, \widetilde\gamma}$ has a unique critical point $\theta^*$ satisfying $\left|\theta^* - e^{\frac{i\beta}{2} - \frac{3\pi i}{4}}\sin\frac{\beta}{2}\sqrt{2\widetilde\gamma}\right| \leq \mathcal{O}\left(\widetilde\gamma^{3/2}\right)$.
\begin{proof}
By an earlier observation, $\theta$ is a critical point of $\Phi_{\beta, \widetilde\gamma}$ iff. it is a fixed point of $F$ defined in lemma \ref{lemma:hamming_weight_squared_fixed_point}. Applying the Banach fixed point theorem to $F$, for small enough $\widetilde\gamma$, $F$ has a unique fixed point $\theta^*$ which can be obtained by iteratively applying $F$ to $0$. The first iterate is $F(0) = \frac{-i\sqrt{-2i\widetilde{\gamma}}\sin\frac{\beta}{2}}{\cos\frac{\beta}{2} - i\sin\frac{\beta}{2}} = e^{\frac{i\beta}{2} - \frac{3\pi i}{4}}\sin\frac{\beta}{2}\sqrt{2\widetilde\gamma}$. We now bound the distance between this and $\theta^*$:
\begin{align*}
    \left|\theta^* - F(0)\right| & = \left|F(\theta^*) - F(0)\right|\\
    & \leq \left|F(\theta^*) - F(F(0))\right| + \left|F(F(0)) - F(0)\right|\\
    & \leq \widetilde\gamma\left|\theta^* - F(0)\right| + \left|F(F(0)) - F(0)\right| \qquad \textrm{(lemma \ref{lemma:hamming_weight_squared_fixed_point})}
\end{align*}
so that
\begin{align*}
    \left|\theta^* - F(0)\right| & \leq \frac{\left|F(F(0)) - F(0)\right|}{1 - \widetilde\gamma}\\
    & \leq \frac{\widetilde\gamma\left|F(0) - 0\right|}{1 - \widetilde\gamma}\\
    & \leq \frac{\sqrt{2}\widetilde\gamma^{3/2}}{1 - \widetilde\gamma} = \mathcal{O}(\widetilde\gamma^{3/2})
\end{align*}
\end{proof}
\end{prop}
We now provide several additional estimates of $\Phi_{\beta, \widetilde\gamma}$ itself and its higher-order derivatives that will make the Gaussian approximation rigorous:
\begin{lem}
\label{lemma:hamming_weight_squared_bounds_phi}
Let the conditions of proposition \ref{prop:hamming_weight_square_critical_point} be satisfied and $\theta^*$ the unique fixed point of $\Phi_{\beta, \widetilde\gamma}$. The following estimates hold:
\begin{align}
    \forall \theta \in \mathbf{C},\quad\Phi_{\beta, \widetilde\gamma}\left(\theta\right) & = -\frac{\log(2)}{2} - \frac{i\beta}{2} - \frac{\theta^2}{4} - e^{\frac{i\pi}{4}}\sin\frac{\beta}{2}e^{\frac{i\beta}{2}}\sqrt{\frac{\widetilde\gamma}{2}}\theta + \mathcal{O}\left(\widetilde\gamma\theta^2\right),\\
    \forall \theta \in \mathbf{C},\quad\Phi''_{\beta, \widetilde\gamma}\left(\theta\right) & = -\frac{1}{2} + \mathcal{O}\left(\widetilde\gamma\right),\\
    \Phi_{\beta, \widetilde\gamma}(\theta^*) & = -\frac{i\beta}{2} - \frac{\log 2}{2} + \frac{i}{2}\sin^2\frac{\beta}{2}e^{i\beta}\widetilde\gamma + \mathcal{O}\left(\widetilde\gamma^2\right),\\
    \Phi''_{\beta, \widetilde\gamma}(\theta^*) & = -\frac{1}{2} + \mathcal{O}\left(\widetilde\gamma\right),\\
    \forall \theta \in \mathbf{C}, \quad \left|\exp(n\Phi_{\beta, \widetilde\gamma}(\theta))\right| & \leq \exp\left(n\left(\frac{\log 2}{2} - \frac{\Re\left(\theta^2\right)}{4} + |\theta|\sqrt{\frac{\widetilde\gamma}{2}}\right)\right),\\
    \forall \theta \in \mathbf{C}, |\theta| = o\left(\frac{1}{\sqrt{\widetilde\gamma}}\right) \implies \quad \left|\Phi_{\beta, \widetilde\gamma}'''\left(\theta\right)\right| & = \mathcal{O}\left(\widetilde\gamma^{3/2}\right).
\end{align}
\begin{proof}
The first and second result are systematic Taylor expansions. More precisely, an error $\mathcal{O}\left(\widetilde\gamma\theta^2\right)$ is obtained in the first equation since $\Phi_{\beta, \widetilde\gamma}\left(\theta\right)$ can be written as the sum of $-\frac{\log(2)}{2} - \frac{i\beta}{2} - \frac{\theta^2}{4}$ and a function of $\sqrt{\widetilde\gamma}\theta$; the first-order Taylor expansion of the latter gives the stated error term.

The third and fourth results then follow from plugging the estimate of the critical point from proposition
\ref{prop:hamming_weight_square_critical_point} in the Taylor expansions of $\Phi_{\beta, \widetilde\gamma}, \Phi''_{\beta, \widetilde\gamma}$ just derived.

The fifth result follows from a crude bound:
\begin{align*}
    \left|\exp\left(n\Phi_{\beta, \widetilde\gamma}(\theta)\right)\right| & = \left|\frac{1}{\sqrt{2^n}}e^{-n\theta^2/4}\left(\cos\frac{\beta}{2} - i\sin\frac{\beta}{2}e^{\theta\sqrt{-\frac{i\widetilde\gamma}{2}}}\right)^n\right|\\
    & \leq \frac{1}{\sqrt{2^n}}e^{-n\Re(\theta^2)/4}\left(1 + \left|e^{\theta\sqrt{-\frac{i\widetilde\gamma}{2}}}\right|\right)^n\\
    & \leq \frac{1}{\sqrt{2^n}}e^{-n\Re(\theta^2)/4}\left(1 + e^{|\theta|\sqrt{\frac{\widetilde\gamma}{2}}}\right)^n\\
    & \leq \frac{1}{\sqrt{2^n}}e^{-n\Re(\theta^2)/4}\left(2e^{|\theta|\sqrt{\frac{\widetilde\gamma}{2}}}\right)^n\\
    & = \exp\left(n\left(\frac{\log 2}{2} - \frac{\Re(\theta^2)}{4} + |\theta|\sqrt{\frac{\widetilde\gamma}{2}}\right)\right).
\end{align*}

For the fourth result,
\begin{align*}
    \Phi'''_{\beta, \widetilde\gamma}(\theta) & = \frac{e^{-\frac{3\pi i}{4}}}{2^{5/2}}\frac{e^{\theta\sqrt{-\frac{i\widetilde\gamma}{2}}}\widetilde\gamma^{3/2}\sin\beta\left(\cos\frac{\beta}{2} + i\sin\frac{\beta}{2}e^{\theta\sqrt{-\frac{i\widetilde\gamma}{2}}}\right)}{\left(\cos\frac{\beta}{2} - i\sin\frac{\beta}{2}e^{\theta\sqrt{-\frac{i\widetilde\gamma}{2}}}\right)^3}\\
    & = \frac{e^{-\frac{3\pi i}{4}}}{2^{5/2}}\frac{e^{\theta\sqrt{-\frac{i\widetilde\gamma}{2}}}\widetilde\gamma^{3/2}\sin\beta\left(e^{\frac{i\beta}{2}} + o(1)\right)}{\left(e^{-\frac{i\beta}{2}} + o(1)\right)^3} \qquad \left(\theta = o(\widetilde\gamma^{-1/2})\right)\\
    & = \mathcal{O}\left(\widetilde\gamma^{3/2}\right).
\end{align*}
\end{proof}
\end{lem}

\begin{prop}
\label{prop:hamming_weight_square_scaling_success_probability}
For sufficiently small $\widetilde\gamma > 0$ and $\gamma := \frac{\widetilde\gamma}{n}$, the leading exponential contribution of $\braket{0|e^{-\frac{i\beta H_B}{2}}e^{-\frac{i\gamma H_C}{2}}|+}$ as $n \to \infty$ is given by:
\begin{align}
    \lim_{n \to \infty}\frac{\log\left(e^{\frac{i\beta n}{2}}\braket{0|e^{-\frac{i\beta H_B}{2}}e^{-\frac{i\gamma H_C}{2}}|+}\right)}{n} & = \frac{i\beta}{2} + \Phi_{\beta, \widetilde\gamma}(\theta^*),
\end{align}
where $\theta^*$ is the unique critical point of $\Phi_{\beta, \widetilde\gamma}$ defined in proposition \ref{prop:hamming_weight_square_critical_point}.
\begin{proof}
Recall $\braket{0|e^{-\frac{i\beta H_B}{2}}e^{-\frac{i\gamma H_C}{2}}|+} = \sqrt{\frac{n}{4\pi}}\int_{\mathbf{R}}\mathrm{d}\theta\,\exp\left(n\Phi_{\beta, \widetilde\gamma}(\theta)\right) =: \sqrt{\frac{n}{4\pi}}I$. We now decompose integral $I$ into 3 contributions:
\begin{align*}
    I & = \int_{\theta^* - \frac{\widetilde\gamma^{-1/4}}{\sqrt{-\Phi''_{\beta, \widetilde\gamma}(\theta^*)}} - \infty}^{\theta^* - \frac{\widetilde\gamma^{-1/4}}{\sqrt{-\Phi''_{\beta, \widetilde\gamma}(\theta^*)}}}\!\mathrm{d}\theta\,\exp\left(n\Phi_{\beta, \widetilde\gamma}(\theta)\right) + \int_{\theta^* - \frac{\widetilde\gamma^{-1/4}}{\sqrt{-\Phi''_{\beta, \widetilde\gamma}(\theta^*)}}}^{\theta^* + \frac{\widetilde\gamma^{-1/4}}{\sqrt{\Phi''_{\beta, \widetilde\gamma}(\theta^*)}}}\!\mathrm{d}\theta\,\exp\left(n\Phi_{\beta, \widetilde\gamma}(\theta)\right)\\
    & \hspace*{30px} + \int_{\theta^* + \frac{\widetilde\gamma^{-1/4}}{\sqrt{-\Phi''_{\beta, \widetilde\gamma}(\theta^*)}}}^{\theta^* + \frac{\widetilde\gamma^{-1/4}}{\sqrt{-\Phi''_{\beta, \widetilde\gamma}(\theta^*)}} + \infty}\!\mathrm{d}\theta\,\exp\left(n\Phi_{\beta, \widetilde\gamma}(\theta)\right)\\
    & =: I_1 + I_2 + I_3.
\end{align*}
Here, the integrals are understood as complex contour integrals; $I_1$ and $I_3$ are along horizontal half-lines, while $I_2$ is along an oblique segment. Deforming $\mathbf{R}$ to this path is licit by analyticity and sufficiently fast decrease (cf. estimates from lemma \ref{lemma:hamming_weight_squared_bounds_phi} for the last point) of the integrand. We will now show that the dominant integral is $I_2$, with $I_1$ and $I_3$ being negligible.

Let us then first consider $I_2$. Introducing a parametrization of the integration path,
\begin{align*}
    I_2 & = \frac{\widetilde\gamma^{-1/4}}{\sqrt{-\Phi''_{\beta, \widetilde\gamma}(\theta^*)}}\int_{-1}^1\!\mathrm{d}t\,\exp\left(n\Phi_{\beta, \widetilde\gamma}\left(\theta^* + t\frac{\widetilde\gamma^{-1/4}}{\sqrt{-\Phi''_{\beta, \widetilde\gamma}(\theta^*)}}\right)\right).
\end{align*}
By applying the Taylor expansion formula to second order:
\begin{align*}
    f(z + h) & = f(z) + hf'(z) + \frac{h^2}{2}f''(z) + h^3\int_0^1\!\mathrm{d}u\,\frac{(1 - u)^2}{2}f'''(z + uh),
\end{align*}
we can estimate the integrand as follows:
\begin{align*}
    & \Phi_{\beta, \widetilde\gamma}\left(\theta^* + t\frac{\widetilde\gamma^{-1/4}}{\sqrt{-\Phi''(\theta^*)}}\right)\\
    & = \Phi_{\beta, \widetilde\gamma}(\theta^*) + t\frac{\widetilde\gamma^{-1/4}}{\sqrt{-\Phi''_{\beta, \widetilde\gamma}(\theta^*)}}\Phi'_{\beta, \widetilde\gamma}(\theta^*) + \frac{t^2}{2}\frac{\widetilde\gamma^{-1/2}}{-\Phi''_{\beta, \widetilde\gamma}(\theta^*)}\Phi''_{\beta, \gamma}(\theta^*)\\
    & \hspace*{30px} + \frac{t^3}{2}\frac{\widetilde\gamma^{-3/4}}{\left(-\Phi''_{\beta, \widetilde\gamma}(\theta^*)\right)^{3/2}}\int_0^1\!\mathrm{d}u\,(1 - u)^2\Phi'''_{\beta, \widetilde\gamma}\left(\theta^* + ut\frac{\widetilde\gamma^{-1/4}}{\sqrt{-\Phi''_{\beta, \widetilde\gamma}(\theta^*)}}\right)\\
    & = \Phi_{\beta, \widetilde\gamma}(\theta^*) - \frac{t^2\widetilde\gamma^{-1/4}}{2} + \frac{t^3}{2}\frac{\widetilde\gamma^{-3/4}}{\left(-\Phi''_{\beta, \widetilde\gamma}(\theta^*)\right)^{3/2}}\int_0^1\!\mathrm{d}u\,(1 - u)^2\Phi'''_{\beta, \widetilde\gamma}\left(\theta^* + ut\frac{\widetilde\gamma^{-1/4}}{\sqrt{-\Phi''_{\beta, \widetilde\gamma}(\theta^*)}}\right).
\end{align*}
Now, according to the estimate on $\Phi_{\beta, \widetilde\gamma}'''$ from lemma \ref{lemma:hamming_weight_squared_bounds_phi} and the estimate on $\theta^*$ from proposition \ref{prop:hamming_weight_square_critical_point}, the function in the last integral is uniformly bounded by $\mathcal{O}\left(\widetilde{\gamma}^{3/2}\right)$. Therefore,
\begin{align*}
    & \int_{-1}^1\!\mathrm{d}t\,\exp\left(n\Phi_{\beta, \widetilde\gamma}\left(\theta^* + t\frac{\widetilde\gamma^{-1/4}}{\sqrt{-\Phi''(\theta^*)}}\right)\right)\\
    & = \int_{-1}^1\!\mathrm{d}t\,\exp\left(n\left(\Phi_{\beta, \widetilde\gamma}(\theta^*) - \frac{t^2\widetilde\gamma^{-1/4}}{2} + \mathcal{O}\left(\widetilde\gamma^{3/4}t^3\right)\right)\right)\\
    & = n^{-1/2}\exp\left(n\Phi_{\beta, \widetilde\gamma}(\theta^*)\right)\int_{-\sqrt{n}}^{\sqrt{n}}\!\mathrm{d}t\,\exp\left(-\frac{t^2\widetilde\gamma^{-1/4}}{2} + \frac{1}{\sqrt{n}}\mathcal{O}\left(\widetilde\gamma^{3/4}t^3\right)\right) \qquad t \longrightarrow \sqrt{n}t.
\end{align*}
Assuming $\widetilde\gamma$ small enough, the integrand is dominated by $\exp\left(-\frac{t^2\widetilde\gamma^{-1/4}}{4}\right)$; therefore, dominated convergence applies and the integral converges to $\int_{\mathbf{R}}\!\mathrm{d}t\,\exp\left(-\frac{\widetilde\gamma^{-1/4}t^2}{2}\right) = \sqrt{2\pi\widetilde\gamma^{1/4}}$. In summary,
\begin{align*}
    I_2 & \underset{n \to \infty}{\sim} \left(\frac{4\pi}{n} \right)^{1/2}\widetilde\gamma^{-1/8}\left(1 + \mathcal{O}\left(\widetilde\gamma\right)\right)\exp\left(n\Phi_{\beta, \widetilde\gamma}(\theta^*)\right).
\end{align*}

We now bound $I_3$ (the reasoning for $I_1$ being similar). For that, we use the bound on $\exp\left(n\Phi_{\beta, \widetilde\gamma}(\theta)\right)$ provided by lemma \ref{lemma:hamming_weight_squared_bounds_phi} along the integration path $t \geq 0 \longmapsto \theta^* + \frac{\widetilde\gamma^{-1/4}}{\sqrt{-\Phi''_{\beta, \widetilde\gamma}(\theta^*)}} + t$:
\begin{align*}
    & \left|\exp\left(n\Phi_{\beta, \widetilde\gamma}\left(\theta^* + 
    \frac{\widetilde\gamma^{-1/4}}{\sqrt{-\Phi''_{\beta, \widetilde\gamma}(\theta^*)}} + t\right)\right)\right|\\
    & \leq \exp\left(n\left(\frac{\log 2}{2} - \frac{1}{4}\Re\left\{\left(\theta^* + 
    \frac{\widetilde\gamma^{-1/4}}{\sqrt{-\Phi''_{\beta, \widetilde\gamma}(\theta^*)}} + t\right)^2\right\} + \left|\theta^* + 
    \frac{\widetilde\gamma^{-1/4}}{\sqrt{-\Phi''_{\beta, \widetilde\gamma}(\theta^*)}} + t\right|\sqrt{\frac{\widetilde\gamma}{2}}\right)\right).
\end{align*}
In fact, one can show that the real part squared dominates from the following simple estimates:
\begin{align*}
    \Re\left\{\left(\theta^* + \frac{\widetilde\gamma^{-1/4}}{\sqrt{-\Phi''_{\beta, \widetilde\gamma}(\theta^*)}} + t\right)^2\right\} & = \Re\left\{\left(\mathcal{O}\left(\sqrt{\widetilde\gamma}\right) + \sqrt{2}\widetilde{\gamma}^{-1/4} + \mathcal{O}\left(\widetilde\gamma^{3/4}\right) + t\right)^2\right\}\\
    & = \Re\left\{\left(\sqrt{2}\widetilde\gamma^{-1/4} + t\right)^2\left(1 + \mathcal{O}\left(\sqrt{\widetilde\gamma}\right)\right)\right\}\\
    & = \left(\sqrt{2}\widetilde\gamma^{-1/4} + t\right)^2\left(1 + \mathcal{O}\left(\sqrt{\widetilde\gamma}\right)\right)\\
    \left|\theta^* + 
    \frac{\widetilde\gamma^{-1/4}}{\sqrt{-\Phi''_{\beta, \widetilde\gamma}(\theta^*)}} + t\right|\sqrt{\frac{\widetilde\gamma}{2}} & = \left|\mathcal{O}\left(\sqrt{\widetilde\gamma}\right) + \sqrt{2}\widetilde\gamma^{-1/4} + 
    \mathcal{O}\left(\widetilde\gamma^{3/4}\right) + t\right|\sqrt{\frac{\widetilde\gamma}{2}}\\
    & = \left|\sqrt{2}\widetilde\gamma^{-1/4} + t\right|\sqrt{\frac{\widetilde\gamma}{2}}\left(1 + \mathcal{O}\left(\sqrt{\widetilde\gamma}\right)\right).
\end{align*}
It is important that in the expressions above, the error terms are uniform in $t \geq 0$. Therefore, it holds
\begin{align*}
    \frac{1}{8}\left|\Re\left\{\left(\theta^* + \frac{\widetilde\gamma^{-1/4}}{\sqrt{-\Phi''\left(\theta^*\right)}} + t\right)^2\right\}\right| & \geq \left|\theta^* + \frac{\widetilde\gamma^{-1/4}}{\sqrt{-\Phi''\left(\theta^*\right)}} + t\right|\sqrt{\frac{\widetilde\gamma}{2}}
\end{align*}
as long as
\begin{align*}
    \left|\sqrt{2}\widetilde\gamma^{-1/4} + t\right| & \geq \sqrt{32\widetilde\gamma}\left(1 + \mathcal{O}\left(\sqrt{\widetilde\gamma}\right)\right),
\end{align*}
which is indeed verified for all $t \geq 0$ for sufficiently small $\widetilde\gamma > 0$. Using that the quadratic part dominates, one can simplify the bound as follows:
\begin{align*}
    \left|\exp\left(n\Phi_{\beta, \widetilde\gamma}\left(\theta^* + 
    \frac{\widetilde\gamma^{-1/4}}{\sqrt{-\Phi''_{\beta, \widetilde\gamma}(\theta^*)}} + t\right)\right)\right| & \leq \exp\left(-\frac{n}{8}\left(\sqrt{2}\widetilde\gamma^{-1/4} + t\right)^2\right)\\
    & \leq \exp\left(-\frac{n\widetilde\gamma^{-1/2}}{4} - \frac{n\widetilde\gamma^{-1/4}t}{2^{3/2}}\right).
\end{align*}
It follows that $I_3 = \mathcal{O}\left(\widetilde\gamma^{1/4}\right)\exp\left(-\frac{\widetilde\gamma^{-1/2}n}{4}\right)$. Comparing it with the leading exponential contribution of $I_2$: $\exp\left(n\Phi_{\beta, \widetilde\gamma}(\theta^*)\right)$, using lemma \ref{lemma:hamming_weight_squared_bounds_phi} to estimate $\Phi_{\beta, \widetilde\gamma}(\theta^*)$ there, we conclude that for $\widetilde\gamma$ sufficiently small, $I_3$ is exponentially negligible compared to $I_2$ as $n \to \infty$. The same holds for $I_1$. Putting all this together yields the leading exponential contribution of $\braket{0|e^{-\frac{i\beta H_B}{2}}e^{-\frac{i\gamma H_C}{2}}|+}$ stated in the proposition.
\end{proof}
\end{prop}

Combining the last proposition with estimates on $\widetilde\Phi$ given in lemma \ref{lemma:hamming_weight_squared_bounds_phi} leads to the following corollary:
\begin{cor}
\label{cor:hamming_weight_squared_success_probability_lowest_order}
For sufficiently small $\widetilde\gamma > 0$ and $\gamma := \frac{\widetilde\gamma}{n}$, the leading exponential contribution of $\braket{0|e^{-\frac{i\beta H_B}{2}}e^{-\frac{i\gamma H_C}{2}}|+}$ as $n \to \infty$ is given, up to first order in $\widetilde\gamma$, by:
\begin{align}
    \lim_{n \to \infty}\frac{\log\left(e^{\frac{i\beta n}{2}}\braket{0|e^{-\frac{i\beta H_B}{2}}e^{-\frac{i\gamma H_C}{2}}|+}\right)}{n} & = -\frac{\log 2}{2} + \frac{i}{2}\sin^2\frac{\beta}{2}e^{i\beta}\widetilde\gamma + \mathcal{O}\left(\widetilde\gamma^2\right).
\end{align}
\end{cor}
Given $\beta$, the last result informs on how to choose $\widetilde\gamma$ to increase the success probability (compared to the trivial random sampling case $\widetilde\gamma = 0$). This is equivalent to increasing the real part of the scaling exponent just derived, which is $-\frac{\log(2)}{2} - \frac{1}{2}\widetilde\gamma\sin^2\frac{\beta}{2}\sin\beta + \mathcal{O}\left(\widetilde\gamma^2\right)$. Therefore, to first order in $\widetilde\gamma > 0$, one needs $\beta \in [-\pi, 0]$ for the exponent to be greater than $-\frac{\log(2)}{2}$, and the optimal choice to maximize the growth rate in $\widetilde\gamma$ is $\beta = -\frac{2\pi}{3}$, which yields $-\frac{\log(2)}{2} + \frac{3^{3/2}}{16}\widetilde\gamma + \mathcal{O}\left(\widetilde\gamma^2\right)$ for the exponent previously derived.

In fact, from exact evaluation of the amplitude at finite size $n$ (which only requires to compute the $n$-terms sum in equation \ref{eq:hamming_weight_squared_binomial_sum}), we conjecture the optimal parameters rather converge to $\beta = -\frac{\pi}{2}$ and $\widetilde\gamma = \pi$ in the limit $n \to \infty$. Besides, we conjecture that for this choice of parameters, the success probability decays at most polynomially in $n$. Remarkably, the analytic formula from proposition \ref{prop:hamming_weight_square_scaling_success_probability} agrees with these observations (although we do not rigorously know it to apply for these parameters). Indeed, for $\beta = -\frac{\pi}{2}$ and $\widetilde\gamma = \pi$, the critical point equation is easily checked to admit the simple solution $\theta^* = \sqrt{-\frac{i\pi}{2}}$. One can check that for this solution,
\begin{align}
    \Phi_{\beta, \widetilde\gamma}\left(\theta^*\right) & = \frac{i\pi}{8}.
\end{align}
Since this has vanishing real part, the saddle-point method, if correct, would predict the amplitude to decay only polynomially in $n$, in agreement with empirical results at finite size.

\subsection{Generalized multinomial sums}

Next we generalize the arguments of the previous section in order to evaluate success probabilities for QAOA applied to random $2^q$-SAT instances.

\subsubsection{Statement of the problem}
\begin{defn}
\label{defn:generalized_binomial_sum}
Let $p \geq 1$ an integer, $q \geq 1$ an integer and $\bm{A} = \left(A_{\alpha s}\right)_{\substack{\alpha \in \mathcal{A}\\s \in \mathcal{S}}}, \bm{b} = \left(b_s\right)_{s \in \mathcal{S}}, \bm{c} = \left(c_{\alpha}\right)_{\alpha \in \mathcal{A}}$ families of complex numbers indexed by two arbitrary index sets $\mathcal{S}$ and $\mathcal{A}$. Besides, assume
\begin{align}
    \sum_{s \in \mathcal{S}}b_s = 1.
\end{align}
We consider the problem of estimating:
\begin{align}
\label{eq:generalized_binomial_sum}
    S_q\left(\bm{A}, \bm{b}, \bm{c}\,;\,n\right) & := \sum_{\substack{\left(n_s\right)_{s \in \mathcal{S}}\\\sum_sn_s = n}}\hspace*{-5px}\binom{n}{\left(n_s\right)_s}\left(\prod_{s}b_s^{n_s}\right)\exp\left(n\sum_{\alpha}c_{\alpha}\left(\frac{1}{n}\sum_{s}A_{\alpha s}n_s\right)^{2^q}\right)
\end{align}
as $n \to \infty$ while $\bm{A}, \bm{b}, \bm{c}$ are kept constant.
\end{defn}

\begin{eg}[Random $2^q$-SAT QAOA]
\label{example:random_2q_sat_as_generalized_binomial_sum}
The expression obtained in proposition \ref{prop:qaoa_ksat_expected_success_probability} for the expected fixed-parameters success probability of QAOA on random $k$-SAT is of the desired form when $k = 2^q$ up to the trivial factor $e^{-2^{-2^q}rn\left(1 + 4\sum_{j \in [p]}\sin^2\frac{\gamma_j}{4}\right)}$. In this particular case, the parameters of the generalized multinomial sum can be taken as follows:
\begin{align}
\label{eq:generalized_binomial_sum_parameters_qaoa}
    \mathcal{S} & := \{0, 1\}^{2p + 1},\\
    \mathcal{A} & := \left\{J \subset [2p + 1]\,:\,|J| \geq 2\right\},\\
    b_s & := \frac{B_{\bm\beta, s}}{2},\\
    c_{\alpha} & := r(-1)^{\mathbf{1}[p \in \alpha]}\prod_{j \in \alpha,\,j < p}\left(e^{-i\gamma_j/2} - 1\right)\prod_{j \in \alpha,\,j > p}\left(e^{i\gamma_{2p - j}/2} - 1\right),\\
    A_{\alpha s} & := \frac{1}{2}\mathbf{1}\left[\forall j, j' \in \alpha,\,s_j = s_{j'}\right].
\end{align}
\end{eg}

\subsubsection{The quadratic case}
\label{sec:quadratic_case}
In the case $q = 1$ of definition \ref{defn:generalized_binomial_sum}, the generalized multinomial sum in equation \ref{eq:generalized_binomial_sum} can be given an integral representation by Gaussian integration, similar to the toy model example considered in section \ref{sec:hamming_weight_squared_example}. Explicitly,
\begin{align*}
    & S_1\left(\bm{A}, \bm{b}, \bm{c}\,;\,n\right)\\
    & = \sum_{\substack{\left(n_s\right)_{s \in \mathcal{S}}\\\sum_sn_s = n}}\hspace*{-5px}\binom{n}{\left(n_s\right)_s}\left(\prod_{s}b_s^{n_s}\right)\exp\left(n\sum_{\alpha}c_{\alpha}\left(\frac{1}{n}\sum_{s}A_{\alpha s}n_s\right)^{2}\right)\\
    & = \int_{\mathbf{R}^{|\mathcal{A}|}}\prod_{\alpha \in \mathcal{A}}\mathrm{d}y_{\alpha}\,\frac{e^{-y_{\alpha}^2/4}}{\sqrt{4\pi}}\sum_{\substack{\left(n_s\right)_{s \in \mathcal{S}}\\\sum_sn_s = n}}\hspace*{-5px}\binom{n}{\left(n_s\right)_s}\left(\prod_{s}b_s^{n_s}\right)\exp\left(\sqrt{n}\sum_{\alpha}\sqrt{c_{\alpha}}\left(\frac{1}{n}\sum_{s}A_{\alpha s}n_s\right)y_{\alpha}\right)\\
    & = n^{|\mathcal{A}|/2}\int_{\mathbf{R}^{|\mathcal{A}|}}\prod_{\alpha \in \mathcal{A}}\mathrm{d}y_{\alpha}\,\frac{e^{-ny_{\alpha}^2/4}}{\sqrt{4\pi}}\sum_{\substack{\left(n_s\right)_{s \in \mathcal{S}}\\\sum_sn_s = n}}\hspace*{-5px}\binom{n}{\left(n_s\right)_s}\left(\prod_{s}b_s^{n_s}\right)\exp\left(\sum_{\alpha}\sqrt{c_{\alpha}}\left(\sum_{s}A_{\alpha s}n_s\right)y_{\alpha}\right)\\
    & \hspace*{270px} \textrm{(change variables } y_{\alpha} := \sqrt{n}y'_{\alpha} \quad \forall \alpha \in \mathcal{A} \textrm{)}\\
    & = n^{|\mathcal{A}|/2}\int_{\mathbf{R}^{|\mathcal{A}|}}\prod_{\alpha \in \mathcal{A}}\mathrm{d}y'_{\alpha}\,\frac{e^{-n\left(y'_{\alpha}\right)^2/4}}{\sqrt{4\pi}}\sum_{\substack{\left(n_s\right)_{s \in \mathcal{S}}\\\sum_sn_s = n}}\hspace*{-5px}\binom{n}{\left(n_s\right)_s}\prod_s\left(b_s\exp\left(\sum_{\alpha}\sqrt{c_{\alpha}}A_{\alpha s}y'_{\alpha}\right)\right)^{n_s}\\
    & = n^{|\mathcal{A}|/2}\int_{\mathbf{R}^{|\mathcal{A}|}}\prod_{\alpha \in \mathcal{A}}\mathrm{d}y_{\alpha}\,\frac{e^{-ny_{\alpha}^2/4}}{\sqrt{4\pi}}\sum_{\substack{\left(n_s\right)_{s \in \mathcal{S}}\\\sum_sn_s = n}}\hspace*{-5px}\binom{n}{\left(n_s\right)_s}\prod_s\left(b_s\exp\left(\sum_{\alpha}\sqrt{c_{\alpha}}A_{\alpha s}y_{\alpha}\right)\right)^{n_s}\\
    & = \frac{n^{|\mathcal{A}|/2}}{\left(4\pi\right)^{|\mathcal{A}|/2}}\int_{\mathbf{R}^{|\mathcal{A}|}}\left(\prod_{\alpha \in \mathcal{A}}\mathrm{d}y_{\alpha}\right)\exp\left(-\frac{n}{4}\sum_{\alpha \in \mathcal{A}}y_{\alpha}^2\right)\left(\sum_sb_s\exp\left(\sum_{\alpha}\sqrt{c_{\alpha}}A_{\alpha s}y_{\alpha}\right)\right)^n\\
    & =: \frac{n^{|\mathcal{A}|/2}}{\left(4\pi\right)^{|S|/2}}\int_{\mathbf{R}^{|\mathcal{A}|}}\left(\prod_{\alpha \in \mathcal{A}}\mathrm{d}y_{\alpha}\right)\,\exp\left(n\Phi\left(\left(y_{\alpha}\right)_{\alpha \in \mathcal{A}}\right)\right),
\end{align*}
where
\begin{align}
\label{eq:phi_quadratic}
    \Phi\left(\left(y_{\alpha}\right)_{\alpha \in \mathcal
    A}\right) & := -\frac{1}{4}\sum_{\alpha \in \mathcal{A}}y_{\alpha}^2 + \log\sum_{s \in \mathcal{S}}b_s\exp\left(\sum_{\alpha \in \mathcal{A}}\sqrt{c_{\alpha}}A_{\alpha s}y_{\alpha}\right).
\end{align}
This has the same formal structure as the Hamming weight squared toy example discussed in section \ref{sec:hamming_weight_squared_example}, except the integral is multivariable instead of univariate. Note that the prefactor $\frac{n^{|\mathcal{A}|/2}}{\left(4\pi\right)^{|\mathcal{A}|/2}}$, similar to $\sqrt{\frac{n}{4\pi}}$ in the toy example, is irrelevant to the exponential scaling here since we assume $|\mathcal{A}|$ constant, so that $\frac{n^{|\mathcal{A}|/2}}{\left(4\pi\right)^{|\mathcal{A}|/2}} = \textrm{poly}(n)$. The analysis performed there for small $\widetilde\gamma$ therefore carries over here \textit{for sufficiently small $c_{\alpha}$}; this hypothesis, precisely stated in assumption \ref{assp:c_bounded}, will also be required for the analysis of the $q > 1$ case. In the cases considered in section \ref{sec:qaoa_multinomial_sums} where the generalized multinomial sum arises from a QAOA expectation, this is equivalent to assuming small (yet constant) $\bm\gamma$ angles but places no restriction on the $\bm\beta$\footnote{However, it may be that to each choice of $\bm\beta$ angles corresponds an upper bound on the $\gamma_j$ for the method to work, and that the upper bound be more favourable for some $\bm\beta$ than for others. Still, this upper bound will be constant for each $\bm\beta$ and not e.g. $\mathcal{O}\left(n^{-1}\right)$.}. We then state the following proposition establishing the scaling of $S_q\left(\bm{A}, \bm{b}, \bm{c}\,;\,n\right)$ in the $q = 1$ case. We leave out the proof, which would essentially be a repetition of that of proposition \ref{prop:hamming_weight_square_scaling_success_probability} from section \ref{sec:hamming_weight_squared_example}; besides, the more general case $q \geq 1$ will be investigated in section \ref{sec:power_of_two_case}.
\begin{prop}
Let a general multinomial sum $S_1\left(\bm{A}, \bm{b}, \bm{c}\,;\,n\right)$ be given as in definition \ref{defn:generalized_binomial_sum} with $q = 1$ and recall the definition of $\Phi\left(\cdot\right)$ in equation \ref{eq:phi_quadratic}. Then, for sufficiently small $\left(c_{\alpha}\right)_{\alpha \in \mathcal{A}}$,
\begin{align}
    \lim_{n \to \infty}\frac{\log S_1\left(\bm{A}, \bm{b}, \bm{c}\,;\,n\right)}{n} & = \Phi\left(\bm{y}^*\right),
\end{align}
where $\bm{y}^* = \left(y^*_{\alpha}\right)_{\alpha \in \mathcal{A}}$ is the unique critical point of $\Phi\left(\cdot\right)$, whose existence and uniqueness is indeed guaranteed for sufficiently small $\left(c_{\alpha}\right)_{\alpha \in \mathcal{A}}$. Besides, this critical point can be understood as the fixed point of the function:
\begin{align}
    \begin{array}{ccc}
        \mathbf{C}^{|\mathcal{A}|} & \longrightarrow & \mathbf{C}^{|\mathcal{A}|}\\
        \left(y_{\alpha}\right)_{\alpha \in \mathcal{A}} & \longmapsto & \left(\frac{2\sqrt{c_{\alpha}}\sum\limits_{s \in \mathcal{S}}b_sA_{\alpha s}\exp\left(\sum\limits_{\alpha' \in \mathcal{A}}\sqrt{c_{\alpha'}}A_{\alpha' s}y_{\alpha'}\right)}{\sum\limits_{s \in \mathcal{S}}b_s\exp\left(\sum\limits_{\alpha' \in S}\sqrt{c_{\alpha'}}A_{\alpha' s}y_{\alpha'}\right)}\right)_{\alpha \in \mathcal{A}}.
    \end{array}
\end{align}
In particular, assuming $\left|c_{\alpha}\right| \leq c$ for definiteness, the following estimates hold in the limit $c \to 0$:
\begin{align}
    y^*_{\alpha} & = 2\sqrt{c_{\alpha}}\sum_{s \in \mathcal{S}}b_sA_{s\alpha} + \mathcal{O}\left(c^{3/2}\right),\\
    \lim_{n \to \infty}\frac{\log S_1\left(\bm{A}, \bm{b}, \bm{c}\,;\,n\right)}{n} & = \sum_{\alpha \in \mathcal{A}}c_{\alpha}\left(\sum_{s \in \mathcal{S}}b_sA_{\alpha s}\right)^2 + \mathcal{O}\left(c^{3/2}\right)\label{eq:generalized_binomial_sum_scaling_quadratic_lowest_order}.
\end{align}
\begin{rem}
Note that the leading order contribution of the scaling exponent of $S_q\left(\bm{A}, \bm{b}, \bm{c}\,;\,n\right)$ in equation \ref{eq:generalized_binomial_sum_scaling_quadratic_lowest_order} is what we would have obtained had we (non-rigorously) estimated $S_q\left(\bm{A}, \bm{b}, \bm{c}\,;\,n\right)$ by naively expanding the $\log$ in $\Phi$ to lowest order in $c$ for each value of $\left(y_{\alpha}\right)_{\alpha \in \mathcal{A}}$, before integrating the resulting Gaussian over these variables. That is,
\begin{align*}
    & S_1\left(\bm{A}, \bm{b}, \bm{c}\,;\,n\right)\\
    & \eqpoly \int_{\mathbf{R}^{|\mathcal{A}|}}\!\left(\prod_{\alpha \in \mathcal{A}}\mathrm{d}y_{\alpha}\right)\,\exp\left(n\Phi\left(\left(y_{\alpha}\right)_{\alpha \in \mathcal{A}}\right)\right)\\
    & = \int_{\mathbf{R}^{|\mathcal{A}|}}\!\left(\prod_{\alpha \in \mathcal{A}}\mathrm{d}y_{\alpha}\right)\,\exp\left(n\left(-\frac{1}{4}\sum_{\alpha \in \mathcal{A}}y_{\alpha}^2 + \log\sum_{s \in \mathcal{S}}b_s\exp\left(\sum_{\alpha \in  \mathcal{A}}\sqrt{c_{\alpha}}A_{\alpha s}y_{\alpha}\right)\right)\right)\\
    & \approx \int_{\mathbf{R}^{|\mathcal{A}|}}\!\left(\prod_{\alpha \in \mathcal{A}}\mathrm{d}y_{\alpha}\right)\,\exp\left(n\left(-\frac{1}{4}\sum_{\alpha \in \mathcal{A}}y_{\alpha}^2 + \log\left(\sum_{s \in \mathcal{S}}b_s\left(1 + \sum_{\alpha \in \mathcal{A}}\sqrt{c_{\alpha}}A_{\alpha s}y_{\alpha}\right)\right)\right)\right) \quad \textrm{(not rigorous!)}\\
    & = \int_{\mathbf{R}^{|\mathcal{A}|}}\!\left(\prod_{\alpha \in \mathcal{A}}\mathrm{d}y_{\alpha}\right)\,\exp\left(n\left(-\frac{1}{4}\sum_{\alpha \in \mathcal{A}}y_{\alpha}^2 + \log\left(\sum_{s \in \mathcal{S}}b_s + \sum_{s \in \mathcal{S}}b_s\sum_{\alpha \in \mathcal{A}}\sqrt{c_{\alpha}}A_{\alpha s}y_{\alpha}\right)\right)\right)\\
    & = \int_{\mathbf{R}^{|\mathcal{A}|}}\!\left(\prod_{\alpha \in \mathcal{A}}\mathrm{d}y_{\alpha}\right)\,\exp\left(n\left(-\frac{1}{4}\sum_{\alpha \in \mathcal{A}}y_{\alpha}^2 + \log\left(1 + \sum_{s \in \mathcal{S}}b_s\sum_{\alpha \in \mathcal{A}}\sqrt{c_{\alpha}}A_{\alpha s}y_{\alpha}\right)\right)\right)\\
    & \approx \int_{\mathbf{R}^{|\mathcal{A}|}}\!\left(\prod_{\alpha \in \mathcal{A}}\mathrm{d}y_{\alpha}\right)\,\exp\left(n\left(-\frac{1}{4}\sum_{\alpha \in \mathcal{A}}y_{\alpha}^2 + \sum_{s \in \mathcal{S}}b_s\sum_{\alpha \in \mathcal{A}}\sqrt{c_{\alpha}}A_{\alpha s}y_{\alpha}\right)\right) \quad \textrm{(not rigorous!)}\\
    & \eqpoly \exp\left(n\sum_{\alpha \in \mathcal{A}}c_{\alpha}\left(\sum_{s \in \mathcal{S}}b_sA_{\alpha s}\right)^2\right).
\end{align*}
\end{rem}
\end{prop}

\subsubsection{The power-of-two case}
\label{sec:power_of_two_case}

The case $q \geq 2$ requires a generalization of the Gaussian integration trick used in section \ref{sec:quadratic_case} for $q = 1$. This generalization is detailed in section \ref{sec:iterated_gaussian_integration} below and essentially contained in proposition \ref{prop:iterated_gaussian_integral_generate_higher_order_gaussian}, the statement of which relies on definitions \ref{def:zeta} and \ref{def:d_domain}. Recalling the general expression of $S_q\left(\bm{A}, \bm{b}, \bm{c}\,;\,n\right)$ in definition \ref{defn:generalized_binomial_sum}:
\begin{align*}
    S_q(\bm{A}, \bm{b}, \bm{c}\,;\,n) & = \sum_{\substack{\left(n_s\right)_{s \in \mathcal{S}}\\\sum_sn_s = n}}\hspace*{-5px}\binom{n}{\left(n_s\right)_s}\left(\prod_{s}b_s^{n_s}\right)\exp\left(n\sum_{\alpha}c_{\alpha}\left(\frac{1}{n}\sum_{s}A_{\alpha s}n_s\right)^{2^q}\right)\\
    & = \sum_{\substack{\left(n_s\right)_{s \in \mathcal{S}}\\\sum_sn_s = n}}\hspace*{-5px}\binom{n}{\left(n_s\right)_s}\left(\prod_{s}b_s^{n_s}\right)\prod_{\alpha}\exp\left(nc_{\alpha}\left(\frac{1}{n}\sum_{s}A_{\alpha s}n_s\right)^{2^q}\right),
\end{align*}
we observe that for all fixed $\left(n_s\right)_{s \in \mathcal{S}}$ and $\alpha \in \mathcal{A}$, the exponential can be expressed as the following integral:
\begin{align*}
    & \exp\left(nc_{\alpha}\left(\frac{1}{n}\sum_{s}A_{\alpha s}n_s\right)^{2^q}\right)\\
    & = \int\limits_{\left(y_{\alpha}^{(j)}\right)_{j \in [q]} \in \mathcal{D}_{q, C_{\alpha}}}\prod_{j \in [q]}\frac{\mathrm{d}y^{(j)}_{\alpha}}{\sqrt{4\pi}}e^{-\frac{\left(y^{(j)}_{\alpha}\right)^2}{4}}\,\exp\left(n^{1/2^q - 1}\left(-c_{\alpha}\right)^{1/2^q}\sum_{s \in \mathcal{S}}A_{\alpha s}n_s\prod_{j \in [q]}\left(-iy_{\alpha}^{(j)}\right)^{1/2^{q - 1 - j}}\right)\\
    & = \int\limits_{\left(y_{\alpha}^{(j)}\right)_{j \in [q]} \in \mathcal{D}_{q, C_{\alpha}}}\prod_{j \in [q]}\frac{\mathrm{d}y^{(j)}_{\alpha}}{\sqrt{4\pi}}e^{-\frac{\left(y^{(j)}_{\alpha}\right)^2}{4}}\,\exp\left(n^{1/2^q - 1}\left(-c_{\alpha}\right)^{1/2^q}\sum_{s \in \mathcal{S}}A_{\alpha s}n_s\zeta\left(\left(y^{(j)}_{\alpha}\right)_{j \in [q]}\right)\right)
\end{align*}
for a function $\zeta$ defined in Definition \ref{def:zeta}. Here $C_{\alpha}$ (corresponding to $C$ in proposition \ref{prop:iterated_gaussian_integral_generate_higher_order_gaussian}) is an arbitrary constant $\in \mathbf{C} - \mathbf{R}_-$. In fact, this representation is not unique; for instance, using the fact that for all $\varphi_{\alpha, 0}, \varphi_{\alpha, 1} \in \mathbf{R}$,
\begin{align}
    -c_{\alpha} & = -c_{\alpha}\frac{e^{2^qi\varphi_{\alpha, 0}}}{e^{2^qi\varphi_{\alpha, 0}} + e^{2^qi\varphi_{\alpha, 1}}} - c_{\alpha}\frac{e^{2^qi\varphi_{\alpha, 1}}}{e^{2^qi\varphi_{\alpha, 0}} + e^{2^qi\varphi_{\alpha, 1}}}\\
    & = \left(\left(-c_{\alpha}\right)^{1/2^q}\frac{e^{i\varphi_{\alpha, 0}}}{\left(e^{2^qi\varphi_{\alpha, 0}} + e^{2^qi\varphi_{\alpha, 1}}\right)^{1/2^q}}\right)^{2^q} + \left(\left(-c_{\alpha}\right)^{1/2^q}\frac{e^{i\varphi_{\alpha, 0}}}{\left(e^{2^qi\varphi_{\alpha, 1}} + e^{2^qi\varphi_{\alpha, 1}}\right)^{1/2^q}}\right)^{2^q},
\end{align}
the alternative integral identity can be obtained:
\begin{align*}
    & \int\limits_{\forall i \in \{0, 1\},\,\left(y^{(j)}_{\alpha, i}\right)_{j \in [q]} \in \mathcal{D}_{q, C_{\alpha, i}}}\prod_{\substack{i \in \{0, 1\}\\j \in [q]}}\frac{\mathrm{d}y_{\alpha, i}^{(j)}}{\sqrt{4\pi}}e^{-\frac{\left(y^{(j)}_{\alpha, i}\right)^2}{4}}\exp\left\{n^{1/2^q - 1}\sum_{\substack{\alpha \in \mathcal{A}\\s \in \mathcal{S}}}A_{\alpha s}n_s \times \frac{(-c_{\alpha})^{1/2^q}}{\left(e^{2^qi\varphi_{\alpha, 0}} + e^{2^qi\varphi_{\alpha, 1}}\right)^{1/2^q}}\right.\\
    & \left. \hspace*{60px} \times \sum_{i \in \{0, 1\}}e^{i\varphi_{\alpha, i}}\zeta\left(\left(y^{(j)}_{\alpha, i}\right)_{j \in [q]}\right)\right\}
\end{align*}
Plugging this identity in the original expression for $S_q\left(\bm{A}, \bm{b}, \bm{c}\,;\,n\right)$, one can rewrite the generalized multinomial sum as follows:
\begin{align*}
    & S_q(\bm{A}, \bm{b}, \bm{c}\,;\,n)\\
    & = \sum_{\substack{\left(n_s\right)_{s \in \mathcal{S}}\\\sum_sn_s = n}}\hspace*{-5px}\binom{n}{\left(n_s\right)_s}\left(\prod_{s}b_s^{n_s}\right)\exp\left(n\sum_{\alpha}c_{\alpha}\left(\frac{1}{n}\sum_{s}A_{\alpha s}n_s\right)^{2^q}\right)\\
    & = \int\limits_{\forall \alpha \in \mathcal{A},\,\forall i \in \{0, 1\},\,\left(y^{(j)}_{\alpha, i}\right)_{j \in [q]} \in \mathcal{D}_{q, C_{\alpha, i}}}\prod_{\substack{\alpha \in \mathcal{A},\,i \in \{0, 1\}\\j \in [q]}}\frac{\mathrm{d}y_{\alpha, i}^{(j)}}{\sqrt{4\pi}}e^{-\frac{\left(y^{(j)}_{\alpha, i}\right)^2}{4}}\,\sum_{\substack{\left(n_s\right)_s\\\sum_sn_s = n}}\binom{n}{\left(n_s\right)_s}\left(\prod_sb_s^{n_s}\right)\\
    & \hspace*{30px} \times \exp\left\{n^{1/2^q - 1}\sum_{\substack{\alpha \in \mathcal{A}\\s \in \mathcal{S}}}A_{\alpha s}n_s \times \frac{(-c_{\alpha})^{1/2^q}}{\left(e^{2^qi\varphi_{\alpha, 0}} + e^{2^qi\varphi_{\alpha, 1}}\right)^{1/2^q}}\sum_{i \in \{0, 1\}}e^{i\varphi_{\alpha, i}}\zeta\left(\left(y^{(j)}_{\alpha, i}\right)_{j \in [q]}\right)\right\},
\end{align*}
where $\left(\varphi_{\alpha, i}\right)_{\alpha \in \mathcal{A},\,i \in \{0, 1\}}$ are real parameters to be adjusted later. Besides, when applying the identity from proposition \ref{prop:iterated_gaussian_integral_generate_higher_order_gaussian}, we allowed a different constant $C$, denoted by $C_{\alpha, i}$ for each $q$-tuple of integration variables $\left(y^{(j)}_{\alpha, i}\right)_{j \in [q]}$, $\alpha \in \mathcal{A},\,i \in \{0, 1\}$. The complex $2^q$-th root in the expression above can be taken to be any determination, even inconsistently between distinct $c_{\alpha}$; such inconsistency would be fixed at the time of choosing the $\varphi_{\alpha, i}$ as we will soon show. One can now perform the multinomial sum for each value of the integration variables $\left(y_{\alpha, i}^{(j)}\right)_{\alpha \in \mathcal{A},\,j \in [q],\,i \in \{0, 1\}}$:
\begin{align}
    & S_q\left(\bm{A}, \bm{b}, \bm{c}\,;\,n\right)\nonumber\\
    & = \int\limits_{\forall \alpha \in \mathcal{A},\,\forall i \in \{0, 1\},\,\left(y^{(j)}_{\alpha, i}\right)_{j \in [q]} \in \mathcal{D}_{q, C_{\alpha, i}}}\prod_{\substack{\alpha \in \mathcal{A}\\j \in [q]}}\frac{\mathrm{d}y_{\alpha, 0}^{(j)}}{\sqrt{4\pi}}e^{-\frac{\left(y^{(j)}_{\alpha, 0}\right)^2}{4}}\,\frac{\mathrm{d}y_{\alpha, 1}^{(j)}}{\sqrt{4\pi}}e^{-\frac{\left(y^{(j)}_{\alpha, 1}\right)^2}{4}}\left\{\sum_sb_s\exp\left[n^{1/2^q - 1}\sum_{\alpha \in \mathcal{A}}A_{\alpha s}\right.\right.\nonumber\\
    & \left.\left. \hspace*{100px} \times \frac{(-c_{\alpha})^{1/2^q}}{\left(e^{2^qi\varphi_{\alpha, 0}} + e^{2^qi\varphi_{\alpha, 1}}\right)^{1/2^q}}\sum_{i \in \{0, 1\}}e^{i\varphi_{\alpha, i}}\zeta\left(\left(y^{(j)}_{\alpha, i}\right)_{j \in [q]}\right)\right]\right\}^n\\
    & = n^{q|\mathcal{A}|}\int\limits_{\forall \alpha \in \mathcal{A},\,\forall i \in \{0, 1\},\,\left(y^{(j)}_{\alpha, i}\right)_{j \in [q]} \in \mathcal{D}_{q, C_{\alpha, i}}}\prod_{\substack{\alpha \in \mathcal{A}\\j \in [q]}}\frac{\mathrm{d}y_{\alpha, 0}^{(j)}}{\sqrt{4\pi}}e^{-\frac{\left(y^{(j)}_{\alpha, 0}\right)^2}{4}n}\,\frac{\mathrm{d}y_{\alpha, 1}^{(j)}}{\sqrt{4\pi}}e^{-\frac{\left(y^{(j)}_{\alpha, 1}\right)^2}{4}n}\left\{\sum_sb_s\exp\left[\sum_{\alpha \in \mathcal{A}}A_{\alpha s}\right.\right.\nonumber\\
    & \left.\left. \hspace*{120px} \times \frac{(-c_{\alpha})^{1/2^q}}{\left(e^{2^qi\varphi_{\alpha, 0}} + e^{2^qi\varphi_{\alpha, 1}}\right)^{1/2^q}}\sum_{i \in \{0, 1\}}e^{i\varphi_{\alpha, i}}\zeta\left(\left(y^{(j)}_{\alpha, i}\right)_{j \in [q]}\right)\right]\right\}^n\label{eq:generalized_binomial_sum_integral_representation}.
\end{align}
Forgetting about the $\frac{1}{\sqrt{4\pi}}$, the integrand can be written\footnote{At least, for small enough integration variables for the $\log$ to be defined.}
\begin{align}
    \exp\left\{n\left(-\frac{1}{4}\sum_{\substack{\alpha \in \mathcal{A},\,i \in \{0, 1\}\\j \in [q]}}\left(y_{\alpha, i}^{(j)}\right)^2 + F\left(\left(\zeta\left(\left(y^{(j)}_{\alpha, i}\right)_{j \in [q]}\right)\right)_{\alpha \in \mathcal{A},\,i \in \{0, 1\}}\right)\right)\right\}\label{eq:generalized_binomial_sum_integral_representation_f},
\end{align}
where
\begin{align}
\label{eq:generalized_binomial_sum_parent_function}
    F\left(\left(z_{\alpha, i}\right)_{\alpha \in \mathcal{A},\,i \in \{0, 1\}}\right) & = \log\sum_{s \in \mathcal{S}}b_s\exp\left(\sum_{\alpha \in \mathcal{A}}A_{\alpha s}\frac{(-c_{\alpha})^{1/2^q}}{\left(e^{2^qi\varphi_{\alpha, 0}} + e^{2^qi\varphi_{\alpha, 1}}\right)^{1/2^q}}\sum_{i \in \{0, 1\}}e^{i\varphi_{\alpha, i}}z_{\alpha, i}\right).
\end{align}
Note that $F(\bm{0}) = 0$ since $\sum_sb_s = 1$ by assumption. In addition, the derivatives of $F$:
\begin{align}
\label{eq:generalized_binomial_sum_parent_function_derivative}
    \partial_{\alpha}F\left(z\right) & = (-c_{\alpha})^{1/2^q}\frac{\sum\limits_{s \in\mathcal{S}}b_sA_{\alpha s}\exp\left(\sum\limits_{\alpha' \in \mathcal{A}}A_{\alpha' s}(-c_{\alpha'})^{1/2^q}z_{\alpha'}\right)}{\sum\limits_{s \in \mathcal{S}}b_s\exp\left(\sum\limits_{\alpha' \in \mathcal{A}}A_{\alpha' s}(-c_{\alpha'})^{1/2^q}z_{\alpha'}\right)}
\end{align}
will play an important role in the following.
We now specify how to choose parameters $\varphi_{\alpha, i}$. Briefly, the choice is to satisfy the argument condition from proposition \ref{prop:critical_point_iterative_gaussian}, i.e. $\left|\arg\left(-\partial_{(\alpha, i)}F\left(\left(z_{\alpha, i}\right)_{\alpha \in \mathcal{A},\,i \in \{0, 1\}}\right)\right)\right| < \frac{\pi}{2^{q - 1}}$ for small enough $\left(z_{\alpha, i}\right)_{\alpha \in \mathcal{A},\,i \in \{0, 1\}}$; in fact, we will require the stronger upper bound $< \frac{\pi}{2^q}$. This will ensure that critical points lie in the upper half plane, which will facilitate a full justification of the saddle-point method. We now make the following assumption:
\begin{assp}
\label{assp:c_bounded}
Parameters $\bm{c}$ are individually bounded by a constant $c > 0$:
\begin{align}
    \forall \alpha \in \mathcal{A},\quad \left|c_{\alpha}\right| < c.
\end{align}
\end{assp}
The results we will show will hold for small enough $c$ (yet, constant as $n \to \infty$). More precisely, given unrestricted $\bm{A}, \bm{b}$ parameters, an upper bound will be required on the components of $\bm{c}$. For the application of our results to QAOA, this corresponds to requiring small (yet constant as the problem size goes to infinity) $\bm{\gamma}$ angles but allowing arbitrary $\bm{\beta}$ angles. Now, for fixed $\left(z_{\alpha, i}\right)_{\alpha \in \mathcal{A}, i \in \{0, 1\}}$, $F\left(\left(z_{\alpha, i}\right)_{\alpha \in \mathcal{A}, i \in \{0, 1\}}\right)$ defined in equation \ref{eq:generalized_binomial_sum_parent_function} expands as follows to lowest order in $c$:
\begin{align}
    F\left(\left(z_{\alpha, i}\right)_{\alpha \in \mathcal{A},\,i \in \{0, 1\}}\right) & = \sum_{\substack{\alpha \in \mathcal{A}\\s \in \mathcal{S}}}b_sA_{\alpha s}\frac{(-c_{\alpha})^{1/2^q}}{\left(e^{2^qi\varphi_{\alpha, 0}} + e^{2^qi\varphi_{\alpha, 1}}\right)^{1/2^q}}\sum_{i \in \{0, 1\}}e^{i\varphi_{\alpha, i}}z_{\alpha, i} + \mathcal{O}\left(c^{1/2^{q - 1}}\right).
\end{align}
Informally differentiating with respect to $z_{\alpha, i}$ ($i \in \{0, 1\}$) yields:
\begin{align}
    \partial_{(\alpha, i)}F\left(\left(z_{\alpha, i}\right)_{\alpha \in \mathcal{A},\,i \in \{0, 1\}}\right) \approx \frac{(-c_{\alpha})^{1/2^q}}{\left(e^{2^qi\varphi_{\alpha, 0}} + e^{2^qi\varphi_{\alpha, 1}}\right)^{1/2^q}}e^{i\varphi_{\alpha, i}}\sum_{s \in \mathcal{S}}b_sA_{\alpha s}.
\end{align}
This motivates the following choice for the $\varphi_{\alpha, i}$:
\begin{defn}
\label{def:generalized_binomial_sum_fix_c}
Given families of complex numbers $\bm{A} = \left(A_{\alpha s}\right)_{\substack{\alpha \in \mathcal{A}\\s \in \mathcal{S}}}$, $\bm{b} = \left(b_s\right)_{s \in \{0, 1\}}$, $\bm{c} = \left(c_{\alpha}\right)_{\alpha \in \mathcal{A}}$ indexed by arbitrary sets $\mathcal{A}$ and $\mathcal{S}$ as in definition \ref{defn:generalized_binomial_sum}, let $\theta_{\alpha}$ the unique representative in $\left(-\frac{\pi}{2^q}, \frac{\pi}{2^q}\right]$ of
\begin{align}
    \arg\left\{\left(-c_{\alpha}\right)^{1/2^q}\sum_{s \in \mathcal{S}}b_sA_{\alpha s}\right\} \quad \left(\mathrm{mod}\,\,\frac{\pi}{2^{q - 1}}\right).
\end{align}
Denote by $m_{\alpha}$ the unique integer such that:
\begin{align}
    \arg\left\{\left(-c_{\alpha}\right)^{1/2^q}\sum_{s \in \mathcal{S}}b_sA_{\alpha s}\right\} & = \theta_{\alpha} + m_{\alpha}\frac{\pi}{2^{q - 1}}.
\end{align}
$\varphi_{\alpha, 0}$ and $\varphi_{\alpha, 1}$ are then defined as follows:
\begin{align}
    \varphi_{\alpha, 0} & := \left\{\begin{array}{cc}
        -m_{\alpha}\frac{\pi}{2^{q - 1}} - \frac{\pi}{2^{q + 2}} & \mathrm{if}\,\, \theta_{\alpha} = \frac{\pi}{2^q}\\
        -m_{\alpha}\frac{\pi}{2^{q - 1}} & \mathrm{otherwise}
    \end{array}\right.,\\
    \varphi_{\alpha, 1} & := \left\{\begin{array}{cc}
        -(m_{\alpha} + 1)\frac{\pi}{2^{q - 1}} + \frac{\pi}{2^{q + 2}} & \mathrm{if}\,\, \theta_{\alpha} = \frac{\pi}{2^q}\\
        -m_{\alpha}\frac{\pi}{2^q} & \mathrm{otherwise}
    \end{array}\right..
\end{align}
\end{defn}
Note that in the first case above, $e^{2^qi\varphi_{\alpha, 0}} + e^{2^qi\varphi_{\alpha, 1}} = 2\cos\left(\frac{\pi}{4}\right) = \sqrt{2}$, while $e^{2^qi\varphi_{\alpha, 0}} + e^{2^qi\varphi_{\alpha, 1}} = 2\cos\left(0\right) = 2$ in the second case. In short, we are just handling the edge case where $\theta_{\alpha} = \frac{\pi}{2^q}$. To make it more concrete, we give a simple example of this ``rephasing trick" for $q = 2$:
\begin{eg}
Let $x > 0$ and consider expressing $\exp\left(x^4\right)$ using the integral identity from proposition \ref{prop:iterated_gaussian_integral_generate_higher_order_gaussian}:
\begin{align*}
    & \exp\left(x^4\right)\\
    & = \exp\left(-\left(e^{i\pi/4}x\right)^4\right)\\
    & = \int\limits_{\mathbf{R} + 2iC^{1/2}}\!\mathrm{d}y^{(0)}\hspace*{-20px}\int\limits_{\mathbf{R} + 2iC^{1/4}\left(-iy^{(0)}\right)^{1/2}}\hspace*{-20px}\mathrm{d}y^{(1)}\,\exp\left(-\frac{\left(y^{(0)}\right)^2}{4} - \frac{\left(y^{(1)}\right)^2}{4} - e^{i\pi/4}x\left(-iy^{(1)}\right)\left(-iy^{(0)}\right)^{1/2}\right).
\end{align*}
However, this representation is not unique; another can be obtained by duplicating variables:
\begin{align*}
    & \exp\left(x^4\right)\\
    & = \exp\left(-\left(e^{i\pi/4}x\right)^4\right)\\
    & = \exp\left(-\frac{1}{\sqrt{2}}\left(e^{i\pi/4 + i\pi/16}x\right)^4 - \frac{1}{\sqrt{2}}\left(e^{i\pi/4 - i\pi/16}x\right)^4\right)\\
    & = \exp\left(-\frac{1}{\sqrt{2}}\left(e^{-i\pi/4 + i\pi/16}x\right)^4 - \frac{1}{\sqrt{2}}\left(e^{i\pi/4 - i\pi/16}x\right)^4\right)\\
    & =  \int\limits_{\mathbf{R} + 2iC_0^{1/2}}\!\mathrm{d}y_0^{(0)}\hspace*{-20px}\int\limits_{\mathbf{R} + 2iC_0^{1/4}\left(-iy_0^{(0)}\right)^{1/2}}\hspace*{-20px}\mathrm{d}y_0^{(1)}\int\limits_{\mathbf{R} + 2iC_1^{1/2}}\!\mathrm{d}y_1^{(0)}\hspace*{-20px}\int\limits_{\mathbf{R} + 2iC_1^{1/4}\left(-iy_1^{(0)}\right)^{1/2}}\hspace*{-30px}\mathrm{d}y_1^{(1)}\,\exp\left(-\frac{\left(y_0^{(0)}\right)^2}{4} - \frac{\left(y_0^{(1)}\right)^2}{4} - \frac{\left(y_1^{(0)}\right)^2}{4} - \frac{\left(y_1^{(1)}\right)^2}{4}\right.\\
    & \hspace*{100px} - 2^{-1/8}e^{-3i\pi/16}x\left(-iy_0^{(1)}\right)\left(-iy_0^{(0)}\right)^{1/2} - 2^{-1/8}e^{3i\pi/16}x\left(-iy_1^{(1)}\right)\left(-iy_1^{(0)}\right)^{1/2}\Bigg).
\end{align*}
At the cost of duplicating integration variables, the phases before $x$ in the exponential are now $\pm \frac{3\pi}{16}$, hence smaller than $\frac{\pi}{4}$.
\end{eg}
Now, we observe that the definition of $\varphi_{\alpha, i}$ just stated is formally equivalent to considering another generalized multinomial sum of the form \ref{eq:generalized_binomial_sum} with new $\mathcal{A}$ index set $\mathcal{A}' := \mathcal{A} \times \{0, 1\}$, new $\bm{A}$ matrix $A'_{(\alpha, i), s} := A_{\alpha, s}\frac{e^{i\varphi_{\alpha, i}}}{e^{2^qi\varphi_{\alpha, 0}} + e^{2^qi\varphi_{\alpha, 1}}}$ and unchanged $\bm{c}$ vector $c'_{(\alpha, i)} := c_{\alpha}$ and $\mathcal{S}$ set $\mathcal{S}' := \mathcal{S}$, while taking the same conventions for complex roots. Using this equivalence, one can then make the following additional assumption on parameters $\bm{A}, \bm{b}, \bm{c}$:
\begin{assp}
\label{assp:restriction_argument}
For all $\alpha \in \mathcal{A}$, let $\theta_{\alpha}$ the unique representative in $\left(-\frac{\pi}{2^q}, \frac{\pi}{2^q}\right]$ of
\begin{align}
    \arg\left\{\left(-c_{\alpha}\right)^{1/2^q}\sum_{s \in \mathcal{S}}b_sA_{\alpha s}\right\} \quad \left(\mathrm{mod}\,\,\frac{\pi}{2^{q - 1}}\right).
\end{align}
We may assume $\left|\theta_{\alpha}\right| < \frac{\pi}{2^q}$.
\end{assp}
This assumption allows for some simplification. Namely, the integral representation of $S(\bm{A}, \bm{b}, \bm{c}\,;\,n)$ in equation \ref{eq:generalized_binomial_sum_integral_representation} becomes:
\begin{align}
    & S(\bm{A}, \bm{b}, \bm{c}\,;\,n)\nonumber\\
    & = n^{q|\mathcal{A}|/2}\hspace*{-30px}\int\limits_{\forall \alpha \in \mathcal{A},\,\left(y^{(j)}\right)_{j \in [q]} \in \mathcal{D}_{q, C_{\alpha}}}\left(\prod_{\substack{\alpha \in \mathcal{A}\\j \in [q]}}\frac{\mathrm{d}y_{\alpha}^{(j)}}{\sqrt{4\pi}}e^{-\frac{\left(y_{\alpha}^{(j)}\right)^2}{4}n}\right)\left\{\sum_{s \in \mathcal{S}}b_s\exp\left(\sum_{\alpha \in \mathcal{A}}(-c_{\alpha})^{1/2^q}A_{\alpha s}\zeta\left(\left(y^{(j)}_{\alpha}\right)_{j \in [q]}\right)\right)\right\}^n\label{eq:generalized_binomial_sum_integral_representation_simplified}.
\end{align}
Up to irrelevant $\textrm{poly}(n)$ prefactors, the integrand can now be written:
\begin{align}
\label{eq:generalized_binomial_sum_simplified_integral_representation_f}
    \exp\left\{n\Phi\left(\left(y_{\alpha}^{(j)}\right)_{\substack{\alpha \in \mathcal{A}\\j \in [q]}}\right)\right\} & := \exp\left\{n\left(-\frac{1}{4}\sum_{\substack{\alpha \in \mathcal{A}\\j \in [q]}}\left(y_{\alpha}^{(j)}\right)^2 + F\left(\left(\zeta\left(\left(y_{\alpha}^{(j)}\right)_{j \in [q]}\right)\right)_{\alpha \in \mathcal{A}}\right)\right)\right\},
\end{align}
where now
\begin{align}
\label{eq:generalized_binomial_sum_simplified_parent_function}
    F\left(\left(z_{\alpha}\right)_{\alpha \in \mathcal{A}}\right) & := \log\sum_{s \in \mathcal{S}}b_s\exp\left(\sum_{\alpha \in \mathcal{A}}A_{\alpha s}\left(-c_{\alpha}\right)^{1/2^q}z_{\alpha}\right),
\end{align}
coinciding with equations \ref{eq:generalized_binomial_sum_integral_representation_f} and \ref{eq:generalized_binomial_sum_parent_function} with $i$ index dropped. Having slightly simplified the expression of $\Phi\left(\cdot\right)$ thanks to the ``rephasing trick" described in definition \ref{def:generalized_binomial_sum_fix_c}, we can now more conveniently state a very crude bound on it that will allow to control the iterated Gaussian integral far from the saddle point. This bound is analogous to the one stated for the Hamming weight squared toy model in lemma \ref{lemma:hamming_weight_squared_bounds_phi}: the gist of the result is that $\exp\left(nF\left(\left(z_{\alpha}\right)_{\alpha \in \mathcal{A}}\right)\right)$ grows no faster than exponentially in the $\left(z_{\alpha}\right)_{\alpha \in \mathcal{A}}$. 
\begin{lem}[Crude exponential bound on $\exp\left(nF\left(\cdot\right)\right)$]
\label{lemma:crude_exp_bound}
The following bound holds for $\exp\left(nF\left(\left(z_{\alpha}\right)_{\alpha \in \mathcal{A}}\right)\right)$:
\begin{align}
    \left|\exp\left(nF\left(\left(z_{\alpha}\right)_{\alpha \in \mathcal{A}}\right)\right)\right| & \leq \left(\sum_{s \in \mathcal{S}}|b_s|\right)^n\exp\left(n\sum_{\alpha \in \mathcal{A}}\left(\max_{s \in \mathcal{S}}|A_{\alpha s}|\right)|c_{\alpha}|^{1/2^q}|z_{\alpha}|\right).
\end{align}
\begin{proof}
\begin{align*}
    \left|\exp\left(nF\left(\left(z_{\alpha}\right)_{\alpha \in \mathcal{A}}\right)\right)\right| & = \left|\left(\sum_{s \in \mathcal{S}}b_s\exp\left(\sum_{\alpha \in \mathcal{A}}A_{\alpha s}\left(-c_{\alpha}\right)^{1/2^q}z_{\alpha}\right)\right)^n\right|\\
    & \leq \left(\sum_{s \in \mathcal{S}}|b_s|\exp\left(\sum_{\alpha \in \mathcal{A}}|A_{\alpha s}||c_{\alpha}|^{1/2^q}|z_{\alpha}|\right)\right)^n\\
    & \leq \left(\sum_{s \in \mathcal{S}}|b_s|\right)^n\exp\left(n\sum_{\alpha \in \mathcal{A}}\left(\max_{s \in \mathcal{S}}|A_{\alpha s}|\right)|c_{\alpha}|^{1/2^q}|z_{\alpha}|\right).
\end{align*}
\end{proof}
\end{lem}
Equipped with the final integral representation \ref{eq:generalized_binomial_sum_integral_representation_simplified} of $S\left(\bm{A}, \bm{b}, \bm{c}\,;\,n\right)$ and the crude bound just stated, we are now ready to establish the exponential scaling of this generalized multinomial sum for sufficiently small $c$.
\begin{prop}[Exponential scaling of generalized multinomial sum]
\label{prop:exp_scaling_generalized_binomial_sum}
Let $p \geq 1$, $q \geq 2$ an integer, and $S, \bm{A}, \bm{b}, \bm{c}$ be as in definition \ref{defn:generalized_binomial_sum}. Besides, suppose the components of $\bm{c}$ are bounded by some constant $c > 0$ (assumption \ref{assp:c_bounded}) and $\bm{A}, \bm{b}, \bm{c}$ satisfy the argument restriction assumption \ref{assp:restriction_argument}. [As showed earlier, the latter statement can always been satisfied for small enough $c$ by redefining $\bm{A}, \bm{b}, \bm{c}$ without changing the value of $S(\bm{A}, \bm{b}, \bm{c}\,;\,n)$.] Recall the definition of $F\left(\cdot\right)$ in terms of $\bm{A}, \bm{b}, \bm{c}$ given in equation \ref{eq:generalized_binomial_sum_simplified_parent_function}. Then, for sufficiently small $c$, $S_q\left(\bm{A}, \bm{b}, \bm{c}\,;\,n\right)$ satisfies the following exponential scaling as $n \to \infty$:
\begin{align}
    \lim_{n \to \infty}\frac{1}{n}\log S_q\left(\bm{A}, \bm{b}, \bm{c}\,;\,n\right) & = F\left(z^*\right) - \left(1 - 2^{-q}\right)\sum_{\alpha \in \mathcal{A}}z^*_{\alpha}\partial_{\alpha}F\left(z^*\right)\\
    & = F\left(z^*\right) + \left(2^q - 1\right)\sum_{\alpha \in \mathcal{A}}\left\{\partial_{\alpha}F\left(z^*\right)\right\}^{2^q}.
\end{align}
where $\left(z_{\alpha}^*\right)_{\alpha \in \mathcal{A}}$ is the unique fixed point of:
\begin{align}
\begin{array}{rcl}
    \mathbf{C}^{\mathcal{A}} & \longrightarrow & \mathbf{C}^{\mathcal{A}}\\
     \left(z_{\alpha}\right)_{\alpha \in \mathcal{A}} & \longmapsto & \left[-2^q\left(\partial_{\alpha}F\left(\left(z_{\alpha'}\right)_{\alpha' \in \mathcal{A}}\right)\right)^{2^q - 1}\right]_{\alpha \in \mathcal{A}},
\end{array}
\end{align}
whose existence and uniqueness is guaranteed for sufficiently small $c$.
\begin{proof}
First, we observe, using assumptions \ref{assp:c_bounded} and \ref{assp:restriction_argument}, that $F$ defined in equation \ref{eq:generalized_binomial_sum_simplified_parent_function} satisfies the assumption of proposition \ref{prop:critical_point_iterative_gaussian} by substituting $c \longrightarrow \mathcal{O}\left(c^{1/2^q}\right)$ and $r \longrightarrow \mathcal{O}\left(c^{-1/2^{q + 1}}\right)$ there. Therefore, the critical points of $\Phi$ defined in equation \ref{eq:generalized_binomial_sum_simplified_integral_representation_f} are fixed points of
\begin{align}
    \left(z_{\alpha}\right)_{\alpha \in \mathcal{A}} & \longmapsto -2^q\left(\partial_{\alpha}F\left(\left(z_{\alpha}\right)_{\alpha \in \mathcal{A}}\right)\right)^{2^q - 1}.
\end{align}
Now, using arguments similar to the proof of proposition \ref{prop:hamming_weight_square_critical_point} for the Hamming weight squared toy example, for small enough $c$, there exists a unique such fixed point. Denote this fixed point by $\left(z_{\alpha}^*\right)_{\alpha \in \mathcal{A}}$ and the associated critical point of $\Phi$ by $\left({y^{(j)}_{\alpha}}^*\right)_{\substack{\alpha \in \mathcal{A}\\j \in [q]}}$; according to proposition \ref{prop:critical_point_iterative_gaussian},
\begin{align*}
    \forall \alpha \in \mathcal{A}, \forall j \in [q], \quad {y^{(j)}_\alpha}^* & = 2^{1 + j/2}i\left(\partial_{\alpha}F\left(\left(z_{\alpha}^*\right)_{\alpha \in \mathcal{A}}\right)\right)^{2^{q - 1}}.
\end{align*}
Using the explicit formula \ref{eq:generalized_binomial_sum_simplified_parent_function} for $F$, it can be shown that
\begin{align*}
    \forall \alpha \in \mathcal{A}, \forall j \in [q], \quad  z^*_{\alpha} = \mathcal{O}\left(c^{1 - 1/2^q}\right) \textrm{ and } y^{(j)*}_{\alpha} = \mathcal{O}\left(\sqrt{c}\right).
\end{align*}
The estimate on the critical point $z^*$ can be obtained with a reasoning similar to the proof of proposition \ref{prop:hamming_weight_square_critical_point} for the Hamming weight squared toy example. The estimate on $y^*$ then results directly from $\partial_{\alpha}F(z) = \mathcal{O}\left(c^{1/2^q}|z|\right)$ and the relation between $y^*$ and $z^*$ stated above.

We now partition the integral representation of $S_q\left(\bm{A}, \bm{b}, \bm{c}\,;\,n\right)$:
\begin{align*}
    S_q\left(\bm{A}, \bm{b}, \bm{c}\,;\,n\right) & \eqpoly \hspace*{-30px} \int\limits_{\forall \alpha \in \mathcal{A},\,\left(y^{(j)}_{\alpha}\right)_{j \in [q]} \in \mathcal{D}_{q, C_{\alpha}}}\!\left(\prod_{\substack{\alpha \in \mathcal{A}\\j \in [q]}}\!\mathrm{d}y^{(j)}_{\alpha}\right)\,\exp\left\{n\left(-\frac{1}{4}\sum_{\substack{\alpha \in \mathcal{A}\\j \in [q]}}\left(y^{(j)}_{\alpha}\right)^2 + F\left(\left(\zeta\left(\left(y^{(j)}_{\alpha}\right)_{j \in [q]}\right)\right)_{\alpha \in \mathcal{A}}\right)\right)\right\}
\end{align*}
into a ``small" and ``large" variable region:
\begin{align*}
	& \int\limits_{\forall \alpha \in \mathcal{A},\,\left(y^{(j)}_{\alpha}\right)_{j \in [q]} \in \mathcal{D}_{q, C_{\alpha}}}\left(\prod_{\substack{\alpha \in A\\j \in [q]}}\mathrm{d}y^{(j)}_{\alpha}\right) \ldots\\
	& = \int\limits_{\substack{\forall \alpha \in \mathcal{A},\,\left(y^{(j)}_{\alpha}\right)_{j \in [q]} \in \mathcal{D}_{q, C_{\alpha}}\\\forall \alpha \in \mathcal{A},\,\left|\zeta\left(\left(y^{(j)}_{\alpha}\right)_{j \in [q]}\right)\right| \leq \mathcal{O}\left(c^{-1/2^{q + 1}}\right)}}\prod_{\substack{\alpha \in \mathcal{A}\\j \in [q]}}\mathrm{d}y_{\alpha}^{(j)}\,\ldots + \int\limits_{\substack{\forall \alpha \in \mathcal{A},\,\left(y^{(j)}_{\alpha}\right)_{j \in [q]} \in \mathcal{D}_{q, C_{\alpha}}\\\exists \alpha \in \mathcal{A},\,\left|\zeta\left(\left(y^{(j)}_{\alpha}\right)_{j \in [q]}\right)\right| > \mathcal{O}\left(c^{-1/2^{q + 1}}\right)}}\prod_{\substack{\alpha \in \mathcal{A}\\j \in [q]}}\mathrm{d}y_{\alpha}^{(j)}\,\ldots\\
	& =: I_0 + I_1.
\end{align*}
For now, we will use that:
\begin{align*}
	C_{\alpha} & = \mathcal{O}\left(c\right),
\end{align*}
which we will observe further down in the proof. We will show that at least for sufficiently small $c$, $I_0$ gives the dominant contribution (which will be estimated by a Gaussian approximation), while $I_1$ is negligible. Let us start by controlling integral $I_1$, which is the easiest. This analysis will use $C_{\alpha} = \mathcal{O}\left(c\right)$, which will result from a precise choice for $C_{\alpha}$ further down in the proof. Applying a union bound on $\alpha \in \mathcal{A}$, $I_1$ can be upper-bounded by:
\begin{align*}
	& \sum_{\alpha \in \mathcal{A}}\int\limits_{\substack{\forall \alpha' \in \mathcal{A},\,\left(y^{(j)}_{\alpha'}\right)_{j \in [q]} \in \mathcal{D}_{q, C_{\alpha'}}\\\left|\zeta\left(\left(y^{(j)}_{\alpha}\right)_{j \in [q]}\right)\right| > \mathcal{O}\left(c^{-1/2^{q + 1}}\right)}}\left(\prod_{\substack{\alpha' \in \mathcal{A}\\j \in [q]}}\mathrm{d}y^{(j)}_{\alpha'}\right)\,\exp\left\{n\left(-\frac{1}{4}\sum_{\substack{\alpha' \in \mathcal{A}\\j \in [q]}}\left(y^{(j)}_{\alpha'}\right)^2 + F\left(\left(\zeta\left(\left(y^{(j)}_{\alpha'}\right)_{j \in [q]}\right)\right)_{\alpha' \in \mathcal{A}}\right)\right)\right\}.
\end{align*}
We now crudely bound the absolute value of the integrand using lemma \ref{lemma:crude_exp_bound}. This result asserts the existence of positive constants $K$ and $\left(K'_{\alpha}\right)_{\alpha \in \mathcal{A}}$ such that
\begin{align*}
	\left|\exp\left(nF\left(\left(\zeta\left(\left(y^{(j)}_{\alpha}\right)_{j \in [q]}\right)\right)_{\alpha \in \mathcal{A}}\right)\right)\right| & \leq \exp\left(nK + \sum_{\alpha \in \mathcal{A}}K'_{\alpha}\left|\zeta\left(\left(y^{(j)}_{\alpha}\right)_{j \in [q]}\right)\right|\right)\\
	& = \exp\left(nK\right)\prod_{\alpha \in \mathcal{A}}\exp\left(K'_{\alpha}c^{1/2^{q}}\left|\zeta\left(\left(y^{(j)}_{\alpha}\right)_{j \in [q]}\right)\right|\right).
\end{align*}
Therefore, using this bound, we can factorize the integral over the $\alpha$ index, giving:
\begin{align*}
	I_1 & \leq \exp\left(nK\right)\sum_{\alpha \in \mathcal{A}}\int\limits_{\substack{\mathcal{D}_{q, C_{\alpha}}\\\left|\zeta\left(\left(y^{(j)}_{\alpha}\right)_{j \in [q]}\right)\right| > \mathcal{O}\left(c^{-1/2^{q + 1}}\right)}}\hspace*{-40px}\prod_{j \in [q]}\mathrm{d}y^{(j)}_{\alpha}\,\exp\left(-\frac{n}{4}\sum_{j \in [q]}\left(y^{(j)}_{\alpha}\right)^2 + nK'_{\alpha}c^{1/2^{q}}\left|\zeta\left(\left(y^{(j)}_{\alpha}\right)_{j \in [q]}\right)\right|\right)\\
	& \hspace*{110px} \times \prod_{\alpha' \in \mathcal{A} - \{\alpha\}}\int\limits_{\mathcal{D}_{q, C_{\alpha'}}}\!\prod_{j \in [q]}\mathrm{d}y^{(j)}_{\alpha'}\exp\left(-\frac{n}{4}\sum_{j \in [q]}\left(y^{(j)}_{\alpha'}\right)^2 + nK'_{\alpha'}c^{1/2^{q}}\left|\zeta\left(\left(y^{(j)}_{\alpha'}\right)_{j \in [q]}\right)\right|\right)
\end{align*}
For each $\alpha \in \mathcal{A}$ in the outer sum, we bound the integral over $\left(y^{(j)}_{\alpha}\right)_{j \in [q]}$ by:
\begin{align*}
	\exp\left(-\Omega\left(\frac{n}{c^{1/2^{q + 1}}}\right)\right)
\end{align*}
using proposition \ref{prop:iterated_gaussian_tail} and each integral over $\alpha' \neq \alpha$ by
\begin{align*}
	\mathcal{O}\left(1\right)\exp\left(\mathcal{O}(nc)\right)
\end{align*}
using proposition \ref{prop:iterated_gaussian_integral_eliminate_variables}. Note we regard $q$ as a constant here, hence absorb it in the $\mathcal{O}$. Similarly, constants $K$ and $K'_{\alpha}$ introduced by lemma \ref{lemma:crude_exp_bound} only depend on $\bm{A}$ and $\bm{b}$ which are regarded as constants (while $\bm{c}$ is required to be small enough), so $K$ and $K'_{\alpha}$ can be factored in $\mathcal{O}$ as well. Therefore, finally,
\begin{align*}
	I_1 & \leq \exp\left(-\Omega\left(\frac{n}{c^{1/2^{q + 1}}}\right)\right)
\end{align*}
for sufficiently small $c$. This completes the analysis of $I_1$.

Let us now turn to estimating $I_0$. For that purpose, we now set:
\begin{align*}
	C_{\alpha} & := \left\{\partial_{\alpha}F\left(z^*\right)\right\}^{2^q},
\end{align*}
which by assumption \ref{assp:c_bounded} is $\mathcal{O}(c)$ indeed. From then on, it will be convenient to regard
\begin{align*}
    I_0 & = \hspace*{-40px}\int\limits_{\substack{\forall \alpha \in \mathcal{A},\,\left(y_{\alpha}^{(j)}\right)_{j \in [q]} \in \mathcal{D}_{q, C_{\alpha}}\\\forall \alpha \in \mathcal{A},\,\left|\zeta\left(\left(y^{(j)}_{\alpha}\right)_{j \in [q]}\right)\right| \leq \mathcal{O}\left(c^{-1/2^{q + 1}}\right)}}\hspace*{-20px}\prod_{\substack{\alpha \in \mathcal{A}\\j \in [q]}}\mathrm{d}y^{(j)}_{\alpha}\,\exp\left\{n\left(-\frac{1}{4}\sum_{\substack{\alpha \in \mathcal{A}\\j \in [q]}}\left(y^{(j)}_{\alpha}\right)^2 + F\left(\left(\zeta\left(\left(y^{(j)}_{\alpha}\right)_{j \in [q]}\right)\right)_{\alpha \in \mathcal{A}}\right)\right)\right\}
\end{align*}
as an integral over real variables $\left(\widetilde{y}^{(j)}_{\alpha}\right)$, according to the parametrization defined in definitions \ref{def:y_function} and \ref{def:zeta_tilde}. Recalling these, $I_0$ can be rewritten:
\begin{align*}
	I_0 & = \int\limits_{\substack{\mathbf{R}^{q|\mathcal{A}|}\\\forall \alpha \in \mathcal{A},\,\left|\widetilde\zeta_{\alpha}\left(\left(\widetilde{y}^{(j)}_{\alpha}\right)_{j \in [q]}\right)\right| < \mathcal{O}\left(c^{-1/2^{q + 1}}\right)}}\left(\prod_{\substack{\alpha \in \mathcal{A}\\j \in [q]}}\mathrm{d}\widetilde{y}^{(j)}_{\alpha}\right)\,\exp\left(n\widetilde{\Phi}\left(\left(\widetilde{y}^{(j)}_{\alpha}\right)_{\substack{\alpha \in \mathcal{A}\\j \in [q]}}\right)\right),
\end{align*}
where
\begin{align*}
	\widetilde{\Phi}\left(\left(\widetilde{y}^{(j)}_{\alpha}\right)_{\substack{\alpha \in \mathcal{A}\\j \in [q]}}\right) & = -\frac{1}{4}\sum_{\substack{\alpha \in \mathcal{A}\\j \in [q]]}}\mathcal{Y}^{(j)}_{\alpha}\left(\left(\widetilde{y}^{(j')}_{\alpha}\right)_{0 \leq j' \leq j}\right)^2 + F\left(\left(\widetilde{\zeta}_{\alpha}\left(\left(\widetilde{y}^{(j)}_{\alpha}\right)_{j \in [q]}\right)\right)_{\alpha \in \mathcal{A}}\right).
\end{align*}
In the equation above, we used the notational shortcut $\mathcal{Y}^{(j)}_{\alpha}$ for $\mathcal{Y}^{(j)}_{C_{\alpha}}$. Note that the bound $\left|\zeta\left(\left(y^{(j)}_{\alpha}\right)_{j \in [q]}\right)\right| < \mathcal{O}\left(c^{-1/2^{q + 1}}\right)$ ensure that $\widetilde{\Phi}$ is analytic in the integration region for sufficiently small $c$. In particular, there are no issues with the $\log$ entering the definition of $F$ in equation \ref{eq:generalized_binomial_sum_parent_function}, whose argument remains restricted to $1 + \mathcal{O}\left(c^{1/2^{q + 1}}\right) = 1 + o(1)$. Applying corollary \ref{cor:critical_point_zero_iterative_gaussian} of proposition \ref{prop:critical_point_iterative_gaussian} to $\widetilde{\Phi}$ just defined, one obtains $0$ is a critical point of $\widetilde{\Phi}$. We then perform a second-order (exact) Taylor expansion of $\widetilde{\Phi}$ around $0$ to allow for a Gaussian approximation of $I_0$:
\begin{align*}
	\widetilde{\Phi}\left(\left(\widetilde{y}_{\alpha}^{(j)}\right)_{\substack{\alpha \in \mathcal{A}\\j \in [q]}}\right) & = \widetilde{\Phi}(0) + \sum_{\substack{\alpha_0, \alpha_1 \in \mathcal{A}\\j_0, j_1 \in [q]}}\widetilde{y}^{(j_0)}_{\alpha_0}\widetilde{y}^{(j_1)}_{\alpha_1}\int_0^1\!\mathrm{d}v\,(1 - v)\frac{\partial\widetilde{\Phi}}{\partial\widetilde{y}^{(j_0)}_{\alpha_0}\partial\widetilde{y}^{j_1}_{\alpha_1}}\left(\left(v\widetilde{y}^{(j)}_{\alpha}\right)_{\substack{\alpha \in \mathcal{A}\\j \in [q]}}\right).
\end{align*}
Plugging this expression of $\widetilde\Phi$ into $I_0$ and changing variables $\widetilde{y}^{(j)}_{\alpha} \longrightarrow \frac{\widetilde{y}^{(j)}_{\alpha}}{\sqrt{n}}$ there,
\begin{align*}
	I_0 & \eqpoly \exp\left(n\widetilde\Phi(0)\right)\hspace*{-60px}\int\limits_{\substack{\mathbf{R}^{q|\mathcal{A}|}\\\forall \alpha \in \mathcal{A},\,\left|\widetilde\zeta\left(\left(\frac{\widetilde{y}_{\alpha}^{(j)}}{\sqrt{n}}\right)_{j \in [q]}\right)\right| < \mathcal{O}\left(c^{-1/2^{q + 1}}\right)}}\hspace*{-60px}\left(\prod_{\substack{\alpha \in \mathcal{A}\\j \in [q]}}\mathrm{d}\widetilde{y}^{(j)}_{\alpha}\right)\,\exp\left(\sum_{\substack{\alpha_0, \alpha_1 \in \mathcal{A}\\j_0, j_1 \in [q]}}\widetilde{y}^{(j_0)}_{\alpha_0}\widetilde{y}^{(j_1)}_{\alpha_1}\int_0^1\!\mathrm{d}v\,(1 - v)\frac{\partial^2\widetilde\Phi}{\partial\widetilde{y}^{(j_0)}_{\alpha_0}\partial\widetilde{y}^{(j_1)}_{\alpha_1}}\left(\left(\frac{v}{\sqrt{n}}\widetilde{y}^{(j)}_{\alpha}\right)_{\substack{\alpha \in \mathcal{A}\\j \in [q]}}\right)\right).
\end{align*}
We want to evaluate the limit of this quantity as $n \to \infty$. We will apply dominated convergence for that purpose. First, for each $\left(\widetilde{y}^{(j)}_{\alpha}\right)_{\substack{\alpha \in \mathcal{A}\\j \in [q}} \in \mathbf{R}^{q|\mathcal{A}|}$,
\begin{align*}
    & \exp\left(\sum_{\substack{\alpha_0, \alpha_1 \in \mathcal{A}\\j_0, j_1 \in [q]}}\widetilde{y}^{(j_0)}_{\alpha_0}\widetilde{y}^{(j_1)}_{\alpha_1}\int_0^1\!\mathrm{d}v\,(1 - v)\frac{\partial^2\widetilde\Phi}{\partial\widetilde{y}^{(j_0)}_{\alpha_0}\partial\widetilde{y}^{(j_1)}_{\alpha_1}}\left(\left(\frac{v}{\sqrt{n}}\widetilde{y}^{(j)}_{\alpha}\right)_{\substack{\alpha \in \mathcal{A}\\j \in [q]}}\right)\right)\\
    & \xrightarrow[]{n \to \infty} \exp\left(\sum_{\substack{\alpha_0, \alpha_1 \in \mathcal{A}\\j_0, j_1 \in [q]}}\widetilde{y}^{(j_0)}_{\alpha_0}\widetilde{y}^{(j_1)}_{\alpha_1}\int_0^1\!\mathrm{d}v\,(1 - v)\frac{\partial^2\widetilde\Phi}{\partial\widetilde{y}^{(j_0)}_{\alpha_0}\partial\widetilde{y}^{(j_1)}_{\alpha_1}}\left(0\right)\right) = \exp\left(\frac{1}{2}\sum_{\substack{\alpha_0, \alpha_1 \in \mathcal{A}\\j_0, j_1 \in [q]}}\widetilde{y}^{(j_0)}_{\alpha_0}\widetilde{y}^{(j_1)}_{\alpha_1}\frac{\partial^2\widetilde\Phi}{\partial\widetilde{y}^{(j_0)}_{\alpha_0}\partial\widetilde{y}^{(j_1)}_{\alpha_1}}\left(0\right)\right).
\end{align*}
According to the explicit formula in proposition \ref{prop:quadratic_form_close_0}, the quadratic form in the exponential is definite negative (for sufficiently small $c$), so the function of $\left(\widetilde{y}^{(j)}_{\alpha}\right)_{\substack{\alpha \in \mathcal{A}\\j \in [q]}}$ above is integrable. We now verify domination. For that purpose, defining a constant $\varepsilon$ that will specified later, we distinguish the following two cases to bound $\int_0^1\!\mathrm{d}v\,(1 - v)\sum\limits_{\substack{\alpha_0, \alpha_1 \in \mathcal{A}\\j_0, j_1 \in [q]}}\widetilde{y}^{(j_0)}_{\alpha_0}\widetilde{y}^{(j_1)}_{\alpha_1}\frac{\partial^2\widetilde\Phi}{\partial\widetilde{y}^{(j_0)}_{\alpha_0}\partial\widetilde{y}^{(j_1)}_{\alpha_1}}\left(\left(\frac{v}{\sqrt{n}}\widetilde{y}^{(j)}_{\alpha}\right)_{\substack{\alpha \in \mathcal{A}\\j \in [q]}}\right)$ for all $\left(\widetilde{y}^{(j)}_{\alpha}\right)_{\substack{\alpha \in \mathcal{A}\\j \in [q]}}$ and $v \in [0, 1]$:
\begin{itemize}
    \item $\forall \alpha \in \mathcal{A},\,\forall j \in [q], \quad \left|\frac{\widetilde{y}^{(j}_{\alpha}}{\sqrt{n}}\right| \leq \varepsilon c^{1/2}$\\\\The condition remains true when rescaling $\widetilde{y}^{(j)}_{\alpha}$ by $v \in [0, 1]$, i.e.
    \begin{align*}
        \forall \alpha \in \mathcal{A},\,\forall j \in [q],\,\forall v \in [0, 1],\,\quad \left|\frac{v}{\sqrt{n}}\widetilde{y}^{(j}_{\alpha}\right| \leq \varepsilon c^{1/2}
    \end{align*}
    We will therefore control $\sum\limits_{\substack{\alpha_0, \alpha_1 \in \mathcal{A}\\j_0, j_1 \in [q]}}\widetilde{y}^{(j_0)}_{\alpha_0}\widetilde{y}^{(j_1)}_{\alpha_1}\frac{\partial^2\widetilde\Phi}{\partial\widetilde{y}^{(j_0)}_{\alpha_0}\partial\widetilde{y}^{(j_1)}_{\alpha_1}}\left(\left(\frac{v}{\sqrt{n}}\widetilde{y}^{(j)}_{\alpha}\right)_{\substack{\alpha \in \mathcal{A}\\j \in [q]}}\right)$ for all $v \in [0, 1]$. By applying lemma \ref{lemma:y_tilde_close_0_function_close_0_evaluated} to the expression of $\frac{\partial^2\widetilde\Phi}{\partial\widetilde{y}^{(j_0)}_{\alpha_0}\partial\widetilde{y}^{(j_1)}_{\alpha_1}}$ in lemma \ref{lemma:quadratic_form_general}, we find that for all $\alpha_0, \alpha_1 \in \mathcal{A}$ and $j_0, j_1 \in [q]$,
    \begin{align*}
        \left|\frac{\partial\widetilde\Phi^2}{\partial\widetilde{y}^{(j_0)}_{\alpha_0}\partial\widetilde{y}^{(j_1)}_{\alpha_1}}\left(\left(\widetilde{y}^{(j)}_{\alpha}\right)_{\substack{\alpha \in \mathcal{A}\\j \in [q]}}\right) - \frac{\partial^2\widetilde\Phi}{\partial\widetilde{y}^{(j_0)}_{\alpha_0}\partial\widetilde{y}^{(j_1)}_{\alpha_1}}(0)\right| & \leq \mathcal{O}(1)c\varepsilon.
     \end{align*}
     This bound holds uniformly in the range $\forall \alpha \in \mathcal{A},\,\forall j \in [q],\,\left|\frac{v}{\sqrt{n}}\widetilde{y}^{(j}_{\alpha}\right| \leq \varepsilon c^{1/2}$ as long as $\varepsilon$ is sufficiently small (where the upper bound on $\varepsilon$ is independent of $c$). Now, with $\varepsilon$ fixed and $c$ sufficiently small, propositions \ref{prop:quadratic_form_0} and \ref{prop:quadratic_form_close_0} (see also remark \ref{rem:applying_quadratic_form_close_0}) give
     \begin{align*}
         \Re\left\{\sum_{\substack{\alpha_0, \alpha_1 \in \mathcal{A}\\j_0, j_1 \in [q]}}\widetilde{y}^{(j_0)}_{\alpha_0}\widetilde{y}^{(j_1)}_{\alpha_1}\frac{\partial^2\widetilde\Phi}{\partial\widetilde{y}^{(j_0)}\partial\widetilde{y}^{(j_1)}_{\alpha_1}}(0)\right\} & \leq -\frac{1}{8}\sum_{\substack{\alpha \in \mathcal{A}\\j \in [q]}}\left(\widetilde{y}^{(j)}_{\alpha}\right)^2.
     \end{align*}
     Hence, for sufficiently small $c$ and $\epsilon$,
     \begin{align*}
         \Re\left\{\sum_{\substack{\alpha_0, \alpha_1 \in \mathcal{A}\\j_0, j_1 \in [q]}}\widetilde{y}^{(j_0)}_{\alpha_0}\widetilde{y}^{(j_1)}_{\alpha_1}\frac{\partial^2\widetilde\Phi}{\partial\widetilde{y}^{(j_0)}\partial\widetilde{y}^{(j_1)}_{\alpha_1}}\left(\left(\frac{v}{\sqrt{n}}\widetilde{y}^{(j)}_{\alpha}\right)_{\substack{\alpha \in \mathcal{A}\\j \in [q]}}\right)\right\} & \leq -\frac{1}{16}\sum_{\substack{\alpha \in \mathcal{A}\\j \in [q]}}\left(\widetilde{y}^{(j)}_{\alpha}\right)^2.
     \end{align*}
     \item $\exists \alpha \in \mathcal{A},\,\exists j \in [q], \quad \left|\frac{\widetilde{y}^{(j)}_{\alpha}}{\sqrt{n}}\right| > \varepsilon c^{1/2}$.\\\\In this case, we use the direct upper bound on $\Re\left\{\widetilde\Phi\left(\cdot\right) - \widetilde\Phi(0)\right\}$ given in proposition \ref{prop:quadratic_form_far_0} rather than the Taylor expansion of the same quantity. For sufficiently small $c$, the negative quadratic form on the right-hand side of inequality \ref{eq:quadratic_form_far_0} there dominates the $\mathcal{O}\left(c^{1 + 1/2^q}\right)$ error, and one can also write:
     \begin{align*}
         & \Re\left\{\int_0^1\!\mathrm{d}v\,(1 - v)\sum_{\substack{\alpha_0, \alpha_1 \in \mathcal{A}\\j_0, j_1 \in [q]}}\widetilde{y}^{(j_0)}_{\alpha_0}\widetilde{y}^{(j_1)}_{\alpha_1}\frac{\partial^2\widetilde\Phi}{\partial\widetilde{y}^{(j_0)}\partial\widetilde{y}^{(j_1)}_{\alpha_1}}\left(\left(\frac{v}{\sqrt{n}}\widetilde{y}^{(j)}_{\alpha}\right)_{\substack{\alpha \in \mathcal{A}\\j \in [q]}}\right)\right\}\\
         & = n\Re\left\{\widetilde\Phi\left(\left(\frac{\widetilde{y}^{(j)}_{\alpha}}{\sqrt{n}}\right)_{\substack{\alpha \in \mathcal{A}\\j \in [q]}}\right) - \widetilde\Phi(0)\right\}\\
         & \leq n\left(-\frac{1}{16}\sum_{\substack{\alpha \in \mathcal{A}\\j \in [q]}}\left(\frac{\widetilde{y}^{(j)}_{\alpha}}{\sqrt{n}}\right)^2\right)\\
         & = -\frac{1}{16}\sum_{\substack{\alpha \in \mathcal{A}\\j \in [q]}}\left(\widetilde{y}^{(j)}_{\alpha}\right)^2.
     \end{align*}
\end{itemize}
This proves (for sufficiently small $c$) domination of the integrand in $I_0$. Therefore, by the dominated convergence theorem,
\begin{align*}
    & \int\limits_{\substack{\mathbf{R}^{q|\mathcal{A}|}\\\forall \alpha \in \mathcal{A},\,\left|\widetilde\zeta\left(\left(\frac{\widetilde{y}_{\alpha}^{(j)}}{\sqrt{n}}\right)_{j \in [q]}\right)\right| < \mathcal{O}\left(c^{-1/2^{q + 1}}\right)}}\hspace*{-60px}\left(\prod_{\substack{\alpha \in \mathcal{A}\\j \in [q]}}\mathrm{d}\widetilde{y}^{(j)}_{\alpha}\right)\,\exp\left(\sum_{\substack{\alpha_0, \alpha_1 \in \mathcal{A}\\j_0, j_1 \in [q]}}\widetilde{y}^{(j_0)}_{\alpha_0}\widetilde{y}^{(j_1)}_{\alpha_1}\int_0^1\!\mathrm{d}v\,(1 - v)\frac{\partial^2\widetilde\Phi}{\partial\widetilde{y}^{(j_0)}_{\alpha_0}\partial\widetilde{y}^{(j_1)}_{\alpha_1}}\left(\left(\frac{v}{\sqrt{n}}\widetilde{y}^{(j)}_{\alpha}\right)_{\substack{\alpha \in \mathcal{A}\\j \in [q]}}\right)\right)\\
    & \xrightarrow[]{n \to \infty} \int\limits_{\substack{\mathbf{R}^{q|\mathcal{A}|}}}\left(\prod_{\substack{\alpha \in \mathcal{A}\\j \in [q]}}\mathrm{d}\widetilde{y}^{(j)}_{\alpha}\right)\,\exp\left(\frac{1}{2}\sum_{\substack{\alpha_0, \alpha_1 \in \mathcal{A}\\j_0, j_1 \in [q]}}\widetilde{y}^{(j_0)}_{\alpha_0}\widetilde{y}^{(j_1)}_{\alpha_1}\frac{\partial^2\widetilde\Phi}{\partial\widetilde{y}^{(j_0)}_{\alpha_0}\partial\widetilde{y}^{(j_1)}_{\alpha_1}}(0)\right) = \mathcal{O}(1).
\end{align*}
It follows
\begin{align*}
\lim_{n \to \infty}\frac{\log I_0}{n} & = \widetilde\Phi(0).
\end{align*}
and the same limits holds for $\frac{1}{n}\log S(\bm{A}, \bm{b}, \bm{c}\,;\,n)$ since we showed $I_1$ to be negligible.
\end{proof}
\end{prop}

We now make slightly more explicit the formula given in proposition \ref{prop:exp_scaling_generalized_binomial_sum} for the scaling exponent of a generalized multinomial sum by expanding it to lowest non-trivial order in $c$:
\begin{cor}
\label{cor:exp_scaling_generalized_binomial_sum_small_c}
Let the notation and assumptions be as in proposition \ref{prop:exp_scaling_generalized_binomial_sum}. Then, for sufficiently small $c$, the scaling exponent of $S_q\left(\bm{A}, \bm{b}, \bm{c}\,;\,n\right)$ admits the following approximation:
\begin{align}
    \lim_{n \to \infty}\frac{1}{n}\log S_q\left(\bm{A}, \bm{b}, \bm{c}\,;\,n\right) & = -\sum_{\alpha \in \mathcal{A}}\left(\partial_{\alpha}F\left(0\right)\right)^{2^q} + \mathcal{O}\left(c^2\right)\\
    & = \sum_{\alpha \in \mathcal{A}}c_{\alpha}\left(\sum\limits_{s \in \mathcal{A}}b_sA_{\alpha s}\right)^{2^q} + \mathcal{O}\left(c^2\right)\label{eq:exp_scaling_generalized_binomial_sum_small_c}.
\end{align}
\begin{proof}
For sufficiently small $c$, the scaling exponent of $S_q\left(\bm{A}, \bm{b}, \bm{c}\,;\,n\right)$ is given by
\begin{align*}
    F\left(z^*\right) - (1 - 2^{-q})\sum_{\alpha \in \mathcal{A}}z^*_{\alpha}\partial_{\alpha}F\left(z^*\right)
\end{align*}
according to proposition \ref{prop:exp_scaling_generalized_binomial_sum}. We expand the latter with an error of order $\mathcal{O}(c^2)$ using the estimate $z^* = \mathcal{O}\left(c^{1 - 1/2^q}\right)$ already observed in the proof of proposition \ref{prop:exp_scaling_generalized_binomial_sum}.
\begin{align*}
    & F\left(z^*\right) - (1 - 2^{-q})\sum_{\alpha \in \mathcal{A}}z^*_{\alpha}\partial_{\alpha}F\left(z^*\right)\\
    & = F\left(0\right) + \sum_{\alpha \in \mathcal{A}}z^*_{\alpha}\partial_{\alpha}F(0) + \mathcal{O}\left(\left|z^*\right|^2\sup_{\substack{\alpha, \beta \in \mathcal{A}\\|z| \leq \mathcal{O}\left(c^{1 - 1/2^q}\right)}}\left|\partial_{\alpha\beta}F(z)\right|\right) - (1 - 2^{-q})\sum_{\alpha \in \mathcal{A}}z^*_{\alpha}\partial_{\alpha}F\left(z^*\right)\\
    & = F\left(0\right) + \sum_{\alpha \in \mathcal{A}}z^*_{\alpha}\partial_{\alpha}F(0) + \mathcal{O}\left(c^2\right) - (1 - 2^{-q})\sum_{\alpha \in \mathcal{A}}z^*_{\alpha}\partial_{\alpha}F\left(z^*\right)\\
    & = F\left(0\right) + \sum_{\alpha \in \mathcal{A}}z^*_{\alpha}\partial_{\alpha}F(0) - (1 - 2^{-q})\sum_{\alpha \in \mathcal{A}}z^*_{\alpha}\partial_{\alpha}F\left(0\right) + \mathcal{O}\left(c^2\right)\\
    & = 2^{-q}\sum_{\alpha \in \mathcal{A}}z_{\alpha}^*\partial_{\alpha}F(0) + \mathcal{O}(c^2)\\
    & = -\sum_{\alpha \in \mathcal{A}}\left(\partial_{\alpha}F(0)\right)^{2^q} + \mathcal{O}\left(c^2\right),
\end{align*}
where we used that the second derivatives of $F$ are bounded by $c^{2/2^q}$ in the region of interest (each derivation of $F$ generating a power of $c^{1/2^q}$). This completes the proof.
\end{proof}
\end{cor}

In the case of random $k$-SAT QAOA, whose expected fixed-angles success probability was cast to a generalized multinomial sum in proposition \ref{prop:qaoa_ksat_expected_success_probability}, the ``more explicit" formula from corollary \ref{cor:exp_scaling_generalized_binomial_sum_small_c} admits an even simpler expression to lowest order in $\bm\gamma$.

\begin{prop}[Exponential scaling of $k$-SAT success probability for small $\bm\gamma$ angles]
Let $q \geq 1$ an integer, $p \geq 1$ an integer, $\bm\beta, \bm\gamma \in \mathbf{R}^p$ and consider the success probability of random $k$-SAT QAOA at level $p$ for $k = 2^q$ as introduced in definition \ref{def:random_ksat_qaoa}. Assuming all angles $\bm\gamma$ to be bounded by $\gamma > 0$, the scaling exponent of the success probability of random $2^q$-SAT QAOA admits the following approximation to first order in $\gamma$:
\begin{align}
    & \lim_{n \to \infty}\frac{1}{n}\log\mathbf{E}_{\bm\sigma \sim \mathrm{CNF}(n, k, r)}\left[\left\langle\Psi_{\mathrm{QAOA}}(\bm\sigma, \bm\beta, \bm\gamma)|\mathbf{1}\left\{H[\bm\sigma] = 0\right\}|\Psi_{\mathrm{QAOA}}(\bm\sigma, \bm\beta, \bm\gamma)\right\rangle\right]\nonumber\\
    & = -\log(2) - \frac{1}{2^{2^q}}\sum_{j \in [p]}\gamma_j\sin\left(2^{q - 1}\sum_{j \leq l < p}\beta_{l}\right)\cos\left(\frac{1}{2}\sum_{j \leq l < p}\beta_{l}\right)^{2^q} + \mathcal{O}\left(\gamma^2\right).
\end{align}
\begin{proof}
This results from an application of corollary \ref{cor:exp_scaling_generalized_binomial_sum_small_c} just proven to the representation of the QAOA success probability as a generalized multinomial sum, established in proposition \ref{prop:qaoa_ksat_expected_success_probability}. The definitions of index sets $\mathcal{A}, \mathcal{S}$ as well as matrices $\bm{A}, \bm{b}, \bm{c}$ in this context are given in equation \ref{eq:generalized_binomial_sum_parameters_qaoa}, which we restate here for convenience:
\begin{align*}
    \mathcal{S} & := \{0, 1\}^{2p + 1},\\
    \mathcal{A} & := \left\{J \subset [2p + 1]\,:\,|J| \geq 2\right\},\\
    b_s & := \frac{B_{\bm\beta, s}}{2},\\
    c_{\alpha} & := r(-1)^{\mathbf{1}[p \in \alpha]}\prod_{j \in \alpha,\,j < p}\left(e^{-i\gamma_j/2} - 1\right)\prod_{j \in \alpha,\,j > p}\left(e^{i\gamma_{2p - j}/2} - 1\right),\\
    A_{\alpha s} & := \frac{1}{2}\mathbf{1}\left[\forall j, j' \in \alpha,\,s_j = s_{j'}\right].
\end{align*}
From this definition, it appears one can choose $c = \mathcal{O}\left(\gamma\right)$ (where the implicit constant depends on $p$) as an upper bound to the coordinates of $\bm{c}$. Hence, the condition ``$c$ sufficiently small" can be replaced by ``$\gamma$ sufficiently small". We now observe using the definition of $c_{\alpha}$ that the only $c_{\alpha}$ that are not of order $\gamma^2$ or smaller are those for which $\alpha = \{j, p\}$, where $j \in \{0, \ldots, p - 1\} \sqcup \{p + 1, \ldots, 2p\}$. It is sufficient to consider these $c_{\alpha}$ in the sum of equation \ref{eq:exp_scaling_generalized_binomial_sum_small_c} to obtain an error $\mathcal{O}\left(\gamma^2\right)$. Explicitly,
\begin{align*}
    \lim_{n \to \infty}\frac{1}{n}S_q\left(\bm{A}, \bm{b}, \bm{c}\,;\,n\right) & = \sum_{j \in [p]}c_{\{j, p\}}\left(\sum_{s \in \{0, 1\}^{2p + 1}}b_sA_{\{j, p\}, s}\right)^{2^q} + \sum_{j \in [p]}c_{\{p, 2p - j\}}\left(\sum_{s \in \{0, 1\}^{2p + 1}}b_sA_{\{p, 2p - j\}, s}\right)^{2^q}.
\end{align*}
It therefore remains to evaluate sums $\sum\limits_{s \in \{0, 1\}^{2p + 1}}b_sA_{\{j, p\}, s}$ and $\sum\limits_{s \in \{0, 1\}^{2p + 1}}b_sA_{\{p, 2p - j\}, s}$ for $j \in [p]$. Consider for instance the first one:
\begin{align*}
    & \sum_{s \in \{0, 1\}^{2p + 1}}b_sA_{\{j, p\}, s}\\
    & = \frac{1}{2}\sum_{\substack{s \in \{0, 1\}^{2p + 1}\\s_j = s_p}}b_s\\
    & = \frac{1}{4}\sum_{\substack{s \in \{0, 1\}^{2p + 1}\\s_j = s_p}}\begin{pmatrix}
    \cos\frac{\beta_0}{2} & i\sin\frac{\beta_0}{2}\\
    i\sin\frac{\beta_0}{2} & \cos\frac{\beta_0}{2}
    \end{pmatrix}_{s_{2p}s_{2p - 1}}\ldots\begin{pmatrix}
    \cos\frac{\beta_{p - 1}}{2} & i\sin\frac{\beta_{p - 1}}{2}\\
    i\sin\frac{\beta_{p - 1}}{2} & \cos\frac{\beta_{p - 1}}{2}
    \end{pmatrix}_{s_{p + 1}s_p}\begin{pmatrix}
    \cos\frac{\beta_{p - 1}}{2} & -i\sin\frac{\beta_{p - 1}}{2}\\
    -i\sin\frac{\beta_{p - 1}}{2} & \cos\frac{\beta_{p - 1}}{2}
    \end{pmatrix}_{s_ps_{p - 1}}\\
    & \hspace*{30px} \ldots \begin{pmatrix}
    \cos\frac{\beta_j}{2} & -i\sin\frac{\beta_j}{2}\\
    -i\sin\frac{\beta_j}{2} & \cos\frac{\beta_j}{2}
    \end{pmatrix}_{s_{j + 1}s_j}\begin{pmatrix}
    \cos\frac{\beta_{j - 1}}{2} & -i\sin\frac{\beta_{j - 1}}{2}\\
    -i\sin\frac{\beta_{j - 1}}{2} & \cos\frac{\beta_{j - 1}}{2}
    \end{pmatrix}_{s_js_{j - 1}}\ldots\begin{pmatrix}
    \cos\frac{\beta_0}{2} & -i\sin\frac{\beta_0}{2}\\
    -i\sin\frac{\beta_0}{2} & \cos\frac{\beta_0}{2}
    \end{pmatrix}_{s_1s_0}.
\end{align*}
Using this matrix expression, one can sum over $s_{2p}, s_{2p - 1}, \ldots, s_{p + 1}$, then over $s_0, s_1, \ldots, s_{j - 1}$ to obtain:
\begin{align*}
    & \frac{1}{4}\sum_{\substack{s_j, \ldots, s_p \in \{0, 1\}\\s_j = s_p}}e^{\frac{i}{2}\sum\limits_{l \in [p]}\beta_l}\begin{pmatrix}
        \cos\frac{\beta_{p - 1}}{2} & -i\sin\frac{\beta_{p - 1}}{2}\\
        -i\sin\frac{\beta_{p - 1}}{2} & \cos\frac{\beta_{p - 1}}{2}
    \end{pmatrix}_{s_ps_{p - 1}}\ldots\begin{pmatrix}
        \cos\frac{\beta_j}{2} & -i\sin\frac{\beta_j}{2}\\
        -i\sin\frac{\beta_j}{2} & \cos\frac{\beta_j}{2}
    \end{pmatrix}_{s_{j + 1}s_j}e^{-\frac{i}{2}\sum\limits_{l \in [j]}\beta_l}\\
    & = \frac{1}{4}e^{\frac{i}{2}\sum\limits_{j \leq l < p}\beta_l}\begin{pmatrix}
        1\\
        0
    \end{pmatrix}^T\begin{pmatrix}
        \cos\frac{\beta_{p - 1}}{2} & -i\sin\frac{\beta_{p - 1}}{2}\\
        -i\sin\frac{\beta_{p - 1}}{2} & \cos\frac{\beta_{p - 1}}{2}
    \end{pmatrix}\ldots\begin{pmatrix}
        \cos\frac{\beta_j}{2} & -i\sin\frac{\beta_j}{2}\\
        -i\sin\frac{\beta_j}{2} & \cos\frac{\beta_j}{2}
    \end{pmatrix}\begin{pmatrix}
        1\\
        0
    \end{pmatrix}\\
    & + \frac{1}{4}e^{\frac{i}{2}\sum\limits_{j \leq l < p}\beta_l}\begin{pmatrix}
        0\\
        1
    \end{pmatrix}^T\begin{pmatrix}
        \cos\frac{\beta_{p - 1}}{2} & -i\sin\frac{\beta_{p - 1}}{2}\\
        -i\sin\frac{\beta_{p - 1}}{2} & \cos\frac{\beta_{p - 1}}{2}
    \end{pmatrix}\ldots\begin{pmatrix}
        \cos\frac{\beta_j}{2} & -i\sin\frac{\beta_j}{2}\\
        -i\sin\frac{\beta_j}{2} & \cos\frac{\beta_j}{2}
    \end{pmatrix}\begin{pmatrix}
        0\\
        1
    \end{pmatrix}\\
    & = \frac{1}{2}e^{\frac{i}{2}\sum\limits_{j \leq l < p}\beta_l}\cos\left(\frac{1}{2}\sum\limits_{j \leq l < p}\beta_l\right).
\end{align*}
To evaluate the product of matrices, we used that $\begin{pmatrix}1\\0\end{pmatrix} = \frac{1}{\sqrt{2}}\left(\ket{+} + \ket{-}\right), \begin{pmatrix}
0\\
1
\end{pmatrix} = \frac{1}{\sqrt{2}}\left(\ket{+} - \ket{-}\right)$ and that $\ket{+}, \ket{-}$ are eigenvectors of $\begin{pmatrix}
\cos\frac{\beta}{2} & i\sin\frac{\beta}{2}\\
i\sin\frac{\beta}{2} & \cos\frac{\beta}{2}
\end{pmatrix}$ with eigenvalues $e^{\frac{i\beta}{2}}, e^{-\frac{i\beta}{2}}$. Similarly, one can evaluate
\begin{align*}
    \sum_{s \in \{0, 1\}^{2p + 1}}b_sA_{\{p, 2p - j\}, s} & = \frac{1}{2}e^{-\frac{i}{2}\sum\limits_{j \leq l < p}\beta_l}\cos\left(\frac{1}{2}\sum_{j \leq l < p}\beta_l\right).
\end{align*}
Back to the original equation \ref{eq:exp_scaling_generalized_binomial_sum_small_c},
\begin{align*}
    & \lim_{n \to \infty}\frac{1}{n}S_q\left(\bm{A}, \bm{b}, \bm{c}\,;\,n\right) \\
    & = \sum_{j \in [p]}\frac{i\gamma_j}{2}\frac{1}{2^{2^q}}e^{2^{q - 1}i\sum\limits_{j \leq l < p}\beta_l}\cos\left(\frac{1}{2}\sum_{j \leq l < p}\beta_l\right)^{2^q} - \sum_{j \in [p]}\frac{i\gamma_j}{2}\frac{1}{2^{2^q}}e^{-2^{q - 1}i\sum\limits_{j \leq l < p}\beta_l}\cos\left(\frac{1}{2}\sum_{j \leq l < p}\beta_l\right)^{2^q} + \mathcal{O}(\gamma^2)\\
    & = -\frac{1}{2^{2^q}}\sum_{j \in [p]}\gamma_j\sin\left(2^{q - 1}\sum_{j \leq l < p}\beta_l\right)\cos\left(\frac{1}{2}\sum_{j \leq l < p}\beta_l\right)^{2^q} + \mathcal{O}\left(\gamma^2\right).
\end{align*}
\end{proof}
\end{prop}

\subsubsection{Algorithmic implementation}
\label{sec:algorithmic_implementation}
Proposition \ref{prop:exp_scaling_generalized_binomial_sum} establishes the exponential scaling of the generalized multinomial sum \ref{eq:generalized_binomial_sum} (for sufficiently small $\bm{c}$ parameters) by expressing it as the fixed point of a certain function. It remains to specify how this fixed point is found in practice. For sufficiently small $c$, a generalization of the argument for the toy model from section \ref{sec:hamming_weight_squared_example} shows the fixed point can be approximated to error $\mathcal{O}\left(\varepsilon\right)$ using $\mathcal{O}\left(\log\frac{1}{\varepsilon}\right)$ iterations. Each iteration applies the function to the previous approximation of the fixed point, starting with lowest-order approximation $0$. The procedure is explicited in algorithm \ref{alg:find_fixed_point_small_c}.
\begin{algorithm}[H]
	\caption{Compute an approached fixed point $z^*$ for sufficiently small $c$}
	\label{alg:find_fixed_point_small_c}
	\begin{algorithmic}[1]
		\Require $q$ parameter, matrices/vectors $\bm{A}, \bm{b}, \bm{c}$, maximum number of iterations $N_{\textrm{iter}}$, relative fixed-point error threshold $\varepsilon$.
		\Function{FixedPointSmallC}{$q$, $\bm{A}$, $\bm{b}$, $\bm{c}$, $N_{\textrm{iter}}$, $\varepsilon$}
			\State $z^* \longleftarrow 0$
			\State $i \longleftarrow 0$
			\While{$i < N_{\textrm{iter}}$}
				\State $z_{\alpha}^* \longleftarrow 2^q\left(-\partial_{\alpha}F\left(z^*\right)\right)^{2^q - 1}$
			    \If{$\max_{\alpha \in \mathcal{A}}\left|\frac{z^*_{\alpha} - \partial_{\alpha}F\left(z^*\right)}{z^*_{\alpha}}\right| < \varepsilon$}
			        \State \algorithmicbreak
			    \EndIf
				\State $i \longleftarrow i + 1$
			\EndWhile
		    \State \Return $\left(i, z^*\right)$
		\EndFunction
	\end{algorithmic}
\end{algorithm}

For large $c$, we do not rigorously know whether the fixed point introduced in theorem \ref{prop:exp_scaling_generalized_binomial_sum} exists and is unique. Besides, even if this holds, we have not formally proven that this fixed point prescribes the exponential scaling of the generalized multinomial sum. However, these limitations may be a mere artifact of our proof methods; it would therefore be desirable for the fixed-point finding algorithm to extrapolate to larger $c$. For that purpose, we supplement algorithm \ref{alg:find_fixed_point_small_c} with a heuristic algorithm \ref{alg:find_fixed_point_improved} which attempts to find the critical point for large $c$. Informally, the only change is to introduce some damping, controlled by a parameter $\rho \in [0, 1]$, as the approximation to the fixed point is updated: instead of updating $z^*$ to $F(z^*)$, we update it to a weighted combination of this proposal and the previous value. This simple amendment was empirically observed to solve the convergence problem and yield a fixed point varying smoothly with $\bm{c}$.

\begin{algorithm}[H]
	\caption{Try to compute an approached fixed point $z^*$ with initial suggestion and damping}
	\label{alg:find_fixed_point_improved}
	\begin{algorithmic}[1]
		\Require $q$ parameter, matrices/vectors $\bm{A}, \bm{b}, \bm{c}$, maximum number of iterations $N_{\textrm{iter}}$, relative fixed-point error threshold $\varepsilon$, initial suggestion $z^*_{\textrm{init}}$ for $z^*$, damping coefficient $\rho \in [0, 1)$.
		\Function{FixedPointImproved}{$q$, $\bm{A}$, $\bm{b}$, $\bm{c}$, $N_{\textrm{iter}}$, $\varepsilon$, $z^*_{\textrm{init}}$, $\rho$}
			\State $z^* \longleftarrow z^*_{\textrm{init}}$
			\State $i \longleftarrow 0$
			\While{$i < N_{\textrm{iter}}$}
				\State $z^*_{\alpha} \longleftarrow \rho z^*_{\alpha} + (1 - \rho)2^q\left(-\partial_{\alpha}F\left(z^*\right)\right)^{2^q - 1}$
			    \If{$\max_{\alpha \in \mathcal{A}}\left|\frac{z^*_{\alpha} - \partial_{\alpha}F\left(z^*\right)}{z^*_{\alpha}}\right| < \varepsilon$}
			        \State \algorithmicbreak
			    \EndIf
				\State $i \longleftarrow i + 1$
			\EndWhile
		    \State \Return $\left(i, z^*\right)$
		\EndFunction
	\end{algorithmic}
\end{algorithm}

We now specialize to random $k$-SAT QAOA an describe a more efficient implementation of the algorithm in this case. Precisely, we provide a more efficient implementation of the evaluation of $F$ and its derivatives $\partial_{\alpha}F$ by exploiting the specific structure of matrix $\bm{A}$ of the generalized multinomial sum \ref{eq:generalized_binomial_sum} for the expectation value arising from that problem. In general, recalling definition \ref{eq:generalized_binomial_sum_simplified_parent_function} of $F$ and the expression for its derivative in equation \ref{eq:generalized_binomial_sum_parent_function_derivative}, it is easily seen that naively evaluating $F(z)$ and all its derivatives requires $\mathcal{O}\left(|\mathcal{A}|.|\mathcal{S}|\right)$ multiplications. Applying that to random $2^q$-SAT using the parameters stated in example \ref{example:random_2q_sat_as_generalized_binomial_sum}, this yields a complexity $\mathcal{O}\left(16^p\right)$. We now describe a more efficient naive method reducing this to $\mathcal{O}\left(4^p\right)$. For that purpose, it will be more convenient to slightly amend the definition of the generalized multinomial sum parameters in example \ref{example:random_2q_sat_as_generalized_binomial_sum}; namely, we include all subsets of $[2p + 1]$ in the index set $\mathcal{A}$ (not only those with 2 elements or more), based on the following rewriting of equation \ref{eq:qaoa_ksat_expected_success_probability} for the success probability of $k$-SAT QAOA:
\begin{align}
\label{eq:qaoa_ksat_expected_success_probability_all_sets}
    & \mathbf{E}_{\bm\sigma \sim \mathrm{CNF}(n, k, r)}\left[\left\langle\Psi_{\mathrm{QAOA}}(\bm\sigma, \bm\beta, \bm\gamma)|\mathbf{1}\left\{H[\bm\sigma] = 0\right\}|\Psi_{\mathrm{QAOA}}(\bm\sigma, \bm\beta, \bm\gamma)\right\rangle\right]\nonumber\\
    & = \frac{1}{2^n}e^{-2^{-k}rn}\sum_{\substack{\left(n_s\right)_{s \in \{0, 1\}^{2p + 1}}\\\sum_sn_s = n}}\binom{n}{\left(n_s\right)_s}\prod_sB_{\bm\beta, s}^{n_s}\exp\left(rn\sum_{J \subset [2p + 1]}c_J\left(\frac{1}{2n}\sum_{\substack{s \in \{0, 1\}^{2p + 1}\\\forall j, j' \in J,\,s_j = s_{j'}}}n_s\right)^k\right),
\end{align}
where the definition of $c_J$ in proposition \ref{prop:qaoa_ksat_expected_success_probability} is unchanged (and irrelevant to the argument). With this amended setup, we now show that given a $2^{2p + 1}$-component vector $\left(v_j\right)_{j \in \{0, 1\}^{2p + 1}}$, both vector $\left(\sum\limits_{\alpha \subset [2p + 1]}A_{\alpha s}v_{\alpha}\right)_{s \in \{0, 1\}^{2p + 1}}$ and vector $\left(\sum\limits_{s \in \{0, 1\}^{2p + 1}}A_{\alpha s}v_{s}\right)_{\alpha \subset [2p + 1]}$ can be compute in time $\mathcal{O}\left(p4^p\right)$ instead of the naive $\mathcal{O}\left(16^p\right)$.

In the above, we naturally identify $n$-bit bitstrings and subsets of $[n]$, where $1$ bits denote the elements included in the set. We now observe that both these sums can be reduced to the sum-over-subsets problem~\cite{kostka21} that we recall below:
\begin{defn}[Sum-over-subsets]
Given a vector $\left(v_{\alpha}\right)_{\alpha \subset [n]}$ of size $2^n$, the sum-over-subset problem consists of computing $\sum_{\alpha \subset S}v_{\alpha}$ for all subsets $S$ of $[n]$. There exists an algorithm, based on dynamic programming, performing this calculation in time $\mathcal{O}\left(n2^n\right)$.
\end{defn}
A naive algorithm would be to iterate over subsets $S$ and sum the $v_{\alpha}$ such that $\alpha$ is included in $S$, demanding time $\mathcal{O}\left(4^n\right)$. We now connect the sum-over-subsets problem to the calculation of the sums mentioned above.
Let us start with the sum over $\alpha$, which can be expanded as follows:
\begin{align*}
    & \sum_{\alpha \subset [2p + 1]}A_{\alpha s}v_{\alpha}\\
    & = \frac{1}{2}\sum_{\substack{\alpha \subset [2p + 1]\\\forall j, j' \in \alpha,\,s_j = s_{j'}}}v_{\alpha}\\
    & = \frac{1}{2}\sum_{\substack{\alpha \subset \left\{j\,:\,s_j = 0\right\}}}v_{\alpha} + \frac{1}{2}\sum_{\substack{\alpha \subset \left\{j\,:\,s_j = 1\right\}}}v_{\alpha}.
\end{align*}
Interpeting $s$ as describing a subset of $[2p + 1]$, the last sum can then be understood as a sum over the subsets of $s$, and can therefore be evaluated in time $\mathcal{O}\left((2p + 1)2^{2p + 1}\right) = \mathcal{O}\left(p4^p\right)$ with the DP algorithm for sum-over-subsets. The first sum can be interpreted similarly, still regarding $s$ as a subset of $[2p + 1]$, but with set elements now identified to $0$ bits. Let us now consider the sums over $s$. By the discussion on the sums over $\alpha$, this sum can be considered as sums over \textit{supersets}. These in turn reduce to sums over subsets by passing to the complementary sets. Explicitly,
\begin{align*}
    & \sum_{s \in \{0, 1\}^{2p + 1}}A_{\alpha s}v_s\\
    & = \frac{1}{2}\sum_{\substack{s \in \{0, 1\}^{2p + 1}\\\alpha \subset \left\{j\,:\,s_j = 0\right\}}}v_s + \frac{1}{2}\sum_{\substack{s\,:\,\alpha \subset \left\{j\,:\,s_j = 1\right\}}}v_s\\
    & = \frac{1}{2}\sum_{\substack{s \in \{0, 1\}^{2p + 1}\\\left[2p + 1\right] - \left\{j\,:\,s_j = 0\right\} \subset [2p + 1] - \alpha}}v_s + \frac{1}{2}\sum_{\substack{s \in \{0, 1\}^{2p + 1}\\\left[2p + 1\right] - \left\{j\,:\,s_j = 1\right\} \subset [2p + 1] - \alpha}}v_s\\
    & = \frac{1}{2}\sum_{\substack{s \in \{0, 1\}^{2p + 1}\\\left\{j\,:\,s_j = 1\right\} \subset [2p + 1] - \alpha}}v_s + \frac{1}{2}\sum_{\substack{s \in \{0, 1\}^{2p + 1}\\\left\{j\,:\,s_j = 0\right\} \subset [2p + 1] - \alpha}}v_s.
\end{align*}
The first sum in the last expression can be computed as a sum over subsets of $[2p + 1] - \alpha$. Therefore, to evaluate it for all $\alpha$, it suffices to apply the sum-over-subsets algorithm (interpreting $1$ bits of $s \in \{0, 1\}^{2p + 1}$ as designing the elements included in the set) and reverse the resulting vector to account for the complement $[2p + 1] - \alpha$. The same applies for computing the second sum, except one interprets $0$ bits of $s$ as marking elements of the set. The overall complexity is then also $O(p4^p)$. To make these observations concrete, we provide detailed algorithms \ref{alg:sos_compute_sum_alpha} and \ref{alg:sos_compute_sum_s} to evaluate the desired sums.
\begin{algorithm}[!htbp]
	\caption{Compute sum $\sum\limits_{\alpha}A_{\alpha s}v_{\alpha}$ for random $2^q$-SAT QAOA}
    \label{alg:sos_compute_sum_alpha}
	\begin{algorithmic}
	    \Require Vector $\bm{v} = \left(v_{\alpha}\right)_{\alpha \subset [n]}$.
	    \Function{SumAlpha}{$\bm{v}$}
	        \State $z^{(1)} \longleftarrow v$.
	        \State $z^{(0)} \longleftarrow v$.
	        \For{$i \in \{0, \ldots, n - 1\}$} \Comment{Interpret $1$ bits of $s$ as designing set}
	            \For{$s \in \{0, 1\}^n$}
	                \If{$s_i = 1$}
	                    \State $z^{(1)}_s \longleftarrow z^{(1)}_s + z^{(1)}_{s\,\,\mathrm{XOR}\,\,2^i}$
	                \EndIf
	            \EndFor
	        \EndFor
	        \For{$i \in \{0, \ldots, n - 1\}$} \Comment{Interpret $0$ bits of $s$ as designing set}
	            \For{$s \in \{0, 1\}^n$}
	                \If{$s_i = 0$}
	                    \State $z^{(0)}_s \longleftarrow z^{(0)}_s + z^{(0)}_{s\,\,\mathrm{XOR}\,\,2^i}$
	                \EndIf
	            \EndFor
	        \EndFor
	        \State \Return $z^{(0)} + z^{(1)} - v_{\varnothing}$.
	    \EndFunction
	\end{algorithmic}
\end{algorithm}

\begin{algorithm}[!htbp]
	\caption{Compute sum $\sum\limits_{s}A_{\alpha s}v_{s}$ for random $2^q$-SAT QAOA}
    \label{alg:sos_compute_sum_s}
	\begin{algorithmic}
	    \Require Vector $\bm{v} = \left(v_{s}\right)_{s \in \{0, 1\}^n}$.
	    \Function{SumS}{$\bm{v}$}
	        \State $z^{(0)} \longleftarrow \Call{Reverse}{v}$ \Comment{Reversing vector is equivalent to flipping bits of bitstring coordinates}
	        \State $z^{(1)} \longleftarrow v$
	        \For{$i \in \{0, \ldots, n - 1\}$} \Comment{Indices from $\alpha$ are $0$ in $s$}
	            \For{$\alpha \subset [n]$}
	                \If{$i \notin \alpha$}
	                    \State $z^{(0)}_{\alpha} \longleftarrow z^{(0)}_{\alpha} + z^{(0)}_{\alpha \sqcup \{i\}}$
	                \EndIf
	            \EndFor
	        \EndFor
	        \For{$i \in \{0, \ldots, n - 1\}$} \Comment{Indices from $\alpha$ are $1$ in $s$}
	            \For{$\alpha \subset [n]$}
	                \If{$i \notin \alpha$}
	                    \State $z^{(1)}_{\alpha} \longleftarrow z^{(1)}_{\alpha} + z^{(1)}_{\alpha \sqcup \{i\}}$
	                \EndIf
	            \EndFor
	        \EndFor
	        \State \Return $z^{(0)} + z^{(1)} - v_{\varnothing}$
	    \EndFunction
	\end{algorithmic}
\end{algorithm}

\subsubsection{Iterated Gaussian integration}
\label{sec:iterated_gaussian_integration}

In this section, we introduce results generalizing two central ingredients of the method developed in section \ref{prop:hamming_weight_square_critical_point} for the toy example of the Hamming weight square. We first generalize the integration trick that allowed to generate a quadratic cost function from a linear one, based on the Gaussian integration identity:
\begin{align}
    \int_{\mathbf{R}}\!\mathrm{d}x\,\frac{e^{-x^2/4}}{\sqrt{4\pi}}e^{\alpha x} & = e^{\alpha^2} \qquad \forall \alpha \in \mathbf{C}.
\end{align}
This identity will be generalized by replacing the single-variable Gaussian integral with an \textit{iterated} Gaussian integral. The generalization is precisely stated in proposition \ref{prop:iterated_gaussian_integral_generate_higher_order_gaussian}. Second, we generalize to iterated Gaussian integrals the relationship established between the critical point of a Gaussian integrand and the fixed point of a certain function. This is expressed in proposition \ref{prop:critical_point_iterative_gaussian}. While the last two propositions constitute the main ideas of this section, further technicalities will be required as we introduce a special parametrization of the iterated Gaussian integral. This special parametrization, although slightly tedious to expose, is necessary to rigorously justify the saddle-point method (i.e., show that the integral is indeed dominated by the saddle-point when one integrates over all $\mathbf{R}^d$).

The statement of the following results will be facilitated by introducing the following notation.
\begin{defn}
\label{def:zeta}
Let $q \geq 1$ an integer. Given complex numbers $y^{(0)}, y^{(1)}, \ldots, y^{(q - 1)} \notin -i\mathbf{R}_+$, let
\begin{align}
    \zeta\left(\left(y^{(j)}\right)_{j \in [q]}\right) := \zeta\left(y^{(0)}, \ldots, y^{(q - 1)}\right) := \prod_{j \in [q]}\left(-iy^{(j)}\right)^{1/2^{q - 1 - j}}
\end{align}
\end{defn}
For instance,
\begin{align}
    \zeta\left(y^{(0)}\right) & = -iy^{(0)}\\
    \zeta\left(y^{(0)}, y^{(1)}\right) & = -iy^{(1)}\left(-iy^{(0)}\right)^{1/2}\\
    \zeta\left(y^{(0)}, y^{(1)}, y^{(2)}\right) & = -iy^{(2)}\left(-iy^{(1)}\right)^{1/2}\left(-iy^{(0)}\right)^{1/4}.
\end{align}
Also, by convention:
\begin{align}
    \zeta\left(\varnothing\right) & := 1.
\end{align}
We now introduce the generalization of the ``Gaussian integration trick" used in the toy example section \ref{sec:gaussian_integral_tails}. While the previous trick generated exponentials of perfect squares, this extension generates exponentials of arbitrary perfect powers of $2$:
\begin{prop}
\label{prop:iterated_gaussian_integral_generate_higher_order_gaussian}
Let $q \geq 1$ an integer, $C \in \mathbf{C} - \mathbf{R}_-$ and $x \in \mathbf{C}$. Then the following identity holds:
\begin{align}
\label{eq:iterated_gaussian_integral_generate_higher_order_gaussian}
    \frac{1}{\left(4\pi\right)^{q/2}}\prod_{j \in [q]}\int\limits_{\mathbf{R} + 2iC^{1/2^{j + 1}}\zeta\left(\left(y^{(j')}\right)_{j' \in [j]}\right)^{1/2}}\hspace*{-40px}\mathrm{d}y^{(j)}\,\exp\left(-\frac{1}{4}\sum_{j \in [q]}\left(y^{(j)}\right)^2 - x\zeta\left(\left(y^{(j)}\right)_{j \in [q]}\right)\right) & = \exp\left(-x^{2^q}\right).
\end{align}
\begin{proof}
Consider the terms
\begin{align*}
    -\frac{\left(y^{(q - 1)}\right)^2}{4} - x\zeta\left(\left(y^{(j)}\right)_{j \in [q]}\right) & = -\frac{\left(y^{(q - 1)}\right)^2}{4} + ixy^{(q - 1)}\prod_{j \in [q - 1]}\left(-iy^{(j)}\right)^{1/2^{q - 1 - j}}
\end{align*}
in the exponential and complete the square in $y^{(q - 1)}$:
\begin{align*}
    & -\frac{\left(y^{(q - 1)}\right)^2}{4} + ixy^{(q - 1)}\prod_{\in [q - 1]}\left(-iy^{(j)}\right)^{1/2^{q - 1 - j}}\\
    & = -\frac{1}{4}\left(y^{(q - 1)} - 2ix\prod_{j \in [q - 1]}\left(-iy^{(j)}\right)^{1/2^{q - 1 - j}}\right)^2 - x^2\prod_{j \in [q - 1]}\left(-iy^{(j)}\right)^{1/2^{q - 2 - j}}\\
    & = -\frac{1}{4}\left(y^{(q - 1)} - 2ix\zeta\left(\left(y^{(j)}\right)_{j \in [q - 1]}\right)^{1/2}\right)^2 - x^2\zeta\left(\left(y^{(j)}\right)_{j \in [q - 1]}\right).
\end{align*}
Continuing in the same way, i.e. completing the square for $y^{(q - 2)}$, then $y^{(q - 3)}$, etc., one finally obtains:
\begin{align*}
    -\frac{1}{4}\sum_{j \in [q]}\left(y^{(j)} - 2ix^{2^{q - 1 - j}}\zeta\left(\left(y^{(j')}\right)_{j' \in [j]}\right)^{1/2}\right)^2 - x^{2^q}.
\end{align*}
The integral to compute has now been rewritten
\begin{align*}
    & \frac{1}{\left(4\pi\right)^{q/2}}\prod_{j \in [q]}\int\limits_{\mathbf{R} + 2iC^{1/2^{j + 1}}\zeta\left(\left(y^{\left(j'\right)}\right)_{j' \in [j]}\right)^{1/2}}\hspace*{-40px}\mathrm{d}y^{(j)}\,\exp\left(-\frac{1}{4}\sum_{j \in [q]}\left(y^{(j)} - 2ix^{2^{q - 1 - j}}\zeta\left(\left(y^{\left(j'\right)}\right)_{j' \in [j]}\right)^{1/2}\right)^2 - x^{2^q}\right)\\
    & = \frac{\exp\left(-x^{2^q}\right)}{\left(4\pi\right)^{q/2}}\left(\prod_{j \in [q]}\int\limits_{\mathbf{R} + 2iC^{1/2^{j + 1}}\zeta\left(\left(y^{\left(j'\right)}\right)_{j' \in [j]}\right)^{1/2}}\hspace*{-40px}\mathrm{d}y^{(j)}\hspace*{15px}\right)\prod_{j \in [q]}\exp\left(-\frac{1}{4}\left(y^{(j)} - 2ix^{2^{q - 1 - j}}\zeta\left(\left(y^{\left(j'\right)}\right)_{j' \in [j]}\right)^{1/2}\right)^2\right)\\
    & = \frac{\exp\left(-x^{2^q}\right)}{\left(4\pi\right)^{q/2}}\left(\prod_{j \in [q]}\int\limits_{\mathbf{R} + 2iC^{1/2^{j + 1}}\zeta\left(\left(y^{\left(j'\right)}\right)_{j' \in [j]}\right)^{1/2}}\hspace*{-40px}\mathrm{d}y^{(j)}\hspace*{15px}\right)\prod_{j \in [q - 1]}\exp\left(-\frac{1}{4}\left(y^{(j)} - 2ix^{2^{q - 1 - j}}\zeta\left(\left(y^{\left(j'\right)}\right)_{j' \in [j]}\right)^{1/2}\right)^2\right)\\
    & \hspace*{20px} \times \exp\left(-\frac{1}{4}\left(y^{(q - 1)} - 2ix\zeta\left(\left(y^{\left(j'\right)}\right)_{j' \in [q - 1]}\right)^{1/2}\right)^2\right)\\
    & = \frac{\exp\left(-x^{2^q}\right)}{\left(4\pi\right)^{(q - 1)/2}}\left(\prod_{j \in [q]}\int\limits_{\mathbf{R} + 2iC^{1/2^{j + 1}}\zeta\left(\left(y^{\left(j'\right)}\right)_{j' \in [j]}\right)^{1/2}}\hspace*{-40px}\mathrm{d}y^{(j)}\hspace*{15px}\right)\prod_{j \in [q - 1]}\exp\left(-\frac{1}{4}\left(y^{(j)} - 2ix^{2^{q - 1 - j}}\zeta\left(\left(y^{\left(j'\right)}\right)_{j' \in [j]}\right)^{1/2}\right)^2\right)\\
    & \hspace*{20px} \times \frac{1}{\sqrt{4\pi}}\int\limits_{\mathbf{R} + 2iC\zeta\left(\left(y^{(j)}\right)_{j \in [q - 1]}\right)^{1/2}}\hspace*{-20px}\mathrm{d}y^{(q - 1)}\,\exp\left(-\frac{1}{4}\left(y^{(q - 1)} - 2ix\zeta\left(\left(y^{\left(j'\right)}\right)_{j' \in [q - 1]}\right)^{1/2}\right)^2\right)\\
    & = \frac{\exp\left(-x^{2^q}\right)}{\left(4\pi\right)^{(q - 1)/2}}\left(\prod_{j \in [q]}\int\limits_{\mathbf{R} + 2iC^{1/2^{j + 1}}\zeta\left(\left(y^{\left(j'\right)}\right)_{j' \in [j]}\right)^{1/2}}\hspace*{-40px}\mathrm{d}y^{(j)}\hspace*{15px}\right)\prod_{j \in [q - 1]}\exp\left(-\frac{1}{4}\left(y^{(j)} - 2ix^{2^{q - 1 - j}}\zeta\left(\left(y^{\left(j'\right)}\right)_{j' \in [j]}\right)^{1/2}\right)^2\right),
\end{align*}
where in the last equality, we used that
\begin{align*}
    \frac{1}{\sqrt{4\pi}}\int_{\mathbf{R} + z}\mathrm{d}y\,\exp\left(-\frac{y^2}{4}\right) & = 1
\end{align*}
for arbitrary $z \in \mathbf{C}$. Continuing this way by integrating over $y^{(q - 2)}$, $y^{(q - 3)}$, $\ldots$, up to $y^{(0)}$, all integrals are eliminated, leaving us with $\exp\left(-x^{2^q}\right)$, which is the result.
\end{proof}
\end{prop}
For instance, for $q = 3$, the identity reads:
\begin{align}
    & \frac{1}{\left(4\pi\right)^{3/2}}\int_{\mathbf{R} + 2iC^{1/2}}\!\mathrm{d}y^{(0)}\hspace*{-30px}\int\limits_{\mathbf{R} + 2iC^{1/4}\left(-iy^{(0)}\right)^{1/2}}\!\mathrm{d}y^{(1)}\hspace*{-20px}\int\limits_{\mathbf{R} + 2iC^{1/8}\left(-iy^{(1)}\right)^{1/2}\left(-iy^{(0)}\right)^{1/4}}\hspace*{-30px}\mathrm{d}y^{(2)}\,\exp\left(-\frac{\left(y^{(0)}\right)^2}{4} - \frac{\left(y^{(1)}\right)^2}{4} - \frac{\left(y^{(2)}\right)^2}{4}\right.\nonumber\\
    & \hspace*{20px} \left. - x\left(-iy^{(2)}\right)\left(-iy^{(1)}\right)^{1/2}\left(-iy^{(0)}\right)^{1/4}\right) = \exp\left(-x^8\right).
\end{align}

\begin{rem}
In proposition \ref{prop:iterated_gaussian_integral_generate_higher_order_gaussian}, Fubini's theorem applies, i.e. the $q$-variables integrals can either be viewed as successive single-variable integrals or as an integral over a $q$-dimensional volume. This is because the function in the integrand is integrable, as can e.g. be seen from proposition \ref{prop:iterated_gaussian_tail}.
\end{rem}

\begin{rem}
At the moment, the reason for the choice of integration contours in proposition \ref{prop:iterated_gaussian_integral_generate_higher_order_gaussian} may seem a bit obscure. The motivation becomes clearer in section \ref{sec:power_of_two_case}, with the proof of proposition \ref{prop:exp_scaling_generalized_binomial_sum} giving the exponential scaling of the generalized multinomial sum \ref{eq:generalized_binomial_sum}. Essentially, by introducing Cartesian coordinates to parametrize the integration surface defined in equation \ref{eq:iterated_gaussian_integral_generate_higher_order_gaussian}, the integrand becomes a (real) Gaussian centered about $0$ in this system of coordinates. We may already get partial insight into this fact by considering the following particular case of equation \ref{eq:iterated_gaussian_integral_generate_higher_order_gaussian}. Assume $\left|\arg(x)\right| < \frac{\pi}{2^q}$, so that for all integer $j \geq 0$, $\left(x^{2^q}\right)^{1/2^{j + 1}} = x^{2^{q - 1 - j}}$. Let $C := x^{2^q} \in \mathbf{C} - \mathbf{R}_-$ in equation \ref{eq:iterated_gaussian_integral_generate_higher_order_gaussian}. Then, recalling the proof of the proposition, the integrand can be written:
\begin{align*}
    & \frac{1}{\left(4\pi\right)^{q/2}}\exp\left(-\frac{1}{4}\sum_{j \in [q]}\left(y^{(j)} - 2ix^{2^{q - 1 - j}}\zeta\left(\left(y^{\left(j'\right)}\right)_{j' \in [j]}\right)^{1/2}\right)^2 - x^{2^q}\right)\\
    & = \frac{1}{\left(4\pi\right)^{q/2}}\exp\left(-\frac{1}{4}\sum_{j \in [q]}\left(y^{(j)} - 2iC^{1/2^{j + 1}}\zeta\left(\left(y^{\left(j'\right)}\right)_{j' \in [j]}\right)\right)^2 - C\right).
\end{align*}
Now, introduce a parametrization of the integration surface by real coordinates $\left(\widetilde{y}^{(0)}, \ldots, \widetilde{y}^{(q - 1)}\right) \in \mathbf{R}^q$ defined recursively by:
\begin{align*}
    y^{(0)} & := \widetilde{y}^{(0)} + 2iC^{1/2},\\
    y^{(1)} & := \widetilde{y}^{(1)} + 2iC^{1/4}\zeta\left(y^{(0)}\right)^{1/2},\\
    y^{(2)} & := \widetilde{y}^{(2)} + 2iC^{1/8}\zeta\left(y^{(0)}, y^{(1)}\right)^{1/2},\\
    \ldots\\
    y^{(q - 1)} & := \widetilde{y}^{(q - 1)} + 2iC^{1/2^q}\zeta\left(\left(y^{(j)}\right)_{j \in [q - 1]}\right)^{1/2}.
\end{align*}
Expressed in this new system of coordinates, the integrand reads:
\begin{align*}
    & \frac{1}{\left(4\pi\right)^{q/2}}\exp\left(-\frac{1}{4}\sum_{j \in [q]}\left(\widetilde{y}^{(j)}\right)^2 - C\right),
\end{align*}
which, up to the multiplicative (possibly complex) constant $e^{-C}$, is a centered real Gaussian indeed. In loose terms, the difference between this simple motivating example and the proof of the main result proposition \ref{prop:exp_scaling_generalized_binomial_sum} is that in the latter, we will apply integral identity \ref{eq:iterated_gaussian_integral_generate_higher_order_gaussian} to many different $x$ and choose $C$ not equal to some $x^{2^q}$, but rather to a different special value that will ``collectively" account for all $x$.
\end{rem}

It will be convenient to introduce a notation for the integration domain of proposition \ref{prop:iterated_gaussian_integral_generate_higher_order_gaussian} to simplify formulae in the next results and proofs:
\begin{defn}
\label{def:d_domain}
Let $q \geq 2$ an integer and $C \in \mathbf{C} - \mathbf{R}_-$. We define:
\begin{align}
    \mathcal{D}_{q, C} & := \left\{\left(y^{(0)}, \ldots, y^{(q - 1)}\right) \in \mathbf{C}^q\,:\,\forall j \in [q],\,y^{(j)} \in \mathbf{R} + 2iC^{1/2^{j + 1}}\zeta\left(\left(y^{(j')}\right)_{j' \in [j]}\right)\right\}.
\end{align}
\end{defn}
With this notation, the identity from proposition \ref{prop:iterated_gaussian_integral_generate_higher_order_gaussian} reads
\begin{align}
    \frac{1}{(4\pi)^{q/2}}\int_{\mathcal{D}_{q, C}}\!\prod_{j \in [q]}\mathrm{d}y^{(j)}\,\exp\left(-\frac{1}{4}\sum_{j \in [q]}\left(y^{(j)}\right)^2 - x\zeta\left(\left(y^{(j)}\right)_{j \in [q]}\right)\right) & = \exp\left(-x^{2^q}\right).
\end{align}

The next proposition then characterizes the critical points of a function of a collection of $Nq$ variables $\left(y_l^{(j)}\right)_{\substack{l \in [N]\\j \in [q]}}$ that only depends on $\left(\zeta\left(\left(y_l^{(j)}\right)_{j \in [q]}\right)\right)_{l \in [N]}$. More precisely, it expresses these critical points as the fixed points of a different function of the $\zeta$ variables. The motivation for this result is to find the critical points of integrands resulting from applying the integral identity in proposition \ref{prop:iterated_gaussian_integral_generate_higher_order_gaussian}. As shown in proposition \ref{prop:exp_scaling_generalized_binomial_sum}, finding these critical points (or more precisely, the unique critical point in this case) suffices to control the behaviour of the integral.
\begin{prop}
\label{prop:critical_point_iterative_gaussian}
Let $q \geq 2$ an integer. Suppose $F$ is a function of $N$ complex variables $z_0, \ldots, z_{N - 1}$ such that for some constants $0 < r < 1$ and $c > 0$,
\begin{align}
    |z_0|, \ldots, |z_{N - 1}| < r & \implies \left|\arg\left(-\partial_lF\left(z_0, \ldots, z_{N - 1}\right)\right)\right| < \frac{\pi}{2^{q - 1}} \qquad l \in [N],\label{eq:critical_point_iterative_gaussian_derivative_arg}\\
    |z_0|, \ldots, |z_{N - 1}| < r & \implies 0 < \left|\partial_lF(z_0, \ldots, z_{N - 1})\right| < c \qquad l \in [N],\\
    2^{(q + 1)/2}c^{2^{q - 1}} & < r^{\frac{1}{2} + \frac{1}{2(2^q - 1)}}.
\end{align}
Note that if $F\left(z_0, \ldots, z_{N - 1}\right) = \mathcal{O}\left(\max(|z_0|, \ldots, |z_{N - 1}|)\right)$ the last two conditions can always be satisfied by choosing $c = \mathcal{O}\left(r\right)$ and $r$ small enough: indeed, since $q \geq 2$, $c^{2^{q - 1}} = \mathcal{O}\left(r^{2^{q - 1}}\right) = \mathcal{O}\left(r^2\right) = o\left(r^{\frac{1}{2} + \frac{1}{2(2^q - 1)}}\right)$ as $r \to 0$. For $q \geq 1$ an integer, consider the function of $Nq$ complex variables defined by:
\begin{align}
    \Phi: & \left(\left(y_0^{(j)}\right)_{j \in [q]}, \ldots, \left(y_{N - 1}^{(j)}\right)_{j \in [q]}\right) \longmapsto -\frac{1}{4}\sum_{\substack{l \in [N]\\j \in [q]}}\left(y^{(j)}_l\right)^2 + F\left(\zeta\left(\left(y^{(j)}_0\right)_{j \in [q]}\right), \ldots, \zeta\left(\left(y^{(j)}_{N - 1}\right)_{j \in [q]}\right)\right).
\end{align}
Then $\left(y_l^{(j)*}\right)_{\substack{l \in [N]\\j \in [q]}}$ is a critical point of $\Phi$ satisfying
\begin{align}
    0 & < \left|y_l^{(j)*}\right| < r^{\frac{1}{2} + \frac{1}{2(2^q - 1)}},\label{eq:critical_point_iterative_gaussian_module_condition}\\
    y_l^{(j)*} & \notin -i\mathbf{R}_+ \qquad l \in [N], j \in [q].\label{eq:critical_point_iterative_gaussian_arg_condition}
\end{align}
iff. there exist $z_0, \ldots, z_{N - 1} \in \mathbf{C}$, $|z_l| < r$ such that for all $l \in [N]$,
\begin{align}
    z_l = -2^q\left\{\partial_lF\left(z_0, \ldots, z_{N - 1}\right)\right\}^{2^q - 1},\label{eq:critical_point_iterative_gaussian_z}
\end{align}
and for all $l \in [N], j \in [q]$,
\begin{align}
    y_l^{(j)*} & = 2^{1 + j/2}i\left\{\partial_lF\left(z_0, \ldots, z_{N - 1}\right)\right\}^{2^{q - 1}}\label{eq:critical_point_iterative_gaussian_y}.
\end{align}
\begin{proof}
First, assume the existence of a critical point $\left(y_l^{(j)*}\right)_{\substack{l \in [N]\\j \in [q]}}$ of $\Phi$ whose individual components satisfy the module and argument conditions \ref{eq:critical_point_iterative_gaussian_module_condition}, \ref{eq:critical_point_iterative_gaussian_arg_condition}. Let us fix $l_0 \in [N]$ and consider the critical point equations for coordinates $\left(y^{(j)}_{l_0}\right)_{j \in [q]}$:
\begin{align*}
    & -\frac{y_{l_0}^{(j)*}}{2} + \frac{1}{2^{q - 1 - j}}\frac{-i}{-iy_{l_0}^{(j)*}}\prod_{j' \in [q]}\left(-iy_{l_0}^{(j')*}\right)^{1/2^{q - 1 - j'}}\nonumber\\
    & \hspace*{80px} \times \partial_{l_0}F\left(\prod_{j \in [q]}\left(-iy_0^{(j)*}\right)^{1/2^{q - 1 - j}}, \ldots, \prod_{j \in [q]}\left(-iy_{N - 1}^{(j)*}\right)^{1/2^{q - 1 - j}}\right) & = 0 \qquad j \in [q].
\end{align*}
The $y_{l_0}^{(j)*} \notin -i\mathbf{R}_+$ assumption guarantees analyticity of the complex roots. In the following, we will abbreviate the arguments of $F$ by $\ldots$ to simplify the notation. For $q = 2$, these equations read:
\begin{align*}
    -\frac{y_{l_0}^{(1)*}}{2} + (-i)\left(-iy^{(0)*}_{l_0}\right)^{1/2}\partial_{l_0}F\left(\ldots\right) & = 0,\\
    -\frac{y_{l_0}^{(0)*}}{2} + (-i)\frac{1}{2}\left(-iy_{l_0}^{(1)*}\right)\left(-iy_{l_0}^{(0)*}\right)^{-1/2}\partial_{l_0}F\left(\ldots\right) & = 0.
\end{align*}
Substituting the expression for $y_{l_0}^{(1)*}$ in terms of $\partial_{l_0}F\left(\ldots\right)$ given by the first equation into the second equation, we obtain:
\begin{align*}
    -iy_{l_0}^{(0)*} & = 2\left(\partial_{l_0}F\left(\ldots\right)\right)^2.
\end{align*}
Now, we observe that the arguments of $\partial_{l_0}F\left(\ldots\right)$ are all of the form $\prod_{j \in [q]}\left(-iy_{l}^{(j)*}\right)^{1/2^{q - 1 - j}}$; in particular, they have moduli bounded by $\prod_{j \in [q]}\left(r^{\frac{1}{2} + \frac{1}{2(2^q - 1)}}\right)^{1/2^{q - 1 - j}} = r$. Therefore, by assumption \ref{eq:critical_point_iterative_gaussian_derivative_arg}, $\Re\partial_{l_0}F\left(\ldots\right) < 0$, so that $\left(-iy_{l_0}^{(0)*}\right)^{1/2} = -\sqrt{2}\partial_{l_0}F\left(\ldots\right)$. Substituting this expression in the first equation yields
\begin{align*}
    -iy_{l_0}^{(1)*} & = 2^{3/2}\left(\partial_{l_0}F(\ldots)\right)^2.
\end{align*}
This is the required equation \ref{eq:critical_point_iterative_gaussian_y} for $q = 2$. For $q > 2$, we proceed by recursion. Namely, consider the equation for the largest index coordinate $j = q - 1$:
\begin{align*}
    -\frac{y_{l_0}^{(q - 1)*}}{2} + (-i)\prod_{j' \in [q - 1]}\left(-iy_{l_0}^{(j')*}\right)^{1/2^{q - 1 - j'}}\partial_{l_0}F\left(\ldots\right) = 0.
\end{align*}
Substituting the expression given by this relation for $y_{l_0}^{(q - 1)*}$ into the equations for coordinates $j \in [q - 1]$ yields:
\begin{align*}
    -\frac{y_{l_0}^{(j)*}}{2} + \frac{1}{2^{q - 2 - j}}\frac{-i}{-iy_{l_0}^{(j)*}}\prod_{j' \in [q - 1]}\left(-iy_{l_0}^{\left(j'\right)*}\right)^{1/2^{q - 2 - j}} \times \left(-\left(\partial_{l_0}F(\ldots)\right)^2\right) \qquad j \in [q - 1].
\end{align*}
Now, observe that the equation obtained above for all $j \leq q - 2$ is formally identical to the critical point equations with substitutions:
\begin{align*}
    q & \longrightarrow q - 1,\\
    \partial_{l_0}F\left(\ldots\right) & \longrightarrow -\left(\partial_{l_0}F\left(\ldots\right)\right)^2.
\end{align*}
Besides, the argument condition \ref{eq:critical_point_iterative_gaussian_derivative_arg} on $\partial_{l_0}F\left(\ldots\right)$ at level $q$: $\left|\arg\left(-\partial_{l_0}F\left(\ldots\right)\right)\right| < \frac{\pi}{2^{q - 1}}$ implies the same condition at level $q - 1$ on $\left(\partial_{l_0}F\left(\ldots\right)\right)^2$: $\left|\arg\left\{-\left[-\left(\partial_{l_0}F\left(\ldots\right)\right)^2\right]\right\}\right| < \frac{\pi}{2^{q - 2}}$. Therefore, one can continue the recursion until one reaches the (already solved) simple case $q = 2$. This finally yields the following expression for $y_{l_0}^{(j)*}$:
\begin{align*}
    -iy_{l_0}^{(j)*} & = 2^{1 + j/2}\left(\partial_{l_0}F(\ldots)\right)^{2^{q - 1}}.
\end{align*}
Therefore, defining
\begin{align*}
    z_l & := \prod_{j \in [q]}\left(-iy_l^{(j)*}\right)^{1/2^{q - 1 - j}},
\end{align*}
the required equations \ref{eq:critical_point_iterative_gaussian_z} and \ref{eq:critical_point_iterative_gaussian_y} hold. Conversely, given $\left(z_l\right)_{l \in [N]}$ satisfying equation \ref{eq:critical_point_iterative_gaussian_z} and defining $y^{(j)*}_l$ from $z_l$ according to equation \ref{eq:critical_point_iterative_gaussian_y}, $\left(y_l^{(j)*}\right)_{\substack{l \in [N]\\j \in [q]}}$ is a critical point of $\Phi$ by the same calculations and satisfies the required assumptions \ref{eq:critical_point_iterative_gaussian_module_condition} and \ref{eq:critical_point_iterative_gaussian_arg_condition}.
\end{proof}
\end{prop}
We now describe a special parametrization for the integration domain $\mathcal{D}_{q, C}$ introduced in definition \ref{def:d_domain}. The idea is simply to introduce for each $j \in [q]$ a real variable $\widetilde{y}^{(j)}$ and parametrize $y^{(j)} \longleftarrow \widetilde{y}_j + 2iC^{1/2^{j + 1}}\zeta\left(\left(y^{(j')}\right)_{j' \in [j]}\right)^{1/2}$ (since $y^{(j)} \in \mathbf{R} + 2iC^{1/2^{j + 1}}\zeta\left(\left(y^{(j')}\right)_{j' \in [j]}\right)^{1/2}$). More formally:
\begin{defn}
\label{def:y_function}
Let $q \geq 1$ an integer and $C \in \mathbf{C} - \mathbf{R}_-$. Given $\left(\widetilde{y}^{(j)}\right) \in \mathbf{R}^q$, define $\mathcal{Y}_C^{(j)}\left(\left(\widetilde{y}^{(j')}\right)_{0 \leq j' \leq j}\right)$ recursively for  all $j \in [q]$ as follows:
\begin{align}
    \mathcal{Y}_C^{(0)}\left(\widetilde{y}^{(0)}\right) & := \widetilde{y}^{(0)} + 2iC^{1/2}
\end{align}
and for $1 \leq j < q$,
\begin{align}
    \mathcal{Y}_C^{(j)}\left(\left(\widetilde{y}^{(j')}\right)_{0 \leq j' \leq j}\right) & := \widetilde{y}^{(j)} + 2iC^{1/2^{j + 1}}\zeta\left(\left(\mathcal{Y}_C^{(j')}\left(\left(\widetilde{y}^{(j'')}\right)_{0 \leq j'' \leq j'}\right)\right)_{j' \in [j]}\right)^{1/2}.
\end{align}
\end{defn}
Back to informal, one may say $\mathcal{Y}_C^{(j)}$ maps parameters $\widetilde{y}^{(0)}, \ldots, \widetilde{y}^{(j)}$ to coordinate $y^{(j)}$. This parametrization of $\left(y^{(j)}\right)_{j \in [q]}$ as a function of $\left(\widetilde{y}^{(j)}\right)_{j \in [q]}$ in turn allows to consider $\zeta$ as a function of $\left(\widetilde{y}^{(j)}\right)_{j \in [q]}$; more formally:
\begin{defn}
\label{def:zeta_tilde}
Let $q \geq 1$ an integer and $C \in \mathbf{C} - \mathbf{R}_-$. For $\left(\widetilde{y}^{(j)}\right)_{j \in [q]} \in \mathbf{R}^q$, let
\begin{align}
    \widetilde\zeta_C\left(\left(\widetilde{y}^{(j)}\right)_{j \in [q]}\right) & = \zeta\left(\left(\mathcal{Y}^{(j)}\left(\left(\widetilde{y}^{(j)}\right)_{0 \leq j' \leq j}\right)\right)_{j \in [q]}\right).
\end{align}
\end{defn}
To illustrate, the identity in proposition \ref{prop:iterated_gaussian_integral_generate_higher_order_gaussian} can now be written as:
\begin{align}
    \int_{\mathbf{R}^q}\!\prod_{j \in [q]}\mathrm{d}\widetilde{y}^{(j)}\,\exp\left(-\frac{1}{4}\sum_{j \in [q]}\mathcal{Y}_C^{(j)}\left(\left(\widetilde{y}^{(j')}\right)_{0 \leq j' \leq j}\right)^2 - x\widetilde{\zeta}_C\left(\left(\widetilde{y}^{(j)}\right)_{j \in [q]}\right)\right) & = \exp\left(-x^{2^q}\right).
\end{align}

We will need the following formulae to convert between derivatives in the variables $y^{(j)}$ and derivatives in the variables $\widetilde{y}^{(j)}$:
\begin{lem}
	\label{lemma:d_y_function_z_tilde}
	Let $q \geq 1$ an integer, $C \in \mathbf{C} - \mathbf{R}_-$ and let functions $\left(\mathcal{Y}^{(j)}_C\right)_{j \in [q]}$ and $\widetilde{\zeta}_C$ be as in definitions \ref{def:y_function} and \ref{def:zeta_tilde}. Then the following identity on their derivatives hold for all $j, j_0, j_1 \in [q]$:
	\begin{align}
		\partial_{j_0}\mathcal{Y}_C^{(j)}\left(\left(\widetilde{y}^{(j')}\right)_{0 \leq j' \leq j}\right) & = \left\{\begin{array}{cc}
			0 & \textrm{if } j_0 > j\\
			1 & \textrm{if } j_0 = j\\
			2iC^{1/2^{j + 1}}\sum\limits_{j' \in [j]}\frac{\widetilde{\zeta}_C\left(\left(\widetilde{y}^{\left(j''\right)}\right)_{j'' \in [j]}\right)^{1/2}}{2^{j - j'}}\frac{\partial_{j_0}\mathcal{Y}_C^{\left(j'\right)}\left(\left(\widetilde{y}^{\left(j''\right)}\right)_{0 \leq j'' \leq j'}\right)}{\mathcal{Y}_C^{\left(j'\right)}\left(\left(\widetilde{y}^{\left(j''\right)}\right)_{0 \leq j'' \leq j'}\right)} & \textrm{otherwise}
		\end{array}\label{eq:d_y_function}\right.\\
		\partial_{j_0}\widetilde{\zeta}_C\left(\left(\widetilde{y}^{\left(j'\right)}\right)_{j' \in [q]}\right) & = \sum_{0 \leq j \leq j_0}\frac{\widetilde{\zeta}_C\left(\left(\widetilde{y}^{\left(j'\right)}\right)_{j' \in [q]}\right)}{2^{q - 1 - j}}\frac{\partial_{j_0}\mathcal{Y}_C^{\left(j\right)}\left(\left(\widetilde{y}^{\left(j'\right)}\right)_{0 \leq j' \leq j}\right)}{\mathcal{Y}_C^{\left(j\right)}\left(\left(\widetilde{y}^{\left(j'\right)}\right)_{0 \leq j' \leq j}\right)}\label{eq:d_zeta_tilde},\\
		\partial_{j_0j_1}\mathcal{Y}^{(j)}_C\left(\left(\widetilde{y}^{(j')}\right)_{0 \leq j' \leq j}\right) & = \left\{\begin{array}{ll}
			0 & \textrm{if } j_0 \geq j \textrm{ or } j_1 \geq j\\
			2iC^{1/2^{j + 1}}\sum\limits_{j', j'' \in [j]}\left(\frac{\left(\widetilde\zeta_C^{(j)}\right)^{1/2}}{2^{j - j'}2^{j - j''}}\frac{\partial_{j_0}\mathcal{Y}_C^{(j')}\partial_{j_1}\mathcal{Y}_C^{\left(j''\right)}}{\mathcal{Y}_C^{\left(j'\right)}\mathcal{Y}_C^{\left(j''\right)}}\right)\\
			+ 2iC^{1/2^{j + 1}}\sum\limits_{j' \in [j]}\frac{\left(\widetilde\zeta_C^{(j)}\right)^{1/2}}{2^{j - j'}}\left(\frac{\partial_{j_0j_1}\mathcal{Y}^{\left(j'\right)}}{\mathcal{Y}_C^{(j')}} - \frac{\partial_{j_0}\mathcal{Y}_C^{\left(j'\right)}\partial_{j_1}\mathcal{Y}_C^{\left(j'\right)}}{\left(\mathcal{Y}_C^{(j')}\right)^2}\right) & \textrm{otherwise}
		\end{array}\right.\\
		\partial_{j_0j_1}\widetilde\zeta_C\left(\left(\widetilde{y}^{(j)}\right)_{j \in [q]}\right) & = \sum_{0 \leq j, j' \leq \max(j_0, j_1)}\frac{\widetilde\zeta_C}{2^{q - 1 - j}2^{q - 1 - j'}}\frac{\partial_{j_0}\mathcal{Y}_C^{(j')}\partial_{j_1}\mathcal{Y}_C^{(j'')}}{\mathcal{Y}_C^{(j)}\mathcal{Y}_C^{(j')}}\nonumber\\
		&  \hspace*{20px} + \sum_{0 \leq j \leq j_0}\frac{\widetilde\zeta_C}{2^{q - 1 - j}}\left(\frac{\partial_{j_0j_1}\mathcal{Y}_C^{\left(j\right)}}{\mathcal{Y}_C^{\left(j\right)}} - \frac{\partial_{j_0}\mathcal{Y}_C^{(j)}\partial_{j_1}\mathcal{Y}_C^{(j)}}{\left(\mathcal{Y}_C^{(j)}\right)^2}\right).
	\end{align}
	In the last two equations, we abbreviated $\widetilde\zeta_C$ for $\widetilde\zeta_C\left(\left(\widetilde{y}^{(j)}\right)_{j \in [q]}\right)$, $\widetilde\zeta_C^{(j)}$ for $\widetilde\zeta_C\left(\left(\widetilde{y}^{(j')}\right)_{j' \in [j]}\right)$, and similarly omitted to write variables in functions $\mathcal{Y}^{(j)}_C$ and their derivatives.
	\begin{proof}
		These are straightforward applications of the chain rule after recalling definition \ref{def:zeta_tilde} for $\widetilde{\zeta}_C$. Let us for instance prove equation \ref{eq:d_y_function}. Assume $j > 0$ since $j = 0$, where $\mathcal{Y}^{(0)}\left(\widetilde{y}^{(0)}\right) = \widetilde{y}^{(0)} + 2iC^{1/2}$, is trivial. Then
		\begin{align*}
			\mathcal{Y}_C^{(j)}\left(\left(\widetilde{y}^{\left(j''\right)}\right)_{0 \leq j'' \leq j}\right) & = \widetilde{y}^{(j)} + 2iC^{1/2^{j + 1}}\zeta_C\left(\left(\mathcal{Y}_C^{\left(j''\right)}\left(\left(\widetilde{y}^{\left(j'''\right)}\right)_{0 \leq j''' \leq j''}\right)\right)_{j'' \in [j]}\right)^{1/2}\\
			& = \widetilde{y}^{(j)} + 2iC^{1/2^{j + 1}}\prod_{j'' \in [j]}\left(-i\mathcal{Y}_C^{(j'')}\left(\left(\widetilde{y}^{\left(j'''\right)}\right)_{0 \leq j''' \leq j''}\right)\right)^{1/2^{j - j''}}.
		\end{align*}
		The above only depends on $\widetilde{y}^{(0)}, \ldots, \widetilde{y}^{(j)}$, giving the first case of equation \ref{eq:d_y_function}; the second case is similarly easy since the $\zeta_C$ function only depends on $\widetilde{y}^{(0)}, \ldots, \widetilde{y}^{(j - 1)}$. For the third case, i.e. $j' \in [j]$, one can then take the derivative in $\widetilde{y}^{(j')}$ using the chain rule, which gives:
		\begin{align*}
			& \partial_{j'}\mathcal{Y}_C^{(j)}\left(\left(\widetilde{y}^{\left(j''\right)}\right)_{0 \leq j'' \leq j}\right)\\
			& = 2iC^{1/2^{j + 1}}\sum_{j'' \in [j]}\frac{1}{2^{j - j''}}\frac{\partial_{j'}\mathcal{Y}_C^{\left(j''\right)}\left(\left(\widetilde{y}^{\left(j'''\right)}\right)_{0 \leq j''' \leq j''}\right)}{\mathcal{Y}_C^{\left(j''\right)}\left(\left(\widetilde{y}^{\left(j'''\right)}\right)_{0 \leq j''' \leq j''}\right)}\prod_{j'' \in [j]}\left(-i\mathcal{Y}_C^{\left(j''\right)}\left(\left(\widetilde{y}^{\left(j'''\right)}\right)_{0 \leq j''' \leq j''}\right)\right)^{1/2^{j - j''}},
		\end{align*}
		which is the result.
	\end{proof}
\end{lem}

The values of $\mathcal{Y}_C^{(j)}$ and $\widetilde\zeta_C$ at $0$ will be of particular importance in relation to proposition \ref{prop:critical_point_iterative_gaussian}. We give these in the following lemma, and similarly for their derivatives (the latters will be needed in the proof of our main result proposition \ref{prop:exp_scaling_generalized_binomial_sum}).
\begin{prop}
\label{prop:y_function_0}
	Let $q \geq 1$ an integer, $C \in \mathbf{C} - \mathbf{R}_-$ and let coordinate change functions $\left(\mathcal{Y}^{(j)}_C\right)_{j \in [q]}$ be defined as in definition \ref{def:y_function}. Then for all $j, j_0, j_1 \in [q]$,
	\begin{align}
		\mathcal{Y}_C^{(j)}(0_{j + 1}) & = 2^{1 + j/2}iC^{1/2},\label{eq:y_0}\\
		\widetilde\zeta_C(0_q) & = 2^qC^{1 - 1/2^q},\label{eq:zeta_tilde_0}\\
		\partial_{j_0}\mathcal{Y}^{(j)}(0_{j + 1}) & = \left\{\begin{array}{cc}
			0 & \textrm{if } j_0 > j\\
			1 & \textrm{if } j_0 = j\\
			2^{(j - j_0)/2 - 1} & \textrm{otherwise}
		\end{array}\right.,\\
		\partial_{j}\widetilde\zeta_C\left(0_q\right) & = -2^{q - 1 - j/2}iC^{1/2 - 1/2^q}.
	\end{align}
	\begin{proof}
		This is easily proven by induction using the recursive definition \ref{def:y_function} of $\mathcal{Y}^{(j)}_C$.
	\end{proof}
\end{prop}

\begin{rem}
Note that equation \ref{eq:d_y_function} in lemma \ref{lemma:d_y_function_z_tilde} does not exactly give the derivative $\partial_{j'}\mathcal{Y}^{(j)}_C$ explicitly, but rather in a recursive form, i.e. as a function of derivatives $\partial_{j'}\mathcal{Y}^{(j'')}_C$ for $j'' \in [j]$. This function is trivial when $j'' = 0$ (it is $1$ if $j' = 0$ and $0$ otherwise).
\end{rem}
Using these definitions, we can state the following corollary of proposition \ref{prop:critical_point_iterative_gaussian}, allowing to center the critical point ``around zero $\widetilde{y}$ coordinates":
\begin{cor}
	\label{cor:critical_point_zero_iterative_gaussian}
	Let $F$ satisfy the assumptions of proposition \ref{prop:critical_point_iterative_gaussian}. Let $z^* = \left(z^*_0, \ldots, z^*_{N - 1}\right)$ be a fixed point of:
	\begin{align}
		\begin{array}{ccc}
			\mathbf{C}^N & \longrightarrow & \mathbf{C}^N\\
			\left(z_l\right)_{l \in [N]} & \longmapsto & \left(-2^q\left\{\partial_lF\left(z_0, \ldots, z_{N - 1}\right)\right\}^{2^q - 1}\right)_{l \in [N]}
		\end{array}
	\end{align}
	Now, consider for all $l \in [N]$ the coordinate change functions $\left(\mathcal{Y}^{(j)}_{C_l}\right)_{j \in [q]} =: \left(\mathcal{Y}^{(j)}_l\right)_{j \in [q]}$ introduced in definition \ref{def:y_function}, where
	\begin{align}
		C_l & = \left\{\partial_lF\left(z^*\right)\right\}^{2^q},
	\end{align}
	and define $\widetilde{\Phi}$ as ``$\Phi$ in the $\widetilde{y}$ coordinates system":
	\begin{align}
		\widetilde\Phi\left(\left(\widetilde{y}^{(j)}_l\right)_{\substack{l \in [N]\\j \in [q]}}\right) & := \Phi\left(\left(\mathcal{Y}^{(j)}_l\left(\left(\widetilde{y}^{(j')}_l\right)_{j' \in [q]}\right)\right)_{\substack{l \in [N]\\j \in [q]}}\right)\\
		& = -\frac{1}{4}\sum_{\substack{l \in [N]\\j \in [q]}}\mathcal{Y}^{(j)}_l\left(\left(\widetilde{y}^{(j')}_{l}\right)_{j' \in [q]}\right)^2 + F\left(\left(\widetilde{\zeta}_l\left(\left(\widetilde{y}^{(j)}_l\right)_{j \in [q]}\right)\right)_{l \in [N]}\right).
	\end{align}
	Then, $0$ is a fixed point of $\widetilde{\Phi}$.
\begin{proof}
	This is a simple consequence of the expression of $\widetilde{\Phi}$ in terms of $\Phi$, the chain rule from lemma \ref{lemma:d_y_function_z_tilde}, and the choice of constants $C$ for the definition of the $\widetilde{Y}$ functions. Explicitly,
	\begin{align*}
		\frac{\partial\widetilde{\Phi}}{\partial\widetilde{y}^{(j)}_l}\left(0\right) & = \sum_{j \leq j' < q}\frac{\partial\Phi}{\partial y^{(j')}_l}\left(\left(\mathcal{Y}_{l'}^{(j'')}(0)\right)_{\substack{l' \in [N]\\j'' \in [q]}}\right)\partial_j\mathcal{Y}^{(j')}_l\left(0\right).
	\end{align*}
	Now, proposition \ref{prop:y_function_0} gives:
	\begin{align*}
		\mathcal{Y}^{(j)}_l(0) & = 2^{1 + j/2}iC_l^{1/2}\\
		& = 2^{1 + j/2}i\left\{\partial_lF\left(z^*\right)\right\},
	\end{align*}
	which is a critical point of $\Phi$ by proposition \ref{prop:critical_point_iterative_gaussian}. Therefore, the $\Phi$ derivatives vanish and the corollary is proven.
\end{proof}
\end{cor}

After evaluating the coordinate change functions $\mathcal{Y}^{(j)}_C$ and their derivatives at $0$ in proposition \ref{prop:y_function_0}, we show that when their arguments are close enough to $0$, then so are the coordinate change functions. In fact, a similar result holds for their derivatives as well as for $\zeta_C$. This follows from a simple rescaling argument:
\begin{lem}
	\label{lemma:y_tilde_close_0_function_close_0_evaluated}
	Recall the definition \ref{def:y_function} and \ref{def:zeta_tilde} of the coordinate change functions $\left(\mathcal{Y}_C^{(j)}\right)_{j \in [q]}$ and $\widetilde\zeta_C$. 	Let for all $j \in [q]$, $\widetilde{y}^{(j)} = z^{(j)}C^{1/2}$ where $z^{(j)} \in \mathbf{C}, |z^{(j)}| \leq z_{\textrm{max}}$. Then for sufficiently small $z_{\textrm{max}}$ (independent of $C$),
	\begin{align}
		\forall j \in [q],\quad\left|\mathcal{Y}_C^{(j)}\left(\left(\widetilde{y}^{(j')}\right)_{j' \in [q]}\right) - \mathcal{Y}_C^{(j)}(0_{j + 1})\right| & \leq\mathcal{O}(1)|C|^{1/2}z_{\textrm{max}},\label{eq:y_close_to_y_0}\\
		%
		%
		%
		%
		%
		%
		\left|\widetilde\zeta_C\left(\left(\widetilde{y}^{(j)}\right)_{j \in [q]}\right) - \widetilde\zeta_C(0_q)\right| & \leq\mathcal{O}(1)|C|^{1 - 1/2^q}z_{\textrm{max}},\label{eq:zeta_tilde_close_to_zeta_tilde_0}\\
		\forall j, j_0 \in [q],\quad\left|\partial_{j_0}\mathcal{Y}_C^{(j)}\left(\left(\widetilde{y}^{(j')}\right)_{j' \in [q]}\right) - \partial_{j_0}\mathcal{Y}_C^{(j)}(0_{j + 1})\right| & \leq\mathcal{O}(1)z_{\textrm{max}},\label{eq:dy_close_to_dy_0}\\
		\forall j, j_0, j_1 \in [q],\quad\left|\partial_{j_0j_1}\mathcal{Y}_C^{(j)}\left(\left(\widetilde{y}^{(j')}\right)_{j' \in [q]}\right) - \partial_{j_0j_1}\mathcal{Y}_C^{(j)}(0_{j + 1})\right| & \leq\mathcal{O}(1)|C|^{-1/2}z_{\textrm{max}},\label{eq:ddy_close_to_ddy_0}\\
		\forall j_0 \in [q],\quad\left|\partial_{j_0}\widetilde\zeta_C\left(\left(\widetilde{y}^{(j)}\right)_{j \in [q]}\right) - \partial_{j_0}\widetilde\zeta_C(0_q)\right| & \leq\mathcal{O}(1)|C|^{1/2 - 1/2^q}z_{\textrm{max}},\label{eq:d_zeta_tilde_close_to_d_zeta_tilde_0}\\
		\forall j_0, j_1 \in [q],\quad\left|\partial_{j_0j_1}\widetilde\zeta_C\left(\left(\widetilde{y}^{(j)}\right)_{j \in [q]}\right) - \partial_{j_0j_1}\widetilde\zeta_C(0_q)\right| & \leq\mathcal{O}(1)|C|^{-1/2^q}z_{\textrm{max}}.\label{eq:dd_zeta_tilde_close_to_dd_zeta_tilde_zero}
	\end{align}
	The constants hidden in the $\mathcal{O}(1)$ do not depend on $C$ but may depend on $q$.
\begin{proof}
    In the following proofs, we will use the notation $0_d$ to denote an all-zero vector with $d$ components. This choice is motivated by the manipulation of $\zeta, \widetilde\zeta_C$ functions defined in \ref{def:zeta}, \ref{def:zeta_tilde} depending on a variable number of arguments.
    
	These inequalities result from simple rescaling arguments. Let us start by simultaneously proving inequalities \ref{eq:y_close_to_y_0} and \ref{eq:zeta_tilde_close_to_zeta_tilde_0}. We prove them by induction on $j \in [q]$. For $j = 0$ and equality \ref{eq:y_close_to_y_0},
	\begin{align*}
		\left|\mathcal{Y}_C^{(0)}\left(\left(\widetilde{y}^{(j')}\right)_{j' \in [q]}\right) - \mathcal{Y}_C^{(0)}(0)\right| & = \left|\widetilde{y}^{(0)}\right|\\
		& = |C|^{1/2}|z^{(0)}|\\
		& \leq |C|^{1/2}z_{\textrm{max}}.
	\end{align*}
	As for inequality \ref{eq:zeta_tilde_close_to_zeta_tilde_0}, it is trivial at lowest level since $\zeta\left(\varnothing\right) = 1$. Now, assuming inequalities \ref{eq:y_close_to_y_0} and \ref{eq:zeta_tilde_close_to_zeta_tilde_0} to hold below level $j$,
	\begin{align*}
		& \left|\mathcal{Y}_C^{(j)}\left(\left(\widetilde{y}^{(j')}\right)_{j' \in [q]}\right) - \mathcal{Y}_C^{(j)}\left(0_{j + 1}\right)\right|\\
		& = \left|\widetilde{y}^{(j)} + 2iC^{1/2^{j + 1}}\widetilde\zeta_C\left(\left(\widetilde{y}^{(j)}\right)_{j' \in [j]}\right)^{1/2} - 2iC^{1/2^{j + 1}}\widetilde\zeta_C\left(0_{j}\right)^{1/2}\right|\\
		& \leq \left|\widetilde{y}^{(j)}\right| + 2|C|^{1/2^{j + 1}}\left|\widetilde\zeta_C\left(\left(\widetilde{y}^{(j)}\right)_{j' \in [j]}\right)^{1/2} - \widetilde\zeta_C\left(0_j\right)^{1/2}\right|\\
		& =  \left|\widetilde{y}^{(j)}\right| + 2|C|^{1/2^{j + 1}}\left|\widetilde\zeta_C\left(0_j\right)\right|^{1/2}\left|\left(\frac{\widetilde\zeta_C\left(\left(\widetilde{y}^{(j)}\right)_{j' \in [j]}\right)}{\widetilde\zeta_C\left(0_j\right)}\right)^{1/2} - 1\right|\\
		& =  \left|\widetilde{y}^{(j)}\right| + 2|C|^{1/2^{j + 1}}\left|\widetilde\zeta_C\left(0_j\right)\right|^{1/2}\left|\left(1 + \frac{\widetilde\zeta_C\left(\left(\widetilde{y}^{(j)}\right)_{j' \in [j]}\right) - \widetilde\zeta_C\left(0_j\right)}{\widetilde\zeta_C\left(0_j\right)}\right)^{1/2} -1\right|\\
		& \leq \left|\widetilde{y}^{(j)}\right| + 2|C|^{1/2^{j + 1}}\left|\widetilde{\zeta}_C\left(0_j\right)\right|^{1/2}\left(1 - \left|\frac{\widetilde\zeta_C\left(\left(\widetilde{y}^{(j)}\right)_{j' \in [j]}\right) - \widetilde\zeta_C\left(0_j\right)}{\widetilde\zeta_C\left(0_j\right)}\right|\right)^{-1/2}\frac{1}{2}\left|\frac{\widetilde\zeta_C\left(\left(\widetilde{y}^{(j)}\right)_{j' \in [j]}\right) - \widetilde\zeta_C\left(0_j\right)}{\widetilde\zeta_C\left(0_j\right)}\right|\\
		& \leq |C|^{1/2}\left|z^{(j)}\right| + 2|C|^{1/2^{j + 1}}2^{j/2}|C|^{1/2 - 1/2^{j + 1}}\left(1 - \mathcal{O}(1)z_{\textrm{max}}\right)^{-1/2}\mathcal{O}(1)z_{\textrm{max}}\\
		& \leq \mathcal{O}(1)|C|^{1/2}z_{\textrm{max}}.
	\end{align*}
	The second inequality results from
	\begin{align*}
	    \left|(1 + z)^{1/2} - 1\right| & \leq (1 - |z|)^{-1/2}\frac{|z|}{2}
	\end{align*}
	holding for all complex numbers $z$, $|z| < 1$. To obtain the third inequality, we referred to the special values of $\widetilde\zeta_C$ at $0$ given in proposition \ref{prop:y_function_0}. Finally, note we tacitly combined all $\mathcal{O}(1)$ together when going from level $j$ to level $j + 1$ of the induction, where $j \in [q - 1]$, since we allowed this implicit constant to depend on $q$.
	
	Now, for inequality \ref{eq:zeta_tilde_close_to_zeta_tilde_0},
	\begin{align*}
		& \left|\widetilde\zeta_C\left(\left(\widetilde{y}^{(j')}\right)_{j' \in [j]}\right) - \widetilde\zeta_C\left(0_j\right)\right|\\
		& = \left|\left(-i\widetilde{y}^{(j - 1)} + 2C^{1/2^{j}}\widetilde\zeta_C\left(\left(\widetilde{y}^{(j')}\right)_{j' \in [j - 1]}\right)^{1/2}\right)\widetilde\zeta_C\left(\left(\widetilde{y}^{(j')}\right)_{j' \in [j - 1]}\right)^{1/2} - 2^{j}C^{1 - 1/2^j}\right|\\
		& \leq \left|\widetilde{y}^{(j - 1)}\right|\left|\widetilde\zeta_C\left(\left(\widetilde{y}^{(j')}\right)_{j' \in [j - 1]}\right)\right|^{1/2} + \left|2C^{1/2^{j}}\widetilde\zeta_C\left(\left(\widetilde{y}^{(j')}\right)_{j' \in [j - 1]}\right) - 2^jC^{1 - 1/2^j}\right|\\
		& \leq \left|\widetilde{y}^{(j - 1)}\right|\left|\widetilde\zeta_C\left(0_{j - 1}\right)\right|^{1/2}\left|\frac{\widetilde\zeta_C\left(\left(\widetilde{y}^{(j')}\right)_{j' \in [j - 1]}\right) - \widetilde\zeta_C\left(0_{j - 1}\right)}{\widetilde\zeta_C\left(0_{j - 1}\right)} + 1\right|^{1/2}\\
		& \hspace*{20px} + 2|C|^{1/2^{j}}\left|\widetilde\zeta_C\left(\left(\widetilde{y}^{(j')}\right)_{j' \in [j - 1]}\right) - 2^{j - 1}C^{1 - 1/2^{j - 1}}\right|\\
		& = \left|\widetilde{y}^{(j - 1)}\right|\left|\widetilde\zeta_C\left(0_{j - 1}\right)\right|^{1/2}\left|\frac{\widetilde\zeta_C\left(\left(\widetilde{y}^{(j')}\right)_{j' \in [j - 1]}\right) - \widetilde\zeta_C\left(0_{j - 1}\right)}{\widetilde\zeta_C\left(0_{j - 1}\right)} + 1\right|^{1/2}\\
		& \hspace*{20px} + 2|C|^{1/2^{j}}\left|\widetilde\zeta_C\left(\left(\widetilde{y}^{(j')}\right)_{j' \in [j - 1]}\right) - \widetilde\zeta_C\left(0_{j - 1}\right)\right|\\
		& \leq \mathcal{O}(1)|C|^{1/2}z_{\textrm{max}}|C|^{1/2 - 1/2^{j}}\left(1 + \mathcal{O}\left(z_{max}\right)\right) + 2|C|^{1/2^j}\mathcal{O}(1)|C|^{1 - 1/2^{j - 1}}z_{\textrm{max}}\\
		& \leq \mathcal{O}(1)|C|^{1 - 1/2^j}z_{\textrm{max}}.
	\end{align*}
	Inequalities \ref{eq:y_close_to_y_0} and \ref{eq:zeta_tilde_close_to_zeta_tilde_0} have now been proved by induction.

	Now, as an illustration of the recursive identities in lemma \ref{lemma:d_y_function_z_tilde}, let us prove for instance inequality \ref{eq:dy_close_to_dy_0}. We fix $j_0 \in [q]$ and argue by induction on $j_0 \leq j < q$. For $j = j_0$, $\partial_{j_0}\mathcal{Y}_C^{(j_0)} = 1$ is constant, so the inequality holds trivially. Next, assume the inequality to hold at levels below $j$. From lemma \ref{lemma:d_y_function_z_tilde},
	\begin{align*}
		\partial_{j_0}\mathcal{Y}_C^{(j)}\left(\left(\widetilde{y}^{(j')}\right)_{0 \leq j' \leq j}\right) & = 2iC^{1/2^{j + 1}}\sum\limits_{j' \in [j]}\frac{\widetilde{\zeta}_C\left(\left(\widetilde{y}^{\left(j''\right)}\right)_{j'' \in [j]}\right)^{1/2}}{2^{j - j'}}\frac{\partial_{j_0}\mathcal{Y}_C^{\left(j'\right)}\left(\left(\widetilde{y}^{\left(j''\right)}\right)_{0 \leq j'' \leq j'}\right)}{\mathcal{Y}_C^{\left(j'\right)}\left(\left(\widetilde{y}^{\left(j''\right)}\right)_{0 \leq j'' \leq j'}\right)}.
	\end{align*}
	Plugging this into the left-hand side of inequality \ref{eq:dy_close_to_dy_0} to be proven,
	\begin{align*}
		& \left|\partial_{j_0}\mathcal{Y}_C^{(j)}\left(\left(\widetilde{y}^{(j')}\right)_{0 \leq j' \leq j}\right) - \partial_{j_0}\mathcal{Y}_C^{(j)}\left(0_{j + 1}\right)\right|\\
		& \leq 2|C|^{1/2^{j + 1}}\sum_{j' \in [j]}\left|\frac{\widetilde{\zeta}_C\left(\left(\widetilde{y}^{\left(j''\right)}\right)_{j'' \in [j]}\right)^{1/2}}{2^{j - j'}}\frac{\partial_{j_0}\mathcal{Y}_C^{\left(j'\right)}\left(\left(\widetilde{y}^{\left(j''\right)}\right)_{0 \leq j'' \leq j'}\right)}{\mathcal{Y}_C^{\left(j'\right)}\left(\left(\widetilde{y}^{\left(j''\right)}\right)_{0 \leq j'' \leq j'}\right)} - \frac{\widetilde{\zeta}_C\left(0_j\right)^{1/2}}{2^{j - j'}}\frac{\partial_{j_0}\mathcal{Y}_C^{\left(j'\right)}\left(0_{j' + 1}\right)}{\mathcal{Y}_C^{\left(j'\right)}\left(0_{j' + 1}\right)}\right|\\
		& = 2|C|^{1/2^{j + 1}}\sum_{j' \in [j]}\left|\frac{\widetilde\zeta_C\left(0_j\right)^{1/2}\partial_{j_0}\mathcal{Y}^{(j')}(0_{j' + 1})}{2^{j - j'}\mathcal{Y}_C^{(j')}(0_{j' + 1})}\right|\nonumber\\
		& \hspace*{80px} \times \left|\frac{\widetilde\zeta_C\left(\left(\widetilde{y}^{(j'')}\right)_{j'' \in [j]}\right)^{1/2}}{\widetilde\zeta_C\left(0_j\right)^{1/2}}\frac{\partial_{j_0}\mathcal{Y}_C^{(j')}\left(\left(\widetilde{y}^{(j'')}\right)_{0 \leq j'' \leq j'}\right)}{\partial_{j_0}\mathcal{Y}_C^{(j')}\left(0_{j' + 1}\right)}\frac{1}{\frac{\mathcal{Y}_C^{(j')}\left(\left(\widetilde{y}^{(j'')}\right)_{0 \leq j'' \leq j'}\right)}{\mathcal{Y}_C^{(j')}\left(0_{j' + 1}\right)}} - 1\right|\\
		& = 2|C|^{1/2^{j + 1}}\sum_{j' \in [j]}\left|\frac{\widetilde\zeta_C\left(0_j\right)^{1/2}\partial_{j_0}\mathcal{Y}^{(j')}(0_{j' + 1})}{2^{j - j'}\mathcal{Y}_C^{(j')}(0_{j' + 1})}\right|\left|\left(1 + \frac{\widetilde\zeta_C\left(\left(\widetilde{y}^{\left(j''\right)}\right)_{j'' \in [j]}\right) - \widetilde\zeta_C(0_j)}{\widetilde\zeta_C(0_j)}\right)^{1/2}\right.\\
		& \left. \hspace*{20px} \times \left(1 + \frac{\partial_{j_0}\mathcal{Y}_C^{(j')}\left(\left(\widetilde{y}^{(j'')}\right)_{0 \leq j'' \leq j'}\right) - \partial_{j_0} \mathcal{Y}_C^{(j')}(0_{j' + 1})}{\partial_{j_0}\mathcal{Y}_C^{(j')}\left(0_{j' + 1}\right)}\right)\frac{1}{1 + \frac{\mathcal{Y}_C^{(j')}\left(\left(\widetilde{y}^{(j'')}\right)_{0 \leq j'' \leq j'}\right) -  \mathcal{Y}_C^{(j')}(0_{j' + 1})}{\mathcal{Y}_C^{(j')}\left(0_{j' + 1}\right)}} - 1\right|\\
		& \leq 2|C|^{1/2^{j + 1}}\sum_{j' \in [j]}\mathcal{O}(1)|C|^{-1/2^{j + 1}}\left|\left(1 + \mathcal{O}(1)z_{\textrm{max}}\right)\left(1 + \mathcal{O}(1)z_{\textrm{max}}\right)\frac{1}{1 + \mathcal{O}(1)z_{\textrm{max}}} - 1\right|\\
		& \leq \mathcal{O}(1)z_{\textrm{max}}.
	\end{align*}
	The induction is complete.
\end{proof}
\end{lem}

\subsubsection{Technical results on Gaussian integral tails}
\label{sec:gaussian_integral_tails}
In this section, we state and prove technical results bounding the tails of iterated Gaussian integrals that will be required in the proof of the main result proposition \ref{prop:exp_scaling_generalized_binomial_sum}. They will allow to upper bound the integrand arising from the integral representation \ref{eq:generalized_binomial_sum_integral_representation_simplified} of $S\left(\bm{A}, \bm{b}, \bm{c}\,;\,n\right)$ sufficiently far from its critical point, where we will show the quadratic Gaussian contributions to dominate. This generalizes the method applied to the Hamming weight squared toy example in section \ref{sec:hamming_weight_squared_example}, where the single-variable Gaussian integrand in $\theta$ was shown to be negliglible: $\exp\left(-\Omega\left(\frac{n}{\sqrt{\widetilde\gamma}}\right)\right)$ for $\theta$ bounded away from $\theta^*$ by $\Omega\left(\widetilde\gamma^{-1/4}\right)$ (and assuming $\widetilde\gamma$ sufficiently small).

The promised tail bound is ultimately proven in proposition \ref{prop:iterated_gaussian_tail} and relies on several preliminary lemmas. The following lemma starts by showing that the points $\left(y^{(0)}, \ldots, y^{(q - 1)}\right)$ of the hypersurface of $\mathbf{C}^q$ defined by the multivariate integral from proposition \ref{prop:iterated_gaussian_integral_generate_higher_order_gaussian} all lie in the upper complex half-plane. This technical assumption will be needed in the proof of the next proposition.
\begin{lem}
\label{lemma:iterated_gaussian_integration_positive_imaginary_parts}
Let $q \geq 1$ an integer and $C \in \mathbf{C} - \mathbf{R}_-$. Let $y^{(0)}, \ldots, y^{(q - 1)} \in \mathbf{C}$ satisfying the conditions in the integral bounds of proposition \ref{prop:iterated_gaussian_integral_generate_higher_order_gaussian}:
\begin{align}
    \forall j \in [q],\,y^{(j)} \in \mathbf{R} + 2iC^{1/2^{j + 1}}\zeta\left(\left(y^{(j')}\right)_{j' \in [j]}\right)^{1/2}.
\end{align}
Then the following holds for all $j \in [q]$:
\begin{align}
    \Re\left(-iy^{(j)}\right) & > 0,\\
    \Re\left(2C^{1/2^{j + 1}}\zeta\left(\left(y^{(j')}\right)_{j' \in [j]}\right)^{1/2}\right) & > 0.
\end{align}
\begin{proof}
This can be proved by explicitly writing $\zeta\left(\left(y^{(j')}\right)_{j' \in [j]}\right)^{1/2} = \prod_{j' \in [j]}\left(-iy^{(j')}\right)^{1/2^{j - j'}}$ and showing by induction on $j \in [q]$ that $\Re\left(-iy^{(j)}\right) > 0$. Namely, for $j = 0$, $-iy^{(0)} = 2C^{1/2} - i\widetilde{y}^{(0)}$ for some $\widetilde{y}^{(0)} \in \mathbf{R}$, so that $\Re\left(-iy^{(0)}\right) > 0$. Now, assuming $\Re\left(-iy^{(j')}\right) > 0$ for all $j' < j$, $-iy^{(j)} = 2C^{1/2^{j + 1}}\prod_{j' \in [j]}\left(-iy^{(j')}\right)^{1/2^{j - j'}} - i\widetilde{y}^{(j)}$ for some $\widetilde{y}^{(j)} \in \mathbf{R}$. By the induction hypothesis,
\begin{align*}
    & \left|\arg\left(2C^{1/2^{j + 1}}\prod_{j' \in [j]}\left(-iy^{(j')}\right)^{1/2^{j - j'}}\right)\right|\\
    & \leq \left|\arg\left(C^{1/2^{j + 1}}\right)\right| + \sum_{j' \in [j]}\left|\arg\left(\left(-iy^{(j')}\right)^{1/2^{j - j'}}\right)\right|\\
    & < \frac{\pi}{2^{j + 1}} + \sum_{j' \in [j]}\frac{\pi/2}{2^{j - j'}}\\
    & = \frac{\pi}{2},
\end{align*}
so $\Re\left(2C^{1/2^{j + 1}}\prod\limits_{j' \in [j]}\left(-iy^{(j')}\right)^{1/2^{j - j'}}\right) > 0$ and the same holds for $-iy^{(j)} = 2C^{1/2^{j + 1}}\prod\limits_{j' \in [j]}\left(-iy^{(j')}\right)^{1/2^{j - j'}} - i\widetilde{y}^{(j)}$.
\end{proof}
\end{lem}
Armed with this lemma, we can state the following upper bound on an iterated Gaussian integral performed over all the domain introduced in proposition \ref{prop:iterated_gaussian_integral_generate_higher_order_gaussian}. Note this is not a tail bound yet since we are integrating over the whole domain. Anticipating the final proposition \ref{prop:iterated_gaussian_tail}, the intuitive reason why we need this result to prove tail bounds is because condition $\left|\zeta\left(\left(y^{(j)}\right)_{j \in [q]}\right)\right| \geq \max\left(c, |C|\right)^{-1/2^{q + 1}}$ in the integral there will imply that at least one, but not necessarily all, $y^{(j)}$ needs to be large. We will have no choice but to integrate over the whole domain for the $y^{(j)}$ that may be small, hence the need for the next result.
\begin{prop}
\label{prop:iterated_gaussian_integral_eliminate_variables}
Let $q \geq 1$ an integer; let $c > 0$ and $C \in \mathbf{C} - \mathbf{R}_-$ constants. Then the following bound holds:
\begin{align}
    & \frac{1}{\left(4\pi\right)^{q/2}}\int\limits_{\forall j \in [q],\,y^{(j)} \in \mathbf{R} + 2iC^{1/2^{j + 1}}\zeta\left(\left(y^{(j')}\right)_{j' \in [j]}\right)^{1/2}}\!\prod_{j \in [q]}\mathrm{d}y^{(j)}\,\exp\left(-\frac{1}{4}\sum_{j \in [q]}\Re\left\{\left(y^{(j)}\right)^2\right\} + c^{1/2^q}\prod_{j \in [q]}\left|y^{(j)}\right|^{1/2^{q - 1 - j}}\right)\nonumber\\
    & \leq 2^q\exp\left(2^{2^{q + 1} - 2}\max\left\{c, |C|\right\}\right).\label{eq:iterated_gaussian_eliminate_variables}
\end{align}
As a corollary, for any constant $K > 0$,
\begin{align}
	& \frac{1}{\left(4\pi\right)^{q/2}}\hspace*{-30px}\int\limits_{\forall j \in [q],\,y^{(j)} \in \mathbf{R} + 2iC^{1/2^{j + 1}}\zeta\left(\left(y^{(j')}\right)_{j' \in [j]}\right)^{1/2}}\!\prod_{j \in [q]}\mathrm{d}y^{(j)}\,\exp\left\{K\left(-\frac{1}{4}\sum_{j \in [q]}\Re\left\{\left(y^{(j)}\right)^2\right\} + c^{1/2^q}\prod_{j \in [q]}\left|y^{(j)}\right|^{1/2^{q - 1 - j}}\right)\right\}\nonumber\\
	& \leq \left(2/\sqrt{K}\right)^q\exp\left(2^{2^{q + 1} - 2}\max\left\{c, |C|\right\}K\right).\label{eq:iterated_gaussian_eliminate_variables_rescaling}
\end{align}
\begin{proof}
This can be proven by induction. Indeed, integrating first over $y^{(q - 1)}$,
\begin{align*}
    & \int\limits_{\mathbf{R} + 2iC^{1/2^q}\zeta\left(\left(y^{(j')}\right)_{j' \in [q]}\right)^{1/2}}\hspace*{-40px}\mathrm{d}y^{(q - 1)}\,\exp\left(-\frac{1}{4}\Re\left\{\left(y^{(q - 1)}\right)^2\right\} - \frac{1}{4}\sum_{j \in [q - 1]}\Re\left\{\left(y^{(j)}\right)^2\right\} + c^{1/2^q}\left|y^{(q - 1)}\right|\prod_{j \in [q - 1]}\left|y^{(j)}\right|^{1/2^{q - 1 - j}}\right)\\
    & =\int_{\mathbf{R}}\!\mathrm{d}y^{(q - 1)}\,\exp\left(-\frac{1}{4}\Re\left\{\left(y^{(q - 1)} + 2iC^{1/2^q}\zeta\left(\left(y^{(j')}\right)_{j' \in [q - 1]}\right)^{1/2}\right)^2\right\} - \frac{1}{4}\sum_{j \in [q - 1]}\Re\left\{\left(y^{(j)}\right)^2\right\}\right.\nonumber\\
    & \left. \hspace*{100px} + c^{1/2^q}\left|y^{(q - 1)} + 2iC^{1/2^q}\zeta\left(\left(y^{(j')}\right)_{j' \in [q - 1]}\right)^{1/2}\right|\prod_{j \in [q - 1]}\left|y^{(j)}\right|^{1/2^{q - 1 - j}}\right)\\
    & = \int_{\mathbf{R}}\!\mathrm{d}y^{(q - 1)}\,\exp\left(-\frac{1}{4}\Re\left\{\left(y^{(q - 1)} + 2i\Re\left[C^{1/2^q}\zeta\left(\left(y^{(j')}\right)_{j' \in [q - 1]}\right)^{1/2}\right]\right)^2\right\} - \frac{1}{4}\sum_{j \in [q - 1]}\Re\left\{\left(y^{(j)}\right)^2\right\}\right.\nonumber\\
    & \left. \hspace*{100px} + c^{1/2^q}\left|y^{(q - 1)} + 2i\Re\left[C^{1/2^q}\zeta\left(\left(y^{(j')}\right)_{j' \in [q - 1]}\right)^{1/2}\right]\right|\prod_{j \in [q - 1]}\left|y^{(j)}\right|^{1/2^{q - 1 - j}}\right)\\
    & = \exp\left(-\frac{1}{4}\sum_{j \in [q  - 1]}\left(y^{(j)}\right)^2\right)\int_{\mathbf{R}}\!\mathrm{d}y^{(q - 1)}\,\exp\left(-\frac{\left(y^{(q - 1)}\right)^2 - t^2}{4} + c^{1/2^q}\left(\left|y^{(q - 1)}\right| + t\right)\prod_{j \in [q - 1]}\left|y^{(j)}\right|^{1/2^{q - 1 - j}}\right),
\end{align*}
where we temporarily let
\begin{align*}
    t & := 2\Re\left(C^{1/2^q}\zeta\left(\left(y^{(j')}\right)_{j' \in [q - 1]}\right)^{1/2}\right)
\end{align*}
to simplify the notation. We observe that $t > 0$ by lemma \ref{lemma:iterated_gaussian_integration_positive_imaginary_parts}. We then separate the integration between $\mathbf{R}_+$ and $\mathbf{R}_-$. For instance, for $\mathbf{R}_+$,
\begin{align*}
    & \int_{\mathbf{R}_+}\!\mathrm{d}y^{(q - 1)}\,\exp\left(-\frac{\left(y^{(q - 1)}\right)^2 - t^2}{4} + c^{1/2^q}\left(\left|y^{(q - 1)}\right| + t\right)\prod_{j \in [q - 1]}\left|y^{(j)}\right|^{1/2^{q - 1 - j}}\right)\\
    & = \int_{\mathbf{R}+}\!\mathrm{d}y^{(q - 1)}\,\exp\left(-\frac{\left(y^{(q - 1)} + t\right)^2}{4} + \left(y^{(q - 1)} + t\right)\left(\frac{t}{2} + c^{1/2^q}\prod_{j \in [q - 1]}\left|y^{(j)}\right|^{1/2^{q - 1 - j}}\right)\right)\\
    & \leq \int_{\mathbf{R}}\!\mathrm{d}y\,\exp\left(-\frac{y^2}{4} + y\left(\frac{t}{2} + c^{1/2^q}\prod_{j \in [q - 1]}\left|y^{(j)}\right|^{1/2^{q - 1 - j}}\right)\right)\\
    & = \sqrt{4\pi}\exp\left\{\left(\frac{t}{2} + c^{1/2^q}\prod_{j \in [q - 1]}\left|y^{(j)}\right|^{1/2^{q - 1 - j}}\right)^2\right\}\\
    & \leq \sqrt{4\pi}\exp\left(\frac{t^2}{2} + 2c^{1/2^{q - 1}}\prod_{j \in [q - 1]}\left|y^{(j)}\right|^{1/2^{q - 2 - j}}\right).
\end{align*}
Proceeding similarly for $\mathbf{R}_-$, we finally obtain:
\begin{align*}
    & \int\limits_{\mathbf{R} + 2iC^{1/2^q}\zeta\left(\left(y^{(j')}\right)_{j' \in [q]}\right)^{1/2}}\hspace*{-40px}\mathrm{d}y^{(q - 1)}\,\exp\left(-\frac{1}{4}\Re\left\{\left(y^{(q - 1)}\right)^2\right\} - \frac{1}{4}\sum_{j \in [q - 1]}\Re\left\{\left(y^{(j)}\right)^2\right\} + c^{1/2^q}\left|y^{(q - 1)}\right|\prod_{j \in [q - 1]}\left|y^{(j)}\right|^{1/2^{q - 1 - j}}\right)\\
    & \leq 2\sqrt{4\pi}\exp\left(\frac{t^2}{2} - \frac{1}{4}\sum_{j \in [q - 1]}\Re\left\{\left(y^{(j)}\right)^2\right\} + 2c^{1/2^{q - 1}}\prod_{j \in [q - 1]}\left|y^{(j)}\right|^{1/2^{q - 2 - j}}\right).\\
    & \leq 2\sqrt{4\pi}\exp\left(2|C|^{1/2^{q - 1}}\left|\zeta\left(\left(y^{(j')}\right)_{j' \in [q - 1]}\right)\right| - \frac{1}{4}\sum_{j \in [q - 1]}\Re\left\{\left(y^{(j)}\right)^2\right\} + 2c^{1/2^{q - 1}}\prod_{j \in [q - 1]}\left|y^{(j)}\right|^{1/2^{q - 2 - j}}\right)\\
    & = 2\sqrt{4\pi}\exp\left(-\frac{1}{4}\sum_{j \in [q - 1]}\Re\left\{\left(y^{(j)}\right)^2\right\} + 2\left(|C|^{1/2^{q - 1}} + c^{1/2^{q - 1}}\right)\prod_{j \in [q - 1]}\left|y^{(j)}\right|^{1/2^{q - 2 - j}}\right)\\
    & \leq 2\sqrt{4\pi}\exp\left(-\frac{1}{4}\sum_{j \in [q - 1]}\Re\left\{\left(y^{(j)}\right)^2\right\} + 4\max\left(c, |C|\right)^{1/2^{q - 1}}\prod_{j \in [q - 1]}\left|y^{(j)}\right|^{1/2^{q - 2 - j}}\right).
\end{align*}
Observe that the same final bound would have held had we replaced $c$ by $\max\left(c, |C|\right)$ from the onset. Therefore, one may assume $c \geq |C|$. In this case, the integration over $y^{(q - 1)}$ reduced the problem at level $q$ to the one at level $q - 1$ up to the following transformation of constant $c$:
\begin{align*}
    c & \longrightarrow 4c^2.
\end{align*}
After $q \geq 0$ iterations, this transformation gives $2^{2^{q + 1} - 2}c^{2^q}$. The result in equation \ref{eq:iterated_gaussian_eliminate_variables} then follows. To deduce the corollary in equation \ref{eq:iterated_gaussian_eliminate_variables_rescaling}, change variables $y^{(j)} \longrightarrow \frac{y^{(j)}}{\sqrt{K}}$ and apply the former result with $c \longrightarrow Kc$ and $C \longrightarrow KC$ to bound the resulting integral.
\end{proof}
\end{prop}
We have now proven proposition \ref{prop:iterated_gaussian_integral_eliminate_variables} which bounds an iterative Gaussian integral performed over the entire domain. The only missing ingredient for the proof of the final proposition \ref{sec:gaussian_integral_tails} is a simple bound on a similar integral where a variable is constrained to be large. In fact, the following bound on a the tail of a single-variable Gaussian (integrated along a horizontal line in the complex upper-half plane) will suffice for that purpose:
\begin{lem}
\label{lemma:complex_gaussian_tail_integral}
Let $c > 0$, $t > 0$ and $y_{\mathrm{min}} > 0$ constants satisfying:
\begin{align}
    y_{\mathrm{min}} & > 4t + 8c.
\end{align}
Then the following bound holds:
\begin{align}
   \int_{\substack{y \in \mathbf{R} + it\\|y| > y_{\mathrm{min}}}}\!\mathrm{d}y\,\exp\left(-\frac{\Re\left(y^2\right)}{4} + c|y|\right) & \leq \frac{2}{3\left(c + t/2\right)}\exp\left(-\frac{y_{\mathrm{min}}^2}{8}\right).
\end{align}
\begin{proof}
\begin{align*}
    & \int_{\substack{y \in \mathbf{R} + it\\|y| > y_{\mathrm{min}}}}\!\mathrm{d}y\,\exp\left(-\frac{\Re\left(y^2\right)}{4} + c|y|\right)\\
    & = \int_{\substack{y \in \mathbf{R}\\\left|y + it\right| > y_{\mathrm{min}}}}\!\mathrm{d}y\,\exp\left(-\frac{y^2 - t^2}{4} + c\left|y + it\right|\right)\\
    & \leq \int_{\substack{y \in \mathbf{R}\\\left|y\right| > y_{\mathrm{min}} - t}}\!\mathrm{d}y\,\exp\left(-\frac{y^2 - t^2}{4} + c\left(\left|y\right| + t\right)\right).
\end{align*}
We now separate the integral over one on $\mathbf{R}_+$ and one on $\mathbf{R}_-$. For instance, for the $\mathbf{R}_+$ one,
\begin{align*}
    & \int_{\substack{y \in \mathbf{R}_+\\\left|y\right| > y_{\mathrm{min}} - t}}\!\mathrm{d}y\,\exp\left(-\frac{y^2 - t^2}{4} + c\left(\left|y\right| + t\right)\right)\\
    & = \int_{\substack{y \in \mathbf{R}_+\\y > y_{\mathrm{min}} - t}}\!\mathrm{d}y\,\exp\left(-\frac{(y + t)^2}{4} + (y + t)\left(\frac{t}{2} + c\right)\right)\\
    & = \int_{y_{\mathrm{min}}}^{+\infty}\!\mathrm{d}y\,\exp\left(-\frac{y^2}{4} + \left(\frac{t}{2} + c\right)y\right).
\end{align*}
Now, using that $y \longmapsto -\frac{y^2}{4} + \left(\frac{t}{2} + c\right)y$ is concave with maximum attained at $y = t + 2c < y_{\mathrm{min}}$, the argument of the exponential for $y > y_{\mathrm{min}}$ can be bounded as follows using the values of the function and its derivative at $y_{\mathrm{min}}$:
\begin{align*}
    & -\frac{y^2}{4} + \left(\frac{t}{2} + c\right)y\\
    & \leq -\frac{y_{\mathrm{min}}^2}{4} + \left(\frac{t}{2} + c\right)y_{\mathrm{min}} + \left(-\frac{y_{\mathrm{min}}}{2} + \frac{t}{2} + c\right)\left(y - y_{\mathrm{min}}\right)\\
    & \leq -\frac{y_{\mathrm{min}}^2}{4}\left(1 - \frac{2t + 4c}{y_{\mathrm{min}}}\right) - \frac{y_{\mathrm{min}}}{2}\left(1 - \frac{t + 2c}{y_{\mathrm{min}}}\right)\left(y - y_{\mathrm{min}}\right)\\
    & \leq -\frac{y_{\mathrm{min}}^2}{8} - \frac{3y_{\mathrm{min}}}{8}\left(y - y_{\mathrm{min}}\right) \qquad \left(y_{\mathrm{min}} > 4t + 8c\right)\\
    & \leq -\frac{y_{\mathrm{min}}^2}{8} - 3\left(c + \frac{t}{2}\right)\left(y - y_{\mathrm{min}}\right).
\end{align*}
Hence,
\begin{align*}
    \int_{y_{\mathrm{min}}}^{+\infty}\!\mathrm{d}y\,\exp\left(-\frac{y^2}{4} + \left(\frac{t}{2} + c\right)y\right) & \leq \frac{1}{3\left(c + t/2\right)}\exp\left(-\frac{y_{\mathrm{min}}^2}{8}\right).
\end{align*}
The integration over $\mathbf{R}_-$ is similar and the final bound is twice the above.
\end{proof}
\end{lem}
Having proven propositions \ref{prop:iterated_gaussian_integral_eliminate_variables} giving a crude bound on an iterated Gaussian integral and lemma \ref{lemma:complex_gaussian_tail_integral} bounding the tail of a single-variable Gaussian integral, we are now in position to prove the main result of this section: a tail bound on an iterated Gaussian integral of variables $\left(y^{(j)}\right)_{j \in [q]}$ where the quantity $\left|\zeta\left(\left(y^{(j}\right)_{j \in [q]}\right)\right|$ is constrained to be sufficiently large. As alluded before the proof of proposition \ref{prop:iterated_gaussian_integral_eliminate_variables}, the main idea of the proof is to show that the last condition implies at least one the variables $y^{(j)}$ is large. We then use a union bound over such events. On the event where $y^{(j)}$ is large, we then apply proposition \ref{prop:iterated_gaussian_integral_eliminate_variables} to eliminate other variables; the resulting upper bound will be a single-variable Gaussian integral in $y^{(j)}$, which can be bounded via lemma \ref{lemma:complex_gaussian_tail_integral}.
\begin{prop}
\label{prop:iterated_gaussian_tail}
Let $q \geq 1$ an integer; let $c \in \left(0, \frac{1}{16}\right)$ and $C \in \mathbf{C} - \mathbf{R}_-$. Then the following bound holds:
\begin{align}
    & \frac{1}{\left(4\pi\right)^{q/2}}\hspace*{-50px}\int\limits_{\substack{
    \mathcal{D}_{q, C}\\\left|\zeta\left(\left(y^{(j)}\right)_{j \in [q]}\right)\right| \geq \max\left(c, |C|\right)^{-1/2^{q + 1}}}}\hspace*{-50px}\prod_{j \in [q]}\mathrm{d}y^{(j)}\,\exp\left(-\frac{1}{4}\sum_{j \in [q]}\Re\left\{\left(y^{(j)}\right)^2\right\} + c^{1/2^q}\prod_{j \in [q]}\left|y^{(j)}\right|^{1/2^{q - 1 - j}}\right)\nonumber\\
    & \leq \frac{2^qq}{3\max\left(c, |C|\right)^{3/2^{q + 2}}\sqrt{4\pi}}\exp\left(-\frac{1}{16\max\left(c, |C|\right)^{1/2^{q + 1}}}\right).
\end{align}
\begin{proof}
We first decompose the integration domain
\begin{align*}
    \mathcal{D} & := \left\{\bm{y} := \left(y^{(0)}, \ldots, y^{(q - 1)}\right) \in \mathcal{D}_{q, C}\,:\,\left|\zeta\left(\left(y^{(j)}\right)_{j \in [q]}\right)\right| \geq \max\left(c, |C|\right)^{-1/2^{q + 1}}\right\}
\end{align*}
into disjoint domains as follows. Given $\bm{y} \in \mathcal{D}$, we denote by $j_0(\bm{y})$ the smallest $j \in [q]$ such that $\left|\zeta\left(\left(y^{(j')}\right)_{0 \leq j' \leq j}\right)\right| \geq \max\left(c, |C|\right)^{-1/2^{q + 1}}$. We partition $\mathcal{D}$ according to the value of $j_0(\bm{y})$: $\mathcal{D} := \bigsqcup_{j \in [q]}\mathcal{D}_j$, where $\mathcal{D}_j := \left\{\bm{y} \in \mathcal{D}\,:\,j_0(\bm{y}) = j\right\}$.

Let us now bound the integral on $\mathcal{D}_j$, first assuming $j > 0$ for simplicity:
\begin{align*}
    & \frac{1}{(4\pi)^{q/2}}\int_{\mathcal{D}_j}\!\prod_{j' \in [q]}\mathrm{d}y^{(j')}\,\exp\left(-\frac{1}{4}\sum_{j' \in [q]}\Re\left\{\left(y^{(j')}\right)^2\right\} + c^{1/2^q}\prod_{j' \in [q]}\left|y^{(q)}\right|^{1/2^{q - 1 - j'}}\right)\\
    & \leq \frac{2^{q - 1 - j}}{(4\pi)^{(j + 1)/2}}\exp\left(2^{2^{q - j} - 2}\max\{c, |C|\}^{1/2^{j + 1}}\right)\hspace*{-50px}\int\limits_{\substack{\mathcal{D}_{j + 1, C}\\\left|\zeta\left(\left(y^{(j')}\right)_{0 \leq j' \leq j}\right)\right| \geq \max\left(c, |C|\right)^{-1/2^{q + 1}}\\\left|\zeta\left(\left(y^{(j')}\right)_{0 \leq j' < j}\right)\right| < \max\left(c, |C|\right)^{-1/2^{q + 1}}}}\hspace*{-60px}\prod_{0 \leq j' \leq j}\mathrm{d}y^{(j')}\exp\left(-\frac{1}{4}\sum_{0 \leq j' \leq j}\Re\left\{\left(y^{(j')}\right)^2\right\}\right.\\
    & \left. \hspace*{30px} + 2^{2^{q - j} - 2}\max\{c, |C|\}^{1/2^{j + 1}}\prod_{0 \leq j' \leq j}\left|y^{(j')}\right|^{1/2^{j - j'}}\right),
\end{align*}
where we used proposition \ref{prop:iterated_gaussian_integral_eliminate_variables} to eliminate variables $y^{(j + 1)}, \ldots, y^{(q - 1)}$. Next, using the condition
\begin{align*}
    \left|\zeta\left(\left(y^{(j')}\right)_{0 \leq j' < j}\right)\right| = \prod_{0 \leq j' < j}\left|y^{(j')}\right|^{1/2^{j - 1 - j'}} < \max\left(c, |C|\right)^{-1/2^{q + 1}}
\end{align*}
in the integral, we bound the integrand by:
\begin{align*}
    \exp\left(-\frac{1}{4}\sum_{0 \leq j' \leq j}\Re\left\{\left(y^{(j')}\right)^2\right\} + 2^{2^{q - j} - 2}\max\{c, |C|\}^{1/2^{j + 1} - 1/2^{q + 2}}\left|y^{(j)}\right|\right).
\end{align*}
Besides, we enlarge the integration domain to:
\begin{align*}
    \left\{\left(y^{(0)}, \ldots, y^{(j)}\right) \in 
    \mathcal{D}_{j + 1, C}\,:\,\left|y^{(j)}\right| > \max\left(c, |C|\right)^{-1/2^{q + 2}}\right\}.
\end{align*}
The last constraint for $y^{(j)}$ results from combining constraints
\begin{align*}
    \left|\zeta\left(\left(y^{(j')}\right)_{0 \leq j' < j}\right)\right| = \prod_{0 \leq j' < j}\left|y^{(j')}\right|^{1/2^{j - 1 - j'}} < \max\left(c, |C|\right)^{-1/2^{q + 1}}
\end{align*}
and
\begin{align*}
    \left|\zeta\left(\left(y^{(j')}\right)_{0 \leq j' \leq j}\right)\right| = \prod_{0 \leq j' \leq j}\left|y^{(j')}\right|^{1/2^{j - 1 - j'}} \geq \max\left(c, |C|\right)^{-1/2^{q + 1}}
\end{align*}
from the original integral. Next, we integrate over $y^{(0)}, \ldots, y^{(j - 1)}$ using e.g. proposition \ref{prop:iterated_gaussian_integral_eliminate_variables} with $c = 0$. This gives the bound:
\begin{align*}
    & \leq \frac{2^{q - 1}}{\sqrt{4\pi}}\exp\left(2^{2^{j + 1} - 2}|C|^{1/2^{q - j}}\right)\hspace*{-40px}\int\limits_{\substack{y^{(j)} \in \mathbf{R} + 2iC^{1/2^{j + 1}}\zeta\left(\left(y^{(j')}\right)_{j' \in [j]}\right)^{1/2}\\|y^{(j)}| \geq \max\left(c, |C|\right)^{-1/2^{q + 2}}}}\hspace*{-40px}\mathrm{d}y^{(j)}\exp\left(-\frac{\Re\left\{\left(y^{(j)}\right)^2\right\}}{4} + 2^{2^{q - j} - 2}\max\{c, |C|\}^{1/2^{j + 1} - 1/2^{q + 2}}\left|y^{(j)}\right|\right)\\
    & \leq \frac{2^{q - 1}}{\sqrt{4\pi}}\exp\left(2^{2^{j + 1} - 2}|C|^{1/2^{q - j}}\right)\hspace*{-30px}\int\limits_{\substack{y^{(j)} \in \mathbf{R} + 2iC^{1/2^{j + 1}}\zeta\left(\left(y^{(j')}\right)_{j' \in [j]}\right)^{1/2}\\|y^{(j)}| \geq \max\left(c, |C|\right)^{-1/2^{q + 2}}}}\hspace*{-40px}\mathrm{d}y^{(j)}\exp\left(-\frac{\Re\left\{\left(y^{(j)}\right)^2\right\}}{4} + \max\{c, |C|\}^{1/2^q - 1/2^{q + 2}}\left|y^{(j)}\right|\right).
\end{align*}
In the last inequality, we used $\max\left\{c, |C|\right\}^{1/2^q} < \frac{1}{4} \implies 2^{2^{q - j}}\left(\max\{c, |C|\}^{1/2^q}\right)^{2^{q - 1 - j}} = \left(4\max\{c, |C|\}^{1/2^q}\right)^{2^{q - 1 - j}} < 4\max\left\{c, |C|\right\}^{1/2^q}$. Now, we can apply lemma \ref{lemma:complex_gaussian_tail_integral} to bound the single-variable Gaussian tail integral, giving the final estimate:
\begin{align*}
    & \frac{1}{(4\pi)^{q/2}}\int_{\mathcal{D}_j}\!\prod_{j' \in [q]}\mathrm{d}y^{(j')}\,\exp\left(-\frac{1}{4}\sum_{j' \in [q]}\Re\left\{\left(y^{(j')}\right)^2\right\} + c^{1/2^q}\prod_{j' \in [q]}\left|y^{(q)}\right|^{1/2^{q - 1 - j'}}\right)\\
    & \leq \frac{2^q}{3\max\left(c, |C|\right)^{1/2^q - 1/2^{q + 2}}\sqrt{4\pi}}\exp\left(\max\{c, |C|\}^{1/2^q} - \frac{1}{8\max(c, |C|)^{1/2^{q + 1}}}\right)\\
    & \leq \frac{2^q}{3\max\left(c, |C|\right)^{1/2^q - 1/2^{q + 2}}\sqrt{4\pi}}\exp\left(-\frac{1}{16\max(c, |C|)^{1/2^{q + 1}}}\right).
\end{align*}
The condition for applying lemma \ref{lemma:complex_gaussian_tail_integral}, which is here equivalent to
\begin{align*}
    \max\left(c, |C|\right)^{-1/2^{q + 2}} & > 8\Re\left\{C^{1/2^{j + 1}}\zeta\left(\left(y^{(j')}\right)_{j' \in [j]}\right)^{1/2}\right\} + 8\max\left(c, |C|\right)
\end{align*}
is satisfied, since the right-hand side is upper bounded by
\begin{align*}
    & 8\left|C\right|^{1/2^{j + 1}}\left|\zeta\left(\left(y^{(j')}\right)_{j' \in [j]}\right)\right|^{1/2} + 8\max\left(c, |C|\right)\\
    & \leq 8\max\left(c, |C|\right)^{1/2^{j + 1} - 1/2^{q + 2}} + 8\max\left(c, |C|\right),
\end{align*}
which is smaller than the left-hand side indeed for sufficiently small $c, |C|$.

For the integral over $\mathcal{D}_0$, we obtain a similar bound. This completes the proof.
\end{proof}
\end{prop}

\subsubsection{The quadratic form of the Gaussian approximation in the power-of-two case}
\label{sec:quadratic_form_power_two_case}

In this section, we estimate the quadratic form that arises in the second-order Taylor expansion of the integrand around its critical point in the proof of proposition \ref{prop:exp_scaling_generalized_binomial_sum}. These allow to prove that the critical point indeed globally maximizes the exponential over all the integration domain of $I_0$ defined in the proof of the proposition, providing a rigorous justification for the saddle-point method. To achieve this, we essentially need two results: one controlling the quadratic form close (in fact, within distance $\mathcal{O}\left(c^{1/2}\right)$ as we will see) to the critical point, and one controlling the same quantity far from the critical point (up to distance $\Omega\left(c^{1/2^{q + 1}}\right)$).

For convenience, we express in the following lemma the quadratic form evaluated at an arbitrary point:
\begin{lem}[Quadratic form in terms of $\widetilde{\zeta}_{\alpha}$ and $F$ derivatives]
\label{lemma:quadratic_form_general}
Let $\widetilde{\Phi}$ be defined as in the proof of proposition \ref{prop:exp_scaling_generalized_binomial_sum}. The second derivatives of $\widetilde{\Phi}$ can be expressed as follows:
\begin{align}
	& \frac{\partial^2\widetilde{\Phi}}{\partial\widetilde{y}^{(j_0)}_{\alpha_0}\partial\widetilde{y}^{(j_1)}_{\alpha_1}}\left(\left(\widetilde{y}^{(j)}_{\alpha}\right)_{\substack{\alpha \in \mathcal{A}\\j \in [q]}}\right)\nonumber\\
	& = -\frac{\delta_{\alpha_0\alpha_1}\delta_{j_0j_1}}{2} + \delta_{\alpha_0\alpha_1}\partial_{j_0j_1}\widetilde\zeta_{\alpha_0}\left(\left(\widetilde{y}^{(j)}_{\alpha_0}\right)_{j \in [q]}\right)\left[\partial_{\alpha_0}F\left(\left(\widetilde{\zeta}_{\alpha}\left(\left(\widetilde{y}^{(j)}_{\alpha}\right)_{j \in [q]}\right)\right)_{\alpha \in \mathcal{A}}\right) - \partial_{\alpha_0}F(z^*)\right]\nonumber\\
    & \hspace*{20px} + \partial_{j_0}\widetilde{\zeta}_{\alpha_0}\left(\left(\widetilde{y}^{(j)}_{\alpha_0}\right)_{j \in [q]}\right)\partial_{j_1}\widetilde{\zeta}_{\alpha_1}\left(\left(\widetilde{y}^{(j)}_{\alpha_1}\right)_{j \in [q]}\right)\partial_{\alpha_0\alpha_1}F\left(\left(\widetilde{\zeta}_{\alpha}\left(\left(\widetilde{y}^{(j)}_{\alpha}\right)_{j \in [q]}\right)\right)_{\alpha \in \mathcal{A}}\right).\label{eq:quadratic_form_phi_tilde}
\end{align}

\begin{proof}
To simplify the notation, we omit to write variables $\left(\widetilde{y}^{(j)}_{\alpha}\right)_{\substack{\alpha \in \mathcal{A}\\j \in [q]}}$ as function arguments where the context is evident; for instance, it will be implied that $\mathcal{Y}^{(j)}_{\alpha}$ is to be evaluated at variables $\left(\widetilde{y}^{(j)}_{\alpha}\right)_{j \in [q]}$ and $\widetilde{\zeta}_{\alpha}$ is to be evaluated at variables $\left(\widetilde{y}^{(j)}_{\alpha}\right)_{j \in [q]}$.

\begin{align*}
    & \frac{\partial^2\widetilde\Phi}{\partial\widetilde{y}^{(j_0)}_{\alpha_0}\partial\widetilde{y}^{(j_1)}_{\alpha_1}}\left(\left(\widetilde{y}^{(j)}_{\alpha}\right)_{\substack{\alpha \in \mathcal{A}\\j \in [q]}}\right)\\
    & = \frac{\partial^2}{\partial\widetilde{y}^{(j_0)}_{\alpha_0}\partial\widetilde{y}^{(j_1)}_{\alpha_1}}\left\{-\frac{1}{4}\sum_{\substack{\alpha \in \mathcal{A}\\j \in [q]}}\mathcal{Y}_{\alpha}^{(j)}\left(\left(\widetilde{y}^{(j)}_{\alpha}\right)_{\substack{\alpha \in \mathcal{A}\\j \in [q]}}\right)^2 + F\left(\left(\widetilde\zeta_{\alpha}\left(\left(\widetilde{y}^{(j)}_{\alpha}\right)_{j \in [q]}\right)\right)_{\alpha \in \mathcal{A}}\right)\right\}\\
    & = \frac{\partial^2}{\partial\widetilde{y}^{(j_0)}_{\alpha_0}\partial\widetilde{y}^{(j_1)}_{\alpha_1}}\left\{\sum_{\alpha \in \mathcal{A}}\left[-\frac{1}{4}\sum_{\substack{j \in [q]}}\mathcal{Y}_{\alpha}^{(j)}\left(\left(\widetilde{y}^{(j)}_{\alpha}\right)_{\substack{\alpha \in \mathcal{A}\\j \in [q]}}\right)^2 - C_{\alpha}^{1/2^q}\widetilde\zeta_{\alpha}\left(\left(\widetilde{y}^{(j)}_{\alpha}\right)_{j \in [q]}\right) + C_{\alpha}^{1/2^q}\widetilde\zeta_{\alpha}\left(\left(\widetilde{y}^{(j)}_{\alpha}\right)_{j \in [q]}\right)\right]\right.\\
    & \hspace*{100px} + F\left(\left(\widetilde\zeta_{\alpha}\left(\left(\widetilde{y}^{(j)}_{\alpha}\right)_{j \in [q]}\right)\right)_{\alpha \in \mathcal{A}}\right)\Bigg\}\\
    & = \frac{\partial^2}{\partial\widetilde{y}^{(j_0)}_{\alpha_0}\partial\widetilde{y}^{(j_1)}_{\alpha_1}}\left\{\sum_{\alpha \in \mathcal{A}}\left[-\frac{1}{4}\sum_{\substack{j \in [q]}}\left(\widetilde{y}^{(j)}_{\alpha}\right)^2 - C_{\alpha} + C_{\alpha}^{1/2^q}\widetilde\zeta_{\alpha}\left(\left(\widetilde{y}^{(j)}_{\alpha}\right)_{j \in [q]}\right)\right] + F\left(\left(\widetilde\zeta_{\alpha}\left(\left(\widetilde{y}^{(j)}_{\alpha}\right)_{j \in [q]}\right)\right)_{\alpha \in \mathcal{A}}\right)\right\}\\
    & = \frac{\partial^2}{\partial\widetilde{y}^{(j_0)}_{\alpha_0}\partial\widetilde{y}^{(j_1)}_{\alpha_1}}\left\{\sum_{\alpha \in \mathcal{A}}\left[-\frac{1}{4}\sum_{\substack{j \in [q]}}\left(\widetilde{y}^{(j)}_{\alpha}\right)^2 - \left(\partial_{\alpha}F\left(z^*\right)\right)^{2^q} - \partial_{\alpha}F\left(z^*\right)\widetilde\zeta_{\alpha}\left(\left(\widetilde{y}^{(j)}_{\alpha}\right)_{j \in [q]}\right)\right]\right.\\
    & \hspace*{90px} + F\left(\left(\widetilde\zeta_{\alpha}\left(\left(\widetilde{y}^{(j)}_{\alpha}\right)_{j \in [q]}\right)\right)_{\alpha \in \mathcal{A}}\right)\Bigg\}\\
    & = -\frac{\delta_{\alpha_0\alpha_1}\delta_{j_0j_1}}{2} - \delta_{\alpha_0\alpha_1}\partial_{j_0j_1}\widetilde\zeta_{\alpha_0}\left(\left(\widetilde{y}^{(j)}_{\alpha_0}\right)_{j \in [q]}\right)\partial_{\alpha_0}F(z^*)\\
    & \hspace*{20px} + \delta_{\alpha_0\alpha_1}\partial_{j_0j_1}\widetilde{\zeta}_{\alpha_0}\left(\left(\widetilde{y}^{(j)}_{\alpha_0}\right)_{j \in [q]}\right)\partial_{\alpha_0}F\left(\left(\widetilde{\zeta}_{\alpha}\left(\left(\widetilde{y}^{(j)}_{\alpha}\right)_{j \in [q]}\right)\right)_{\alpha \in \mathcal{A}}\right)\\
    & \hspace*{20px} + \partial_{j_0}\widetilde{\zeta}_{\alpha_0}\left(\left(\widetilde{y}^{(j)}_{\alpha_0}\right)_{j \in [q]}\right)\partial_{j_1}\widetilde{\zeta}_{\alpha_1}\left(\left(\widetilde{y}^{(j)}_{\alpha_1}\right)_{j \in [q]}\right)\partial_{\alpha_0\alpha_1}F\left(\left(\widetilde{\zeta}_{\alpha}\left(\left(\widetilde{y}^{(j)}_{\alpha}\right)_{j \in [q]}\right)\right)_{\alpha \in \mathcal{A}}\right)\\
    & = -\frac{\delta_{\alpha_0\alpha_1}\delta_{j_0j_1}}{2} + \delta_{\alpha_0\alpha_1}\partial_{j_0j_1}\widetilde\zeta_{\alpha_0}\left(\left(\widetilde{y}^{(j)}_{\alpha_0}\right)_{j \in [q]}\right)\left[\partial_{\alpha_0}F\left(\left(\widetilde{\zeta}_{\alpha}\left(\left(\widetilde{y}^{(j)}_{\alpha}\right)_{j \in [q]}\right)\right)_{\alpha \in \mathcal{A}}\right) - \partial_{\alpha_0}F(z^*)\right]\nonumber\\
    & \hspace*{20px} + \partial_{j_0}\widetilde{\zeta}_{\alpha_0}\left(\left(\widetilde{y}^{(j)}_{\alpha_0}\right)_{j \in [q]}\right)\partial_{j_1}\widetilde{\zeta}_{\alpha_1}\left(\left(\widetilde{y}^{(j)}_{\alpha_1}\right)_{j \in [q]}\right)\partial_{\alpha_0\alpha_1}F\left(\left(\widetilde{\zeta}_{\alpha}\left(\left(\widetilde{y}^{(j)}_{\alpha}\right)_{j \in [q]}\right)\right)_{\alpha \in \mathcal{A}}\right).
\end{align*}
\end{proof}
\end{lem}
Using this general expression and results from section \ref{sec:iterated_gaussian_integration} on the parametrization of iterated Gaussian integrals, the following proposition then evaluates the quadratic form at the critical point. As detailed in the proof of proposition \ref{prop:exp_scaling_generalized_binomial_sum}, this will also serve as a proxy to estimate the same quadratic form up to a distance $\mathcal{O}\left(c^{1/2}\right)$ from the critical point.
\begin{prop}[The quadratic form at $\widetilde{y} = 0$]
\label{prop:quadratic_form_0}
Let $\widetilde\Phi$ be defined as in the proof of proposition \ref{prop:exp_scaling_generalized_binomial_sum} and recall the notations (in particular the fixed point $z^*$) introduced there. $\widetilde\Phi$ admits the following second derivative at $0$:
\begin{align}
	\frac{\partial^2\widetilde\Phi}{\partial\widetilde{y}^{(j_0)}_{\alpha_0}\partial\widetilde{y}^{(j_1)}_{\alpha_1}}(0) & = -\frac{\delta_{\alpha_0\alpha_1}\delta_{j_0j_1}}{2} + 2^{2q - 2 - (j_0 + j_1)/2}\left(-\partial_{\alpha_0}F\left(z^*\right)\right)^{2^{q - 1} - 1}\left(-\partial_{\alpha_1}F\left(z^*\right)\right)^{2^{q - 1} - 1}\partial_{\alpha_0\alpha_1}F\left(z^*\right).
\end{align}
\begin{proof}
Plugging $\widetilde{y}^{(j)}_{\alpha} = 0$ for all $\alpha \in \mathcal{A}, j \in [q]$ in the general expression for the Hessian in lemma \ref{lemma:quadratic_form_general}, and recalling that $\widetilde\zeta_{\alpha}(0_q) = z^*_{\alpha}$ for all $\alpha \in \mathcal{A}$, one obtains the following expression for the Hessian at $0$:
\begin{align*}
    \frac{\partial^2\widetilde\Phi}{\partial\widetilde{y}^{(j_0)}_{\alpha_0}\partial\widetilde{y}^{(j_1)}_{\alpha_1}}(0) & = -\frac{\delta_{\alpha_0\alpha_1}\delta_{j_0j_1}}{2} + \partial_{j_0}\widetilde\zeta_{\alpha_0}\left(0\right)\partial_{j_1}\widetilde\zeta_{\alpha_1}\left(0\right)\partial_{\alpha_0\alpha_1}F\left(\left(\widetilde\zeta_{\alpha}(0)\right)_{\alpha \in \mathcal{A}}\right)
\end{align*}
To conclude, one uses the explicit expressions for the derivatives of $\widetilde\zeta$ at $0$ given in proposition \ref{prop:y_function_0}.
\end{proof}
\end{prop}
After evaluating the Hessian of $\widetilde\Phi$ at arbitrary $\left(\widetilde{y}^{(j)}_{\alpha}\right)_{\substack{\alpha \in \mathcal{A}\\j \in [q]}}$ in lemma \ref{lemma:quadratic_form_general} and at $0$ in the last proposition, we have yet to estimate the distance between the former and the latter quantities for sufficiently small coordinates. This will result from a simple application of lemma \ref{lemma:y_tilde_close_0_function_close_0_evaluated} bounding the distances of functions $\mathcal{Y}^{(j)}_{\alpha}$ and $\widetilde\zeta_{\alpha}$ from their values at $0$.
\begin{prop}[The quadratic form sufficiently close to $\widetilde{y} = 0$]
\label{prop:quadratic_form_close_0}
Let $\widetilde\Phi$ be defined as in the proof of proposition \ref{prop:exp_scaling_generalized_binomial_sum} and recall the notations there. For sufficiently small $z_{\textrm{max}} > 0$ (independent of $c$), the following holds for all $\alpha_0, \alpha_1 \in \mathcal{A}$ and $j_0, j_1 \in [q]$,
\begin{align}
    \left|\widetilde{y}^{(j)}_{\alpha}\right| \leq c^{1/2}z_{\textrm{max}} \quad \forall \alpha \in \mathcal{A},\,\forall j \in [q] & \implies \left|\frac{\partial^2\widetilde\Phi}{\partial\widetilde{y}^{\left(j_0\right)}_{\alpha_0}\partial\widetilde{y}^{\left(j_1\right)}_{\alpha_1}}\left(\left(\widetilde{y}^{(j)}_{\alpha}\right)_{\substack{\alpha \in \mathcal{A}\\j \in [q]}}\right) - \frac{\partial^2\widetilde\Phi}{\partial\widetilde{y}^{\left(j_0\right)}_{\alpha_0}\partial\widetilde{y}^{\left(j_1\right)}_{\alpha_1}}\left(0\right)\right| \leq \mathcal{O}(1)z_{\textrm{max}}.
\end{align}
The constant in the $\mathcal{O}(1)$ is independent of $c$ as soon as $c$ is small enough. In particular, this means one can satisfy this inequality with fixed constants while $c$ decreases.
\begin{proof}
Recalling the general expression for the Hessian of $\widetilde\Phi$ given in lemma \ref{lemma:quadratic_form_general}, and applying the general inequality:
\begin{align*}
    \left|\prod_{j \in [m]}z_j - \prod_{j \in [m]}w_j\right| & \leq \sum_{j \in [m]}\left(\prod_{j' \in [j]}\left|w_{j'}\right|\right)\left|z_j - w_j\right|\left(\prod_{j < j' < m}\left|z_{j'}\right|\right)
\end{align*}
holding for arbitrary complex numbers $z_0, \ldots, z_{m - 1}, w_0, \ldots, w_{m - 1}$,
\begin{align*}
    & \left|\frac{\partial^2\widetilde\Phi}{\partial\widetilde{y}^{\left(j_0\right)}_{\alpha_0}\partial\widetilde{y}^{\left(j_1\right)}_{\alpha_1}}\left(\left(\widetilde{y}^{(j)}_{\alpha}\right)_{\substack{\alpha \in \mathcal{A}\\j \in [q]}}\right) - \frac{\partial^2\widetilde\Phi}{\partial\widetilde{y}^{\left(j_0\right)}_{\alpha_0}\partial\widetilde{y}^{\left(j_1\right)}_{\alpha_1}}\left(0\right)\right|\\
    & \leq \delta_{\alpha_0\alpha_1}\left|\partial_{j_0j_1}\widetilde\zeta_{\alpha_0}\left(\left(\widetilde{y}^{(j)}_{\alpha_0}\right)_{j \in [q]}\right)\right|\left|\partial_{\alpha_0}F\left(\left(\widetilde\zeta_{\alpha}\left(\left(\widetilde{y}^{(j)}_{\alpha}\right)_{j \in [q]}\right)\right)_{\alpha \in \mathcal{A}}\right) - \partial_{\alpha_0}F(z^*)\right|\\
    & \hspace*{20px} + \left|\partial_{j_0}\widetilde{\zeta}_{\alpha_0}\left(\left(\widetilde{y}^{(j)}_{\alpha_0}\right)_{j \in [q]}\right) - \partial_{j_0}\widetilde\zeta_{\alpha_0}\left(0\right)\right|\left|\partial_{j_1}\widetilde{\zeta}_{\alpha_1}\left(\left(\widetilde{y}^{(j)}_{\alpha_1}\right)_{j \in [q]}\right)\right|\left|\partial_{\alpha_0\alpha_1}F\left(\left(\widetilde{\zeta}_{\alpha}\left(\left(\widetilde{y}^{(j)}_{\alpha}\right)_{j \in [q]}\right)\right)_{\alpha \in \mathcal{A}}\right)\right|\\
    & \hspace*{20px} + \left|\partial_{j_0}\widetilde{\zeta}_{\alpha_0}\left(0\right)\right|\left|\partial_{j_1}\widetilde{\zeta}_{\alpha_1}\left(\left(\widetilde{y}^{(j)}_{\alpha_1}\right)_{j \in [q]}\right) - \partial_{j_1}\widetilde\zeta_{\alpha_1}(0)\right|\left|\partial_{\alpha_0\alpha_1}F\left(\left(\widetilde{\zeta}_{\alpha}\left(\left(\widetilde{y}^{(j)}_{\alpha}\right)_{j \in [q]}\right)\right)_{\alpha \in \mathcal{A}}\right)\right|\\
    & \hspace*{20px} + \left|\partial_{j_0}\widetilde\zeta_{\alpha_0}(0)\right|\left|\partial_{j_1}\widetilde\zeta_{\alpha_1}(0)\right|\left|\partial_{\alpha_0\alpha_1}F\left(\left(\widetilde\zeta_{\alpha}\left(\left(\widetilde{y}^{(j)}_{\alpha}\right)_{j \in [q]}\right)\right)_{\alpha \in \mathcal{A}}\right) - \partial_{\alpha_0\alpha_1}F\left(\left(\widetilde\zeta_{\alpha}(0)\right)_{\alpha \in \mathcal{A}}\right)\right|.
\end{align*}
We now invoke lemma \ref{lemma:y_tilde_close_0_function_close_0_evaluated} to bound each of the terms appearing in this long sum. We keep track of the layout of the previous equation for clarity.
\begin{align*}
    & \left|\frac{\partial^2\widetilde\Phi}{\partial\widetilde{y}^{(j_0)}_{\alpha_0}\partial\widetilde{y}^{(j_1)}_{\alpha_1}}\left(\left(\widetilde{y}^{(j)}_{\alpha}\right)_{\substack{\alpha \in \mathcal{A}\\j \in [q]}}\right) - \frac{\partial^2\widetilde\Phi}{\partial\widetilde{y}^{(j_0)}_{\alpha_0}\partial\widetilde{y}^{(j_1)}_{\alpha_1}}\left(0\right)\right|\\
    & = \delta_{\alpha_0\alpha_1}\mathcal{O}(1)c^{-1/2^q} \times \mathcal{O}(1)\underbrace{c^{1/2^{q - 1}}}_{\textrm{bound on the } \partial_{\alpha_0\alpha}F \textrm{ in ball on radius } c^{-1/2^{q + 1}}}\underbrace{c^{1 - 1/2^q}z_{\textrm{max}}}_{\textrm{bound on variation of the } \widetilde\zeta_{\alpha}}\\
    & \hspace*{20px} + \mathcal{O}(1)c^{1/2 - 1/2^q}z_{\textrm{max}} \times \mathcal{O}(1)c^{1/2 - 1/2^q} \times \mathcal{O}(1)c^{1/2^{q - 1}}\\
    & \hspace*{20px} + \mathcal{O}(1)c^{1/2 - 1/2^q} \times \mathcal{O}(1)c^{1/2 - 1/2^q}z_{\textrm{max}} \times \mathcal{O}(1)c^{1/2^{q - 1}}\\
    & \hspace*{20px} + \mathcal{O}(1)c^{1/2 - 1/2^q} \times \mathcal{O}(1)c^{1/2 - 1/2^q} \times \mathcal{O}(1)\underbrace{c^{3/2^q}}_{\textrm{bound on the } \partial_{\alpha_0\alpha_1\alpha}F \textrm{ in a ball of radius } c^{-1/2^{q + 1}}}\underbrace{c^{1 - 1/2^q}z_{\textrm{max}}}_{\textrm{bound on variation of the } \widetilde\zeta_{\alpha}}.
\end{align*}
These can be combined to a simpler final bound $\mathcal{O}(1)cz_{\textrm{max}}$
\end{proof}
\end{prop}

\begin{rem}
\label{rem:applying_quadratic_form_close_0}
According to proposition \ref{prop:quadratic_form_close_0}, the Hessian of $\widetilde\Phi$ at $0$ is the sum of $-\frac{1}{2}\mathbf{1}_{|\mathcal{A}|} \otimes \mathbf{1}_q$ and a contribution depending on $F$, which, as observed in the proof of proposition \ref{prop:exp_scaling_generalized_binomial_sum}, vanishes as $c \downarrow 0$. Therefore, the real part of this Hessian converges to $-\frac{1}{2}\mathbf{1}_{|\mathcal{A}|} \otimes \mathbf{1}_{q}$ in this limit. It follows that for sufficiently small $c$, the real part of the Hessian is $\preceq -\frac{1}{4}\mathbf{1}_{|\mathcal{A}|} \otimes \mathbf{1}_q$ (where the inequality is in the operator sense) by continuity of eigenvalues.
\end{rem}

Now we have obtained control of the quadratic form at the critical point, hence sufficiently close to it, let us complement this result with the following bound further from the critical point:
\begin{prop}[The quadratic form sufficiently far from $\widetilde{y} = 0$]
\label{prop:quadratic_form_far_0}
Let definitions be as in the statement and proof of proposition \ref{prop:exp_scaling_generalized_binomial_sum}. Then for sufficiently small $c$, the following estimates holds uniformly in $\left(\widetilde{y}^{(j)}_{\alpha}\right)_{\substack{\alpha \in \mathcal{A}\\j \in [q]}}$ as long as $\left|\widetilde{\zeta}_{\alpha}\left(\left(\widetilde{y}^{(j)}_{\alpha}\right)_{j \in [q]}\right)\right| \leq \mathcal{O}\left(c^{-1/2^{q + 1}}\right)$ for all $\alpha \in \mathcal{A}$:
\begin{align}
	\Re\left\{\widetilde{\Phi}\left(\left(\widetilde{y}^{(j)}_{\alpha}\right)_{\substack{\alpha \in \mathcal{A}\\j \in [q]}}\right) - \widetilde\Phi\left(0\right)\right\} & \leq  -\frac{1}{8}\sum_{\substack{\alpha \in \mathcal{A}\\j \in [q]}}\left(\widetilde{y}^{(j)}_{\alpha}\right)^2 + \mathcal{O}\left(c^{1 + 1/2^{q}}\right).\label{eq:quadratic_form_far_0}
\end{align}
\begin{proof}
We only need a first-order Taylor expansion of $F(z) = F\left(\left(z_{\alpha}\right)_{\alpha \in \mathcal{A}}\right)$ around $z = z^*$:
\begin{align*}
	F(z) & = F(z^*) + \sum_{\alpha \in \mathcal{A}}\left(z_{\alpha} - z^*_{\alpha}\right)\int_0^1\!\mathrm{d}v\,\partial_{\alpha}F\left(z^* + v(z - z^*)\right)\\
	& = F(z^*) + \sum_{\alpha \in \mathcal{A}}\left(z_{\alpha} - z^*_{\alpha}\right)\partial_{\alpha}F\left(z^*\right) + \sum_{\alpha \in \mathcal{A}}\left(z_{\alpha} - z^*_{\alpha}\right)\varepsilon_{\alpha}\left(z\right),
\end{align*}
where
\begin{align*}
	\varepsilon_{\alpha}\left(z\right) & := \int_0^1\!\mathrm{d}v\,\left(\partial_{\alpha}F\left(z^* + v\left(z - z^*\right)\right) - \partial_{\alpha}F\left(z^*\right)\right)\\
	& = \mathcal{O}\left(\sup_{v \in [0, 1], \beta \in \mathcal{A}}\left|\partial_{\alpha\beta}F\left(z^* + v\left(z - z^*\right)\right)\right|\right)\\
	& = \mathcal{O}\left(c^{1/2^{q - 1}}\right),
\end{align*}
where the last equality uses that $z^* = \mathcal{O}\left(c^{1/2}\right)$ and $z = \mathcal{O}\left(c^{-1/2^{q + 1}}\right)$.

One now writes $F$ according to this Taylor expansion in the expression of $\widetilde{\Phi}\left(\left(\widetilde{y}^{(j)}_{\alpha}\right)_{\substack{\alpha \in \mathcal{A}\\j \in [q]}}\right)$:
\begin{align*}
	\widetilde{\Phi}\left(\left(\widetilde{y}^{(j)}_{\alpha}\right)_{\substack{\alpha \in \mathcal{A}\\j \in [q]}}\right) & = -\frac{1}{4}\sum_{\substack{\alpha \in \mathcal{A}\\j \in [q]}}\mathcal{Y}^{(j)}_{\alpha}\left(\left(\widetilde{y}^{(j)}_{\alpha}\right)_{j \in [q]}\right)^2 + F\left(\left(\widetilde{\zeta}_{\alpha}\left(\left(\widetilde{y}^{(j)}_{\alpha}\right)_{j \in [q]}\right)\right)_{\alpha \in \mathcal{A}}\right)\\
	& = F\left(z^*\right) - \sum_{\alpha \in \mathcal{A}}z^*_{\alpha}\left(\partial_{\alpha}F\left(z^*\right) + \varepsilon_{\alpha}(z)\right) - \frac{1}{4}\sum_{\substack{\alpha \in \mathcal{A}\\j \in [q]}}\mathcal{Y}^{(j)}_{\alpha}\left(\left(\widetilde{y}^{(j)}_{\alpha}\right)_{j \in [q]}\right)^2\\
	& \hspace*{20px} + \sum_{\alpha \in \mathcal{A}}\widetilde{\zeta}_{\alpha}\left(\left(\widetilde{y}^{(j)}_{\alpha}\right)_{j \in [q]}\right)\partial_{\alpha}F\left(z^*\right) + \sum_{\alpha \in \mathcal{A}}\widetilde{\zeta}_{\alpha}\left(\left(\widetilde{y}^{(j)}_{\alpha}\right)_{j \in [q]}\right)\varepsilon_{\alpha}\left(\left(\widetilde{\zeta}_{\alpha'}\left(\left(\widetilde{y}^{(j)}_{\alpha'}\right)_{j \in [q]}\right)\right)_{\alpha' \in \mathcal{A}}\right).
\end{align*}
Now, by repeating the calculations in the proof of proposition \ref{prop:iterated_gaussian_integral_generate_higher_order_gaussian}, one has for all $\alpha \in \mathcal{A}$:
\begin{align*}
	-\frac{1}{4}\sum_{j \in [q]}\mathcal{Y}^{(j)}_{\alpha}\left(\left(\widetilde{y}^{(j)}_{\alpha}\right)_{j \in [q]}\right)^2 & = -\frac{1}{4}\sum_{j \in [q]}\left(\widetilde{y}^{(j)}_{\alpha}\right)^2 - C_{\alpha} + C_{\alpha}^{1/2^q}\widetilde{\zeta}_{\alpha}\left(\left(\widetilde{y}^{(j)}_{\alpha}\right)_{j \in [q]}\right)\\
	& = -\frac{1}{4}\sum_{j \in [q]}\left(\widetilde{y}^{(j)}_{\alpha}\right)^2 - \left\{\partial_{\alpha}F\left(z^*\right)\right\}^{2^q} + \left(\left\{\partial_{\alpha}F\left(z^*\right)\right\}^{2^q}\right)^{1/2^q}\widetilde{\zeta}_{\alpha}\left(\left(\widetilde{y}^{(j)}_{\alpha}\right)_{j \in [q]}\right)\\
	& = -\frac{1}{4}\sum_{j \in [q]}\left(\widetilde{y}^{(j)}_{\alpha}\right)^2 - \left\{\partial_{\alpha}F\left(z^*\right)\right\}^{2^q} + \left(\left\{-\partial_{\alpha}F\left(z^*\right)\right\}^{2^q}\right)^{1/2^q}\widetilde{\zeta}_{\alpha}\left(\left(\widetilde{y}^{(j)}_{\alpha}\right)_{j \in [q]}\right)\\
	& = -\frac{1}{4}\sum_{j \in [q]}\left(\widetilde{y}^{(j)}_{\alpha}\right)^2 - \left\{\partial_{\alpha}F\left(z^*\right)\right\}^{2^q} + \partial_{\alpha}F\left(z^*\right)\widetilde{\zeta}_{\alpha}\left(\left(\widetilde{y}^{(j)}_{\alpha}\right)_{j \in [q]}\right).
\end{align*}
In the last equality, we used assumption \ref{assp:restriction_argument}, which implies $\left|\arg\left(-\partial_{\alpha}F\left(z^*\right)\right)\right| < \frac{\pi}{2^{q}}$ for sufficiently small $c$. Hence,
\begin{align*}
	& \widetilde{\Phi}\left(\left(\widetilde{y}^{(j)}_{\alpha}\right)_{\substack{\alpha \in \mathcal{A}\\j \in [q]}}\right)\\
	& = F\left(z^*\right) + \left(2^q - 1\right)\sum_{\alpha \in \mathcal{A}}\left\{\partial_{\alpha}F\left(z^*\right)\right\}^{2^q} - \frac{1}{4}\sum_{\substack{\alpha \in \mathcal{A}\\j \in [q]}}\left(\widetilde{y}^{(j)}_{\alpha}\right)^2 + \sum_{\alpha \in \mathcal{A}}\left(\widetilde\zeta_{\alpha}\left(\left(\widetilde{y}^{(j)}_{\alpha}\right)_{\alpha \in \mathcal{A}}\right) - z_{\alpha}^*\right)\varepsilon_{\alpha}\left(\left(\widetilde\zeta_{\alpha'}\left(\left(\widetilde{y}^{(j)}_{\alpha'}\right)_{j \in [q]}\right)\right)_{\alpha' \in \mathcal{A}}\right)\\
	& = F\left(z^*\right) + \left(2^q - 1\right)\left\{\partial_{\alpha}F\left(z^*\right)\right\}^{2^q} - \frac{1}{4}\sum_{\substack{\alpha \in \mathcal{A}\\j \in [q]}}\left(\widetilde{y}^{(j)}_{\alpha}\right)^2 + \sum_{\alpha \in \mathcal{A}}\widetilde\zeta_{\alpha}\left(\left(\widetilde{y}^{(j)}_{\alpha}\right)_{j \in [q]}\right)\varepsilon_{\alpha}\left(\left(\widetilde\zeta_{\alpha'}\left(\left(\widetilde{y}^{(j)}_{\alpha'}\right)_{j \in [q]}\right)\right)_{\alpha' \in \mathcal{A}}\right)\nonumber\\
	& \hspace*{10px} + \mathcal{O}\left(c^{1 + 1/2^q}\right),
\end{align*}
where the last equality uses $z^*_{\alpha} = \mathcal{O}\left(c^{1 - 1/2^q}\right)$, which follows from the self-consistent equation $z^*_{\alpha} = -2^q\left\{\partial_{\alpha}F\left(z^*\right)\right\}^{2^q - 1}$ and $\varepsilon_{\alpha}(z) = \mathcal{O}(c^{1/2^{q - 1}})$. We now focus on bounding:
\begin{align*}
	-\frac{1}{4}\sum_{\substack{\alpha \in \mathcal{A}\\j \in [q]}}\left(\widetilde{y}^{(j)}_{\alpha}\right)^2 + \sum_{\alpha \in \mathcal{A}}\widetilde\zeta_{\alpha}\left(\left(\widetilde{y}^{(j)}_{\alpha}\right)_{j \in [q]}\right)\varepsilon_{\alpha}(z) & = -\frac{1}{4}\sum_{\substack{\alpha \in \mathcal{A}\\j \in [q]}}\left(\widetilde{y}^{(j)}_{\alpha}\right)^2 + \sum_{\alpha \in \mathcal{A}}\widetilde\zeta_{\alpha}\left(\left(\widetilde{y}^{(j)}_{\alpha}\right)_{j \in [q]}\right)\mathcal{O}\left(c^{1/2^{q - 1}}\right).
\end{align*}
To achieve that, we rewrite it as:
\begin{align*}
	& -\frac{1}{8}\sum_{\substack{\alpha \in \mathcal{A}\\j \in [q]}}\left(\widetilde{y}^{(j)}_{\alpha}\right)^2 - \frac{1}{8}\sum_{\substack{\alpha \in \mathcal{A}\\j \in [q]}}\left(\widetilde{y}^{(j)}_{\alpha}\right)^2 + \sum_{\alpha \in \mathcal{A}}\widetilde\zeta_{\alpha}\left(\left(\widetilde{y}^{(j)}_{\alpha}\right)_{j \in [q]}\right)\mathcal{O}\left(c^{1/2^{q - 1}}\right)
\end{align*}
and bound 
\begin{align*}
	& \Re\left\{- \frac{1}{8}\sum_{\substack{j \in [q]}}\left(\widetilde{y}^{(j)}_{\alpha}\right)^2 + \widetilde\zeta_{\alpha}\left(\left(\widetilde{y}^{(j)}_{\alpha}\right)_{j \in [q]}\right)\mathcal{O}\left(c^{1/2^{q - 1}}\right)\right\}
\end{align*}
by an absolute constant for each $\alpha \in \mathcal{A}$. For that purpose, we partially and successively maximize over variables $\widetilde{y}^{(q - 1)}_{\alpha}, \ldots, \widetilde{y}^{(0)}_{\alpha}$. To maximize over $\widetilde{y}^{(q - 1)}_{\alpha}$, observe:
\begin{align*}
	& -\frac{1}{8}\left(\widetilde{y}^{(q - 1)}_{\alpha}\right)^2 + \widetilde\zeta_{\alpha}\left(\left(\widetilde{y}^{(j)}_{\alpha}\right)_{j \in [q]}\right)\mathcal{O}\left(c^{1/2^{q - 1}}\right)\\
	& = -\frac{1}{8}\left(\widetilde{y}^{(q - 1)}_{\alpha}\right)^2 - i\mathcal{Y}^{(q - 1)}\left(\left(\widetilde{y}^{(j)}_{\alpha}\right)_{j \in [q]}\right)\widetilde{\zeta}_{\alpha}\left(\left(y^{(j)}_{\alpha}\right)_{j \in [q - 1]}\right)^{1/2}\mathcal{O}\left(c^{1/2^{q - 1}}\right)\\
	& = -\frac{1}{8}\left(\widetilde{y}^{(q - 1)}_{\alpha}\right)^2 - i\left(\widetilde{y}^{(q - 1)}_{\alpha} + 2iC_{\alpha}^{1/2^{q}}\widetilde{\zeta}_{\alpha}\left(\left(\widetilde{y}^{(j)}_{\alpha}\right)_{j \in [q - 1]}\right)^{1/2}\right)\widetilde{\zeta}_{\alpha}\left(\left(y^{(j)}_{\alpha}\right)_{j \in [q - 1]}\right)^{1/2}\mathcal{O}\left(c^{1/2^{q - 1}}\right)\\
	& =  -\frac{1}{8}\left(\widetilde{y}^{(q - 1)}_{\alpha}\right)^2 - i\widetilde{y}^{(q - 1)}_{\alpha}\widetilde{\zeta}_{\alpha}\left(\left(\widetilde{y}^{(j)}_{\alpha}\right)_{j \in [q]}\right)^{1/2}\mathcal{O}\left(c^{1/2^{q - 1}}\right) + 2\partial_{\alpha}F\left(z^*\right)\widetilde{\zeta}_{\alpha}\left(\left(\widetilde{y}^{(j)}_{\alpha}\right)_{j \in [q - 1]}\right)\mathcal{O}\left(c^{1/2^{q - 1}}\right)
\end{align*}
Therefore, using the simple result:
\begin{align*}
	\sup_{y \in \mathbf{R}}\Re\left\{-\frac{y^2}{8} + z y\right\} & = 2\Re(z)^2 \qquad \forall z \in \mathbf{C},
\end{align*}
one can bound:
\begin{align*}
	& \Re\left\{-\frac{1}{8}\left(\widetilde{y}^{(q - 1)}_{\alpha}\right)^2 + \widetilde\zeta_{\alpha}\left(\left(\widetilde{y}^{(j)}_{\alpha}\right)_{j \in [q]}\right)\mathcal{O}\left(c^{1/2^{q - 1}}\right)\right\}\\
	& \leq 2\left|\widetilde{\zeta}_{\alpha}\left(\left(\widetilde{y}^{(j)}_{\alpha}\right)_{j \in [q - 1]}\right)\right|\mathcal{O}\left(c^{1/2^{q - 2}}\right) + 2\left|\partial_{\alpha}F\left(z^*\right)\right|\left|\widetilde{\zeta}_{\alpha}\left(\left(\widetilde{y}^{(j)}_{\alpha}\right)_{j \in [q - 1]}\right)\right|\mathcal{O}\left(c^{1/2^{q - 1}}\right)\\
	& = \left|\widetilde{\zeta}_{\alpha}\left(\left(\widetilde{y}^{(j)}_{\alpha}\right)_{j \in [q - 1]}\right)\right|\mathcal{O}\left(c^{\frac{1}{2^{q - 1}} + \frac{1}{2^q}}\right).
\end{align*}
We then partially maximize over $\widetilde{y}^{(q - 1)}_{\alpha}$ using the last upper bound for the partial maximization over $\widetilde{y}^{(q - 1)}_{\alpha}$:
\begin{align*}
	& \Re\left\{-\frac{1}{8}\left(\widetilde{y}^{(q - 2)}_{\alpha}\right)^2 - \frac{1}{8}\left(\widetilde{y}^{(q - 1)}_{\alpha}\right)^2 + \widetilde{\zeta}_{\alpha}\left(\left(\widetilde{y}^{(j)}_{\alpha}\right)_{j \in [q]}\right)\mathcal{O}\left(c^{1/2^{q - 1}}\right)\right\}\\
	& \leq \Re\left\{-\frac{1}{8}\left(\widetilde{y}^{(q - 2)}_{\alpha}\right)^2 + \left|\widetilde{\zeta}_{\alpha}\left(\left(\widetilde{y}^{(j)}_{\alpha}\right)_{j \in [q - 1]}\right)\right|\mathcal{O}\left(c^{\frac{1}{2^{q - 1}} + \frac{1}{2^q}}\right)\right\}\\
	& \leq \Re\left\{-\frac{1}{8}\left(\widetilde{y}^{(q - 2)}_{\alpha}\right)^2 + \left|\widetilde{y}^{(q - 2)}_{\alpha}\right|\left|\widetilde\zeta_{\alpha}\left(\left(\widetilde{y}^{(j)}_{\alpha}\right)_{j \in [q - 2]}\right)\right| + 2|\partial_{\alpha}F\left(z^*\right)|^2\left|\widetilde\zeta_{\alpha}\left(\left(\widetilde{y}^{(j)}_{\alpha}\right)_{j \in [q - 2]}\right)\right|\right\}\\
	& \leq \left|\widetilde\zeta_{\alpha}\left(\left(\widetilde{y}^{(j)}_{\alpha}\right)_{j \in [q - 2]}\right)\right|\mathcal{O}\left(c^{\frac{1}{2^{q - 2}} + \frac{1}{2^{q - 1}}}\right) + \left|\widetilde\zeta_{\alpha}\left(\left(\widetilde{y}^{(j)}_{\alpha}\right)_{j \in [q - 2]}\right)\right|\mathcal{O}\left(c^{\frac{1}{2^{q - 1}} + \frac{1}{2^{q}} + \frac{1}{2^{q - 1}}}\right)\\
	& \leq \left|\widetilde\zeta_{\alpha}\left(\left(\widetilde{y}^{(j)}_{\alpha}\right)_{j \in [q - 2]}\right)\right|\mathcal{O}\left(c^{\frac{1}{2^{q - 2}} + \frac{1}{2^q}}\right).
\end{align*}
Continuing this way, one finally obtains
\begin{align*}
	& \Re\left\{- \frac{1}{8}\sum_{\substack{j \in [q]}}\left(\widetilde{y}^{(j)}_{\alpha}\right)^2 + \widetilde\zeta_{\alpha}\left(\left(\widetilde{y}^{(j)}_{\alpha}\right)_{j \in [q]}\right)\mathcal{O}\left(c^{1/2^{q - 1}}\right)\right\}\\
	& \leq \left|\widetilde\zeta_{\alpha}\left(\varnothing\right)\right|\mathcal{O}\left(c^{\frac{1}{2^0} + \frac{1}{2^q}}\right) = \mathcal{O}\left(c^{1 + 1/2^q}\right).
\end{align*}
It follows
\begin{align*}
	- \frac{1}{4}\sum_{\substack{\alpha \in \mathcal{A}\\j \in [q]}}\left(\widetilde{y}^{(j)}_{\alpha}\right)^2 + \sum_{\alpha \in \mathcal{A}}\widetilde\zeta_{\alpha}\left(\left(\widetilde{y}^{(j)}_{\alpha}\right)_{j \in [q]}\right)\varepsilon_{\alpha}(z) & = -\frac{1}{8}\sum_{\substack{\alpha \in \mathcal{A}\\j \in [q]}}\left(\widetilde{y}^{(j)}_{\alpha}\right)^2 + \mathcal{O}\left(c^{1 + 1/2^{q}}\right),
\end{align*}
and
\begin{align*}
	\Re\left\{\widetilde{\Phi}\left(\left(\widetilde{y}^{(j)}_{\alpha}\right)_{\substack{\alpha \in \mathcal{A}\\j \in [q]}}\right) - \widetilde\Phi\left(0\right)\right\} & \leq  -\frac{1}{8}\sum_{\substack{\alpha \in \mathcal{A}\\j \in [q]}}\left(\widetilde{y}^{(j)}_{\alpha}\right)^2 + \mathcal{O}\left(c^{1 + 1/2^{q}}\right).
\end{align*}
\end{proof}
\end{prop}

\begin{rem}
It may appear unclear at first sight why the estimate from proposition \ref{prop:quadratic_form_close_0} on the quadratic form close to the critical point is needed given the latest proposition, whose estimate is furthermore simpler. The reason can be understood by coming back to the context of the proof of proposition \ref{prop:exp_scaling_generalized_binomial_sum}, where results from this section are invoked to justify dominated convergence. The issue of applying proposition \ref{prop:quadratic_form_far_0} close to the critical point is that the error term $\mathcal{O}\left(c^{1/2 + 1/2^{q + 1}}\right)$ may be positive, hence $n\mathcal{O}\left(c^{1/2 + 1/2^{q + 1}}\right) \to \infty$ as $n \to \infty$, preventing from applying the dominated convergence theorem in this limit in the proof of proposition \ref{prop:exp_scaling_generalized_binomial_sum}.
\end{rem}

\subsection*{Acknowledgements}

This project has received funding from the European Research Council (ERC) under the European Union’s Horizon 2020 research and innovation programme (grant agreement No. 817581). This work was supported by the EPSRC Centre for Doctoral Training in Delivering Quantum Technologies, grant ref.\ EP/S021582/1.

\appendix

\section{Further numerical results}
\label{app:further_figures}

In this appendix we present the full optimization landscape for QAOA at $p=1$ in figure \ref{fig:p1_qaoa_landscapes}. More precisely, for each value of $k$, we report 3 quantities as functions of $\beta$ and $\gamma$, corresponding to the 3 columns of the figures.
\begin{itemize}
    \item The expected success probability, with a logarithm taken and rescaled by the number of variables $n$:
    \begin{align}
        \frac{1}{n}\log\mathbf{E}_{\bm\sigma \sim \mathrm{CNF}(n, k, r)}\left[\left\langle\Psi_{\mathrm{QAOA}}(\bm\sigma, \beta, \gamma)|\mathbf{1}\left\{H[\bm\sigma] = 0\right\}|\Psi_{\mathrm{QAOA}}(\bm\sigma, \beta, \gamma)\right\rangle\right],
    \end{align}
    which, assuming the existence of an exponential scaling, converges to the scaling exponent as $n \to \infty$.
    
    For this set of experiments, $n = 20$.
    \item The empirical scaling exponent determined from an exponential fit. In the results shown, the fit is performed on $11$ points corresponding to problem sizes $20 \leq n \leq 30$.
    \item The correlation coefficient of the exponential fit.
\end{itemize}
From these results, it first appears that the existence of the exponential scaling is not granted for all $\beta$ and $\gamma$, with a correlation coefficient even approaching $0$ in certain regions. This effect is particularly pronounced for larger $k$ and these problematic regions exhibit a complex pattern that cannot be only explained by the magnitude of the angles. However, in the regions where an exponential scaling exists, the landscapes in the first two columns should coincide for sufficiently large $n$. We observe this is indeed the case in the figures. Interestingly, the regions where the success probability is maximized always exhibit an exponential scaling, as shown by the correlation coefficient close to unity. As for extrema, while $2$-SAT and $4$-SAT possess a single local maximum (up to the central symmetry), $8$- and $16$-SAT have spurious local maxima.
\begin{figure}
    \centering
    \begin{subfigure}{\textwidth}
        \centering
        \begin{subfigure}{0.32\textwidth}
            \includegraphics[width=\textwidth]{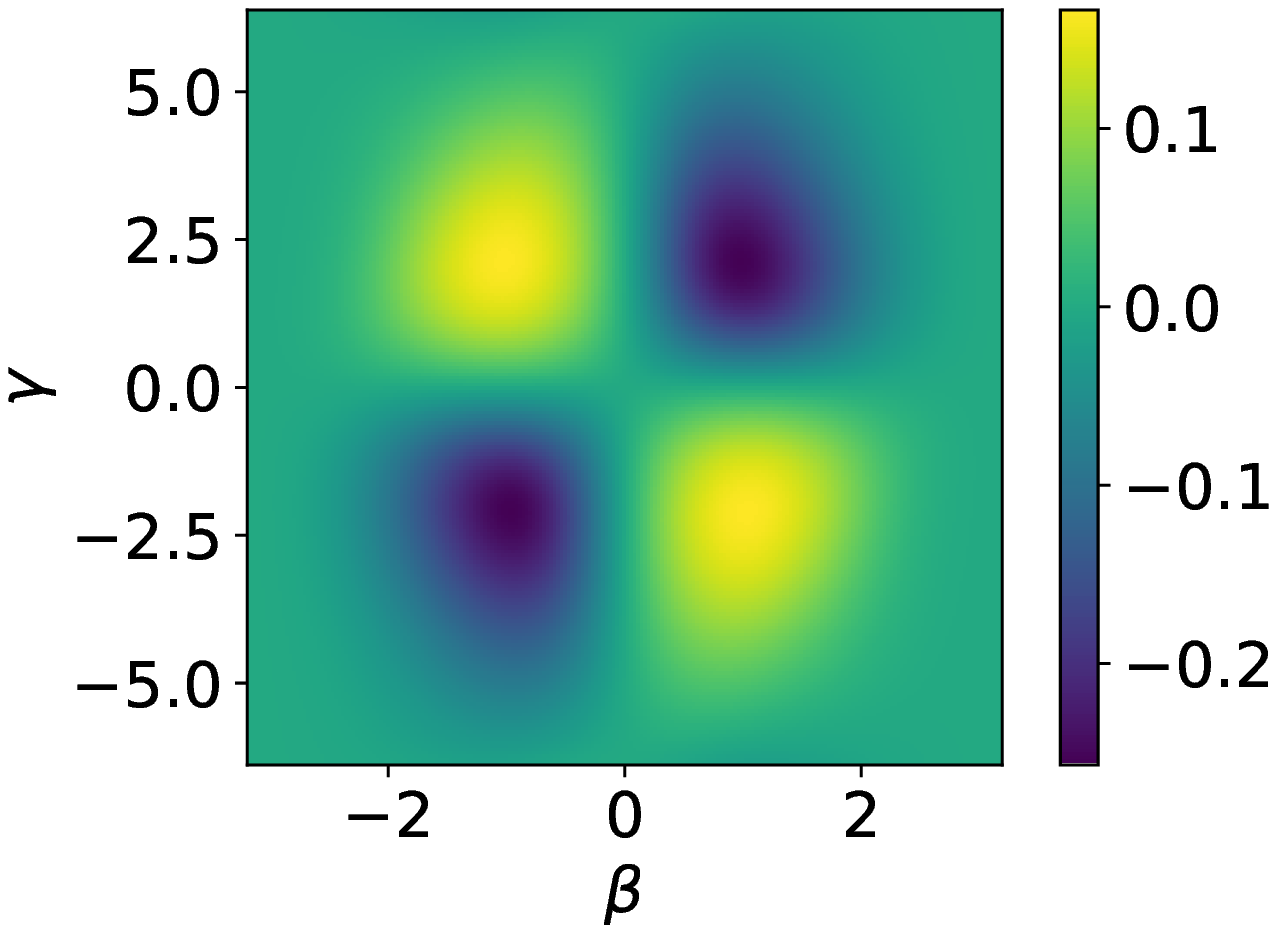}
            \renewcommand\thesubfigure{\alph{subfigure}1}
            \caption{Logarithmic success probability}
        \end{subfigure}
        \begin{subfigure}{0.32\textwidth}
            \includegraphics[width=\textwidth]{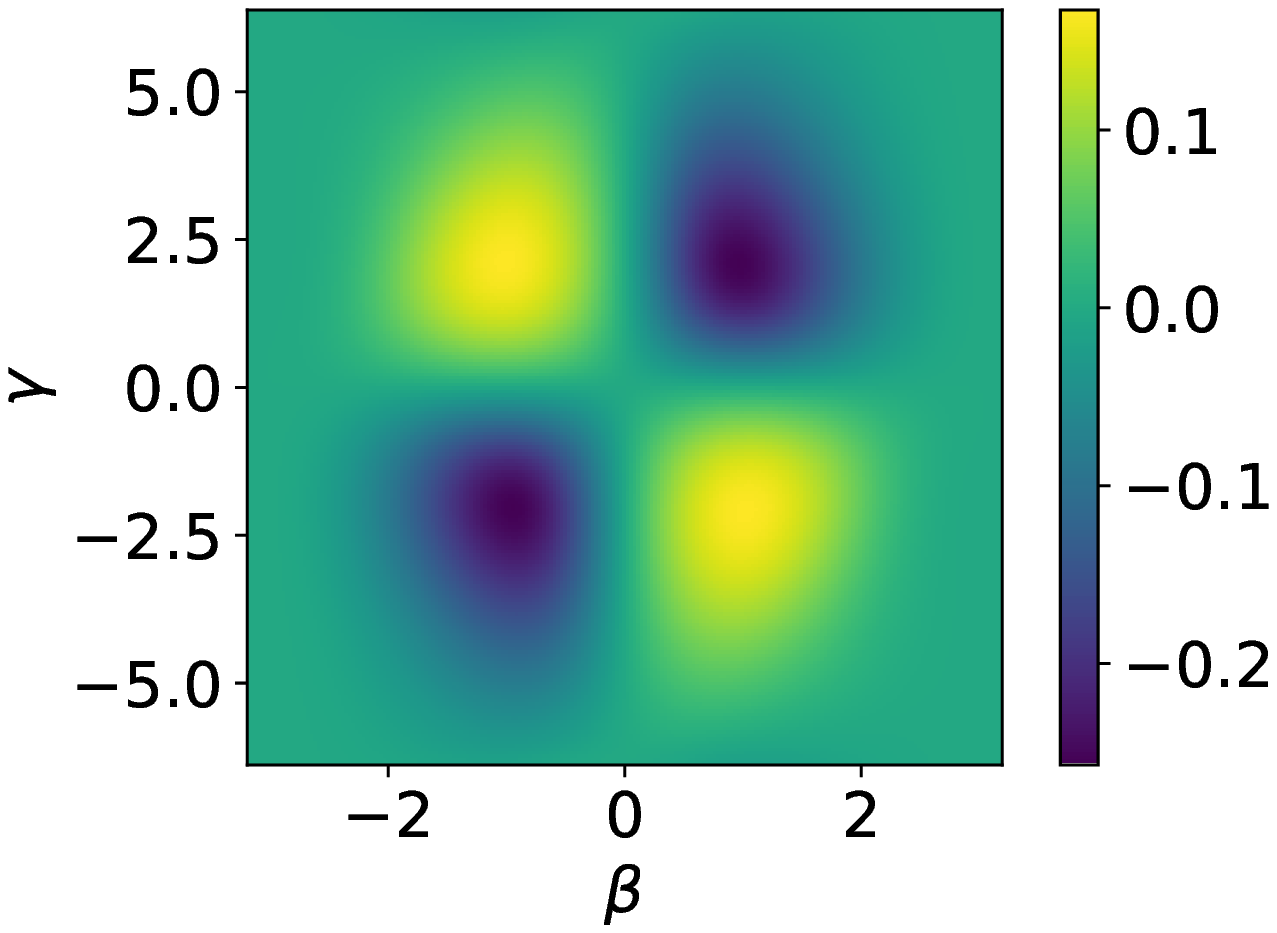}
            \addtocounter{subfigure}{-1}
            \renewcommand\thesubfigure{\alph{subfigure}2}
            \caption{Exponential fit}
        \end{subfigure}
        \begin{subfigure}{0.32\textwidth}
            \includegraphics[width=\textwidth]{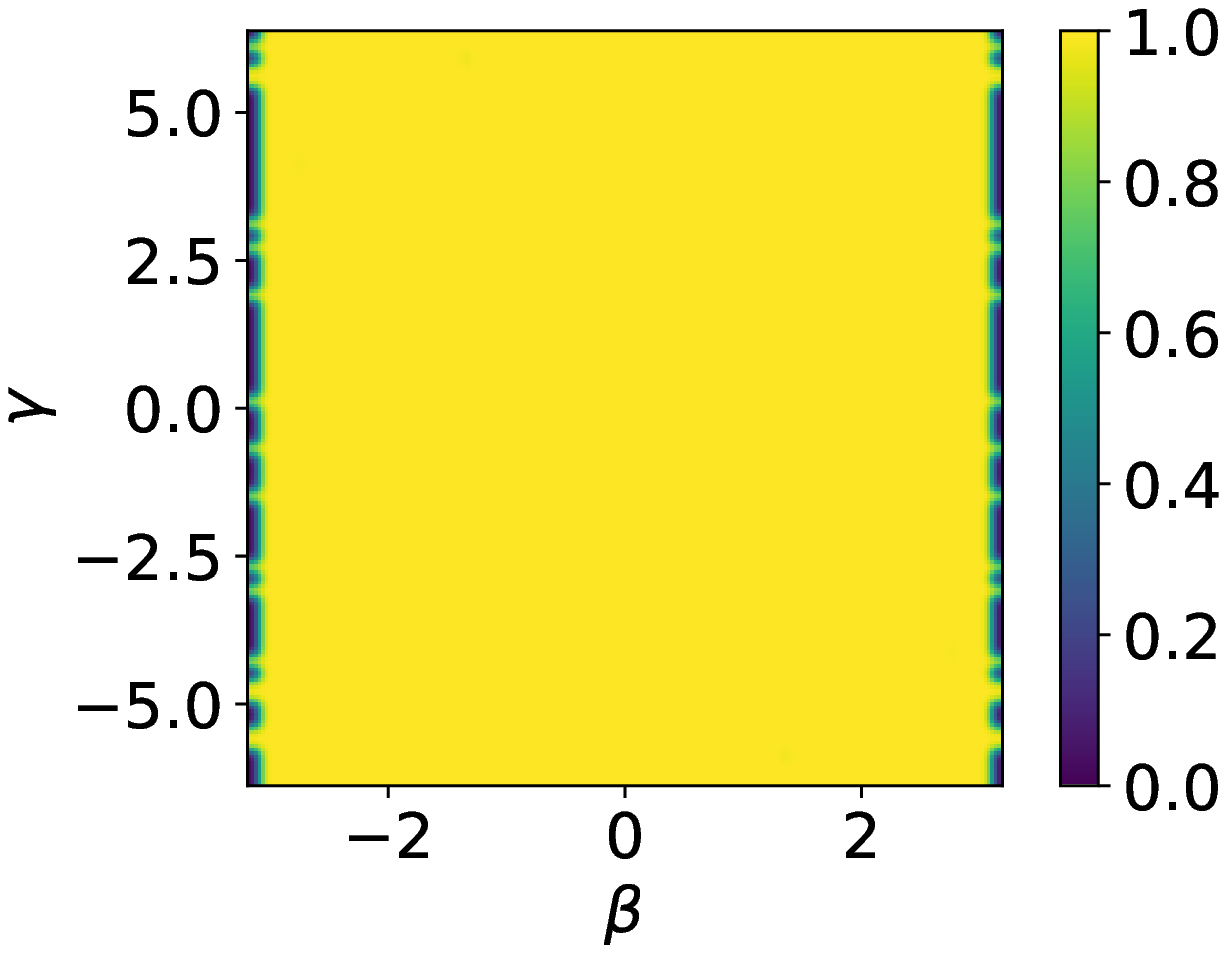}
            \addtocounter{subfigure}{-1}
            \renewcommand\thesubfigure{\alph{subfigure}3}
            \caption{Correlation coefficient}
        \end{subfigure}
        \addtocounter{subfigure}{-1}
        \caption{$k = 2$ (zoom: $1 \times$)}
    \end{subfigure}\\
    \vspace*{30px}
    \begin{subfigure}{\textwidth}
        \centering
        \begin{subfigure}{0.32\textwidth}
            \includegraphics[width=\textwidth]{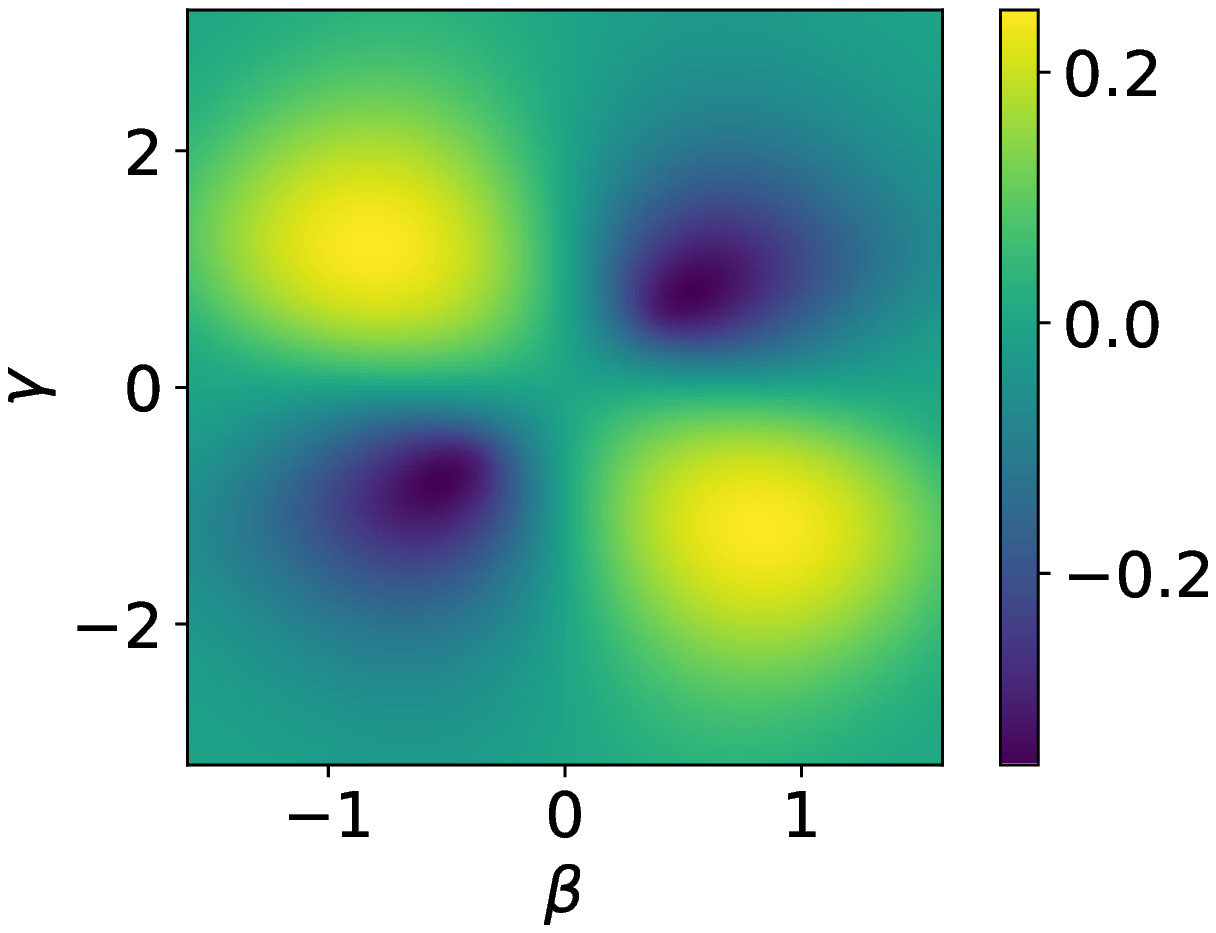}
            \renewcommand\thesubfigure{\alph{subfigure}1}
            \caption{Logarithmic success probability}
        \end{subfigure}
        \begin{subfigure}{0.32\textwidth}
            \includegraphics[width=\textwidth]{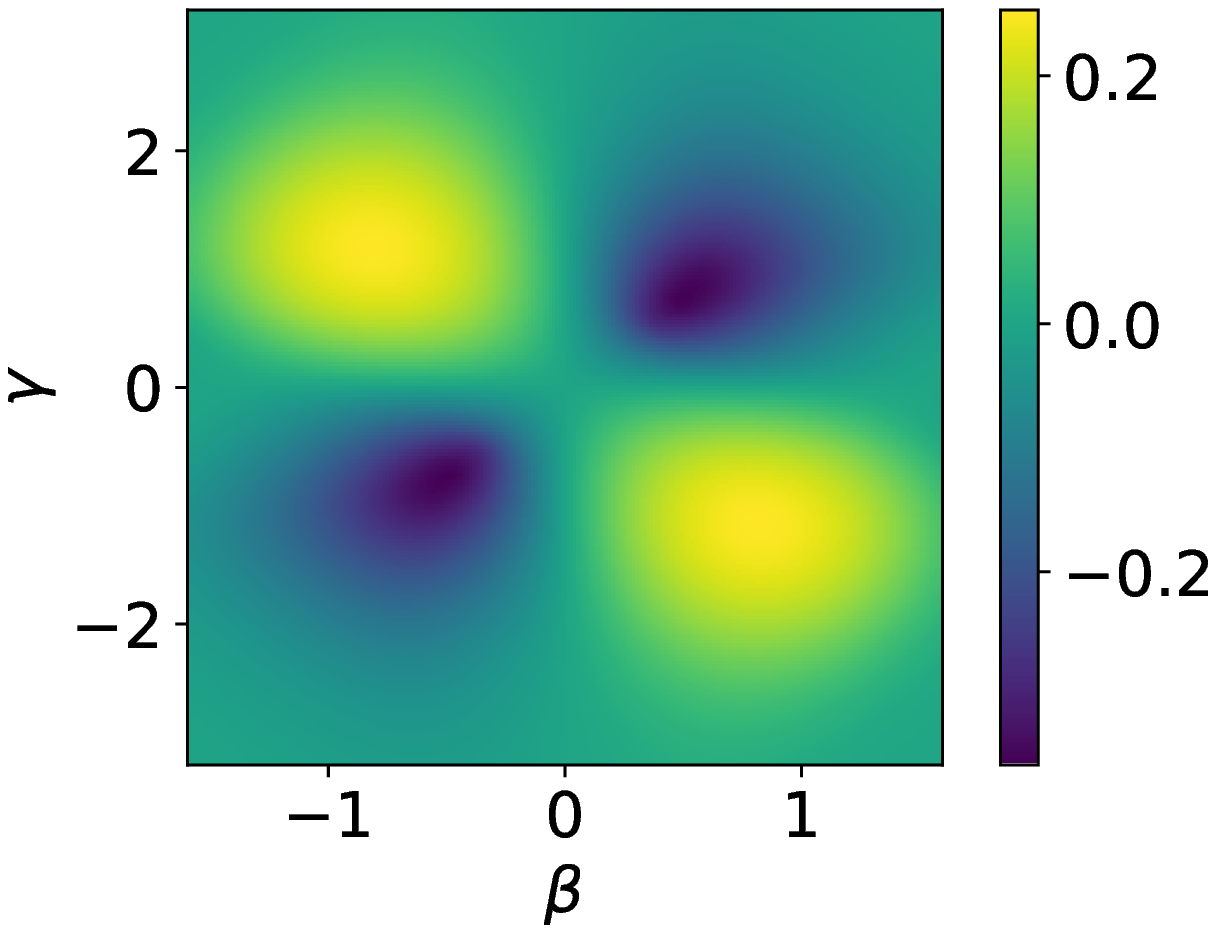}
            \addtocounter{subfigure}{-1}
            \renewcommand\thesubfigure{\alph{subfigure}2}
            \caption{Exponential fit}
        \end{subfigure}
        \begin{subfigure}{0.32\textwidth}
            \includegraphics[width=\textwidth]{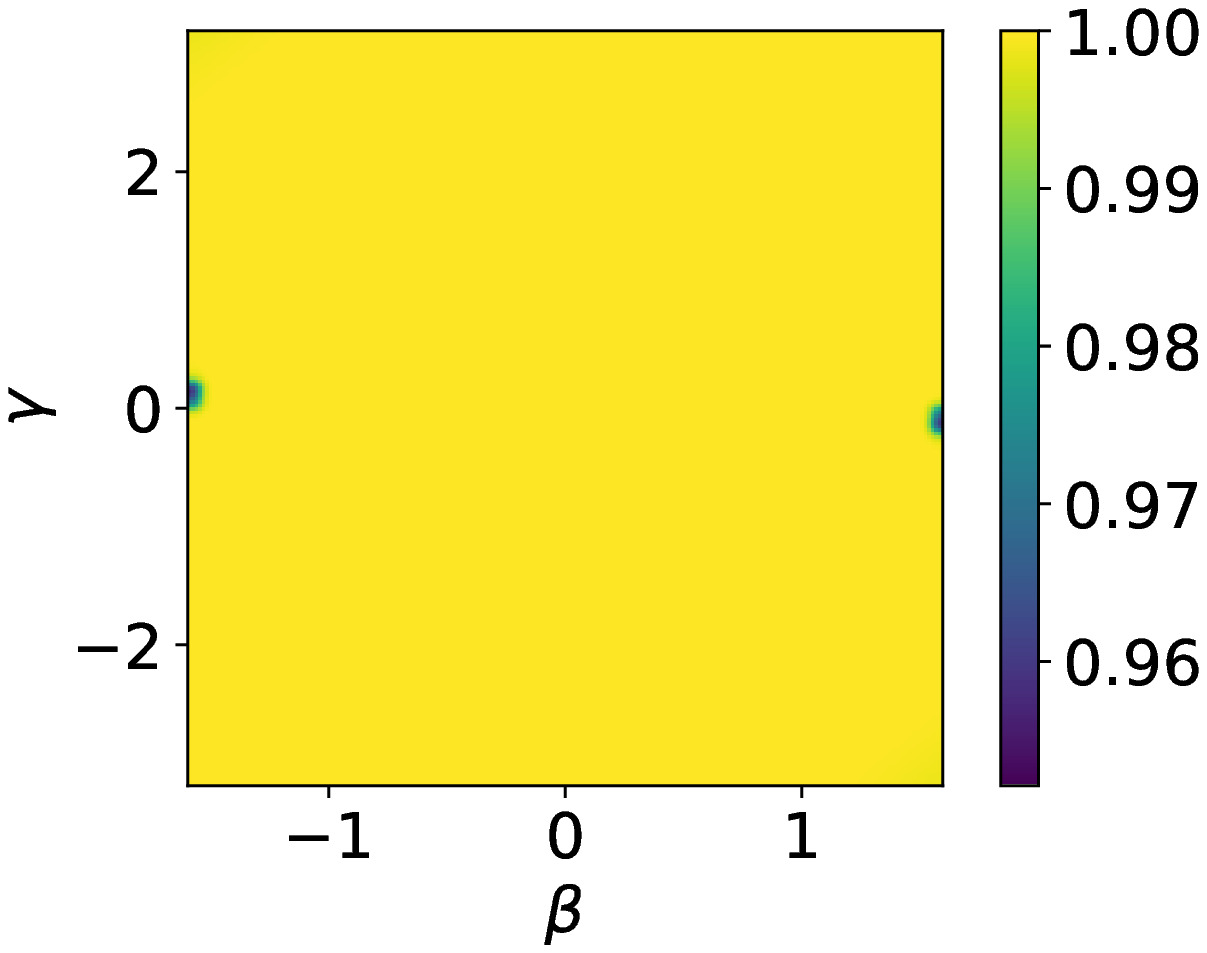}
            \addtocounter{subfigure}{-1}
            \renewcommand\thesubfigure{\alph{subfigure}3}
            \caption{Correlation coefficient}
        \end{subfigure}
        \addtocounter{subfigure}{-1}
        \caption{$k = 4$ (zoom: $2\times$)}
    \end{subfigure}
\end{figure}
\begin{figure}
    \ContinuedFloat
    \begin{subfigure}{\textwidth}
        \centering
        \begin{subfigure}{0.32\textwidth}
            \includegraphics[width=\textwidth]{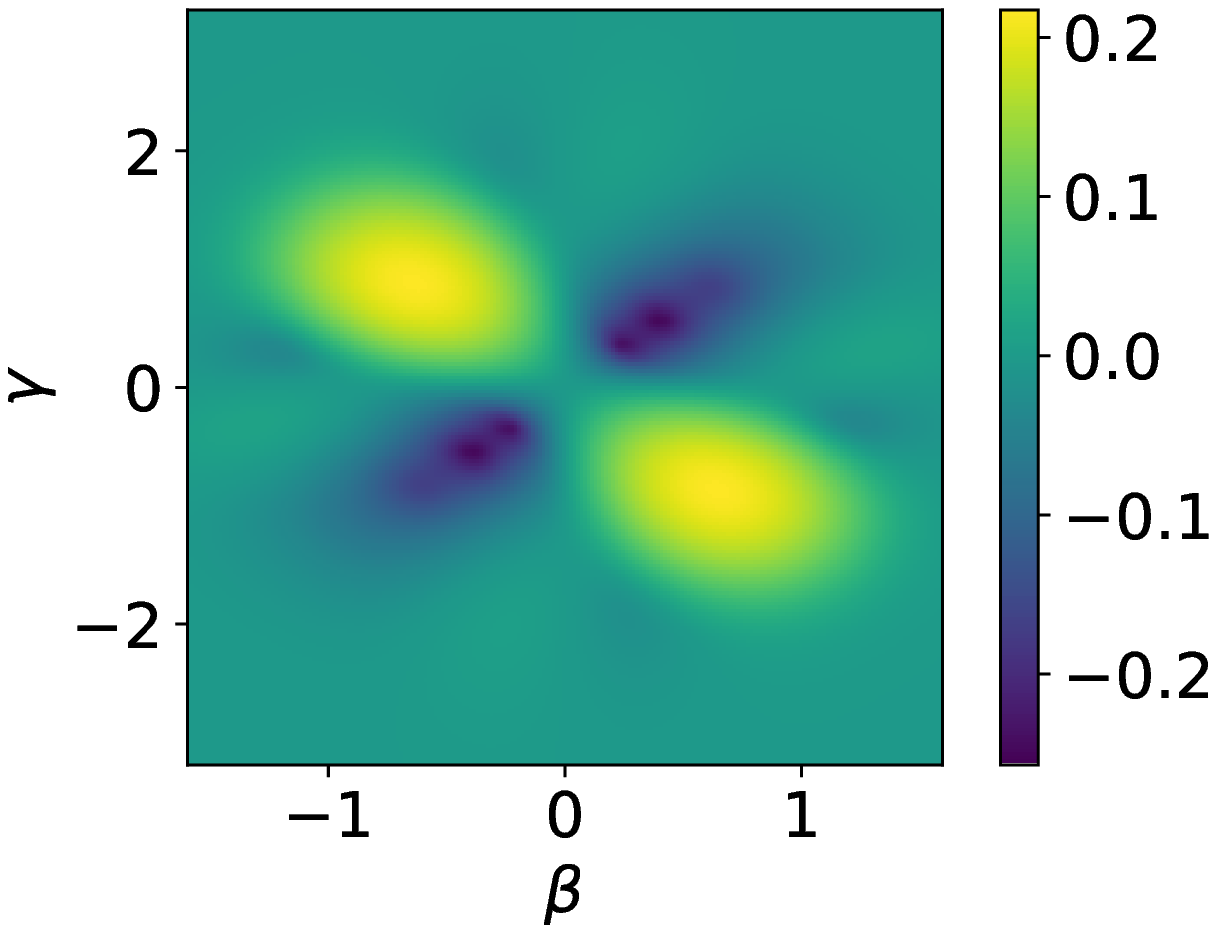}
            \renewcommand\thesubfigure{\alph{subfigure}1}
            \caption{Logarithmic success probability}
        \end{subfigure}
        \begin{subfigure}{0.32\textwidth}
            \includegraphics[width=\textwidth]{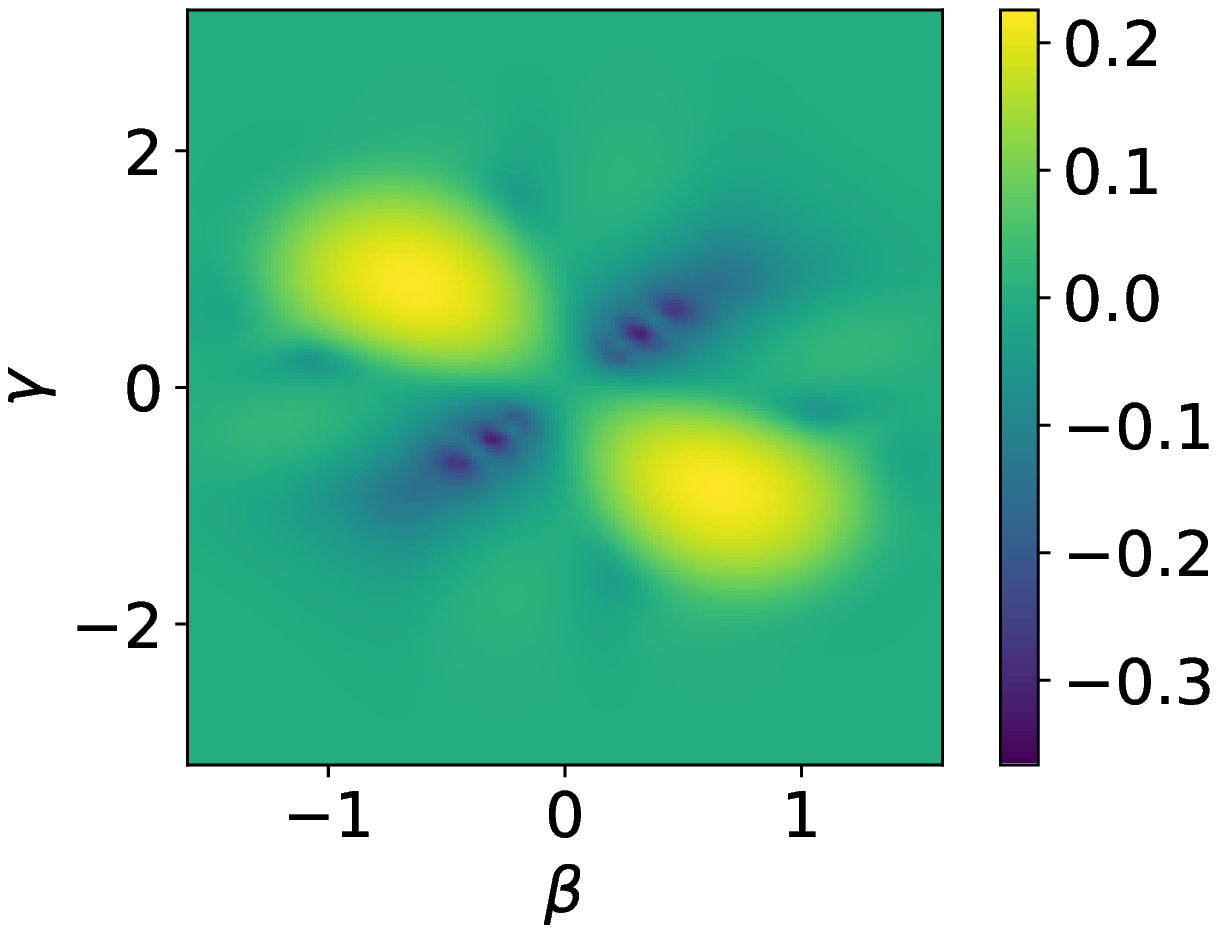}
            \addtocounter{subfigure}{-1}
            \renewcommand\thesubfigure{\alph{subfigure}2}
            \caption{Exponential fit}
        \end{subfigure}
        \begin{subfigure}{0.32\textwidth}
            \includegraphics[width=\textwidth]{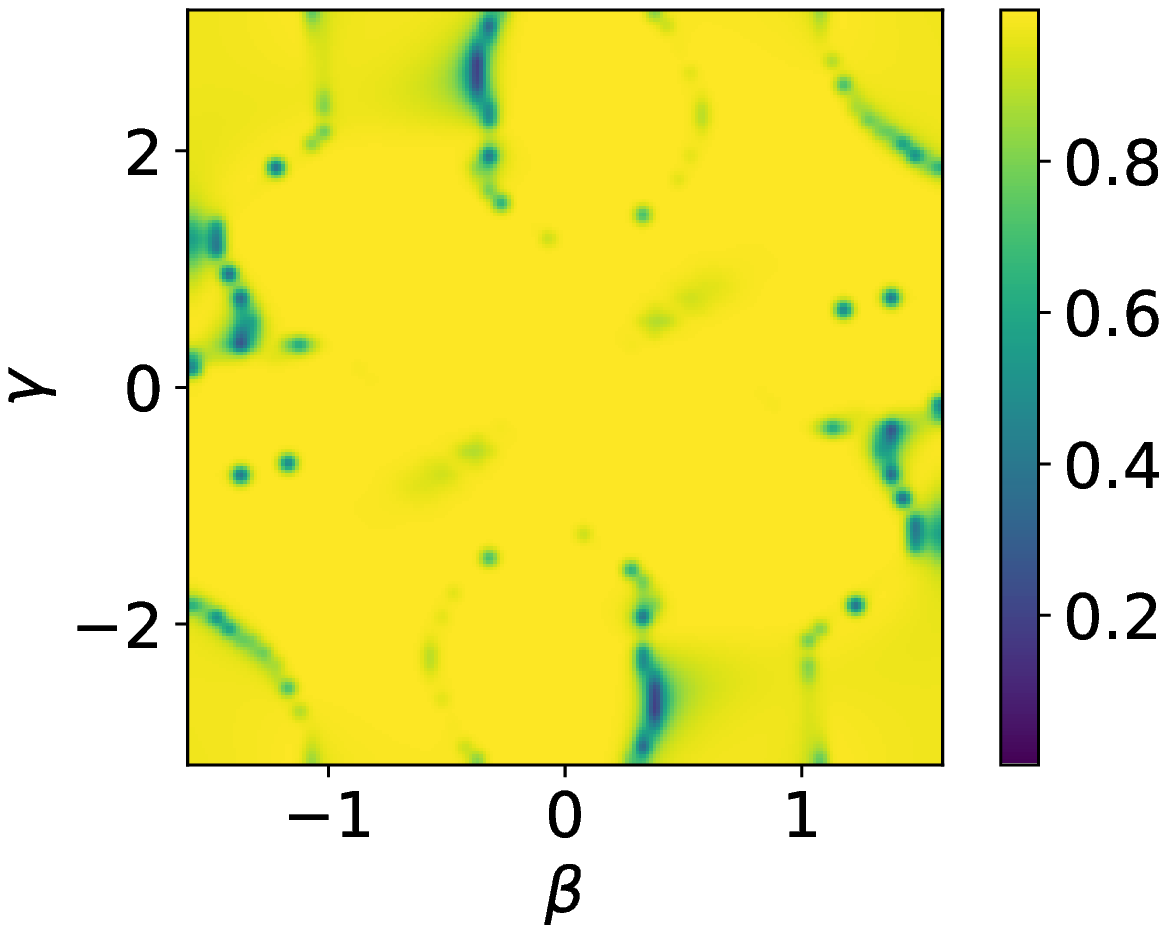}
            \addtocounter{subfigure}{-1}
            \renewcommand\thesubfigure{\alph{subfigure}3}
            \caption{Correlation coefficient}
        \end{subfigure}
        \addtocounter{subfigure}{-1}
        \caption{$k = 8$ (zoom: $2\times$)}
    \end{subfigure}
    \begin{subfigure}{\textwidth}
    \vspace*{20px}
        \centering
        \begin{subfigure}{0.32\textwidth}
            \includegraphics[width=\textwidth]{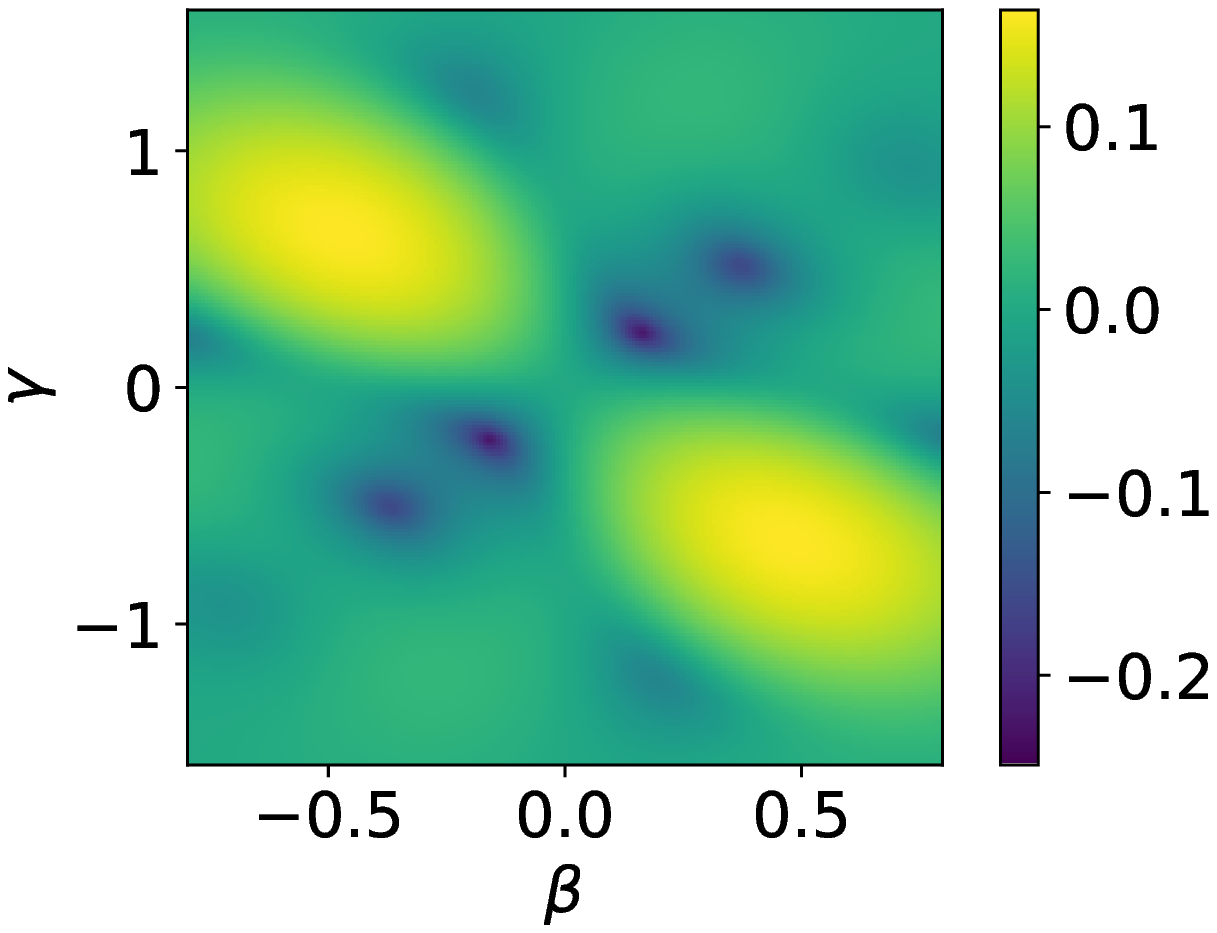}
            \renewcommand\thesubfigure{\alph{subfigure}1}
            \caption{Logarithmic success probability}
        \end{subfigure}
        \begin{subfigure}{0.32\textwidth}
            \includegraphics[width=\textwidth]{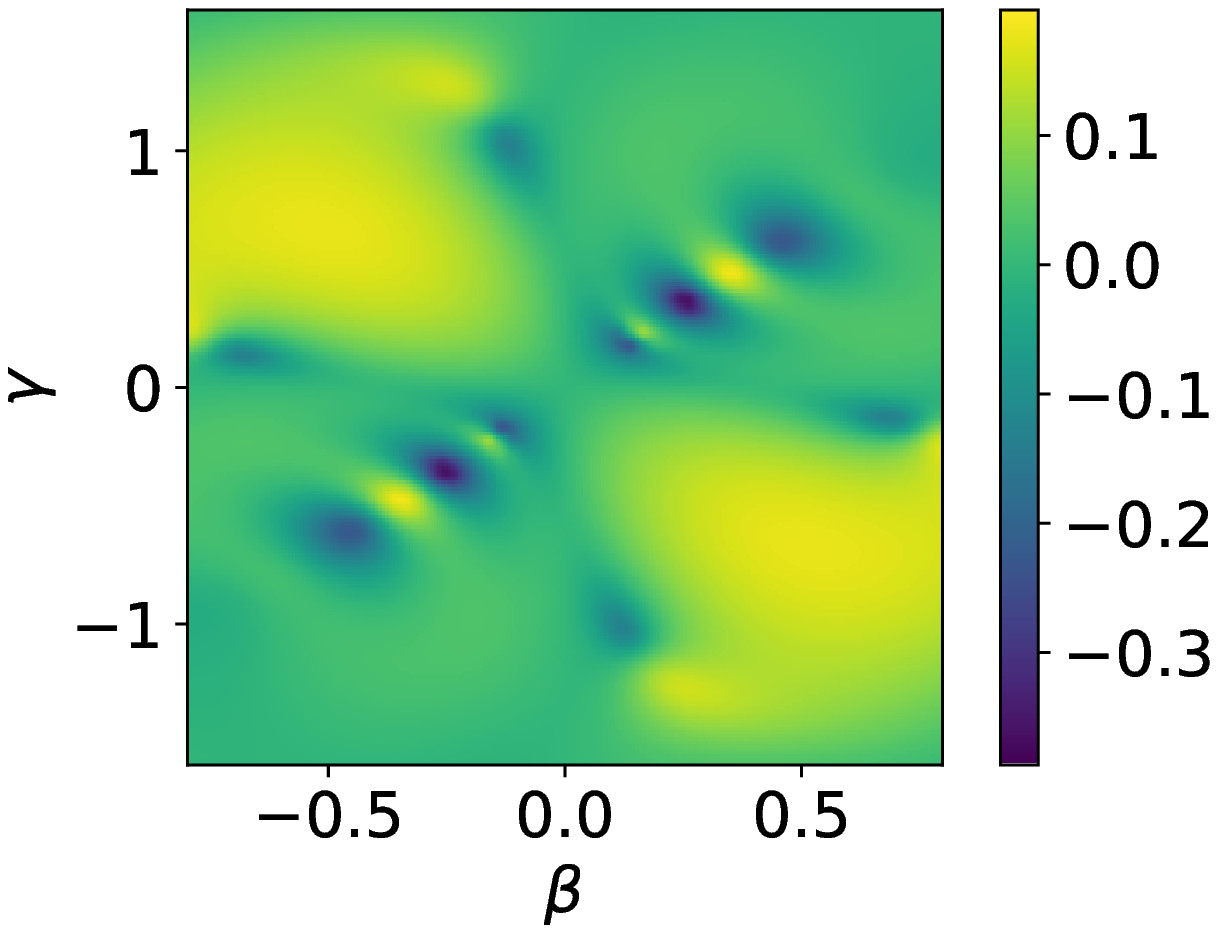}
            \addtocounter{subfigure}{-1}
            \renewcommand\thesubfigure{\alph{subfigure}2}
            \caption{Exponential fit}
        \end{subfigure}
        \begin{subfigure}{0.32\textwidth}
            \includegraphics[width=\textwidth]{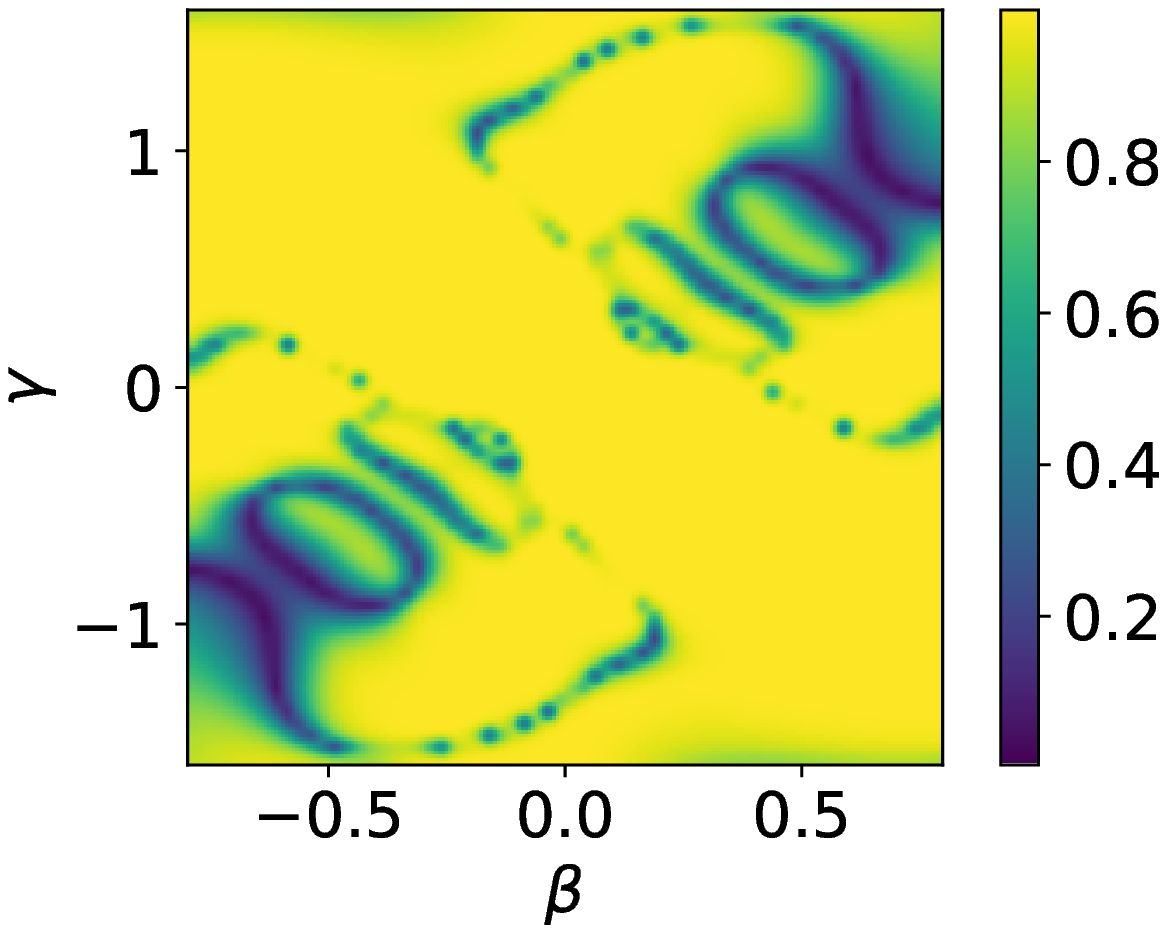}
            \addtocounter{subfigure}{-1}
            \renewcommand\thesubfigure{\alph{subfigure}3}
            \caption{Correlation coefficient}
        \end{subfigure}
        \addtocounter{subfigure}{-1}
        \caption{$k = 16$ (zoom: $4\times$)}
    \end{subfigure}
    \caption{QAOA optimization landscapes at $p = 1$. By periodicity (following from the integrality of the cost function), $\beta$ and $\gamma$ can be respectively restricted to $[-\pi, \pi]$ and $[-2\pi, 2\pi]$. In case the represented function is negligible except on a small part of this domain, we choose to zoom in on the rectangle while keeping centered at $0$; for instance, when zooming by a factor of $2$, the represented domain is $\left[-\frac{\pi}{2}, \frac{\pi}{2}\right] \times \left[\pi, \pi\right]$. The central symmetry is a general feature of QAOA, applying to all cost functions and diagonal unitaries.}
    \label{fig:p1_qaoa_landscapes}
\end{figure}

\subsection{Quality of empirically determined optimal variational angles}

In the main text, we compared exact (for $p = 1$ only) or empirical (for $p \in [1, 10]$) finite-size results to the prediction made by proposition \ref{prop:success_probability_random_ksat_qaoa_saddle_point} for the infinite-size limit. These comparisons helped us build confidence that the analytic method, which is in principle restricted to sufficiently small angles, remains correct when evaluated with the pseudo-optimal angles we chose. Assuming the analytic method remains operational in a neighbourhood of these variational parameters, allowing for an \textit{exact} evaluation of the exact success probability in this region, one may now wonder how far from optimal these angles are.

\begin{figure}
    \centering
    \begin{subfigure}{0.45\textwidth}
        \centering
        \includegraphics[width=\textwidth]{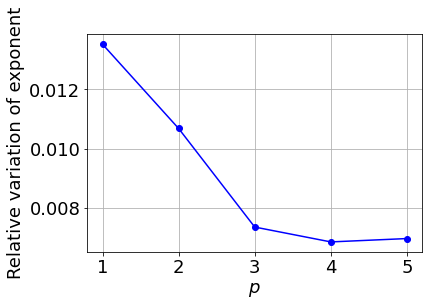}
        \caption{Exponent}
    \end{subfigure}
    \begin{subfigure}{0.45\textwidth}
        \centering
        \includegraphics[width=\textwidth]{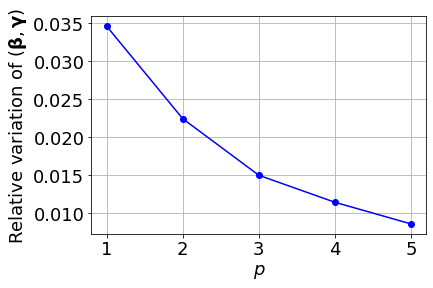}
        \caption{Angles}
    \end{subfigure}
    \caption{Relative variation of scaling exponent and optimized angles after reoptimization from analytic method. The relative error for the exponent is defined as the ratio between the new and old value, minus 1. For angles, the distance between the old angles $\left(\bm{\beta^{(i)}}, \bm{\gamma^{(i)}}\right)$ and the new ones $\left(\bm{\beta^{(f)}}, \bm{\gamma^{(f)}}\right)$ is calculated by considering the representative of $\left(\bm{\beta^{(f)}}, \bm{\gamma^{(f)}}\right)$ closest to the $\left(\bm{\beta^{(i)}}, \bm{\gamma^{(i)}}\right)$ in order to account for $2\pi$-periodicity in the $\beta_j$ and $4\pi$-periodicity in the $\gamma_j$. The $\bm\beta$ components of the difference vector $\left(\bm{\beta^{(f)}} - \bm{\beta^{(i)}}, \bm{\gamma^{(f)}} - \bm{\gamma^{(i)}}\right)$ are then rescaled by $\frac{1}{\pi\sqrt{2p}}$, and the $\gamma$ components by $\frac{1}{2\pi\sqrt{2p}}$, mapping the $2$-norm of the resulting vector into $[0, 1]$.}
    \label{fig:reoptimize_relative_variation}
\end{figure}

To investigate this, we reoptimize the success probability using the analytic formula, with the initial angles (empirically optimised for $n=12$) as an educated guess. As usual, a simple gradient descent algorithm is used. We then consider the relative variation of the analytically determined scaling exponent along the optimization, together with the relative variation of the angles. Figure \ref{fig:reoptimize_relative_variation} reports these results for 8-SAT and precisely defines the metrics just evoked. The $p$ values used here are more restricted than in the main text due to the overhead of evaluations required by optimization as opposed to simple evaluation. These results show the angles determined by the simple empirical method described in section \ref{sec:numerical_methods} are very close to optimal, or at least to a local optimum. Relative variations for both the angles and the exponent along the optimization are of order a few percent. Surprisingly, the relative variation even appears to decrease with $p$, but we do not know how robust the last conclusion is due to the limited number of values of $p$ considered and the dependence of these results on the small dataset used to determine approximately optimal angles.

\subsection{Further experimental results for classical solvers for $k=8$}
\label{app:k8results}

We include all the classical experimental results on the complexity of solvers for 8-SAT in table \ref{tab:8sat_scaling_exponents}.

\begin{table}
	\begin{tabular}{c|c|c|c}
		Solver & Fit & Correlation coefficient & Estimated exponent error\\
		\hline
		\verb|walksat_qaoa| & $-3.232 + 0.295n$ & $0.963143$ & $0.011$\\
		\hline
		\verb|eval_qaoa| ($p=14$) & $-1.064 + 0.326n$ & $0.999422$ & $0.005$\\
		\hline
		\verb|eval_qaoa| ($p=60$) &  $-2.842 + 0.302n$ & $0.998161$ & $0.007$\\
		\hline
		\verb|walksatlm| & $-0.309 + 0.325n$ & $0.997503$ & $0.007$\\
		\hline
		\verb|maplesat| & $1.531 + 0.461n$ & $0.999626$ & $0.004$\\
		\hline
		\verb|glucose3| & $2.998 + 0.498n$ & $0.999826$ & $0.004$\\
		\hline
		\verb|glucose4| & $2.998 + 0.498n$ & $0.999826$ & $0.004$\\
		\hline
		\verb|gluecard3| & $2.998 + 0.498n$ & $0.999826$ & $0.006$\\
		\hline
		\verb|gluecard4| & $2.998 + 0.498n$ & $0.999826$ & $0.005$\\
		\hline
		\verb|mergesat3| & $2.974 + 0.500n$ & $0.999846$ & $0.004$\\
		\hline
		\verb|lingeling| & $2.681 + 0.505n$ & $0.999700$ & $0.005$\\
		\hline
		\verb|cadical| & $2.702 + 0.518n$ & $0.999645$ & $0.005$\\
		\hline
		\verb|minicard| & $2.725 + 0.523n$ & $0.999689$ & $0.004$\\
		\hline
		\verb|minisat22| & $2.725 + 0.523n$ & $0.999689$ & $0.005$\\
		\hline
		\verb|minisatgh| & $2.725 + 0.523n$ & $0.999689$ & $0.005$\\
		\hline
		\verb|maple_chrono| & $2.557 + 0.533n$ & $0.999814$ & $0.005$\\
		\hline
		\verb|maple_cm| & $-0.713 + 0.581n$ & $0.989826$ & $0.005$\\
		\hline
		\verb|schoning| & $-2.657 + 0.649n$ & $0.999826$ & $0.006$
	\end{tabular}
	\caption{Empirical exponential fits for all SAT solvers for 8-SAT. We further report the correlation coefficient of the fit. Besides, we estimate the error on the scaling exponent as described in section \ref{sec:numerical_methods}.}
	\label{tab:8sat_scaling_exponents}
\end{table}

\bibliographystyle{hapalike}
\bibliography{bibliography}

\end{document}